\documentclass{article}

\usepackage[preprint]{neurips_2026}

\usepackage{xcolor}
\usepackage[most]{tcolorbox}
\usepackage{algorithm}
\usepackage{algorithmic}
\usepackage{algorithm}
\usepackage{amsmath}
\usepackage{graphicx}
\usepackage{subcaption}
\usepackage[utf8]{inputenc} 
\usepackage[T1]{fontenc}    
\usepackage{hyperref}       
\usepackage{url}            
\usepackage{booktabs}       
\usepackage{multirow}       
\usepackage{amsfonts}       
\usepackage{nicefrac}       
\usepackage{microtype}      
\usepackage{xcolor}         

\title{GReFEM: Multimodal LLMs as Zero-Shot Semantic Assistants for Physics-Guided 3D Mesh Refinement }

\author{%
  Kartik Bali\\
  Institute of Material Systems Modeling\\
  Helmholtz Zentrum Hereon\\
  Max-Planck-Straße 1, 21502 Geesthacht \\
  \texttt{kartik.bali@hereon.de} \\
  \And
  Mahish K.Guru \\
  Institute of Material and Process Design \\
  Helmholtz Zentrum Hereon\\
  Max-Planck-Straße 1, 21502 Geesthacht \\
  \texttt{mahish.guru@hereon.de} \\
  \And
  Christian J Cyron\\
  Institute of Material Systems Modeling\\
  Helmholtz Zentrum Hereon\\
  Max-Planck-Straße 1, 21502 Geesthacht \\
  \texttt{christian.cyron@hereon.de} \\
  \And
  Roland Aydin \\
  Data-Driven Simulation and Analysis in Materials Science \\
  German Research Center for Artificial Intelligence \\
  Campus D3 2, 66123 Saarbrücken \\
  \texttt{roland.aydin@dfki.de} \\
}

\begin{document}

\maketitle

\begin{abstract}
Adaptive volumetric finite element meshing is a critical step in computer-aided engineering and analysis that dictates the computational budget of a given problem. It traditionally requires iterative PDE solvers or heavily supervised, data-driven surrogates trained on large-scale simulation data. While Multimodal Large Language Models (MLLMs) excel in 2D visual tasks, their zero-shot capability to semantically ground regions based on geometric understanding and physics remains an open question. This work explores a significant question: can the high-level semantic understanding of off-the-shelf MLLMs serve as a viable, zero-shot geometric proxy for finite element mesh refinement?
We introduce \textbf{GReFEM} (Geometric Reasoning Enhanced Multimodal LLMs for Finite Element Meshing), a framework that utilizes MLLMs to visually localize stress-critical regions based on physics-guided textual prompts. To bridge the gap between 2D MLLM pre-training and 3D geometries, we introduce \textit{orthoViews}, a trained view-selection module that assists these MLLMs by maximizing the observability of key geometric features. We evaluate \textbf{GReFEM} by computing finite-element error at stress concentrations, on diverse CAD mesh geometries under multiple loading cases, with MLLMs and prompt variants, against blind and load-informed geometric heuristics, solver-informed Zienkiewicz--Zhu (ZZ) refinement, and a retrained learned adaptive-meshing baseline, all at a matched element budget. Without any FEM solve or simulation training data, \textbf{GReFEM} lowers the critical-region peak von Mises stress error relative to a blind geometric heuristic and to refinement driven by a coarse FEM solve, closing $58\%$ of the gap between the unrefined mesh and a reference-solve oracle, while a hand-derived analytical load-informed heuristic remains stronger and the learned baseline performs slightly better overall. With this we investigate their performance as semantic assistants in automated simulation workflows. Code is available at \url{https://github.com/kbali1297/GReFEM}.
\end{abstract}

\section{Introduction}
Recent large language models (LLMs) have demonstrated strong capabilities in reasoning and visual perception, enabling their application across scientific domains that require structured spatial understanding. In particular, recent work has explored the use of LLMs for 3D CAD modeling, object generation, and geometric reasoning, leading to notable advances in design automation and shape synthesis~\cite{kienle2025querycad, zhou2025cadialogue, yang2023sam3d}.

Despite these advances, the role of LLMs in fine-grained numerical simulation workflows remains largely unexplored. High-fidelity finite element simulation fundamentally relies on discretization of the problem domain, where mesh resolution must be adaptively concentrated in regions of high physical variation. While existing approaches adapt LLMs to reason about surface meshes~\cite{wang2024llama, fang2025meshllm} and CAD representations~\cite{wu2023cad, alrashedy2024generating} of 3D objects, comparatively little work investigates their effectiveness in guiding volumetric discretization required to resolve localized physical phenomena while controlling computational cost.

Traditional adaptive volumetric meshing methods rely on PDE solvers to compute error indicators that identify regions requiring refinement (\cite{zienkiewicz1992superconvergent, babuvska1978posteriori}). This tightly couples meshing to solver availability and computational expense. In practice, however, experienced engineers frequently anticipate regions requiring refinement directly from geometric structure, loading directions, and boundary conditions, without explicitly solving for field quantities. Several works exploit such geometry-driven heuristics~\cite{bronson2012particle, gkrispanis2024enhancing}, but typically assume structured CAD representations that explicitly encode high-level geometric features.

\begin{figure*}
\centering
\includegraphics[width=1.0\columnwidth]{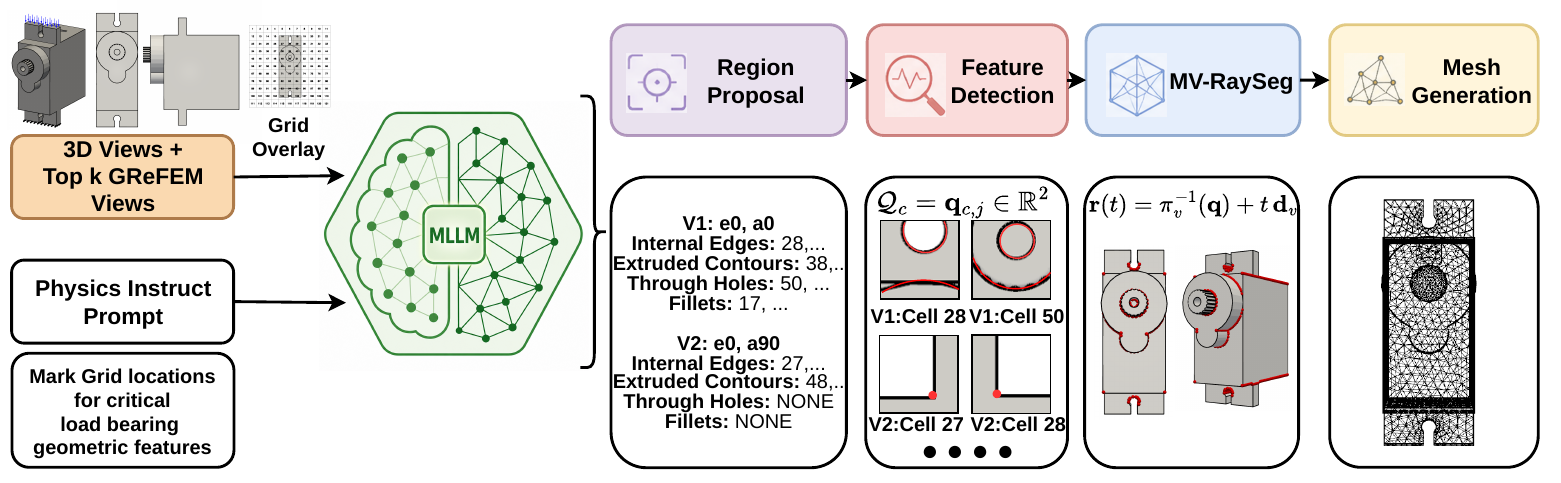}
\caption{\textbf{GReFEM Pipeline}: We process the top views \textbf{$\mathcal{V}^*$}, via the \textbf{Region Proposal} and \textbf{Feature Detection} Pipelines. Once the points \textbf{$Q$} are obtained on the views, we use our \textbf{Surface Projection (2D to 3D)} pipeline to obtain anchor points $\mathcal{P}$ for mesh generation.}
\label{fig:Mesh_Inference}    
\end{figure*}

\begin{figure}[t]
\centering
\begin{subfigure}[t]{0.42\columnwidth}
    \centering
    \includegraphics[width=\linewidth]{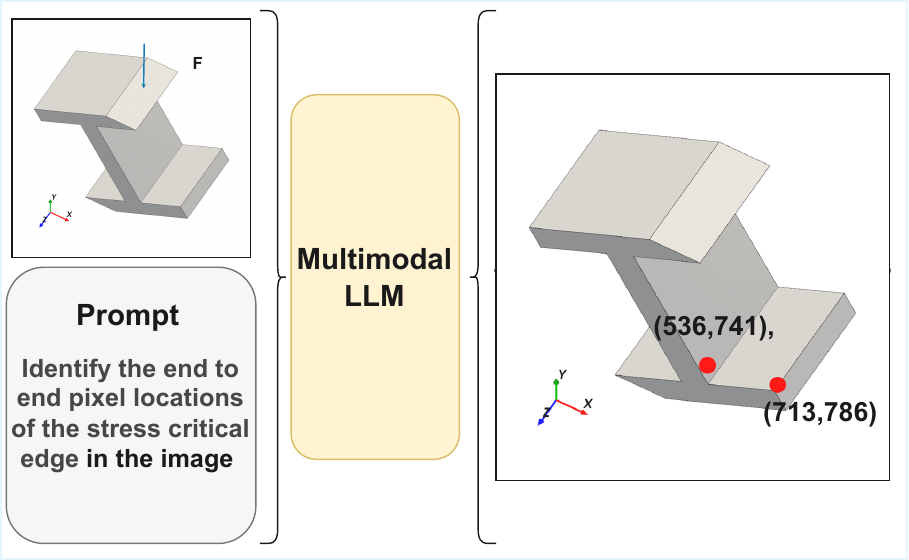}
    \caption{Pixel localization.}
    \label{fig:Pixel_localization_gemini}
\end{subfigure}
\hfill
\begin{subfigure}[t]{0.56\columnwidth}
    \centering
    \includegraphics[width=\linewidth]{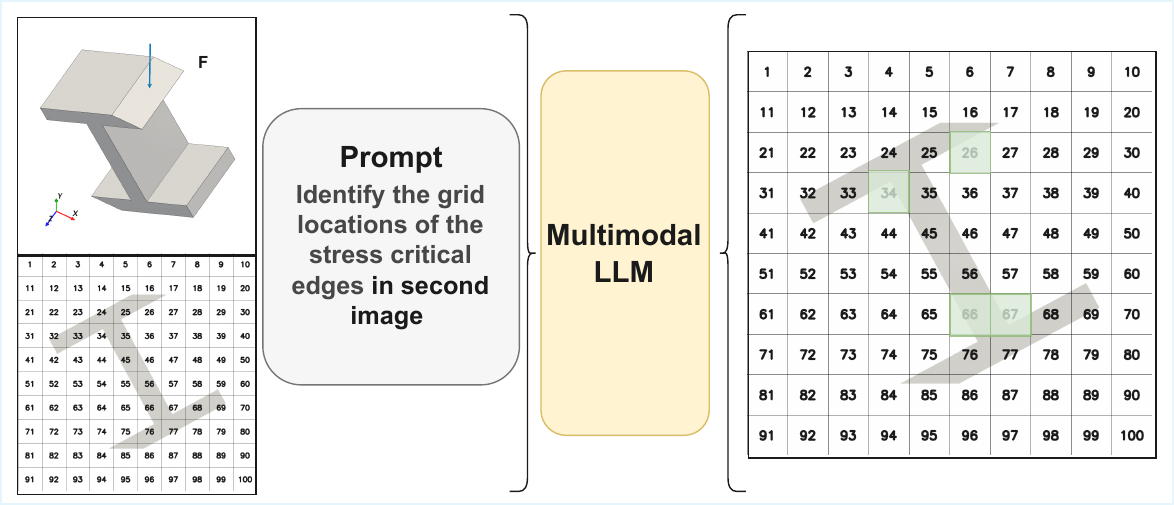}
    \caption{Region (grid-cell) localization.}
    \label{fig:Region_localization_gemini}
\end{subfigure}

\caption{\textbf{Localization behavior of MLLMs}
\emph{Left:} Direct pixel-level localization is unreliable, even when the model correctly identifies the stress-critical feature semantically.
\emph{Right:} Grid-based region localization in \textbf{GReFEM} views yields slightly more accurate results: Predicted green grid cells are concentrated around stress-critical re-entrant edges.
}
\label{fig:localization_comparison}
\end{figure}

In this work, we investigate a fundamental question: \textit{can off-the-shelf MLLMs act as zero-shot semantic assistants to emulate expert, geometry-driven meshing intuition?} To systematically explore this, we introduce  \textbf{GReFEM} as a framework to evaluate MLLM spatial grounding in 3D engineering environments. Because MLLMs often struggle with precise pixel-level localization (Fig.~\ref{fig:localization_comparison}), we propose grid-based region localization as a solution and integrate \textit{orthoViews}, a trained view-scoring module to maximize geometric observability. This enables us to reliably project coarse 2D MLLM inferences back to the 3D domain, steering volumetric mesh refinement without PDE solvers or explicit CAD annotations.

Through rigorous experiments across CAD geometries, multiple loading conditions, and state-of-the-art MLLMs, we present the following contributions:

(1) \textbf{A Semantic Evaluation Framework:} We introduce \textbf{GReFEM} (with \textit{orthoViews}) as a CAD-agnostic probing pipeline that translates geometric instructions grounded in problem physics into localized 3D volumetric mesh refinement anchors. We score every refinement policy by the downstream finite-element error it produces at a matched element budget, namely the peak von Mises stress and the strain energy inside the stress-critical region of a high-fidelity reference solve.

(2) \textbf{Analysis of Instruction-Driven Grounding:} Prompt content measurably steers where MLLMs refine. Describing only the load (\textit{Load-Only}) raises recall and anchor count but lowers anchor precision for all four MLLMs, and at a matched element budget enumerating the stress-critical feature types lowers peak-stress error relative to \textit{Load-Only} for Gemini-3-flash and GPT-5.4-mini.

(3) \textbf{Downstream Accuracy at a Matched Budget:} Across multiple geometry-load cases, zero-shot \textbf{GReFEM} yields lower critical-region peak-stress error than a blind geometric heuristic and ZZ refinement from a coarse FEM solve, with no solve of its own, and performs comparably overall to a learned-AMR baseline AMBER (~\cite{freymuth2025amber}) retrained on 440 expert meshes without requiring any of them. A load-informed heuristic built from hand-derived beam-theory formulas remains stronger, but relies on hand-derived analytical form for every separate loading case contrary to \textbf{GReFEM}, which treats loading cases as different only at the prompt level.

\section{Related Works}

\textbf{Adaptive Meshing}
Adaptive meshing non-uniformly allocates resolution to minimize finite element simulation costs. Traditional pipelines rely on computationally expensive, solver-in-the-loop error estimators tailored to the governing PDEs (~\cite{babuvska1978posteriori, babuvvska1978error, zienkiewicz1992superconvergent}), though some geometry-driven heuristics avoid explicit PDE solves using mathematical criteria (~\cite{rivara1984algorithms, Shewchuk2002, bronson2012particle}). Recently, deep learning has emerged as a surrogate, using Graph Neural Networks (GNNs) to learn h-adaptive sizing fields (~\cite{khan2024graphmesh, freymuth2025amber}) or r-adaptive topological optimizations (~\cite{pfaff2020learning, rowbottom2024g}). While these supervised architectures achieve high accuracy, their strict reliance on massive datasets of converged FEM solutions prevents zero-shot generalization. Rather than proposing a data-hungry supervised model,  \textbf{GReFEM} explores a new solver-free paradigm: evaluating whether the semantic reasoning of off-the-shelf MLLMs can act as a natural language-guided proxy for mesh refinement, significantly circumventing the need for offline FEM training data. Furthermore, supervised GNNs fundamentally solve a different problem: they act as end-to-end surrogates for a specific, pre-defined PDE and boundary condition distribution. Adapting a GNN to a new physical phenomenon (e.g., thermal stress instead of linear elasticity) or a fundamentally different load type requires complete model retraining and a newly generated FEM dataset. In contrast, \textbf{GReFEM} acts as a domain-agnostic semantic interface. By shifting the refinement logic from learned network weights to expert-provided natural language prompts, \textbf{GReFEM} can dynamically adapt to entirely new loading scenarios without a single parameter update.

\textbf{3D Spatial Reasoning with MLLMs}
Extending vision-language models to 3D reasoning remains challenging due to limited 3D training data. Several approaches address this by directly training LLMs on structured 3D representations, such as 3D feature fields~\cite{hong20233d}, point clouds (~\cite{xu2024pointllm, batra2025spatialthinker}), or radiance fields ~\cite{kerr2023lerf}. Alternatively, other works infer 3D understanding from 2D projections without explicit 3D structures. This includes predicting 3D bounding boxes ~\cite{cho2024language}, optimizing view selection for 3D QA ~\cite{wang20253d}, regressing geometric attributes via 3D query tokens ~\cite{yang2026llmi3d}, estimating absolute distances from synthetic datasets ~\cite{chen2024spatialvlm}, progressive prompting for trajectory imagination ~\cite{ma2024spatialpin}, and leveraging video sequences for 3D grounding ~\cite{zheng2025video}. Our work follows this latter 2D-to-3D projection paradigm; however, rather than targeting general semantic scene understanding, we uniquely leverage intelligently ranked 2D orthographic views to infer mechanically relevant 3D features for downstream engineering discretization.

\section{Preliminaries}
\paragraph{Governing Equations} We consider linear elastic deformation of a solid body occupying a bounded domain
$\Omega \subset \mathbb{R}^3$, subject to prescribed boundary displacements.
Let $\mathbf{u} : \Omega \rightarrow \mathbb{R}^3$ denote the displacement field, which satisfies
\begin{equation}
\begin{aligned}
-\nabla \cdot \boldsymbol{\sigma}(\mathbf{u}) &= \mathbf{0}
&& \text{in } \Omega, \\
\mathbf{u} &= \bar{\mathbf{u}}
&& \text{on } \partial \Omega .
\end{aligned}
\label{eq:equilibrium}
\end{equation}
where $\boldsymbol{\sigma}$ is the Cauchy stress tensor and $\bar{\mathbf{u}}$
denotes the prescribed boundary displacements.
In all experiments, loads are induced by imposing fixed displacements on opposing boundaries of the domain. 

\paragraph{Downstream Error Metrics}
Closed-form solutions are unavailable for these geometries, so we assess a candidate mesh by how closely its finite-element solution $\mathbf{u}_h$ reproduces the high-resolution reference $\mathbf{u}_{\mathrm{ref}}$ on the design-governing quantities, evaluated \emph{inside} the stress-critical region that the refinement budget targets rather than over the smooth bulk.

\textbf{Critical region.} Let $\eta_{ZZ}(\mathbf{x})$ be the Zienkiewicz--Zhu error indicator (Appendix~\ref{app:fem}) of the reference solution and $q_{p}$ its $p$-th percentile over $\Omega$. The reference's most stress-critical points are the top $0.1\%$ of this field, $\mathcal{C}=\{\mathbf{x}\in\Omega:\eta_{ZZ}(\mathbf{x})\ge q_{99.9}\}$ (after excluding a $2\%$ slab of the bounding box at the loaded boundaries, which removes load-application singularities), and the critical region $\Omega_{\mathrm{crit}}$ is their volumetric neighbourhood of radius $r_{\mathrm{crit}}=4\,h_{\min}$:
\begin{equation}
\Omega_{\mathrm{crit}} = \Big\{\,\mathbf{x}\in\Omega \;:\; \min_{\mathbf{c}\in\mathcal{C}} \|\mathbf{x}-\mathbf{c}\|_2 \le r_{\mathrm{crit}}\,\Big\}.
\label{eq:omega_crit}
\end{equation}

\textbf{Peak von Mises stress.} The failure-relevant quantity is the von Mises stress, built from the deviatoric part $\mathbf{s}=\boldsymbol{\sigma}-\tfrac{1}{3}\operatorname{tr}(\boldsymbol{\sigma})\,\mathbf{I}$ of the Cauchy stress $\boldsymbol{\sigma}=\mathbb{C}:\boldsymbol{\varepsilon}(\mathbf{u})$,
\begin{equation}
\sigma_{vM}(\mathbf{u}) = \sqrt{\tfrac{3}{2}\,\mathbf{s}(\mathbf{u}):\mathbf{s}(\mathbf{u})}.
\end{equation}
To remain robust to the integrable singularities at re-entrant corners we summarise it by its $99$-th percentile over the critical region, $\sigma_{vM}^{p99}(\mathbf{u})=\mathrm{P}_{99}\{\sigma_{vM}(\mathbf{u}(\mathbf{x})):\mathbf{x}\in\Omega_{\mathrm{crit}}\}$, and report the relative error against the reference,
\begin{equation}
\delta_{vM}^{p99} = \frac{\big|\,\sigma_{vM}^{p99}(\mathbf{u}_h)-\sigma_{vM}^{p99}(\mathbf{u}_{\mathrm{ref}})\,\big|}{\sigma_{vM}^{p99}(\mathbf{u}_{\mathrm{ref}})}.
\label{eq:vm_err}
\end{equation}

\textbf{Critical-region strain energy.} As the natural norm of the elasticity operator we also report the strain energy stored in $\Omega_{\mathrm{crit}}$ and its relative error,
\begin{equation}
E_{\mathrm{crit}}(\mathbf{u}) = \tfrac{1}{2}\!\int_{\Omega_{\mathrm{crit}}}\! \boldsymbol{\varepsilon}(\mathbf{u}):\mathbb{C}:\boldsymbol{\varepsilon}(\mathbf{u})\,\mathrm{d}\Omega,
\qquad
\delta_{E}^{\mathrm{crit}} = \frac{\big|\,E_{\mathrm{crit}}(\mathbf{u}_h)-E_{\mathrm{crit}}(\mathbf{u}_{\mathrm{ref}})\,\big|}{E_{\mathrm{crit}}(\mathbf{u}_{\mathrm{ref}})}.
\label{eq:energy_err}
\end{equation}

$(\delta_{vM}^{p99},\,\delta_{E}^{\mathrm{crit}})$ (both relative, lower is better) are the two quantities reported throughout (Table~\ref{tab:downstream_main}); restricting them to $\Omega_{\mathrm{crit}}$ prevents the large, well-resolved smooth region from masking the differences at the concentrations that refinement must resolve. $\Omega_{\mathrm{crit}}$ is a fixed region of space, identical for every candidate mesh, and each quantity is evaluated natively on the candidate's own mesh, so no cross-mesh interpolation enters. Because $\sigma_{vM}^{p99}$ sits at stress concentrations that do not converge under $h$-refinement, absolute errors stay high for every policy; the numbers are meaningful for \emph{ranking} policies at a common budget, not as convergence estimates. We aggregate over cases by the mean and median and compare policies within each geometry and load case. 


\begin{figure}[t]
\centering
\subfloat[Compression Loading Case]{
    \includegraphics[width=0.3\textwidth]{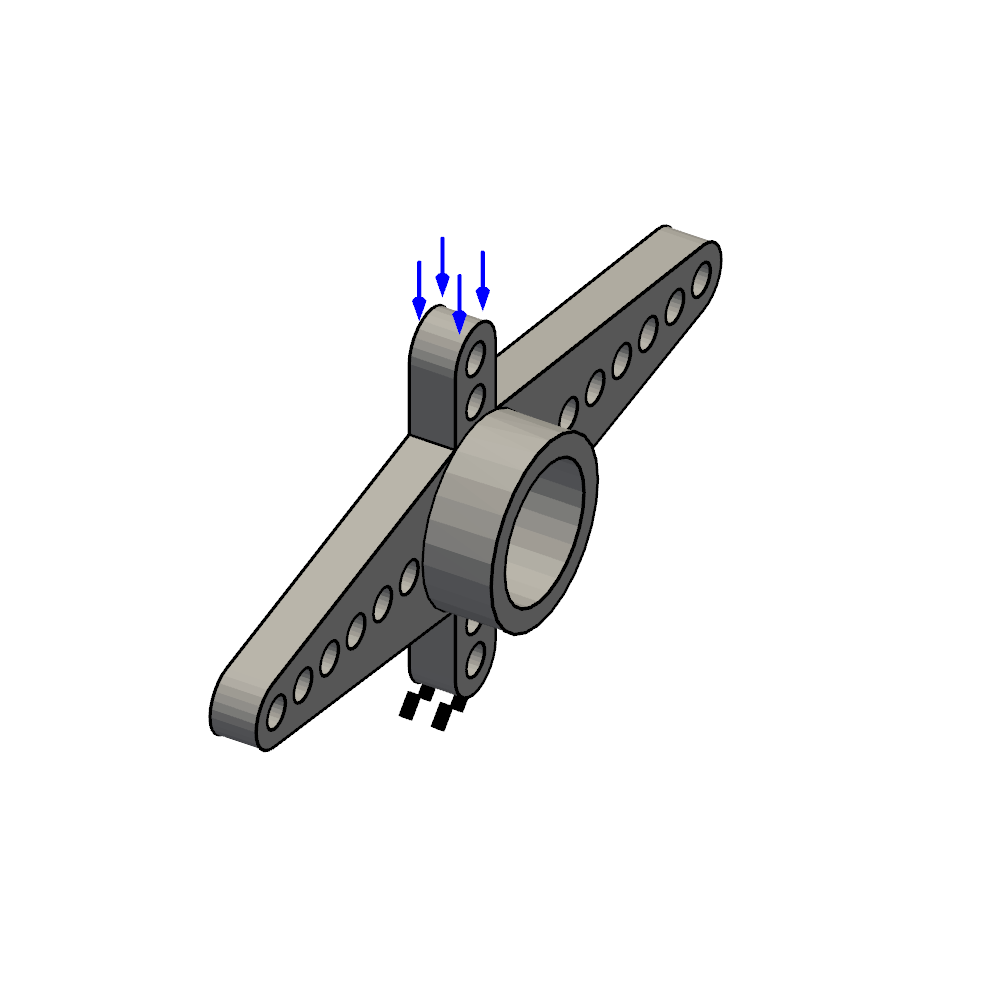}
    \label{fig:GReFEM_points1}
}
\hfill
\subfloat[ \textbf{GReFEM} Predicted refinement anchors]{
    \includegraphics[width=0.3\textwidth]{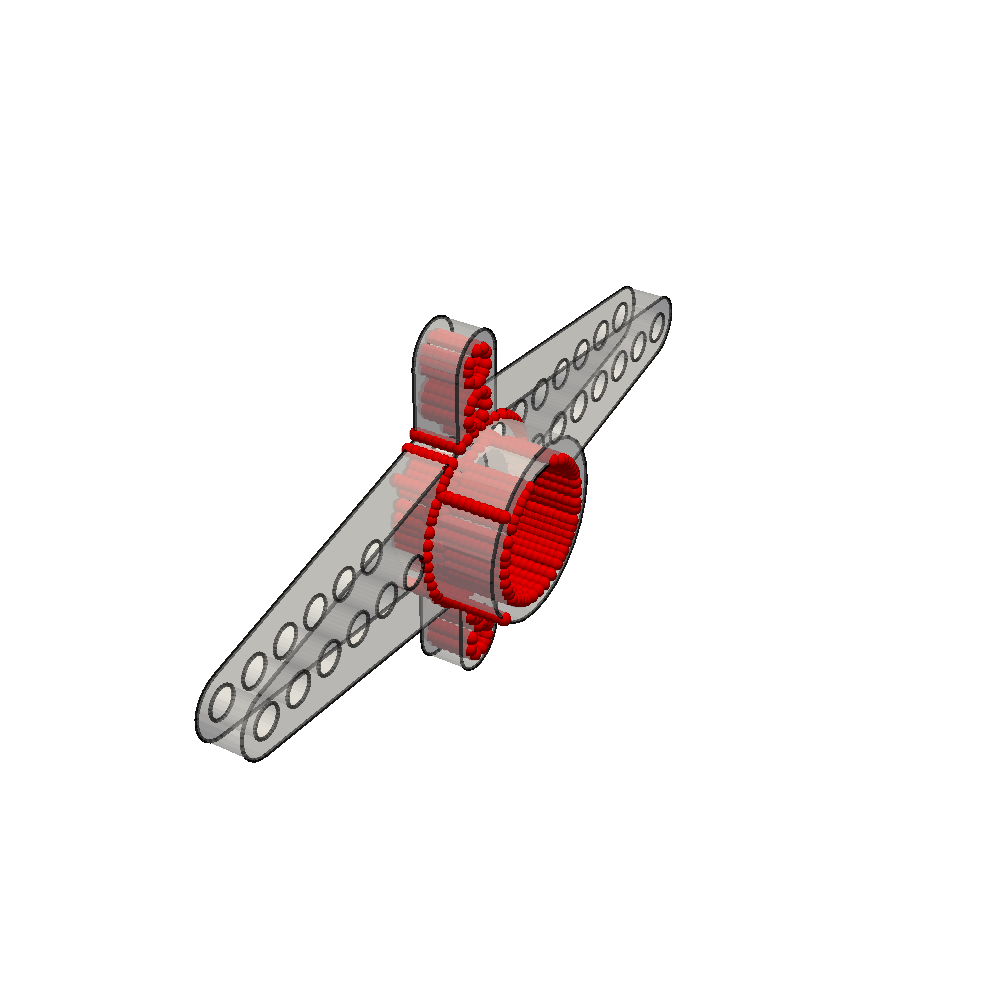}
    \label{fig:GReFEM_points2}
}
\hfill
\subfloat[Ground-truth ZZ stress indicator.]{
    \includegraphics[width=0.3\textwidth]{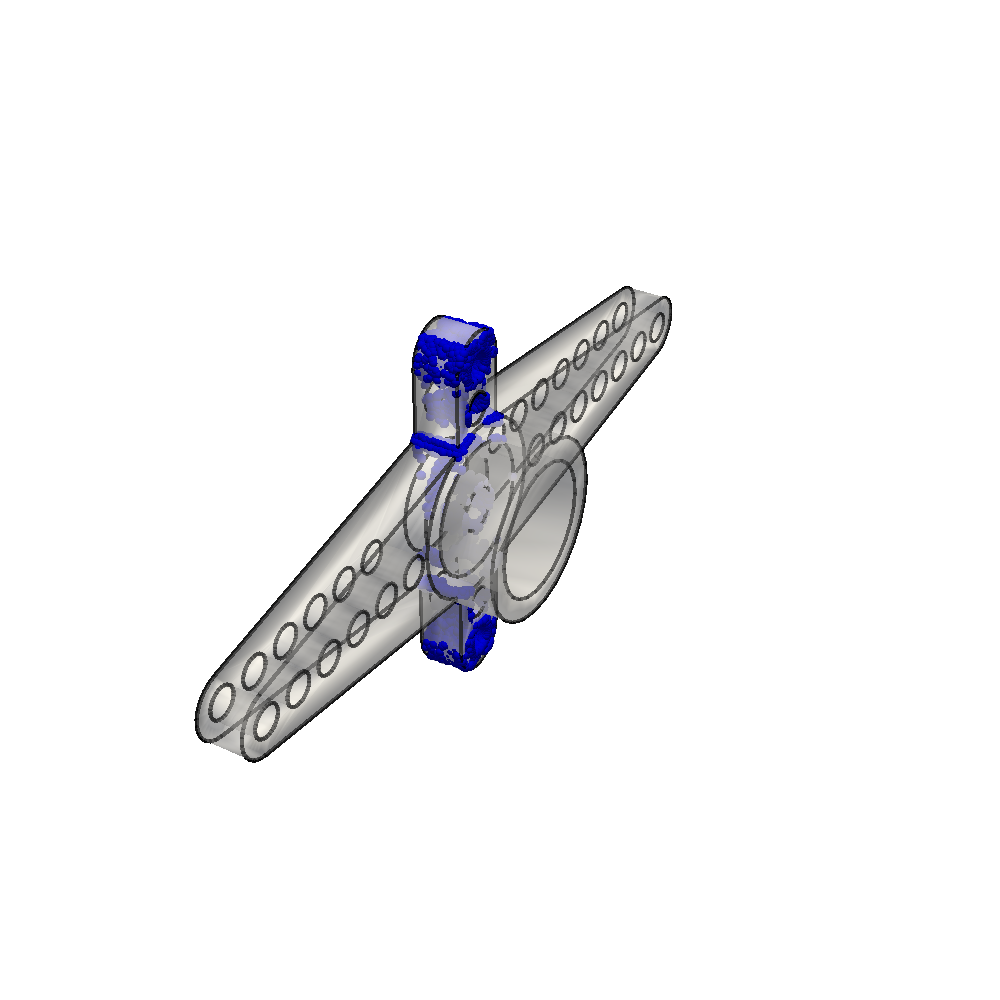}
    \label{fig:GReFEM_zz}
}

\caption{
Qualitative evaluation of \textbf{GReFEM}.
Predicted refinement points (Red) align with stress-critical regions (Blue) identified by the ground truth ZZ stress error indicator field computed using the fine reference solution $u_{ref}$. Blue points within the geometry denote stress critical areas above the 99th percentile of the $\eta_{ZZ}$ field.
}
\label{fig:GReFEM_overall_evaluation}
\end{figure}


\paragraph{Finite Element Meshing}
We employ tetrahedral volumetric meshes due to their ability to conform to complex 3D geometries while maintaining element quality under local refinement.
Mesh generation is performed using Gmsh (\cite{GeuzaineRemacle2009}), where spatially varying element sizes are prescribed via a distance-based sizing field. Given a set of refinement anchor points $\mathcal{P} \subset \mathbb{R}^3$ located on the surface, we define a distance field $d(\mathbf{x}) = \min_{\mathbf{p} \in \mathcal{P}} \| \mathbf{x} - \mathbf{p} \|_2$, 
which measures the proximity of any point $\mathbf{x} \in \Omega$ to the closest refinement location. This distance field is mapped to a spatially varying linear target element size function $h(\mathbf{x})$,
\begin{equation}
h(\mathbf{x}) =
\begin{cases}
h_{\min}, & d(\mathbf{x}) \leq D_{\min}, \\[2pt]
h_{\max}, & d(\mathbf{x}) \geq D_{\max}, \\[2pt]
\ell\!\left(d(\mathbf{x})\right), & \text{otherwise},
\label{eq:h_field}
\end{cases}
\end{equation}

\begin{equation}
\ell(d) =
h_{\min}
+ \frac{d - D_{\min}}{D_{\max} - D_{\min}}
\left( h_{\max} - h_{\min} \right).
\label{eq:element_field}
\end{equation}

\section{Method}

We propose  \textbf{GReFEM}, a semantic spatial grounding testbed framework applied to adaptive finite element meshing. Rather than relying on solver-based error estimators, \textbf{GReFEM} leverages pretrained MLLMs and problem specific instructions to identify mechanically salient regions using only visual observations of geometry. \textbf{GReFEM} operates solely on triangulated surface meshes and requires no access to B-Rep entities, STEP files, or simulation feedback. 

\subsection{Problem Formulation}

We consider a solid domain $\Omega \subset \mathbb{R}^3$ with boundary $\partial \Omega$ represented by a triangulated surface mesh
$\mathcal{M} = (V, F)$.
The objective is to generate a volumetric finite element mesh that locally refines regions expected to exhibit large stress gradients under mechanical load, while maintaining coarse resolution elsewhere to control computational cost. Given $\mathcal{M}$, the method predicts a sparse set of surface-localized refinement points
$\mathcal{P} = \{ \mathbf{p}_i \in \mathbb{R}^3 \}_{i=1}^{N}$, 
corresponding to geometrically or mechanically important features, such as sharp edges, fillets, holes and load-bearing faces. The resulting size field $h : \Omega \rightarrow \mathbb{R}^+$, as constructed according to equation \ref{eq:h_field}, is provided to a standard tetrahedral meshing algorithm in Gmsh, which deterministically generates a conforming volumetric mesh.

\subsection{orthoViews}

\begin{figure*}[t]
\centering
\includegraphics[width=1\columnwidth]{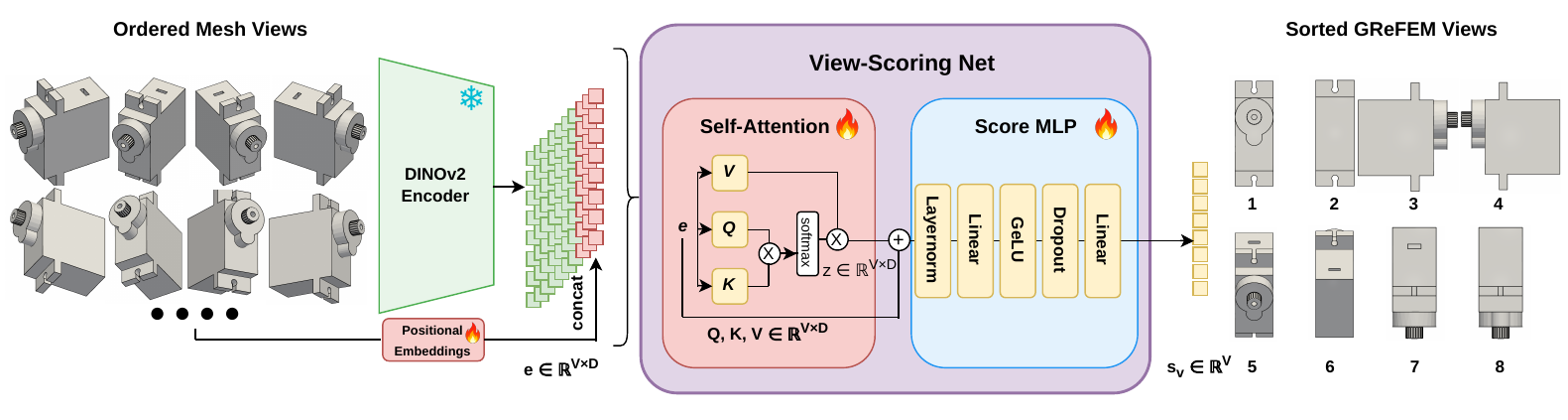}
\caption{\textit{orthoViews} performs self-attention across the V views of each CAD geometry, allowing each view embedding to aggregate information from all other views before being scored by a per-view MLP.}
\label{fig:orthoViews}
\end{figure*}

We render a candidate set of $V$ orthographic views $\mathcal{V} = \{ v_k = (\theta_k,\phi_k) \}_{k=1}^V$ by systematically enumerating azimuth and elevation angles on a fixed angular grid.
Orthographic view rendering preserves metric structure and avoids perspective distortion, which is critical for our projection pipeline. Each view is embedded using a frozen DINOv2 encoder, yielding feature vectors
$\mathbf{e}_k \in \mathbb{R}^D$.
To score views by geometric informativeness, we order these, compute self-attention across the views $V$ and input these view embeddings to the \textit{Score-MLP} network, in Figure \ref{fig:orthoViews}.
This ranking module is trained using a pairwise margin loss to prioritize views that expose a higher number of salient geometric features, given that we have view labels. The top-$K$ views form the final inference set $\mathcal{V}^*$ for \textbf{GReFEM}.

\subsection{Inference Pipeline}

Figure~\ref{fig:Mesh_Inference} summarizes the  \textbf{GReFEM} inference pipeline, which consists of three stages: region proposal, feature detection, and surface projection.

\textbf{Region Proposal}
Each selected view $v\in\mathcal{V}^*$ is overlaid with a uniform grid $\mathcal{G}_v$.
A multimodal LLM predicts a subset of salient grid cells $\mathcal{C}_v \subset \mathcal{G}_v$, corresponding to geometrically or mechanically significant regions.
Proposals are aggregated across views via $\mathcal{C} = \bigcup_{v\in\mathcal{V}^*} \mathcal{C}_v$

\textbf{Feature Detection}
Each cell in $\mathcal{C}$ defines an image patch processed using classical vision operators and mesh-ray projection to recover precise feature locations in the 2D views. We specifically spot features resembling internal corner edges, extruded contours/bosses, holes and fillets that are key geometric discontinuities in a load bearing structure. This stage converts coarse LLM output into accurate 2D feature pixels $\mathcal{Q}_c = \{ \mathbf{q}_{c,j} \in \mathbb{R}^2 \}$.

\textbf{MV-RaySeg (Multi-View Ray Segment Sampling)}
Given a feature point $\mathbf{q} \in \mathbb{R}^2$ detected in a view $v$, we recover its corresponding spatial projection on the 3D surface by ray–mesh intersection under known camera parameters.
Let $\pi_v^{-1}$ denote the inverse projection for view $v$, and let $\mathbf{d}_v \in \mathbb{R}^3$ be the associated view direction.
The corresponding ray is defined as
$\mathbf{r}(t) = \pi_v^{-1}(\mathbf{q}) + t \, \mathbf{d}_v, \quad t \in \mathbb{R}$.

We compute the complete ordered set of ray–mesh intersections with the surface mesh $\mathcal{M}$, $\mathcal{T}_q = \{ t_i \mid \mathbf{r}(t_i) \in \mathcal{M} \}$, with $t_1 < t_2 < \dots < t_{|\mathcal{T}_q|}$.
Under standard assumptions on watertight meshes, consecutive intersection pairs $(t_{2k-1}, t_{2k})$ define ray segments that lie inside the solid. For each such segment, we sample a finite set of 3D points $
\mathcal{P}_q = \bigcup_k \left\{ \mathbf{r}(t) \;\middle|\; t \in [t_{2k-1}, t_{2k}] \right\}$, which yields a spatially extended set of surface-adjacent anchor points corresponding to the 2D feature $\mathbf{q}$. This formulation allows a single 2D detection to influence multiple depth-aligned surface regions, which is essential for internal cavities and edges. 

\section{Experiments}

\paragraph{Training}
We train the \textit{orthoViews} view-selection module on a dataset consisting of 5k diverse CAD parts from ABC CAD dataset \cite{koch2019abc} and FreeCAD \cite{riegel2016freecad} spanning mechanical components with varying geometric complexity (e.g., extrusions, holes, fillets, and load-bearing features).
Each object is represented during training by both its triangulated surface mesh and the corresponding STEP file. Specifically, we leverage high-level geometric primitives available in STEP (e.g., planar faces, cylindrical axes) for training by scoring each view by the number of features aligned in its view direction. During inference, our framework can work \emph{solely} with spherical views of surface meshes of 3D objects. For more information, the reader is encouraged to read the appendix section on \textit{orthoViews}. Training the network on the entire dataset for 4 epochs took a couple of hours on one NVIDIA H100 GPU.

\paragraph{Experimental Setup}
We evaluate on 156 CAD parts from the ABC~\cite{koch2019abc} and FreeCAD~\cite{riegel2016freecad} datasets (Appendix~\ref{app:testset}), disjoint from the \textit{orthoViews} training and validation splits, each under five loading scenarios, uniaxial (compression, bending, torsion) and multi-axial (bending-compression, torsion-compression), giving 780 paired geometry--load cases. 
We use three prompt templates: \textit{Load+Features}, \textit{Load+Features+Heuristics} and \textit{Load-Only}. For the \textit{Load+Features} prompt we provide the model with the loading instruction prompt and the key geometric features that could be stress critical under the load (Internal corner Edges, Corner Edges of extruded or Protruding parts, Holes and Fillets). The model is tested on the ability to correctly identify the cells under each feature enough to localize it using  contour and depth detection CV pipeline. For the \textit{Load+Features+Heuristics} we provide additional load based hints to spot stress critical regions in addition to the input provided in the \textit{Load+Features} prompt. For the \textit{Load-Only} prompt, we merely describe the loading on the object and instruct the MLLM to recommend cells containing the stress-critical geometric features. All the cells obtained are given to contour and depth detection algorithms in feature detection part of the pipeline to compute anchor points in these 2D views that will be later projected onto the 3D mesh. We evaluate four MLLMs, Gemini-3-flash-preview, Claude-Haiku-4.5, GPT-5.4-mini and Qwen3-VL-235B, all accessed through the OpenRouter (\cite{openrouter}) API and all given the top five \textit{orthoViews} views, without domain-specific training or PDE solves. The minimum element size is $h_{\min} = 0.2\,h_{\max}$ for all refined meshes, and $D_{min}=2h_{min}$, $D_{max}=4h_{min}$ define the element size field of Eq.~\ref{eq:h_field}. Anchors within $6h_{\min}$ of the loaded faces are ignored by the mesher for every policy alike, so that error differences arise from interior refinement and not from differences in the imposition of boundary conditions. Every (geometry, load) case is scored against an independent reference solve on a uniform mesh with $h_{\mathrm{fine}}=h_{\max}/16$, capped at $6\times10^6$ tetrahedra ($\approx 5\times10^6$ typical; linear elasticity, $E=210$\,GPa, $\nu=0.3$), through the critical-region errors of Eqs.~\ref{eq:vm_err} and~\ref{eq:energy_err}.

\paragraph{Qualitative Results}
In the qualitative examples, the \textbf{GReFEM} predicted refinement locations align with stress-critical regions. Notably, the predicted points are not uniformly distributed across all geometric features such as holes or internal corners, but are selectively localized to those features that should be mechanically activated under the prescribed loading conditions (for instance Figure~\ref{fig:GReFEM_zz}).
In particular, while multiple concave features may be present in the geometry,  \textbf{GReFEM}, when provided with physics guided prompts about the problem and physics, attempts to strongly highlight those regions that correspond to dominant stress concentrations in the ZZ stress field, rather than marking all such features indiscriminately, as seen in Figure \ref{fig:visual_examples_4x5} in the appendix.

\paragraph{Refinement Policies at a Matched Budget}
Adaptive meshing is fundamentally a resource-constrained optimization problem: introducing too many refinement points severely degrades downstream solver efficiency. We therefore compare seven refinement policies at a \emph{matched element budget}, so that differences reflect \emph{where} smaller elements are placed rather than how many are spent: (i) \textbf{Coarse}, the unrefined mesh produced by the pipeline (a fixed floor, not budget-matched); (ii) \textbf{GReFEM}, our framework with trained \textit{orthoViews} selection and zero-shot MLLM inference; (iii) \textbf{Geometric heuristic (blind)}, the load-agnostic detector of concave sharp edges, fillets, and holes based on dihedral angles and discrete curvature, independent of loading; (iv) \textbf{Load-informed heuristic}, the same geometric candidates re-weighted by a solver-free closed-form nominal-stress estimate (per-layer voxel cross-section properties $A$, $I_z$, $J$ along the loading axis), so the loading case decides which candidates are refined; (v) \textbf{ZZ from coarse solve}, anchors placed at the top-percentile ZZ error points of a single coarse FEM solve; (vi) \textbf{ZZ from fine reference solve}, anchors at the top-percentile ZZ points of the $\approx 5\times10^6$-element reference solution, an oracle reference; and (vii) \textbf{AMBER}~\cite{freymuth2025amber}, a learned sizing-field AMR method retrained from scratch with load-case conditioning on 440 expert meshes derived from reference solves. Each anchor-based baseline is reduced to \textbf{GReFEM}'s anchor count for that object, and the $h_{\min}$ of every refined baseline is calibrated iteratively to \textbf{GReFEM}'s element count ($10\%$ tolerance); AMBER emits a continuous field, matched via a global scaling factor. The load-informed heuristic and AMBER are detailed in Appendix~\ref{app:downstream}.

\subsection{Downstream FEM Error at a Matched Budget}
\label{sec:downstream}
We treat downstream FEM error, not localization overlap, as the primary criterion. Table~\ref{tab:downstream_main} reports the relative error in peak von Mises stress ($\delta_{vM}^{p99}$, the failure-relevant design quantity) and in $\delta_{E}^{crit}$ (the natural finite-element error norm), both inside $\Omega_{\mathrm{crit}}$, pooled over all 780 cases.

\textbf{GReFEM versus solver-free, solver-informed and learned baselines.} With no FEM solve, no simulation training data and no hand-derived mechanics, \textbf{GReFEM} (Gemini, \textit{Load+Features+Heuristics}) attains a lower peak-stress error than the blind geometric heuristic (pooled $0.627$ vs.\ $0.672$) and than ZZ refinement from a coarse FEM solve ($0.671$), closing $58\%$ of the gap between the coarse mesh and the reference-solve oracle at a forward cost of view rendering plus one MLLM call (median $\approx 7$\,s). Strain-energy errors separate the refined policies as well. \textbf{GReFEM}'s energy error is on average lower than blind heuristic and coarse-solve ZZ. AMBER, retrained with load conditioning on 440 expert meshes, and \textbf{GReFEM} are comparable in peak-stress error ($0.624$ vs.\ $0.627$; strain energy $0.869$ vs.\ $0.880$). \textbf{GReFEM} reaches this range with no expert meshes, no reference solves and no meshing failures. While the load-informed heuristic surpasses \textbf{GReFEM} in peak-stress error ($0.567$ vs.\ $0.627$), it requires an expert to hand-derive beam-theory stress formulas for each load family and is valid only where nominal-stress reasoning applies, whereas \textbf{GReFEM} adapts to a new load family by a prompt edit. We report it as the boundary of what semantic localization currently provides.

\textbf{MLLMs and prompts.} With the feature-enumerating prompts all four MLLMs lie within $0.021$ in pooled $\delta_{vM}^{p99}$. At a matched budget, these prompts lower peak-stress error relative to \textit{Load-Only} for Gemini ($0.627$ vs.\ $0.649$) and GPT-5.4-mini ($0.629$ vs.\ $0.664$), whereas adding load-specific heuristics to the feature list (\textit{Load+Features+Heuristics} vs.\ \textit{Load+Features}) decreases MLLM's F1 while slightly increasing precision. For Claude and Qwen the \textit{Load-Only} runs place more anchors and use $10$--$30\%$ more elements than the feature prompts, thus displaying lower localization precision. The sizing sweep (Table~\ref{tab:sizing_sweep}) shows that \textbf{GReFEM}'s advantage over the blind heuristic and coarse-solve ZZ is largest at the full-budget operating point and decays toward parity as elements become scarce, while the load-informed heuristic stays ahead at every level.

\begin{table*}[t]
\centering
\caption{\textbf{Downstream FEM error at a matched element budget} on 156 CAD parts $\times$ 5 load cases. Relative errors in peak von Mises stress ($\delta_{vM}^{p99}$, Eq.~\ref{eq:vm_err}) and in strain energy ($\delta_{E}^{\mathrm{crit}}$, Eq.~\ref{eq:energy_err}) inside $\Omega_{\mathrm{crit}}$ against the $\approx 5\times10^6$-element reference solve (lower is better). The fine-reference ZZ policy is an oracle reference; all other refined policies are matched to GReFEM's anchor count and element budget. $^\dagger$Budget-matched to GReFEM (the native \textit{Load-Only} runs of Gemini and GPT place more anchors). $^\ddagger$Not budget-matched (native element count). AMBER fails to mesh $\approx 6\%$ of cases (730 cases evaluated).}
\label{tab:downstream_main}
\resizebox{\textwidth}{!}{%
\begin{tabular}{llccccc}
\toprule
\textbf{Method} & \textbf{Information used} & \textbf{$\delta_{vM}^{p99}$ mean} & \textbf{$\delta_{vM}^{p99}$ med.} & \textbf{$\delta_{E}^{\mathrm{crit}}$ mean} & \textbf{$\delta_{E}^{\mathrm{crit}}$ med.} & \textbf{cells} \\
\midrule
Coarse (no refinement) & --- & 0.737 & 0.782 & 0.943 & 0.966 & 25\,051 \\
\midrule
GReFEM (Gemini) & Load + Features + Heuristics & 0.627 & 0.650 & 0.880 & 0.920 & 95\,170 \\
GReFEM (GPT-5.4-mini) & Load + Features + Heuristics & 0.629 & 0.650 & 0.872 & 0.915 & 96\,636 \\
GReFEM (Claude-Haiku-4.5) & Load + Features + Heuristics & 0.643 & 0.652 & 0.879 & 0.924 & 87\,316 \\
GReFEM (Qwen3-VL-235B) & Load + Features + Heuristics & 0.648 & 0.670 & 0.884 & 0.923 & 82\,929 \\
GReFEM (Gemini) & Load + Features & 0.635 & 0.666 & 0.872 & 0.920 & 97\,338 \\
GReFEM (GPT-5.4-mini) & Load + Features & 0.642 & 0.664 & 0.878 & 0.921 & 92\,750 \\
GReFEM (Claude-Haiku-4.5) & Load + Features & 0.634 & 0.654 & 0.877 & 0.920 & 95\,762 \\
GReFEM (Qwen3-VL-235B) & Load + Features & 0.645 & 0.674 & 0.887 & 0.930 & 85\,159 \\
GReFEM (Gemini)$^\dagger$ & Load-Only & 0.649 & 0.675 & 0.881 & 0.923 & 93\,832 \\
GReFEM (GPT-5.4-mini)$^\dagger$ & Load-Only & 0.664 & 0.692 & 0.888 & 0.924 & 93\,360 \\
GReFEM (Claude-Haiku-4.5)$^\ddagger$ & Load-Only & 0.635 & 0.658 & 0.870 & 0.913 & 112\,813 \\
GReFEM (Qwen3-VL-235B)$^\ddagger$ & Load-Only & 0.638 & 0.655 & 0.877 & 0.915 & 91\,555 \\
\midrule
Geometric heuristic (blind) & geometry only & 0.672 & 0.696 & 0.891 & 0.927 & 93\,766 \\
Load-informed heuristic & geometry + hand-derived closed-form stress & 0.567 & 0.592 & 0.868 & 0.914 & 87\,846 \\
ZZ from coarse solve & 1 FEM coarse solve + ZZ estimator & 0.671 & 0.699 & 0.886 & 0.924 & 92\,342 \\
AMBER (learned AMR) & supervised expert meshes + load & 0.624 & 0.660 & 0.869 & 0.915 & 84\,994 \\
\midrule
\textit{ZZ from fine reference solve} & 1 FEM reference solve + ZZ estimator & \textit{0.549} & \textit{0.579} & \textit{0.850} & \textit{0.908} & \textit{82\,042} \\
\bottomrule
\end{tabular}%
}
\end{table*}

\subsection{Localization Overlap versus Downstream Accuracy}
\label{sec:localization}

For measuring the degree to which MLLMs localize refinement points based on prompt variations and number of input views, we take the ZZ stress indicator field $\eta_{ZZ}$ of the reference solve and extract the points above its 99th percentile (after the same $2\%$ boundary trim as $\Omega_{\mathrm{crit}}$). \textbf{Recall@$r$} is the fraction of these reference points within radius $r$ of any predicted refinement anchor, and \textbf{Precision@$r$} the fraction of predicted anchors within $r$ of a reference point; \textbf{F1@$r$} is their harmonic mean. We set $r=D_{max}=4h_{min}$ (Eq.~\ref{eq:h_field}), the extent of the refinement transition zone. Metrics are computed per case with the top five \textit{orthoViews} views and averaged over cases; cases in which a variant returned no anchors ($\le 9\%$ per variant) are excluded from that variant's average.

Appendix Table~\ref{tab:exploratory_results} reports anchor-level Precision@$r$, Recall@$r$ and F1@$r$ for all twelve MLLM--prompt variants over the 780 cases. Contrary to common perception, we find the localization overlap metrics are decorrelated with actual downstream FEM accuracy. Reduced to \textbf{GReFEM}'s anchor count, the blind geometric heuristic attains higher recall ($0.766$) and F1 ($0.593$) than every MLLM variant (F1 $0.484$--$0.578$), and its precision ($0.568$) is matched only by Qwen3-VL-235B ($0.573$), yet it has a higher peak-stress error than \textbf{GReFEM} at the same element budget (Table~\ref{tab:downstream_main}). Conversely, the load-informed heuristic has the lowest recall ($0.490$) and F1 ($0.470$) of all policies and the lowest downstream error. Overlap metrics weight every top-percentile ZZ point equally, whereas $\sigma_{vM}^{p99}$ is governed by the few peak concentrations: a policy can cover many mildly critical points while under-resolving the peak. The quantity that does track the downstream ranking is the element size the final mesh places at the hotspots (Appendix Table~\ref{tab:elem_loc}): median $0.42$ (load-informed), $0.45$ (GReFEM) and $0.49$ (blind heuristic), in the same order as their peak-stress errors. We therefore use localization overlap only to characterise \emph{how} the prompt steers each model in the following section.

\subsection{Analysis of Zero-Shot Semantic Grounding}
\paragraph{The Impact of Physics-Guided Prompting}
A central hypothesis of this study is that MLLMs can translate expert-provided textual cues into load-dependent spatial selections. We compare three prompt variants: \textit{Load+Features} (loading conditions plus the stress-critical feature types), \textit{Load+Features+Heuristics} (adding load-specific spatial hints), and \textit{Load-Only} (loading conditions only, asking for ``stress critical'' regions). The effect of the feature list is consistent across all four MLLMs: without it (\textit{Load-Only}) the models propose more anchors ($+0.7$k to $+2.1$k per case), recall rises (e.g.\ GPT-5.4-mini $0.644\rightarrow0.777$) and precision falls by $0.014$--$0.039$ (Appendix Table~\ref{tab:exploratory_results}). Where \textit{Load-Only} can be budget-matched (Gemini, GPT-5.4-mini) the broader selection costs downstream accuracy: peak-stress error rises from $0.627$ to $0.649$ and from $0.629$ to $0.664$. Adding load-specific heuristics to the feature list makes the selection slightly more conservative: anchors per case fall by up to $0.5$k, precision rises marginally ($+0.002$ to $+0.008$), and recall drops for Gemini, Claude, and Qwen ($0.008$--$0.061$; unchanged for GPT-5.4-mini). Downstream, the effect has no consistent sign: peak-stress error falls for Gemini and GPT-5.4-mini ($-0.008$, $-0.013$) and rises for Claude and Qwen ($+0.009$, $+0.003$).


\paragraph{Load Cases}
Per load case (Appendix Table~\ref{tab:load_case_results_comprehensive}, pooled over all 780 cases), \textbf{GReFEM} has lower peak-stress error than the blind heuristic in all five load cases and coarse-solve ZZ in all five. The load-informed heuristic's lead is largest under torsion ($0.512$ vs.\ $0.623$), where the closed-form $\tau=Tr/J$ is most informative. FEM errors vary more across load cases (range $0.062$) than across MLLMs ($0.016$) or prompts ($0.007$) (Appendix Table~\ref{tab:sensitivity_analysis}).



\begin{table}[ht]
\centering
\caption{\textbf{Sizing sweep.} $\delta_{vM}^{p99}$ /\ $\delta_{E}^{\mathrm{crit}}$ in $\Omega_{\mathrm{crit}}$ as a common scale factor is applied to $h_{\min}$ and $h_{\max}$; every baseline is re-matched to \textbf{GReFEM}'s element budget at each level (lower is better; 773--780 cases per level). The \textbf{GReFEM} column pools all four MLLMs (Gemini, GPT-5.4-mini, Claude-Haiku-4.5, Qwen3-VL-235B) per case (\textit{Load+Features+Heuristics} prompt); the MLLMs agree within $\approx0.02$.}
\label{tab:sizing_sweep}
\resizebox{\columnwidth}{!}{%
\begin{tabular}{lccccccc}
\toprule
\textbf{sizing} & \textbf{mean cells} & \textbf{GReFEM} & \textbf{Geom. heur.} & \textbf{Load-inf.} & \textbf{ZZ coarse} & \textbf{AMBER} & \textbf{ZZ fine ref.} \\
\midrule
$h\times1.0$  & 91k & 0.637/0.879 & 0.672/0.891 & 0.567/0.868 & 0.671/0.886 & 0.624/0.869 & \textit{0.549/0.850} \\
$h\times1.25$ & 55k & 0.688/0.903 & 0.688/0.895 & 0.627/0.890 & 0.716/0.906 & 0.657/0.891 & \textit{0.607/0.868} \\
$h\times1.6$  & 32k & 0.727/0.928 & 0.744/0.934 & 0.686/0.914 & 0.727/0.925 & 0.716/0.927 & \textit{0.700/0.916} \\
$h\times2.0$  & 20k & 0.748/0.938 & 0.765/0.943 & 0.734/0.928 & 0.758/0.943 & 0.735/0.931 & \textit{0.727/0.931} \\
\bottomrule
\end{tabular}%
}
\end{table}

\begin{table}[ht]
\centering
\caption{\textbf{Per-case cost}, pooled over all 780 cases. Maximum-displacement error is globally computed and reported for completeness; the decisive column is the anchor-placement cost, where \textbf{GReFEM}'s full forward cost is view rendering plus a single MLLM call with \emph{no} FEM solve. The \textbf{GReFEM} row averages all four MLLMs (Gemini, GPT-5.4-mini, Claude-Haiku-4.5, Qwen3-VL-235B); the cost profile is MLLM-independent.}
\label{tab:cost}
\resizebox{\columnwidth}{!}{%
\begin{tabular}{lcccc}
\toprule
\textbf{method} & \textbf{mean cells} & \textbf{max-disp. err (med.)} & \textbf{solve time (med.)} & \textbf{anchor-placement cost (med.)} \\
\midrule
Coarse & 25\,051 & 0.026 & 0.2\,s & --- \\
\textbf{GReFEM} & 90\,511 & 0.023 & 0.8\,s & $\approx 7$\,s: render + 1 MLLM call; \textbf{no solve} \\
Geometric heuristic & 93\,766 & 0.022 & 0.8\,s & $<1$\,s; no solve \\
Load-informed heuristic & 87\,846 & 0.022 & 1.0\,s & $<1$\,s; no solve \\
ZZ from coarse solve & 92\,342 & 0.022 & 2.5\,s & 1 coarse solve + ZZ \\
ZZ from fine reference solve & 82\,042 & 0.022 & 0.6\,s & reference solve ($\approx 35$--$60$\,min) \\
\bottomrule
\end{tabular}%
}
\end{table}

\section{Conclusion}
This study explored a fundamental question at the intersection of foundation models and computational engineering: can MLLMs act as zero-shot semantic assistants that replace solver-in-the-loop decisions about \emph{where} to refine a mesh? Measured by the finite-element error at stress concentrations, the quantity refinement is meant to improve, the answer on 156 parts and 780 geometry--load cases is a qualified yes. At a matched element budget and without any FEM solve or simulation data, \textbf{GReFEM} lowers critical-region peak-stress error relative to a blind geometric heuristic and to ZZ refinement from a coarse solve, performs comparably overall to a learned baseline AMBER, and is insensitive to the choice among four MLLMs. The gain comes for certain MLLMs (Gemini, GPT) from following explicit instructions about which feature types are stress critical under the given load: removing that list broadens the selection and, at a matched budget, increases error.

\paragraph{Limitations} Our study has several limitations. A load-informed heuristic built from hand-derived beam-theory formulas for each loading case outperforms \textbf{GReFEM}, \textbf{GReFEM} does not outperform the retrained learned baseline AMBER, and $42\%$ of the gap to the reference-solve oracle remains. The MLLMs rely on the classical feature-detection stage to turn coarse grid-cell proposals into anchors, and on the presence of recognisable geometric features. \textbf{GReFEM} decides only \emph{where} to refine; the refinement magnitude ($h_{\min}$, $D_{\min}$, $D_{\max}$) and the final mesh generation follow fixed rules. Our evaluation covers linear elasticity under five displacement-driven load cases on ABC and FreeCAD parts. Within these bounds, the results indicate that off-the-shelf MLLMs, without task-specific training, can serve within \textbf{GReFEM} as solver-free, prompt-conditioned refinement policies and as an interface between human physical intuition and low-level numerical discretization.

\section*{Acknowledgements and Disclosures}
The authors gratefully acknowledge Helmholtz-Zentrum Hereon for providing the GPU computing resources on which all experiments in this work were carried out, and thank Prof.\ Christian Cyron, Head of the Institute of Material Systems Modeling (MSM), Helmholtz-Zentrum Hereon, for providing the platform and support that made this work possible. The authors also thank their colleagues at the Institute of Material Systems Modeling for helpful discussions. The authors have no competing interests to declare that are relevant to the content of this article.

\bibliography{main}
\bibliographystyle{unsrtnat}

\clearpage
\appendix
\onecolumn
\section{Appendix}
This appendix section includes additional details of ZZ stress indicator field, our view selection module \textit{orthoViews}, a table containing performance of various MLLMs for various load cases considered, prompt templates and demonstrations of our \textbf{GReFEM} framework in action via a few test prompts. We also release the complete CAD dataset used for our analyses.

\subsection{ZZ Stress Indicator}
\label{app:fem}

The Cauchy stress tensor is defined via Hooke's law,
\begin{equation}
\boldsymbol{\sigma}(\mathbf{u}) = \mathbb{C} : \boldsymbol{\varepsilon}(\mathbf{u}),
\qquad
\boldsymbol{\varepsilon}(\mathbf{u}) =
\frac{1}{2}
\left(
\nabla \mathbf{u} + \nabla \mathbf{u}^\top
\right),
\end{equation}


where $\mathbb{C}$ denotes the fourth-order elasticity tensor and
$\boldsymbol{\varepsilon}(\mathbf{u})$ the infinitesimal strain tensor.

In standard finite element formulations, the stress field
$\boldsymbol{\sigma}_h$ is computed element-wise and is generally
discontinuous across element boundaries.
The Zienkiewicz--Zhu (ZZ) error estimator addresses this limitation by
constructing a smoothed or \emph{recovered} stress field
$\boldsymbol{\sigma}^\ast$, typically obtained via local least-squares
stress recovery or patch-based averaging of neighboring element stresses
\cite{zienkiewicz1987simple,zienkiewicz1992superconvergent}.

The ZZ error indicator is defined as
\begin{equation}
\eta_{\mathrm{ZZ}} =
\left(
\int_\Omega
(\boldsymbol{\sigma}^\ast - \boldsymbol{\sigma}_h)
:
\mathbb{C}^{-1}
:
(\boldsymbol{\sigma}^\ast - \boldsymbol{\sigma}_h)
\, d\Omega
\right)^{1/2},
\end{equation}
which measures the discrepancy between the recovered and finite element
stress fields in a constitutive-energy-weighted norm.

Regions with large values of $\eta_{\mathrm{ZZ}}$ are commonly interpreted
as stress-critical zones where the finite element approximation is
insufficiently resolved.
Due to its robustness and ease of computation, the ZZ indicator is
widely adopted in industrial adaptive meshing pipelines as a refinement
criterion \cite{babuvska1978posteriori}.

\paragraph{Use as a Reference Stress Signal}
In our experiments, the ZZ stress indicator is computed \emph{exclusively}
on a sufficiently fine reference mesh using the corresponding reference
displacement solution $\mathbf{u}_{\mathrm{ref}}$.
Importantly, the ZZ field is \emph{not} evaluated on the meshes generated
by our refinement strategy, nor does it influence the refinement process
itself.

Instead, the reference ZZ indicator serves as an external, physics-based
validation signal, enabling us to assess whether the refinement points
predicted by our method spatially align with genuinely stress-critical
regions identified from the high-fidelity solution.
This separation ensures that the reported stress localization results
reflect true generalization rather than feedback from the refined mesh
solution.


\begin{figure*}[t] 
    \centering
    \includegraphics[width=0.9\linewidth]{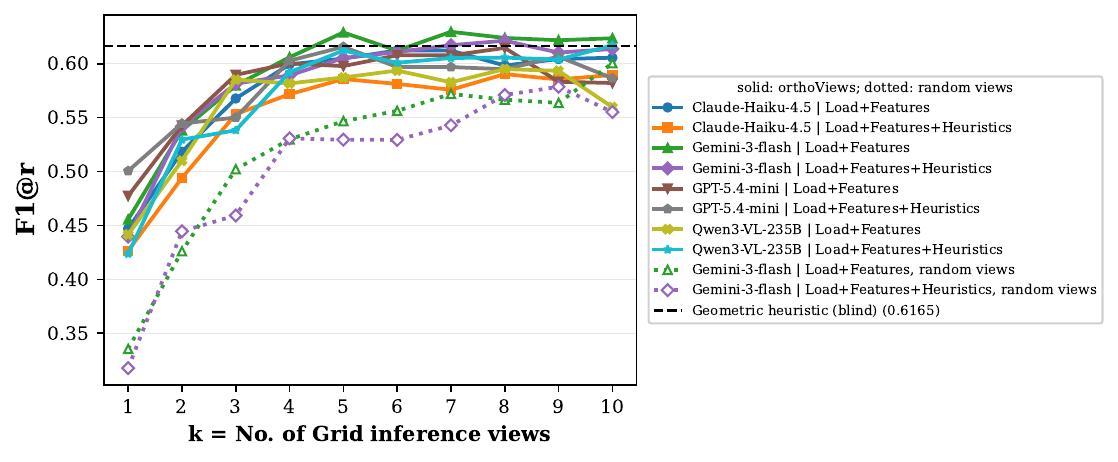}    
       \caption{Anchor-level F1@$r$ versus number of views, inference at 1--10 views. Solid lines use \textit{orthoViews} view selection; dotted lines use randomly selected views (Gemini, both feature prompts). F1 rises steeply up to 4--5 views and plateaus thereafter. With \textit{orthoViews}, Gemini's F1 is higher than with random views at every view count, by $0.024$--$0.123$ ($0.075$--$0.082$ at five views). Gemini with the \textit{Load+Features} prompt exceeds the blind geometric heuristic's F1 at five and seven or more views, and the peak F1 of every other MLLM--prompt variant lies within $0.027$ of it ($0.590$--$0.621$). On the full 156-part benchmark at five views the heuristic attains higher F1@r than most MLLMs (Table~\ref{tab:exploratory_results}), while \textbf{GReFEM} attains lower downstream error (Table~\ref{tab:downstream_main}).}
    \label{fig:combined_evaluation}
\end{figure*}


\subsection{orthoViews} 
We introduced \textit{orthoViews} as a view ranking module that sorts the rendered views in the order of most to least informative views considering the stress critical features like internal edges, holes and notches. We curated a dataset of 5k CAD geometries from open online sources like FreeCAD \cite{riegel2016freecad} and ABC Autodesk CAD dataset \cite{koch2019abc}. 

\paragraph{Label Generation}
To generate supervision for training \textit{orthoViews}, we first render a fixed set of candidate orthographic views for each CAD object. Views are obtained by placing a virtual camera on a sphere whose radius is scaled proportionally to the object’s bounding dimensions, and systematically sweeping the camera over 12 uniformly spaced azimuth angles (0 to 360 degrees) and 11 elevation angles (-90 to 90 degrees in intervals of 18 degrees, with just one azimuth angle for the extremes), yielding a total of 110 views per mesh.

For each rendered view, we analyze the underlying CAD solid to identify stress-relevant geometric primitives, including concave internal edges, cylindrical holes, and filleted surfaces. Each view is then assigned a geometric saliency score based on the alignment of the viewing direction and the principal axes of these features, while accounting for visibility through occlusion checks on the CAD object surface. The resulting ranked view list, along with the corresponding view directions, forms the supervision signal for learning view importance in \textit{orthoViews}. We use a cosine alignment threshold of $\tau = 0.95$ when computing feature-view alignment.

\begin{algorithm}
  \caption{orthoViews Label Generation}
  \label{alg:orthoviews_scoring}
  \begin{algorithmic}
    \STATE {\bfseries Input:} CAD solid $\mathcal{S}$, surface mesh $\mathcal{M}$, view directions $\mathcal{V}$
    \STATE {\bfseries Output:} Ranked views $\mathcal{V}^\star$

    \STATE $\mathcal{E}_c \leftarrow$ ExtractConcaveEdges$(\mathcal{S})$
    \STATE $\mathcal{H} \leftarrow$ ExtractCylindricalHoles$(\mathcal{S})$
    \STATE $\mathcal{F} \leftarrow$ ExtractFillets$(\mathcal{S})$
    \STATE $\mathcal{G} \leftarrow \mathcal{E}_c \cup \mathcal{H} \cup \mathcal{F}$

    \FORALL{$\mathbf{v} \in \mathcal{V}$}
        \STATE $s[\mathbf{v}] \leftarrow 0$
        \FORALL{$g \in \mathcal{G}$}
            \IF{IsOccluded$(g, \mathbf{v}, \mathcal{M})$}
                \STATE {\bfseries continue}
            \ENDIF
            \STATE $\mathbf{a}_g \leftarrow$ PrincipalAxis$(g)$
            \IF{$|\mathbf{a}_g \cdot \mathbf{v}| > \tau$}
                \STATE $s[\mathbf{v}] \leftarrow s[\mathbf{v}] + 1$
            \ENDIF
        \ENDFOR
    \ENDFOR

    \STATE $\mathcal{V}^\star \leftarrow$ SortDescending$(\mathcal{V}, s)$
    \RETURN $\mathcal{V}^\star$
  \end{algorithmic}
\end{algorithm}

\paragraph{Training}
The rendered views are then passed through a vision encoder, and their embeddings are aggregated and passed through a trainable self-attention layer to capture view ranking. We employ a pretrained frozen DINOv2 encoder as it captures both global semantics (e.g., object-level understanding) and local details (e.g., object parts, textures, edges), making it ideal for representing geometric features of objects visible in the views. 

Formally, given $V$ \textbf{GReFEM} views $\{\mathcal{I}_k\}_{k=1}^V$, a frozen DINOv2 encoder $f_{\theta}$ maps each view to an embedding
\begin{equation}
\mathbf{e}_k = f_{\theta}(\mathcal{I}_k) \in \mathbb{R}^D .
\end{equation}
The embeddings are stacked as $\mathbf{e} \in \mathbb{R}^{V \times D}$ and passed through a self-attention module
\begin{equation}
\tilde{\mathbf{Z}} = \mathrm{Attn}(\mathbf{e}, \mathbf{e}, \mathbf{e}),
\end{equation}
followed by an MLP head $g_{\phi}$ to predict scalar view scores $s_v = g_{\phi}(\tilde{\mathbf{z}}_v)$. The self attention layer and the scoring MLP are both kept trainable.

The model is trained using a pairwise ranking loss
\begin{equation}
\mathcal{L} = \mathbb{E}_{(v^+,v^-)} \big[\max(0, m - (s_{v^+} - s_{v^-}))\big],
\end{equation}

$s_{v^+}$ and $s_{v^-}$ denote the geometrically preferred and non-preferred scores, respectively, constructed from oracle view scores in Algorithm \ref{alg:orthoviews_scoring}, $m$ is a margin hyperparameter enforcing a minimum score separation, kept here equal to 1.

\begin{table*}[t]
\centering
\caption{\textbf{Anchor-level localization} on the expanded benchmark (156 parts $\times$ 5 loads, top five \textit{orthoViews} views). Precision@$r$, Recall@$r$ and F1@$r$ against the top-$1\%$ reference ZZ points, $r=4h_{\min}$, averaged over cases; $n$ is the number of cases in which the variant returned anchors. Avg Pts: mean anchors per case. Cell mentions/view: grid-cell indices returned per view, summed over feature categories (a cell listed under several categories counts several times). The blind heuristic is reported at its native anchor count and reduced to \textbf{GReFEM}-Gemini's (\textit{Load+Features+Heuristics}) count per case. Localization overlap does not reproduce the downstream ranking of Table~\ref{tab:downstream_main} (Sec.~\ref{sec:localization}).}
\label{tab:exploratory_results}
\resizebox{\textwidth}{!}{%
\begin{tabular}{llcccccc}
\toprule
\textbf{Model / Baseline} & \textbf{Prompt Type} & \textbf{P@$r$} & \textbf{R@$r$} & \textbf{F1@$r$} & \textbf{Avg Pts (k)} & \textbf{Cell mentions/view} & \textbf{$n$} \\
\midrule
\multirow{3}{*}{Gemini-3-flash-preview} 
 & Load+Features+Heuristics & 0.552 & 0.616 & 0.521 & 4.6 & 18.6 & 756 \\
 & Load+Features   & 0.550 & 0.650 & 0.538 & 4.8 & 22.1 & 758 \\
 & Load-Only   & 0.538 & 0.692 & 0.546 & 5.3 & 9.6 & 765 \\
\midrule
\multirow{3}{*}{Claude-Haiku-4.5}      
 & Load+Features+Heuristics & 0.545 & 0.559 & 0.484 & 4.2 & 15.2 & 748 \\
 & Load+Features   & 0.537 & 0.620 & 0.517 & 4.7 & 19.8 & 767 \\
 & Load-Only   & 0.524 & 0.720 & 0.550 & 5.9 & 15.2 & 774 \\
\midrule
\multirow{3}{*}{GPT-5.4-mini}          
 & Load+Features+Heuristics & 0.549 & 0.644 & 0.525 & 4.3 & 38.6 & 769 \\
 & Load+Features   & 0.546 & 0.644 & 0.525 & 4.3 & 36.8 & 757 \\
 & Load-Only   & 0.520 & 0.777 & 0.578 & 6.4 & 25.5 & 771 \\
\midrule
\multirow{3}{*}{Qwen3-VL-235B}          
 & Load+Features+Heuristics & 0.573 & 0.577 & 0.505 & 3.4 & 59.8 & 735 \\
 & Load+Features   & 0.566 & 0.585 & 0.505 & 3.7 & 50.1 & 718 \\
 & Load-Only   & 0.534 & 0.648 & 0.522 & 4.4 & 18.3 & 735 \\
\midrule
\multirow{2}{*}{Geometric heuristic (blind)} & native count & 0.564 & 0.822 & 0.607 & 6.0 & -- & 780 \\
 & \textbf{GReFEM}-Gemini count & 0.568 & 0.766 & 0.593 & 3.6 & -- & 756 \\
\midrule
Load-informed heuristic & native count & 0.600 & 0.490 & 0.470 & 1.9 & -- & 780 \\
\bottomrule
\end{tabular}%
}
\end{table*}

\begin{table*}[t]
\centering
\caption{\textbf{Downstream FEM error per load case} at a matched element budget, all 156 parts (mean of $\delta_{vM}^{p99}$ /\ $\delta_{E}^{\mathrm{crit}}$; lower is better for both). \textbf{GReFEM} rows use the \textit{Load+Features+Heuristics} prompt. Last block: paired median difference in $\delta_{vM}^{p99}$, \textbf{GReFEM} (Gemini) minus baseline (negative favours \textbf{GReFEM}), with two-sided Wilcoxon $p$-values.}
\label{tab:load_case_results_comprehensive}
\resizebox{\textwidth}{!}{%
\begin{tabular}{lccccc}
\toprule
\textbf{Policy} & \textbf{Compression} & \textbf{Bending} & \textbf{Torsion} & \textbf{Bending + Comp.} & \textbf{Torsion + Comp.} \\
\midrule
Coarse (no refinement) & 0.739/0.948 & 0.730/0.938 & 0.727/0.942 & 0.741/0.941 & 0.740/0.942 \\
\midrule
GReFEM (Gemini) & 0.661/0.900 & 0.625/0.871 & 0.623/0.868 & 0.601/0.850 & 0.620/0.898 \\
GReFEM (GPT-5.4-mini) & 0.653/0.887 & 0.636/0.880 & 0.620/0.869 & 0.607/0.851 & 0.623/0.862 \\
GReFEM (Claude-Haiku-4.5) & 0.670/0.896 & 0.653/0.889 & 0.640/0.862 & 0.610/0.859 & 0.633/0.875 \\
GReFEM (Qwen3-VL-235B) & 0.671/0.900 & 0.657/0.881 & 0.640/0.874 & 0.624/0.857 & 0.639/0.894 \\
\midrule
Geometric heuristic (blind) & 0.696/0.909 & 0.697/0.905 & 0.659/0.878 & 0.646/0.860 & 0.653/0.888 \\
Load-informed heuristic & 0.636/0.889 & 0.574/0.870 & 0.512/0.848 & 0.551/0.831 & 0.561/0.887 \\
ZZ from coarse solve & 0.706/0.917 & 0.684/0.895 & 0.649/0.878 & 0.631/0.838 & 0.672/0.877 \\
AMBER (learned AMR) & 0.616/0.861 & 0.652/0.893 & 0.637/0.863 & 0.610/0.867 & 0.602/0.859 \\
\midrule
\textit{ZZ from fine reference solve} & \textit{0.606/0.858} & \textit{0.573/0.875} & \textit{0.568/0.859} & \textit{0.473/0.791} & \textit{0.521/0.855} \\
\midrule
\multicolumn{6}{l}{\textit{Paired $\Delta \delta_{vM}^{p99}$, \textbf{GReFEM} (Gemini) $-$ baseline (median, $p$)}} \\
vs.\ geometric heuristic & $-0.018$ ($0.023$) & $-0.064$ ($9{\cdot}10^{-7}$) & $-0.026$ ($0.023$) & $-0.031$ ($0.006$) & $-0.004$ ($0.071$) \\
vs.\ load-informed heuristic & $+0.032$ ($0.035$) & $+0.013$ ($0.041$) & $+0.065$ ($8{\cdot}10^{-6}$) & $+0.027$ ($0.056$) & $+0.042$ ($0.003$) \\
vs.\ ZZ from coarse solve & $-0.039$ ($0.003$) & $-0.040$ ($2{\cdot}10^{-4}$) & $-0.039$ ($0.008$) & $-0.033$ ($0.078$) & $-0.047$ ($0.002$) \\
vs.\ AMBER & $+0.052$ ($0.040$) & $-0.016$ ($0.18$) & $-0.021$ ($0.53$) & $-0.031$ ($0.47$) & $+0.031$ ($0.15$) \\
\bottomrule
\end{tabular}%
}
\end{table*}

\begin{table*}[h]
\centering
\caption{\textbf{Sensitivity of \textbf{GReFEM}'s downstream error across experimental dimensions} (all 156 parts). For each dimension, the mean relative error is computed per type, pooling over all other dimensions at $h\times1.0$ (for the sizing level, pooling over MLLMs and prompts at each type), and the standard deviation and range across levels are reported. Prompt types use native element counts. The last row gives, for reference, the spread across the five attainable refinement policies (GReFEM-Gemini, both heuristics, coarse-solve ZZ, AMBER).}
\label{tab:sensitivity_analysis}
\begin{tabular}{lccccc}
\toprule
\textbf{Dimension} & \textbf{Types} & \textbf{$\delta_{vM}^{p99}$ std} & \textbf{$\delta_{vM}^{p99}$ range} & \textbf{$\delta_{E}^{\mathrm{crit}}$ std} & \textbf{$\delta_{E}^{\mathrm{crit}}$ range} \\
\midrule
MLLM & 4 & 0.006 & 0.016 & 0.005 & 0.013 \\
Prompt & 3 & 0.003 & 0.007 & 0.005 & 0.011 \\
Load case & 5 & 0.021 & 0.062 & 0.014 & 0.040 \\
Sizing level & 4 & 0.044 & 0.116 & 0.025 & 0.064 \\
\midrule
\textit{Refinement policy (reference)} & 5 & 0.038 & 0.104 & 0.009 & 0.023 \\
\bottomrule
\end{tabular}
\end{table*}

\begin{table}[H]
\centering
\caption{\textbf{Element-level localization} on the final meshes over all 780 cases, against the top-$0.1\%$ reference ZZ hotspots. ``Predictions'' are the deeply refined elements (size $\le 1.25\,h_{\min}$): element precision \textbf{P} is the fraction of these within $r$ of a hotspot, element recall \textbf{R} the fraction of hotspots covered. Element size at hotspots is the median edge-length proxy (cube root of tetrahedron volume) of elements within $r$ of a hotspot. Element-level F1 is nearly identical across the three policies, but the element size at the hotspots follows the downstream ranking: paired per case, \textbf{GReFEM}'s hotspot elements are smaller than the blind heuristic's in $66\%$ of cases and larger than the load-informed heuristic's in $68\%$.}
\label{tab:elem_loc}
\begin{tabular}{lccccc}
\toprule
\textbf{method} & \textbf{P} & \textbf{R} & \textbf{F1} & \textbf{elem. size at hotspots (med.)} & \textbf{$\delta_{vM}^{p99}$ mean} \\
\midrule
GReFEM & 0.317 & 0.903 & 0.431 & 0.45 & 0.627 \\
Geometric heuristic (blind) & 0.311 & 0.922 & 0.425 & 0.49 & 0.672 \\
Load-informed heuristic & 0.349 & 0.831 & 0.431 & 0.42 & 0.567 \\
\bottomrule
\end{tabular}
\end{table}


\subsection{Expanded Downstream Evaluation}
\label{app:downstream}
This appendix expands the matched-budget downstream study of the main text over the full 156-part ($\times\,5$ load) benchmark. All errors are relative to the $\approx 5\times10^6$-element reference solve and measured inside the critical region $\Omega_{\mathrm{crit}}$.

\paragraph{Load-informed heuristic.}
This policy keeps the \emph{same} geometric candidates as the blind heuristic and changes only \emph{which} of them are refined, using a solver-free estimate of nominal stress. We voxel-fill the part and slice it into thin layers along the loading axis $y$ (clamped at $y_{\min}$, loaded at $y_{\max}$); for each layer we compute purely geometric section properties, area $A(y)$, centroid $(\bar{x},\bar{z})$, second moment of area $I_z(y)$, and polar moment $J(y)$. Each candidate is assigned to its nearest layer and weighted by the textbook nominal stress for the load case: $w=1/A$ (compression, net-section stress), $w=|x-\bar{x}|/I_z$ (bending, $\sigma=Mc/I$), and $w=r/J$ (torsion, $\tau=Tr/J$), with $r$ the distance to the layer centroid; combined loads sum the normalised terms. We keep the top stress-weighted $30\%$ of candidates and density-filter to the matched anchor budget. Geometry thus decides \emph{where} refinement may go and beam-theory stress decides \emph{which} locations matter; this prior wins where closed-form reasoning is valid but must be re-derived by hand for each load family.

\paragraph{Learned-AMR baseline (AMBER).}
We retrained AMBER~\cite{freymuth2025amber}, an iterative graph network that predicts a continuous sizing field, from scratch on this task: 440 expert adaptive meshes (88 ABC training parts $\times$ 5 load cases, disjoint from all 156 evaluation parts), each expert mesh built by projecting the fine reference solve's top-percentile ZZ refinement points onto the coarse-resolution mesh, trained to its validation plateau on H100 GPUs (our 440 expert meshes are $4$--$20\times$ more than the original AMBER datasets), with load-case conditioning added. At test time its sizing field is globally scaled to \textbf{GReFEM}'s element budget at each sweep level; it fails to mesh $\approx 6\%$ of cases. The final-step scale is calibrated by a secant search (up to 10 iterations, $5\%$ tolerance), which brings $95\%$ of AMBER's meshes within $10\%$ of \textbf{GReFEM}'s element count at every sizing level; the remainder are geometries on which the predicted field cannot be meshed. Across the sizing sweep (Table~\ref{tab:sizing_sweep}) the two policies are comparable: their mean errors differ by at most $0.031$ in peak stress and $0.012$ in strain energy at every level. A zero-shot, prompt-based policy thus lands in the same accuracy range as a sizing field supervised on 440 expert meshes; on a narrow part family with abundant expert meshes, a supervised sizing field would likely pull further ahead.

\begin{figure*}[t] 
    \centering

    \begin{subfigure}[t]{0.19\textwidth}
        \centering
        \includegraphics[width=\linewidth]{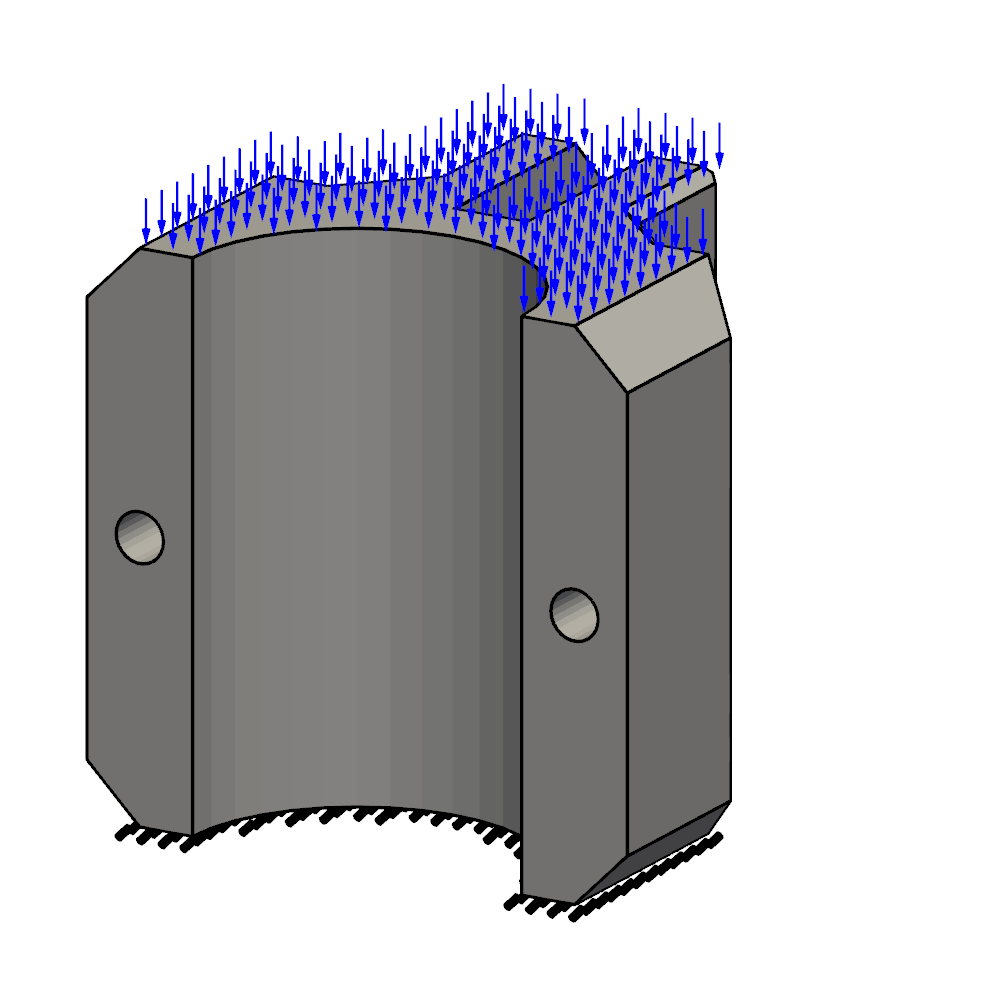}
    \end{subfigure}\hfill
    \begin{subfigure}[t]{0.19\textwidth}
        \centering
        \includegraphics[width=\linewidth]{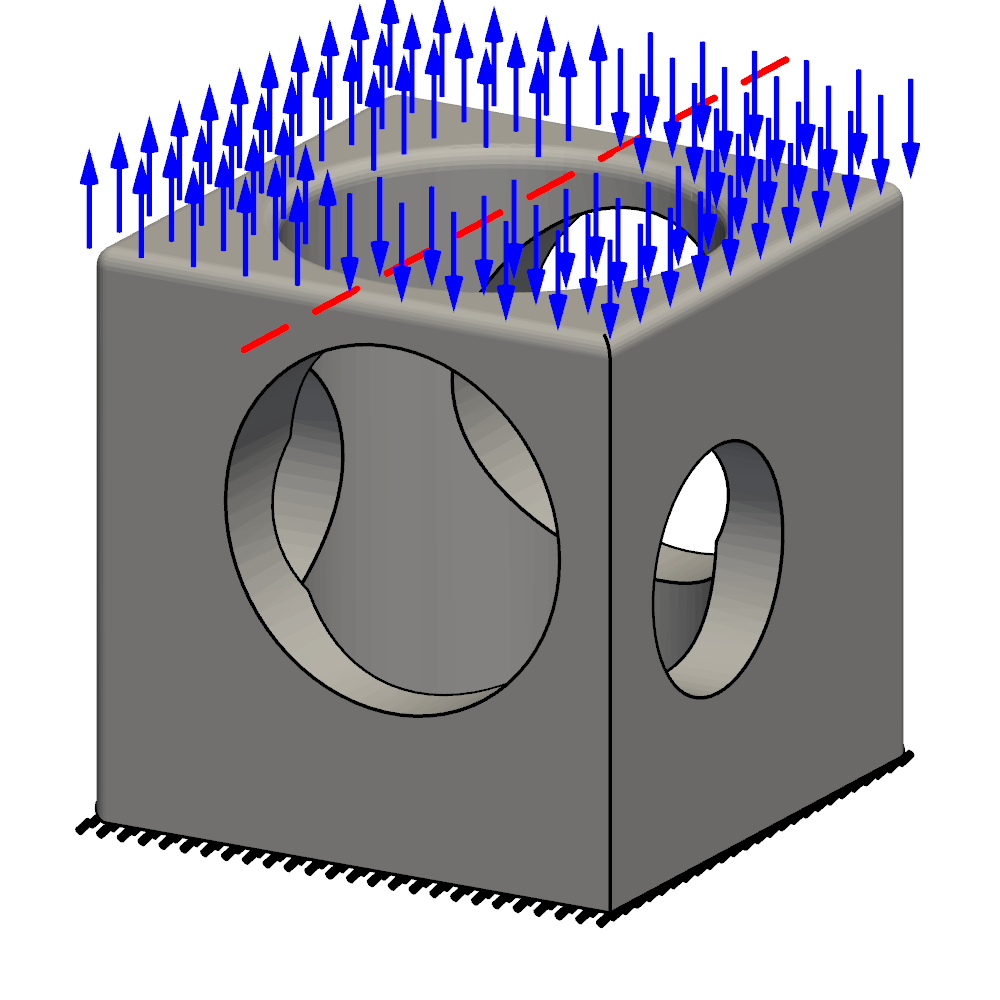}
    \end{subfigure}\hfill
    \begin{subfigure}[t]{0.19\textwidth}
        \centering
        \includegraphics[width=\linewidth]{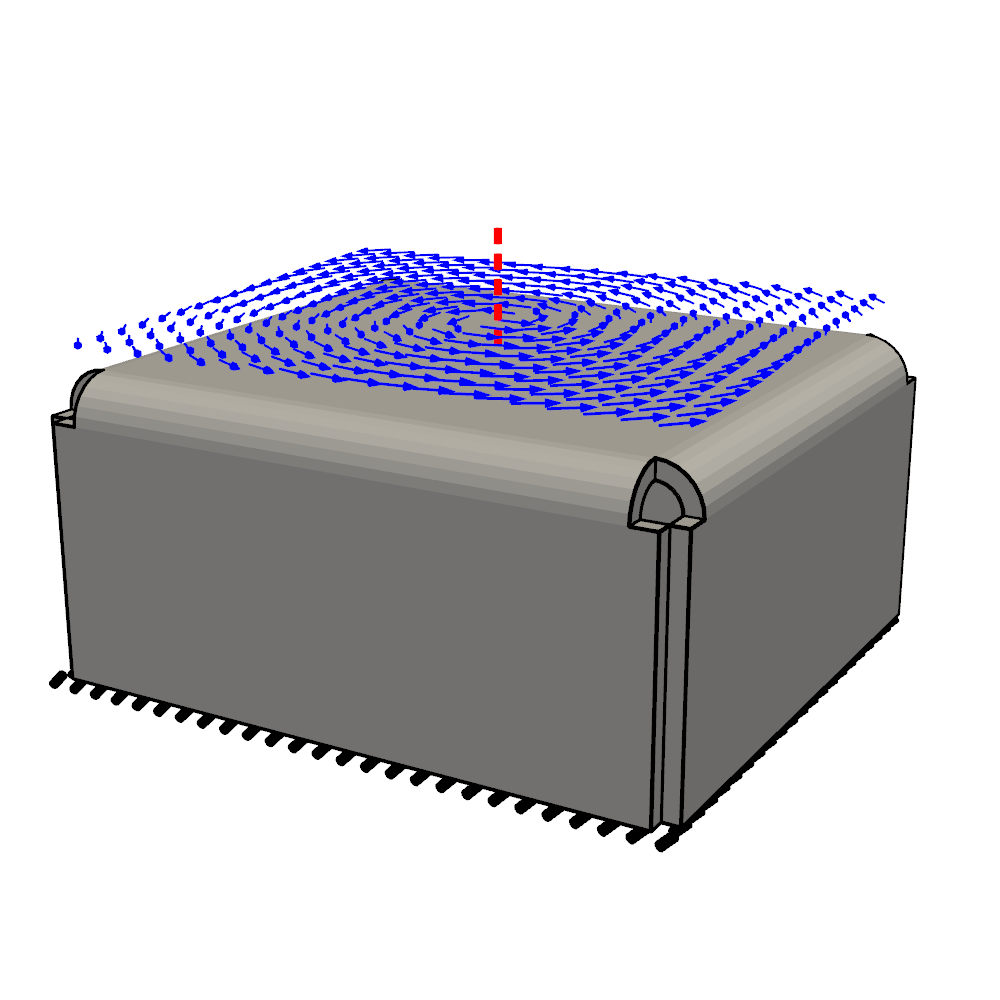}
    \end{subfigure}\hfill
    \begin{subfigure}[t]{0.19\textwidth}
        \centering
        \includegraphics[width=\linewidth]{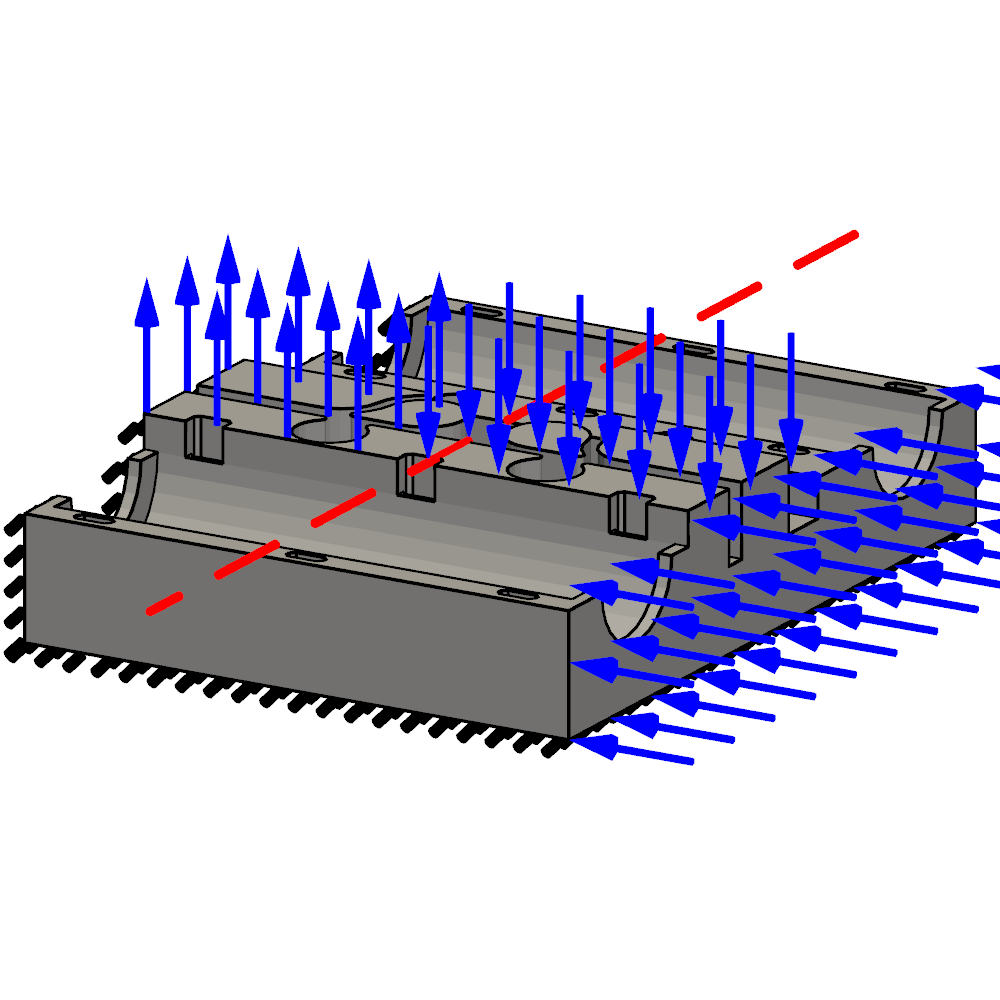}
    \end{subfigure}\hfill
    \begin{subfigure}[t]{0.19\textwidth}
        \centering
        \includegraphics[width=\linewidth]{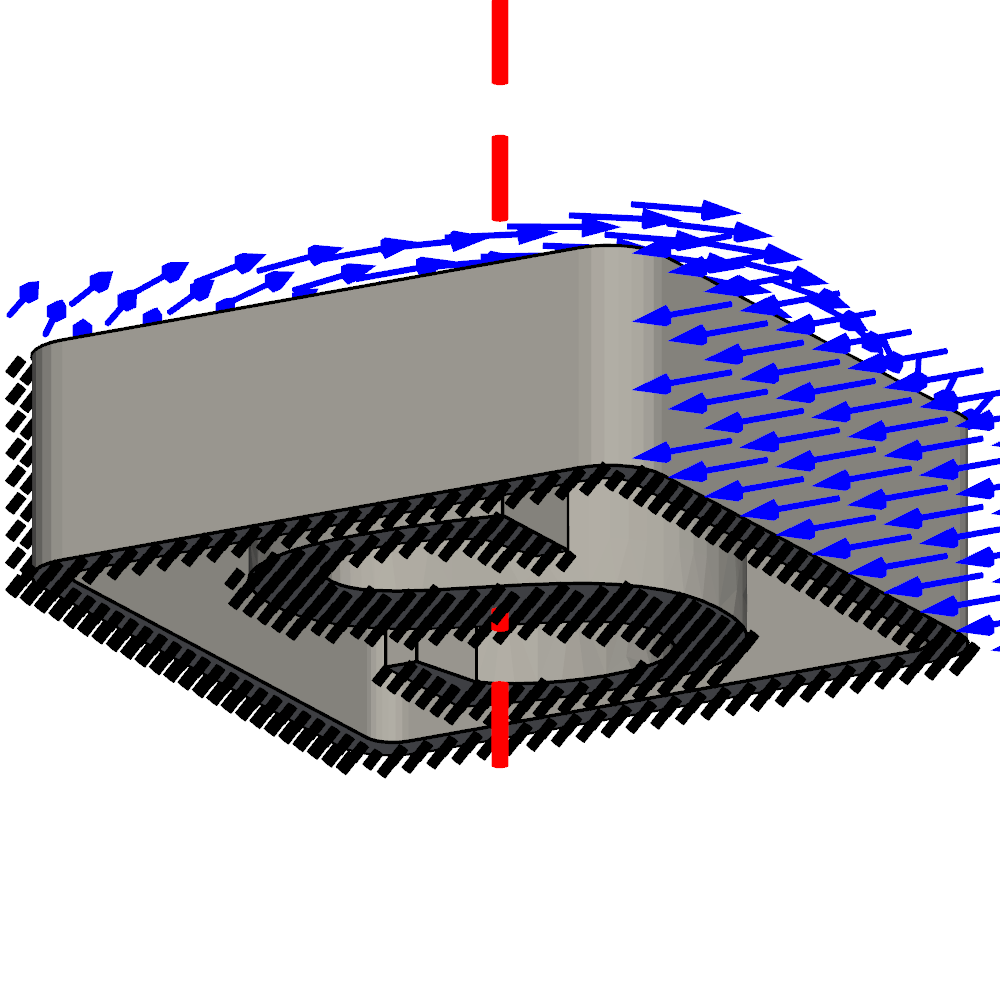}
    \end{subfigure}

    \vspace{0.5em}

    \begin{subfigure}[t]{0.19\textwidth}
        \centering
        \includegraphics[width=\linewidth]{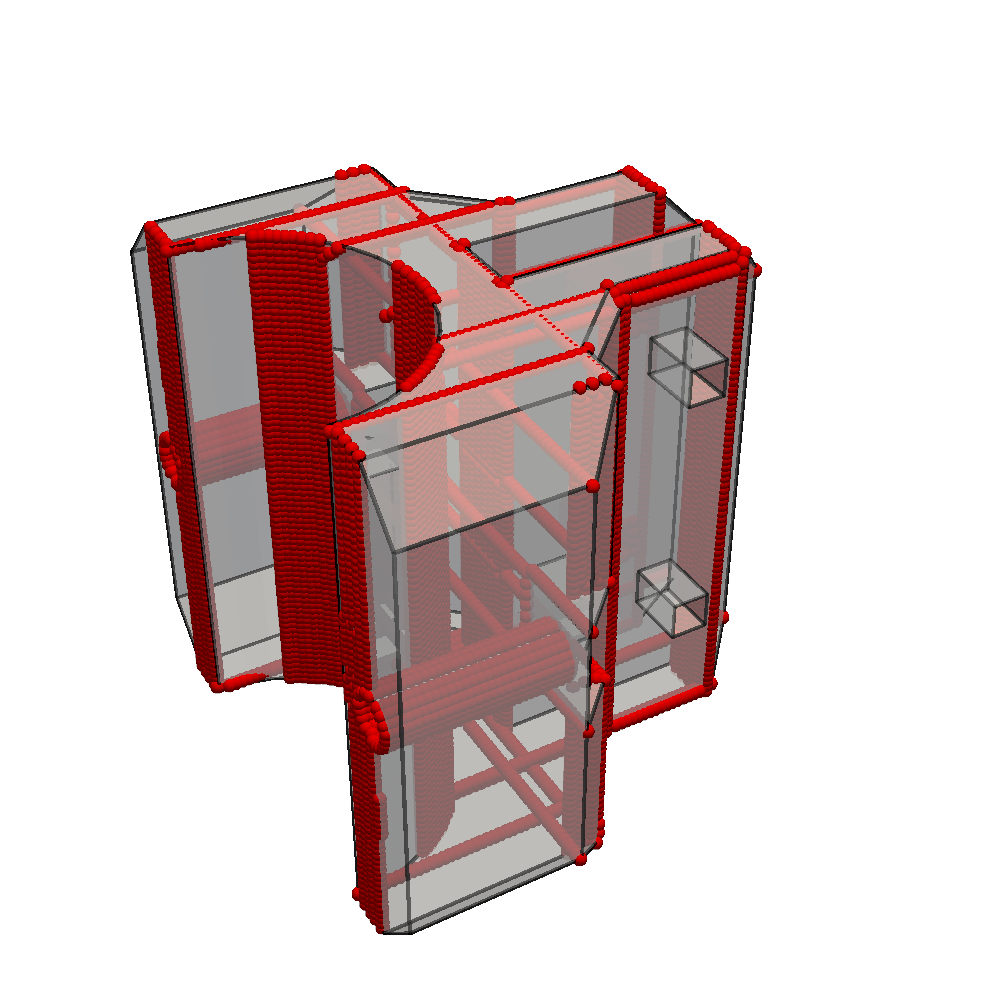}
    \end{subfigure}\hfill
    \begin{subfigure}[t]{0.19\textwidth}
        \centering
        \includegraphics[width=\linewidth]{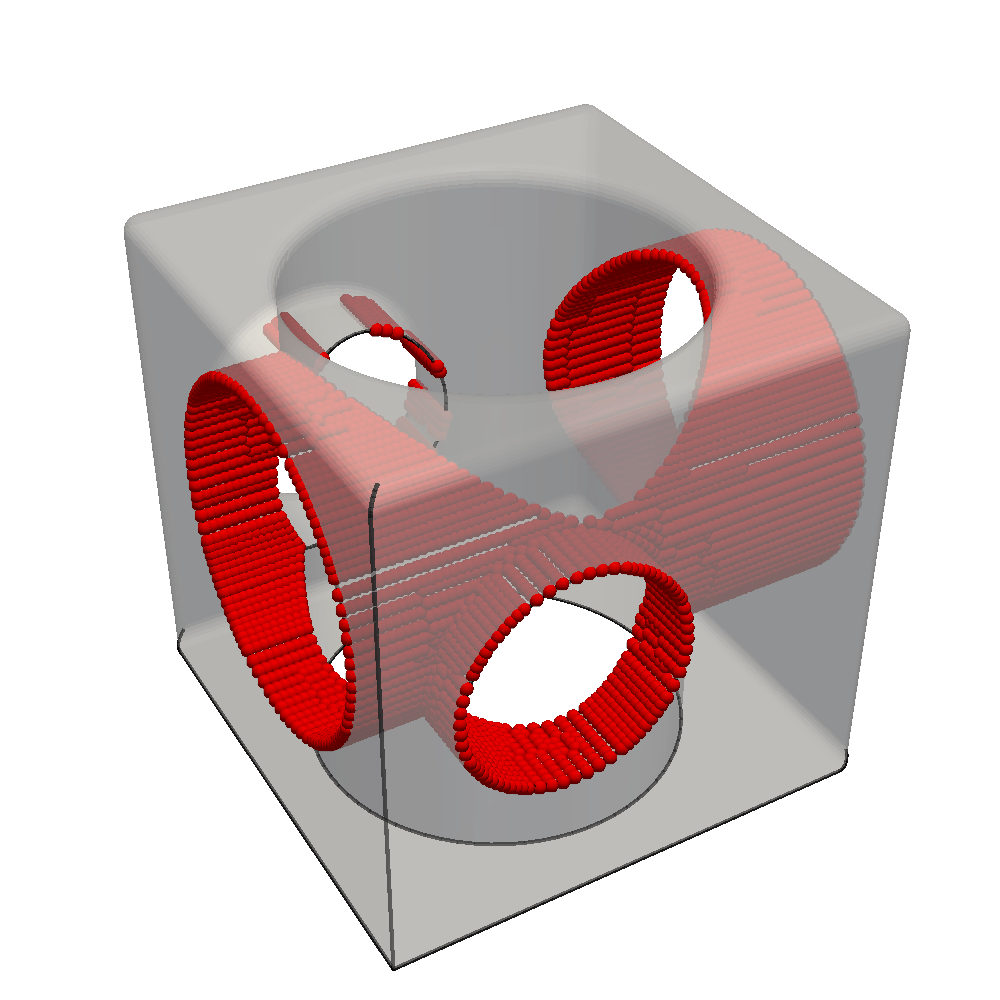}
    \end{subfigure}\hfill
    \begin{subfigure}[t]{0.19\textwidth}
        \centering
        \includegraphics[width=\linewidth]{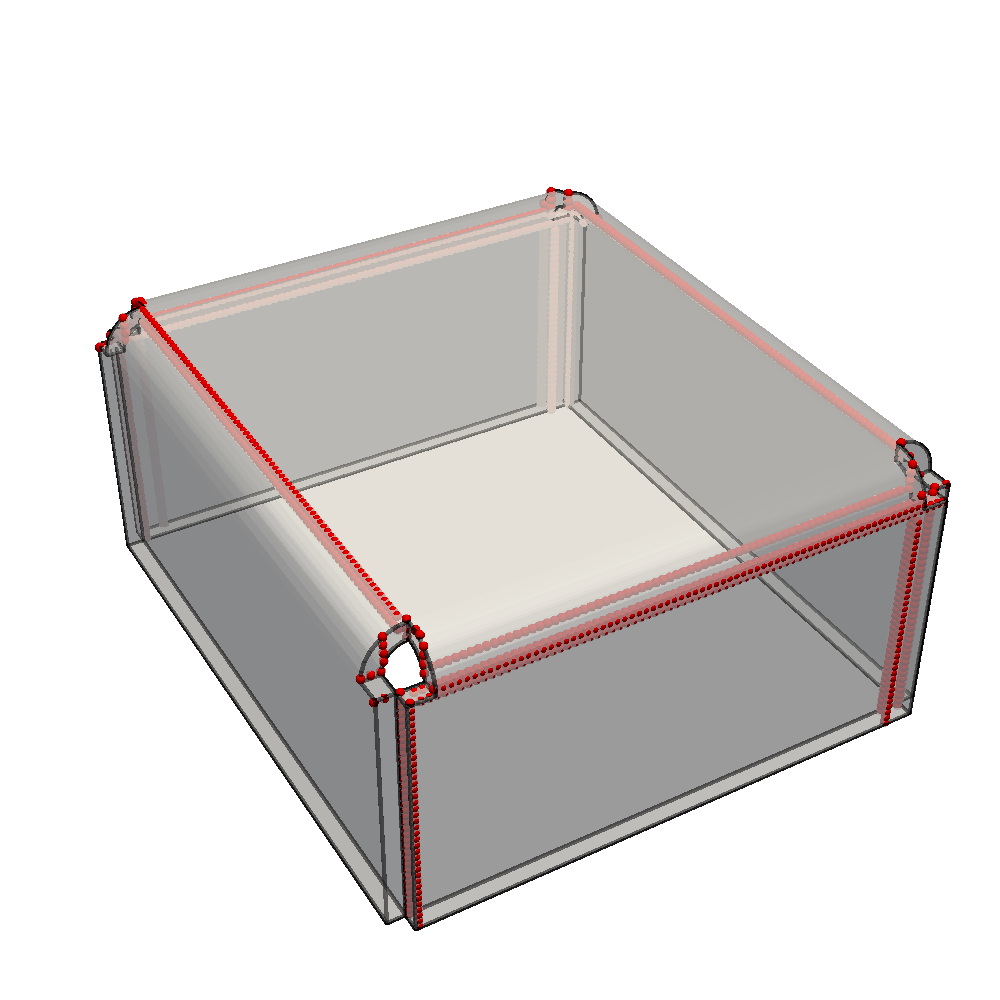}
    \end{subfigure}\hfill
    \begin{subfigure}[t]{0.19\textwidth}
        \centering
        \includegraphics[width=\linewidth]{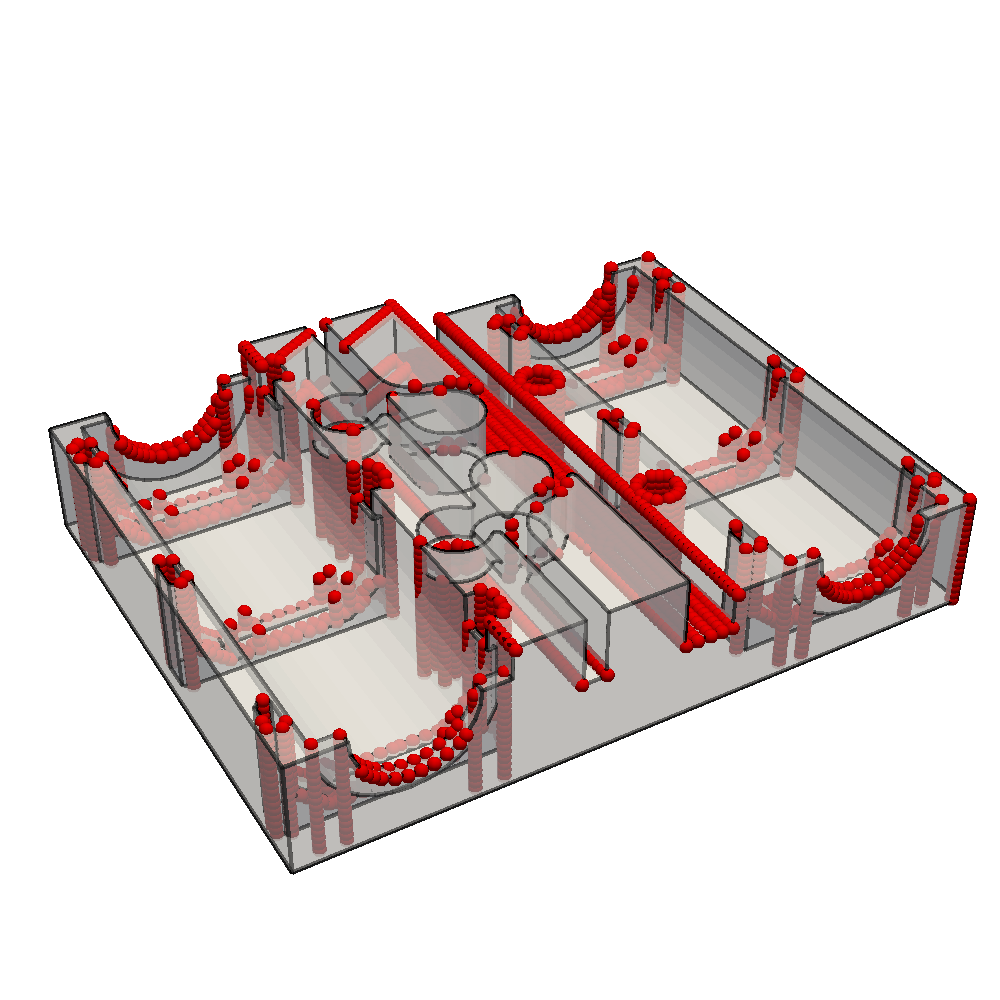}
    \end{subfigure}\hfill
    \begin{subfigure}[t]{0.19\textwidth}
        \centering
        \includegraphics[width=\linewidth]{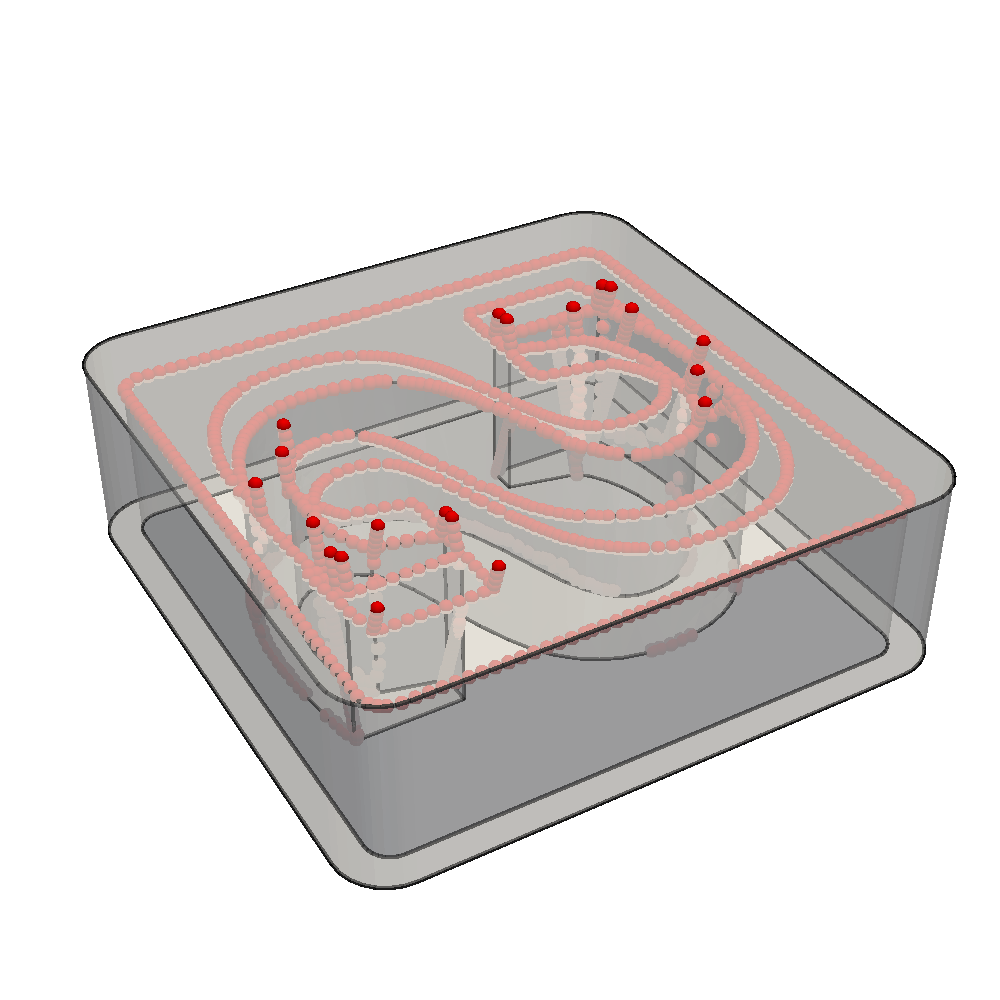}
    \end{subfigure}

    \vspace{0.5em}

    \begin{subfigure}[t]{0.19\textwidth}
        \centering
        \includegraphics[width=\linewidth]{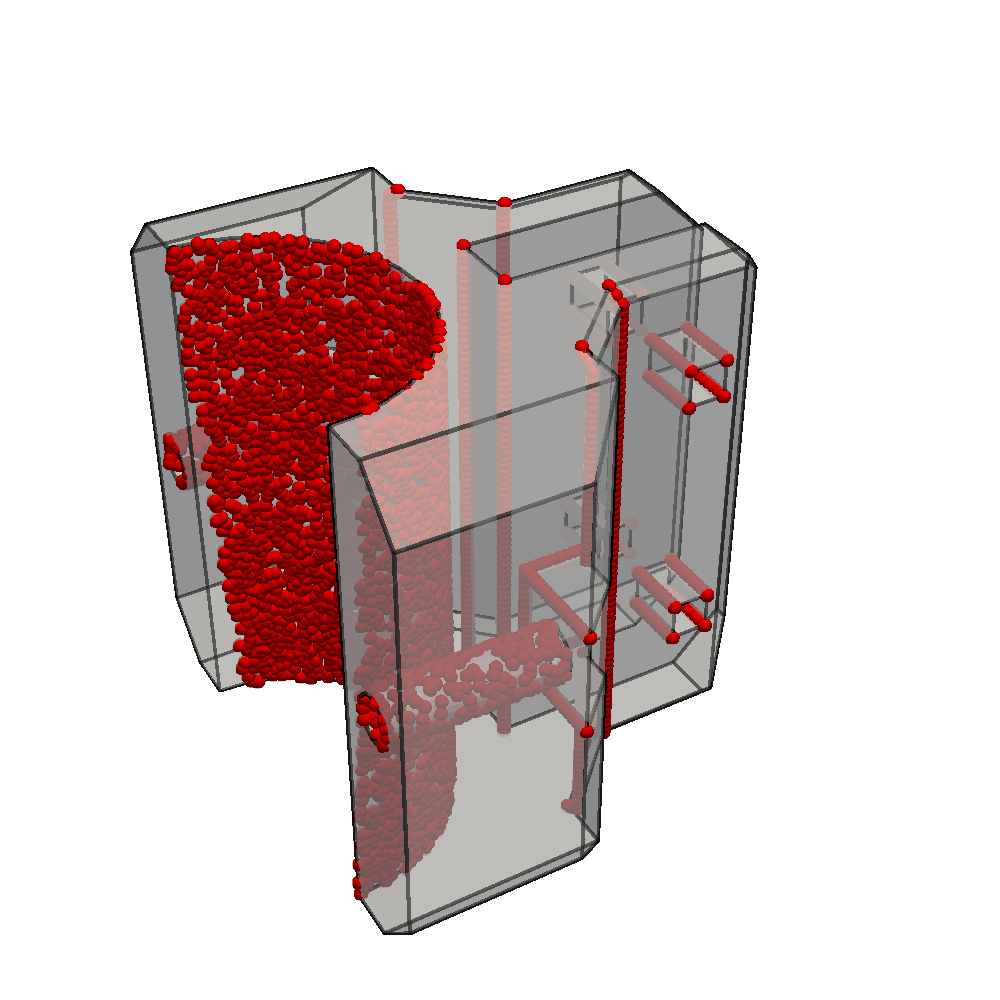}
    \end{subfigure}\hfill
    \begin{subfigure}[t]{0.19\textwidth}
        \centering
        \includegraphics[width=\linewidth]{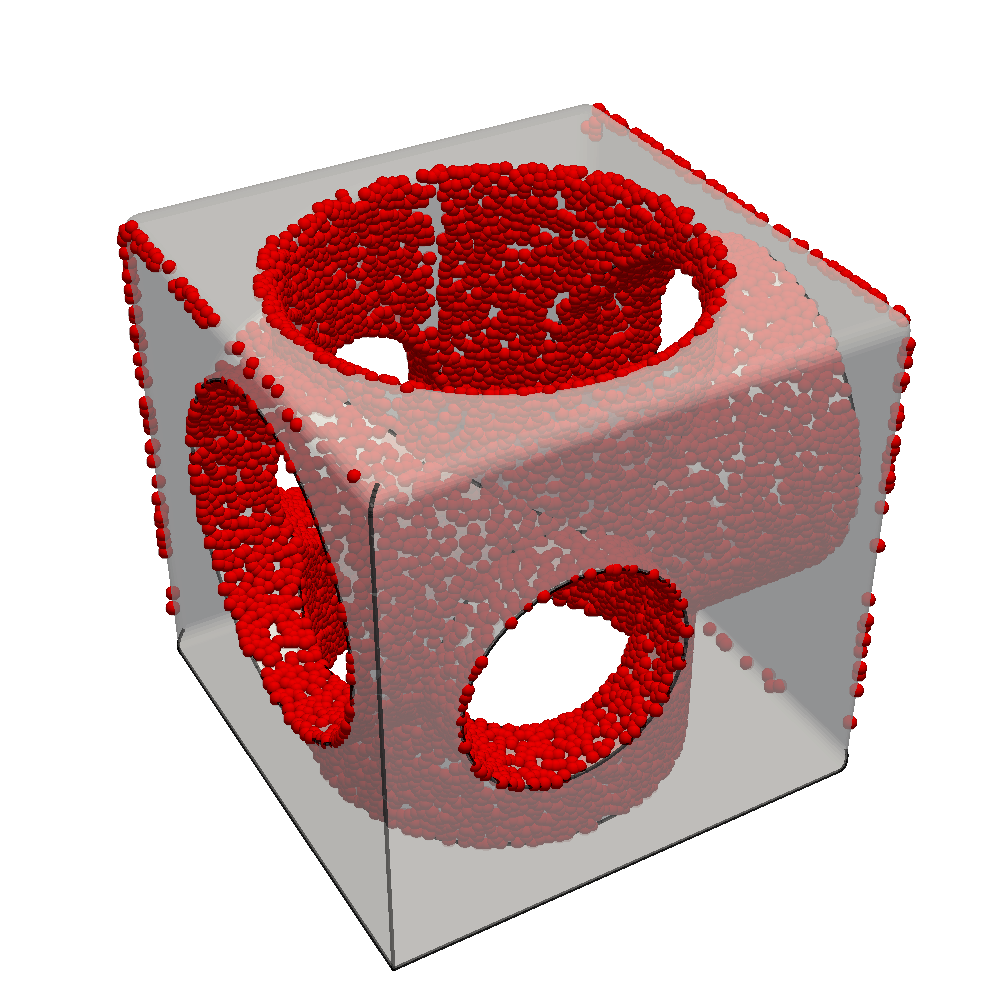}
    \end{subfigure}\hfill
    \begin{subfigure}[t]{0.19\textwidth}
        \centering
        \includegraphics[width=\linewidth]{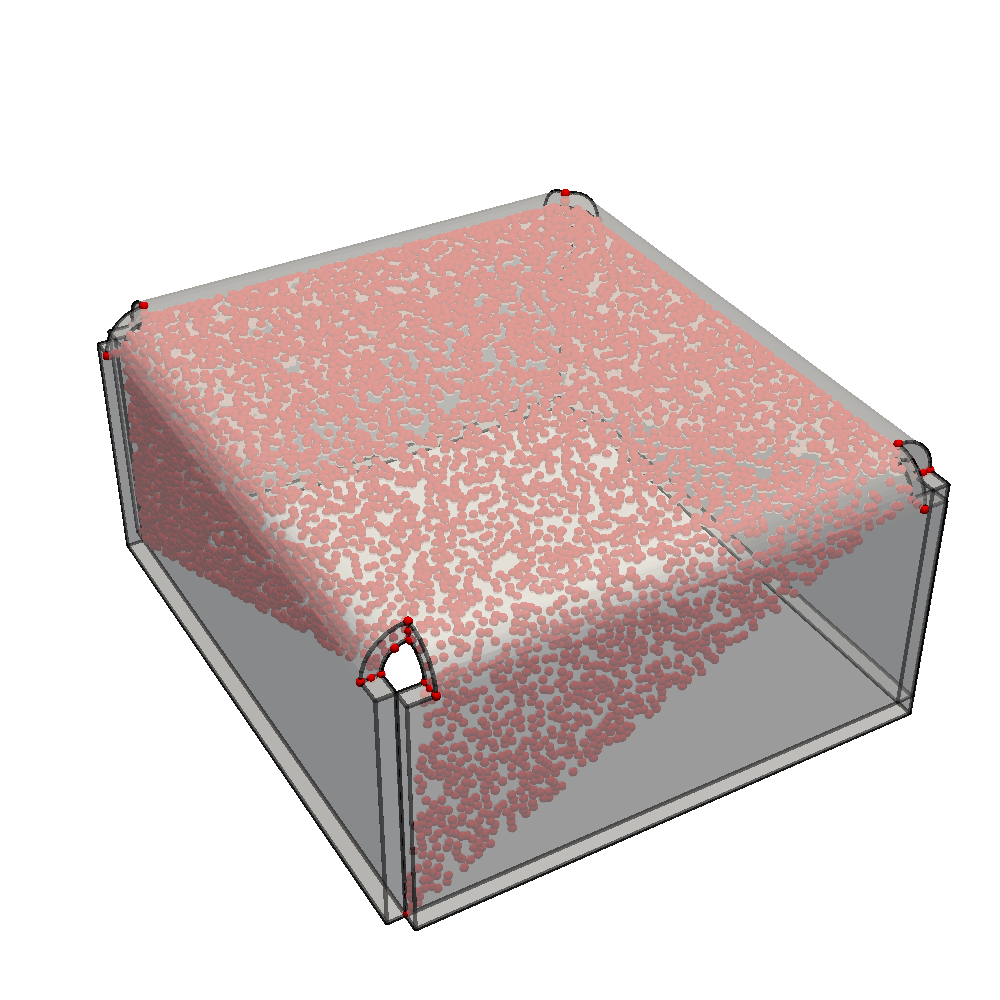}
    \end{subfigure}\hfill
    \begin{subfigure}[t]{0.19\textwidth}
        \centering
        \includegraphics[width=\linewidth]{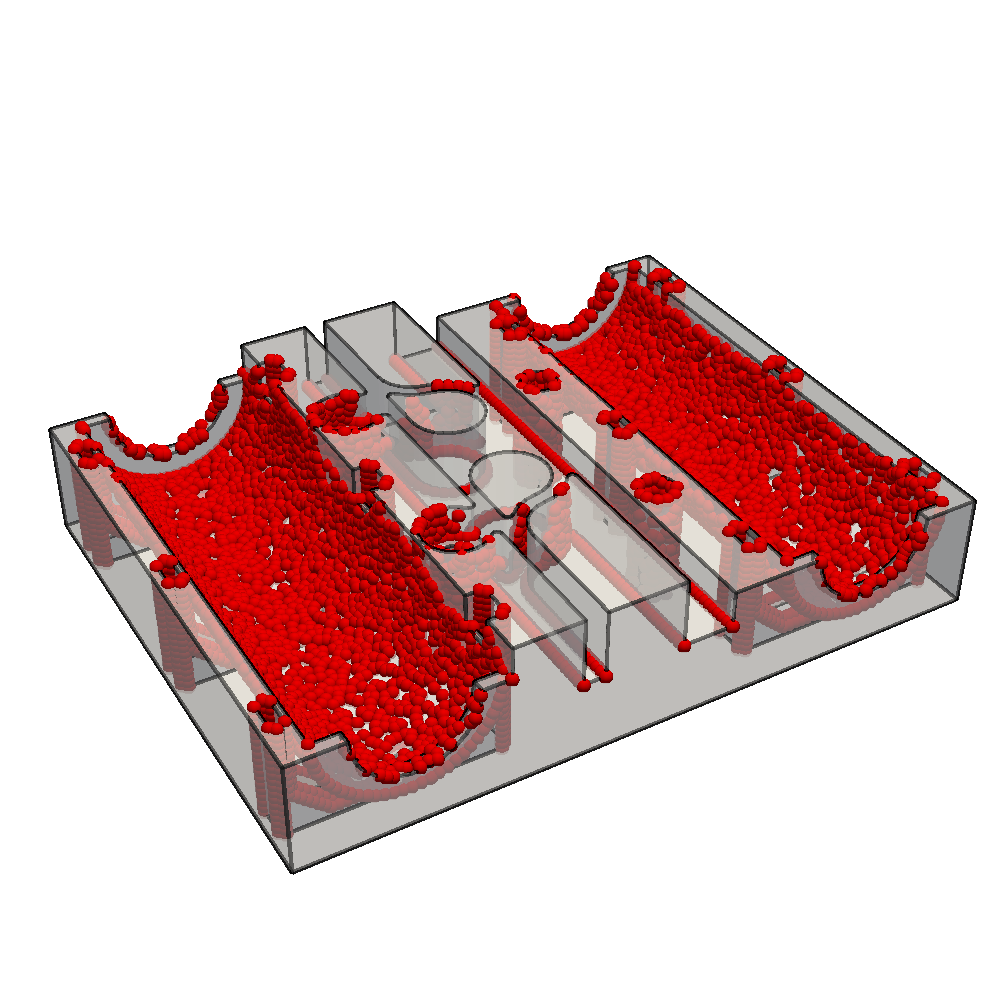}
    \end{subfigure}\hfill
    \begin{subfigure}[t]{0.19\textwidth}
        \centering
        \includegraphics[width=\linewidth]{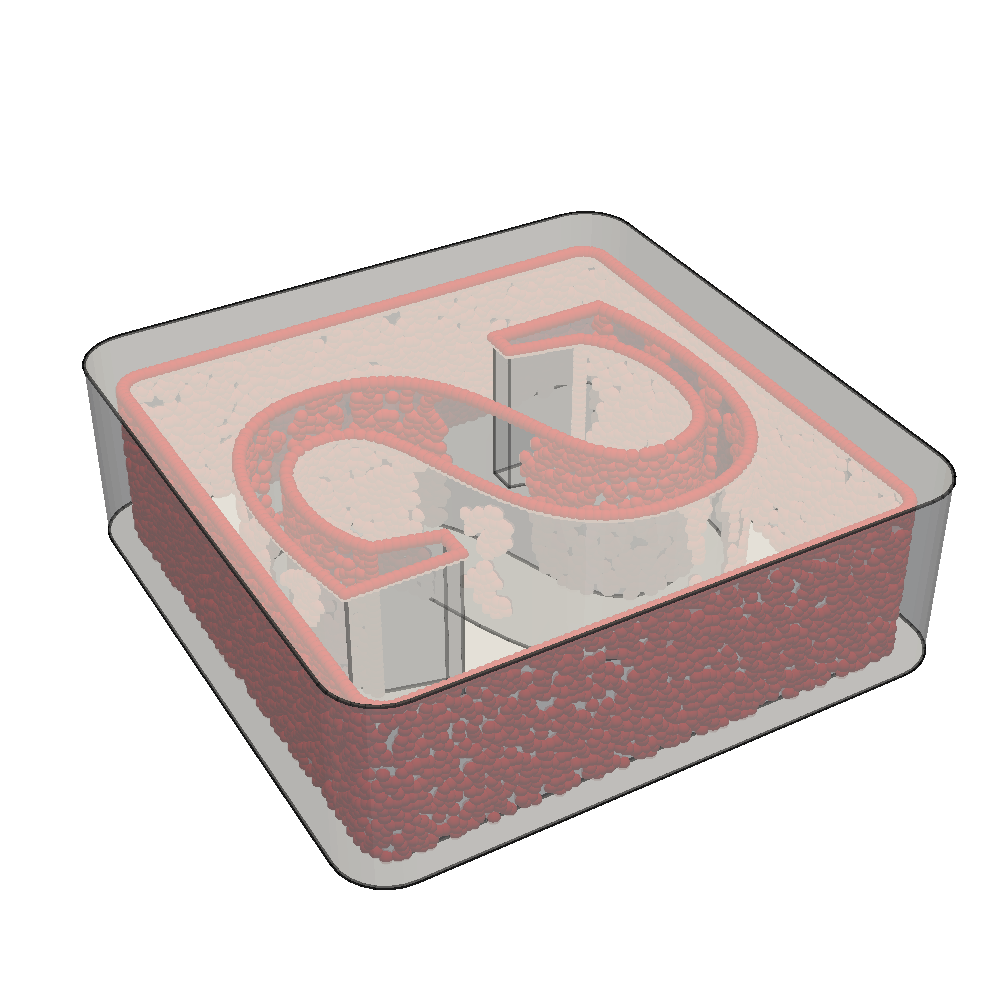}
    \end{subfigure}

    \vspace{0.5em}

    \begin{subfigure}[t]{0.19\textwidth}
        \centering
        \includegraphics[width=\linewidth]{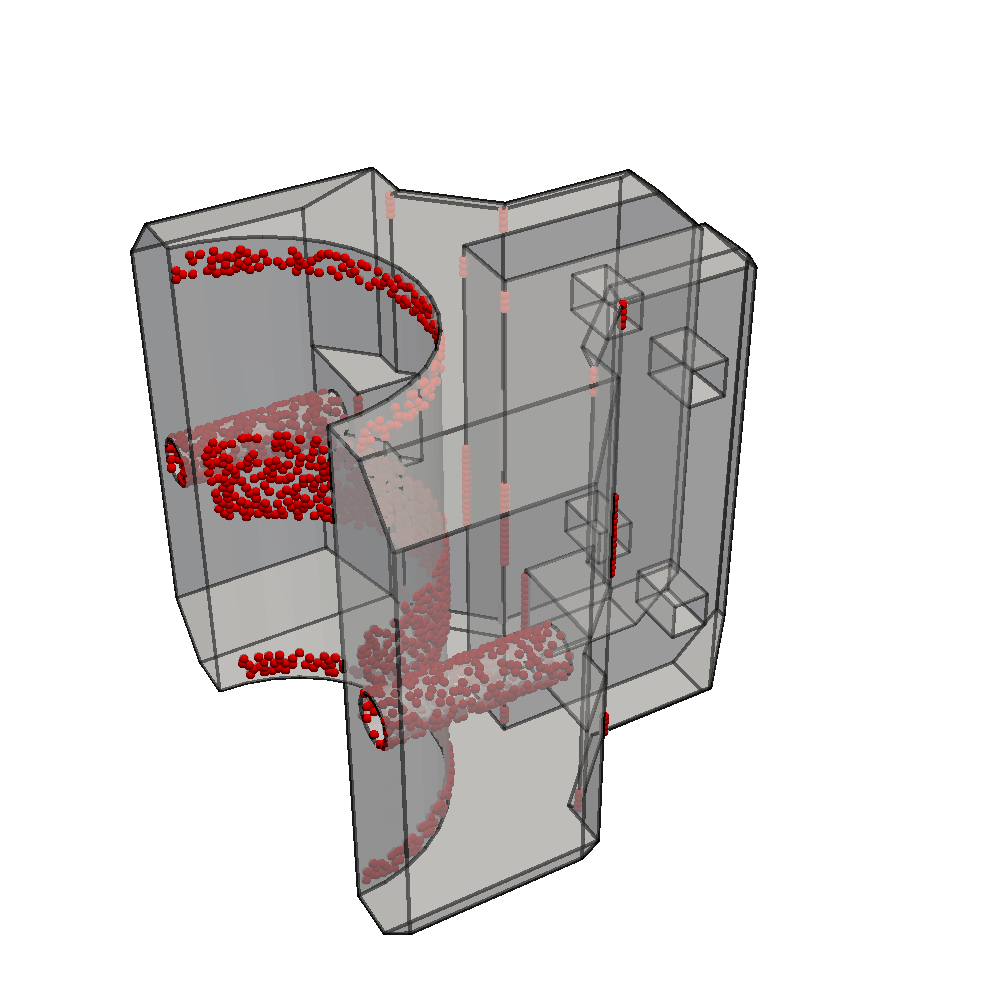}
    \end{subfigure}\hfill
    \begin{subfigure}[t]{0.19\textwidth}
        \centering
        \includegraphics[width=\linewidth]{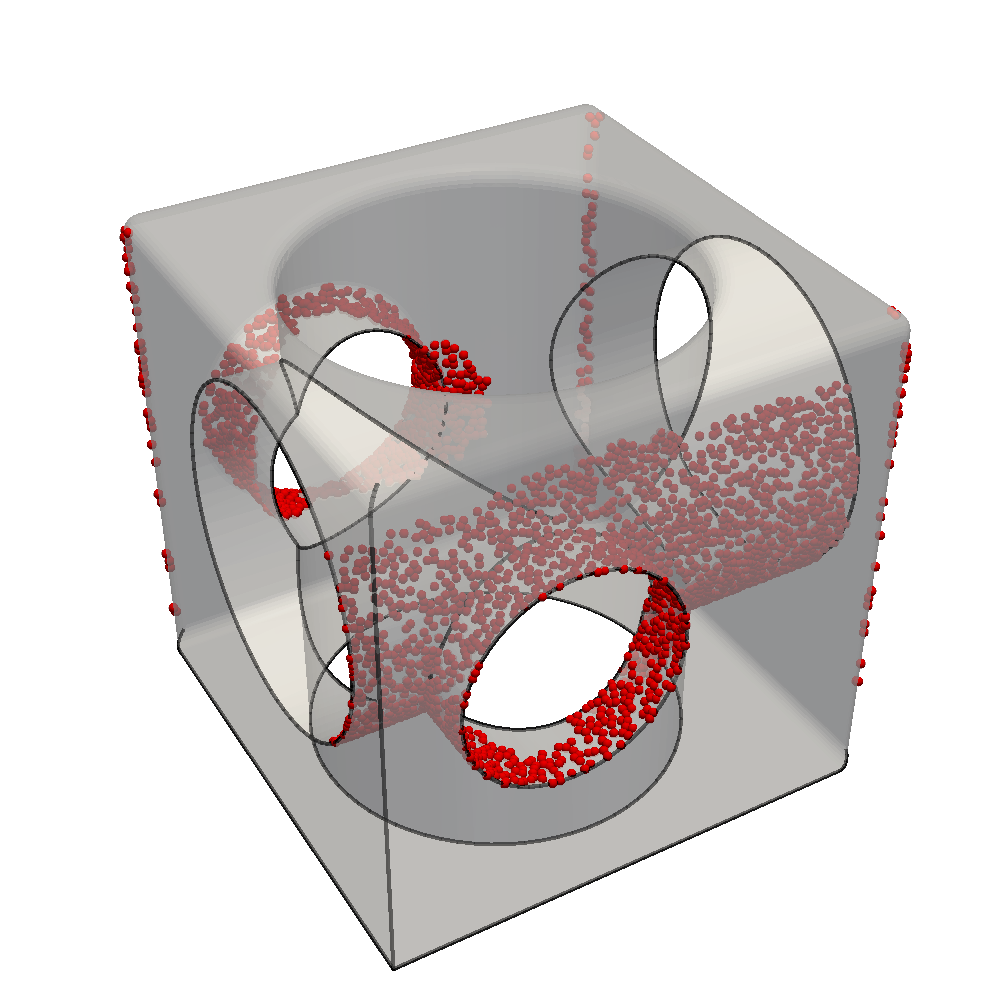}
    \end{subfigure}\hfill
    \begin{subfigure}[t]{0.19\textwidth}
        \centering
        \includegraphics[width=\linewidth]{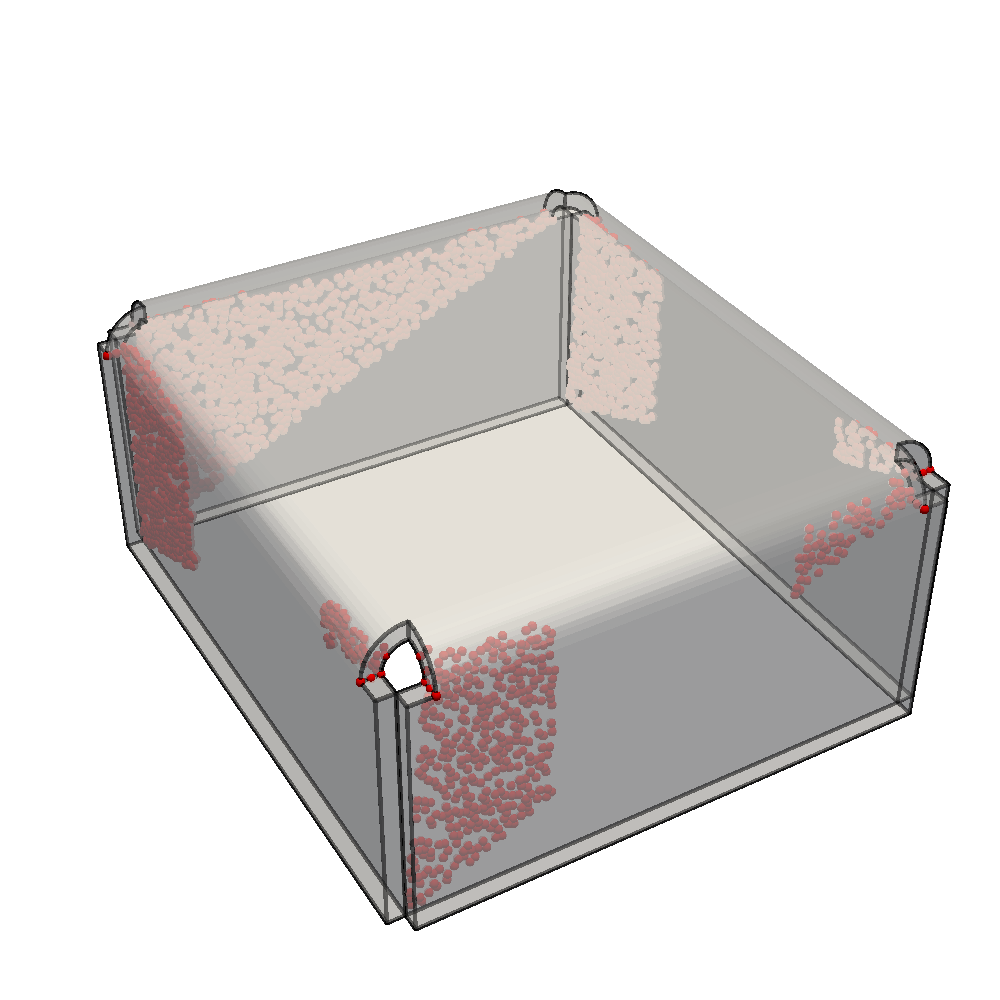}
    \end{subfigure}\hfill
    \begin{subfigure}[t]{0.19\textwidth}
        \centering
        \includegraphics[width=\linewidth]{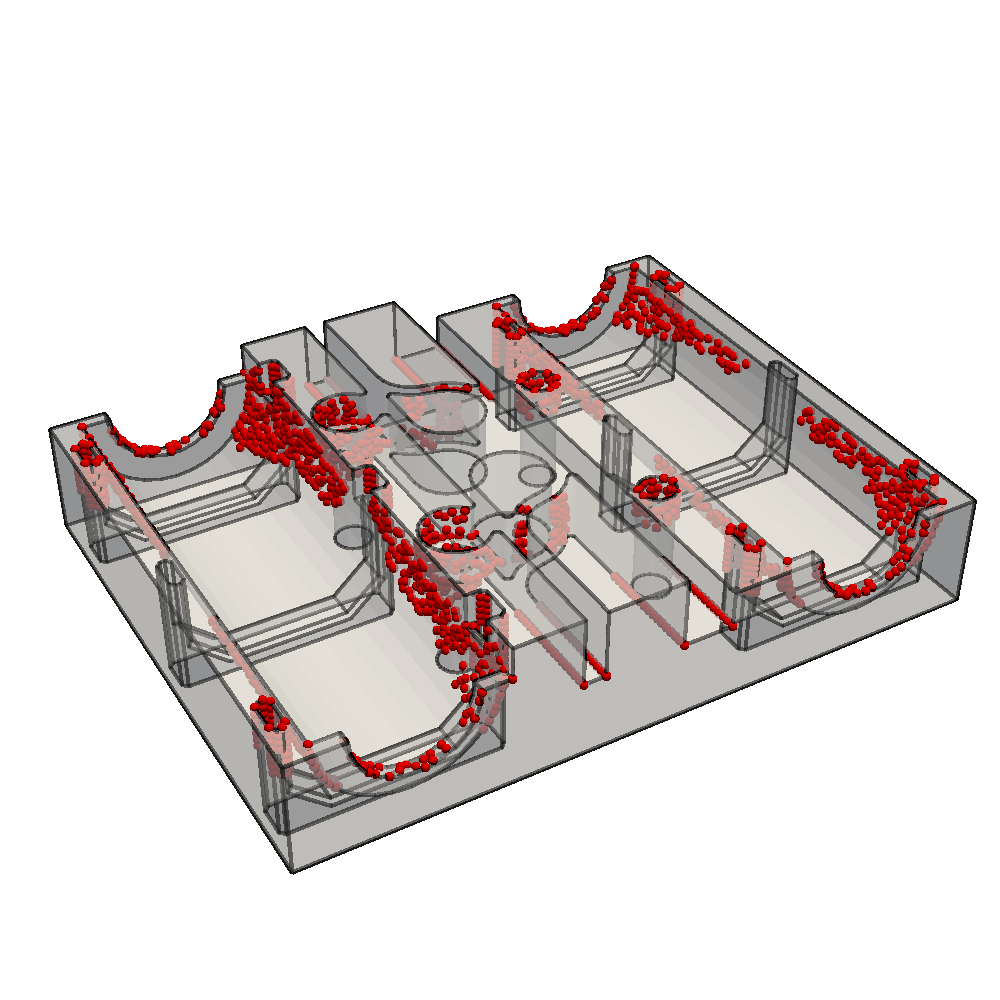}
    \end{subfigure}\hfill
    \begin{subfigure}[t]{0.19\textwidth}
        \centering
        \includegraphics[width=\linewidth]{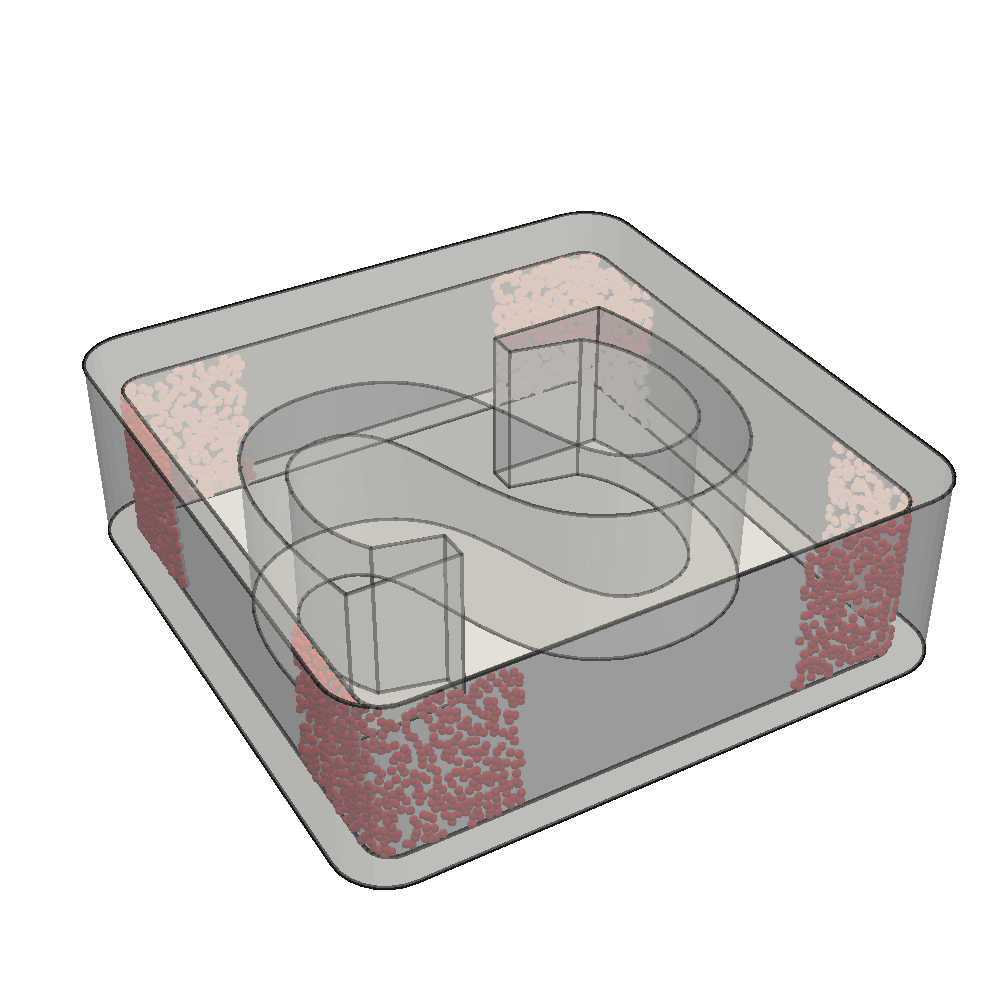}
    \end{subfigure}

    \vspace{0.5em}

    \begin{subfigure}[t]{0.19\textwidth}
        \centering
        \includegraphics[width=\linewidth]{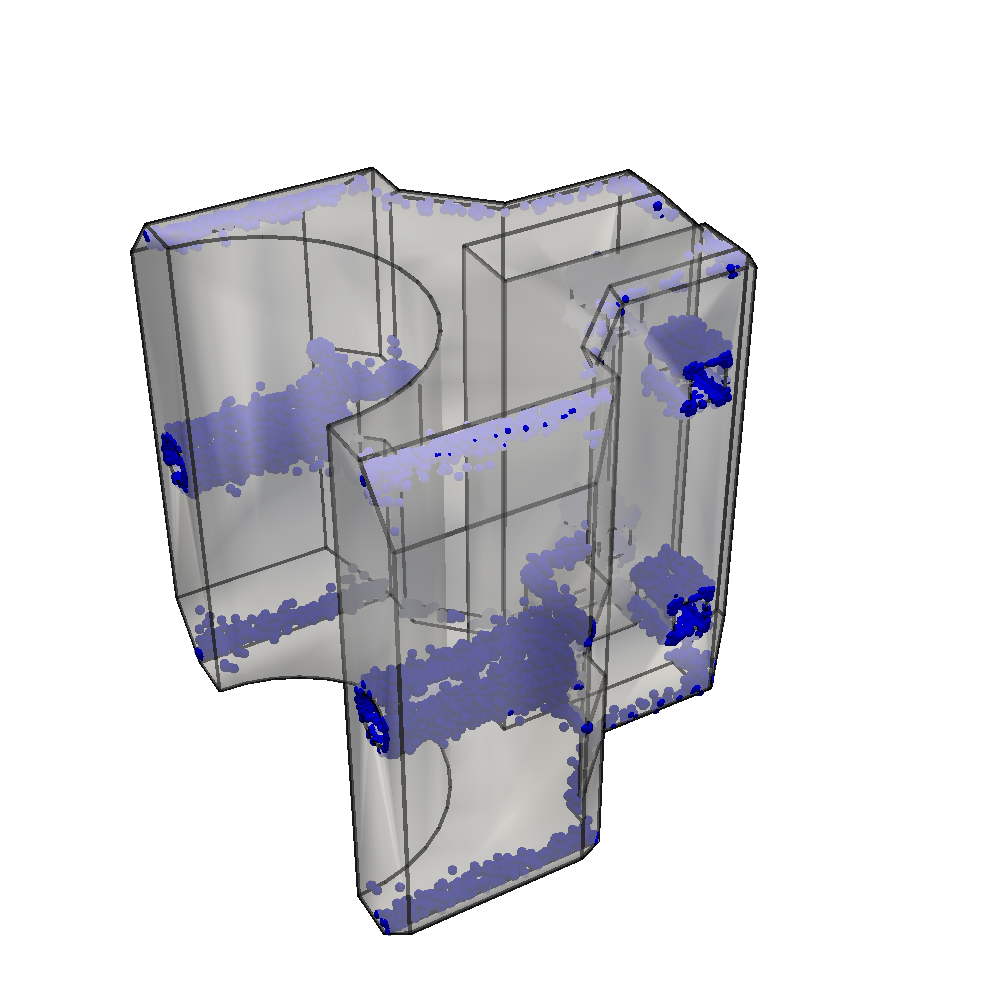}
        \caption*{Compression}
    \end{subfigure}\hfill
    \begin{subfigure}[t]{0.19\textwidth}
        \centering
        \includegraphics[width=\linewidth]{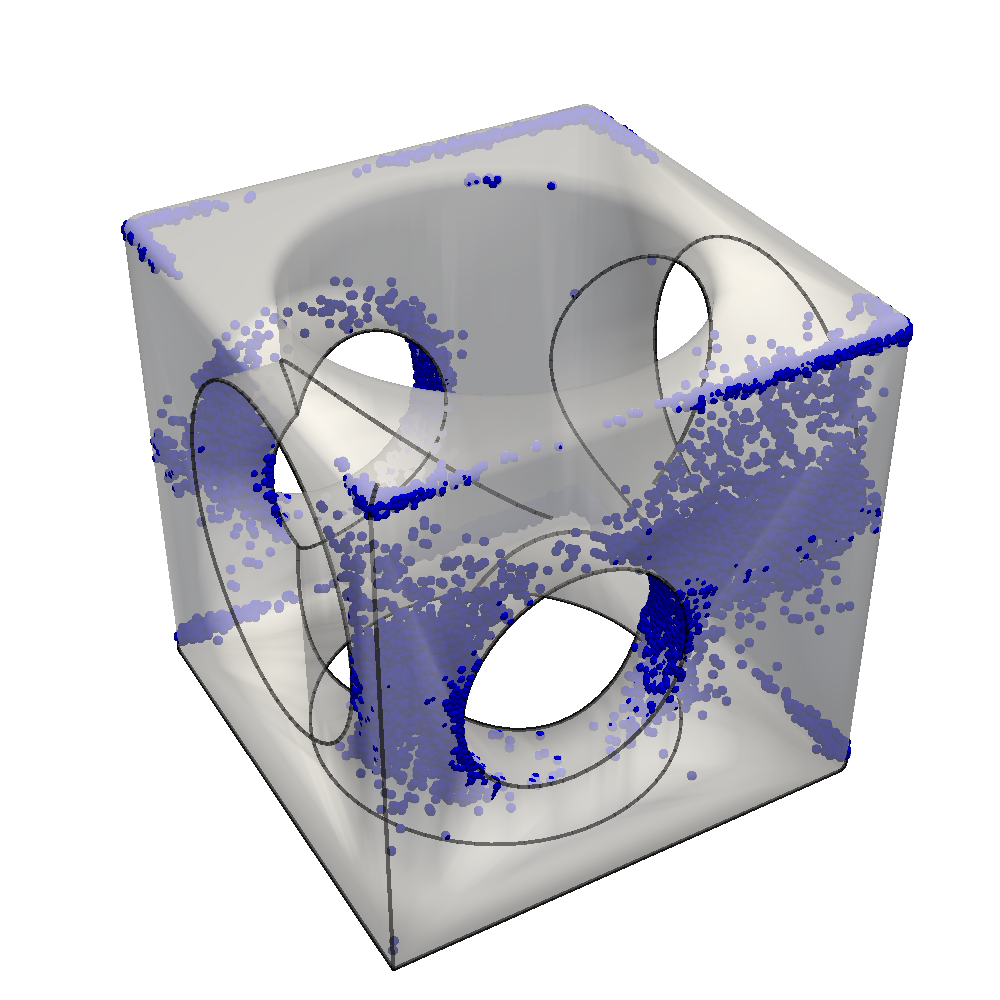}
        \caption*{Bending}
    \end{subfigure}\hfill
    \begin{subfigure}[t]{0.19\textwidth}
        \centering
        \includegraphics[width=\linewidth]{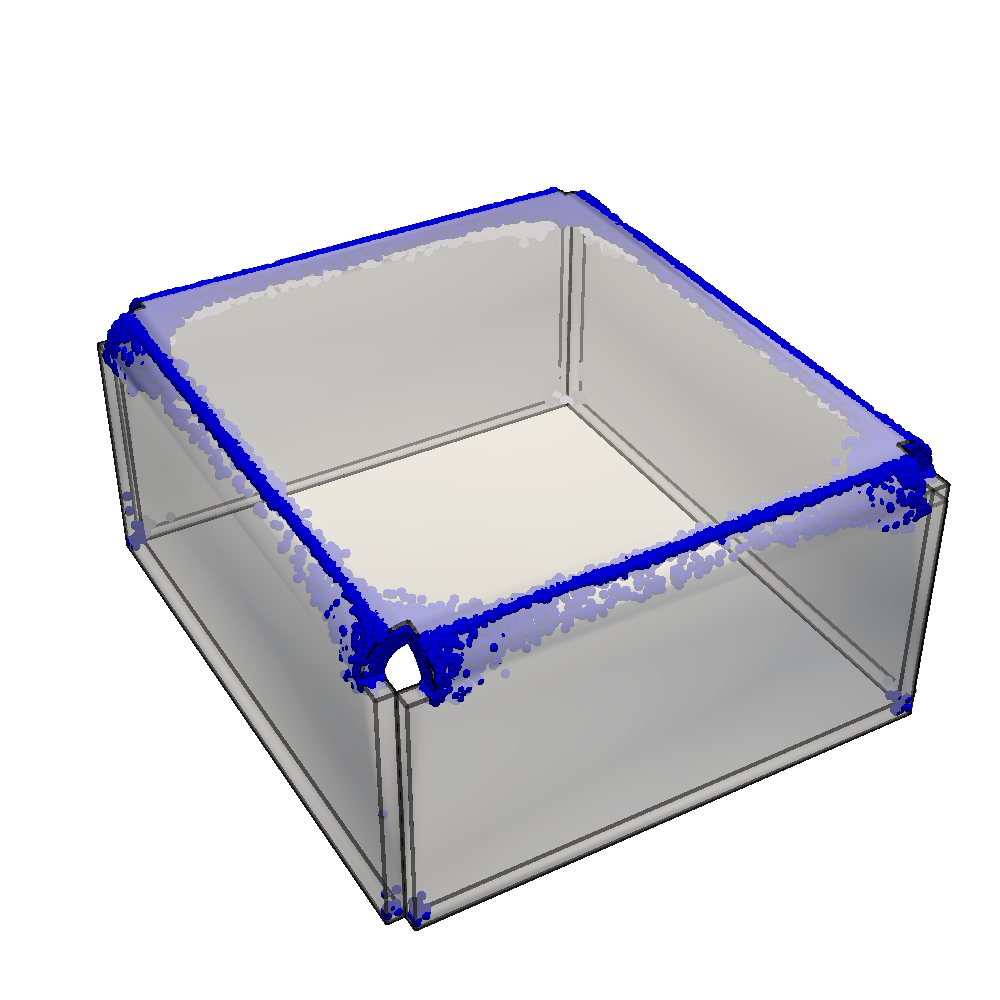}
        \caption*{Torsion}
    \end{subfigure}\hfill
    \begin{subfigure}[t]{0.19\textwidth}
        \centering
        \includegraphics[width=\linewidth]{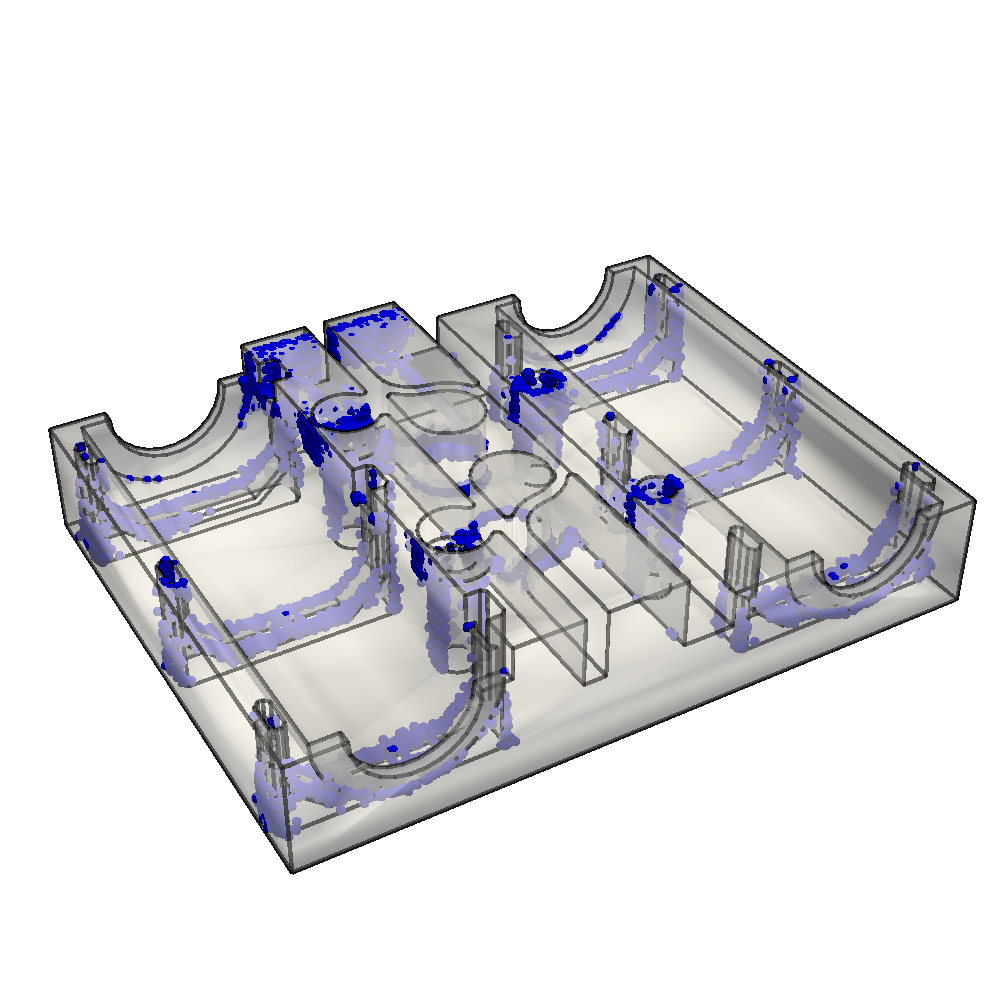}
        \caption*{Bending-Comp.}
    \end{subfigure}\hfill
    \begin{subfigure}[t]{0.19\textwidth}
        \centering
        \includegraphics[width=\linewidth]{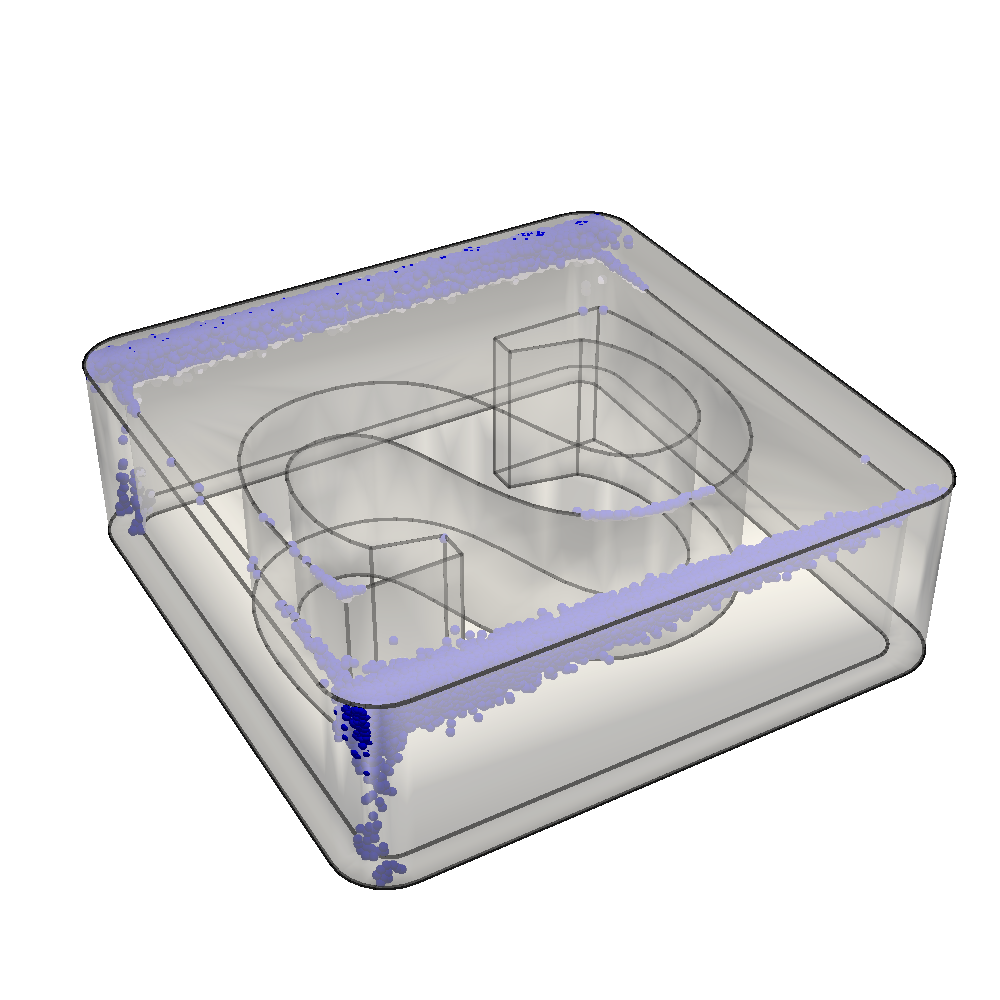}
        \caption*{Torsion-Comp.}
    \end{subfigure}

    \caption{Qualitative visual comparison across experimental loading cases (row 1), predicted refinement points via \textbf{GReFEM} (row 2), blind geometric heuristic points (row 3), load-informed heuristic points (row 4), and ground-truth ZZ stress indicator fields (row 5). All anchor rows show the native anchor sets before matching to \textbf{GReFEM}'s anchor count. The columns represent different object geometries and loading scenarios. All results were generated using a maximum of five \textbf{GReFEM} views selected via the \textit{orthoViews} module using \textbf{GReFEM}-gemini (\textit{Load+Features}). As is apparent from the figures, the blind geometric heuristic spreads refinement points over all concave edges and surfaces that could be stress critical, whereas \textbf{GReFEM} concentrates them on the features activated by the given load, and the load-informed heuristic keeps the subset of the blind candidates with the highest closed-form nominal stress. Quantitatively (Sec.~\ref{sec:downstream}), this selection lowers critical-region peak-stress error at a matched budget, even though anchor-level overlap with the reference ZZ field is not higher than the heuristic's.}
    \label{fig:visual_examples_4x5}
\end{figure*}

\clearpage
\subsection{Prompts}
We prompt the MLLMs in \textbf{GReFEM} to return region proposals from gridded view images. 
We introduce the default prompt template for the two physics informed prompt types introduced in the paper with place holders for perspective views, gridded views and loading information followed by example outputs for each of the model compared in our study. For the \textit{Load+Features+Heuristics} prompt template we append "\{additional\_info\}" in the prompt, while for \textit{Load+Features} prompt we omit it, keeping all else same.

\begin{tcolorbox}[
  title={Region Localization Prompt Template - \textit{Load+Features+Heuristics}/\textit{Load+Features}},
  colback=gray!5,
  colframe=blue,
  fonttitle=\bfseries,
  breakable
]
\textbf{System Prompt:}
\begin{verbatim}
You are an expert in computational geometry and finite element meshing.    
\end{verbatim}

\vspace{0.5em}
\textbf{User Prompt:}
\begin{verbatim}
You are given multiple images of a single CAD part along with their 
camera angles marked in (e: elevation, a: azimuth) format. 
These angles symbolize the location of the camera with respect to the 
objects' center in spherical coordinates.
The first {num_perspective_views} images show general 3D views of the 
object (not gridded) with camera situated at various angles respectively: 
{angles_list_perspective_views}.
The remaining {num_gridded_views} images show orthographic views with 
camera positions in the list respectively 
(e: elevation, a: azimuth): {angles_list_gridded_views}.
Each orthographic view has a visible grid with numbered cells.
{intro_loading[loading_case]}
The object is completely fixed on surfaces as indicated by the 
Protruding dashed black lines
Task:
For EACH gridded orthographic image, identify all grid cells that 
contain inside them these features that will be stress critical for the 
given loading condition:
1) Internal corner edges (concave edges that fold into the object and go 
into the view plane).
2) Extruded Contour/Protruding Contour (extruded closed contour cut 
partially like a blind hole into the object or protruding out of the 
object like a boss)
3) Through Holes (any internal surface with hollow space inside all the 
way).
4) Concave Fillet Surfaces, or arches (Any concave curved surface)
I.C.E: Internal Corner Edge, E.C/P.C: Extruded Contour/Protruding 
Contour, T.H: Through Holes, F: Concave Fillet Surfaces or arches
Use the non-gridded {num_perspective_views} 3D views ONLY to understand 
the overall shape. The grid numbers should be marked only for the 
corresponding gridded orthographic images.
Base all cell predictions ONLY on what is visible in the corresponding 
gridded orthographic image.
For Hollow features mark the boundary cells that contain the hollow 
feature inside them.
Do NOT guess or infer cells that are not clearly visible in that image.
Output format (repeat for each gridded image):
***I.C.E: c1, c2, c3 ***
***E.C/P.C: c4, c5 ***
***T.H: c6, c7 ***
***F: c8, c9 ***
{additional_info[loading_case]}
Rules:
- List only cell numbers.
- If no cells apply, write: NONE.
- DO NOT Add explanations
\end{verbatim}
\end{tcolorbox}

\begin{tcolorbox}[
  title={Region Localization Prompt Template - Load-Only},
  colback=gray!5,
  colframe=blue,
  fonttitle=\bfseries,
  breakable
]
\textbf{System Prompt:}
\begin{verbatim}
You are an expert in computational geometry and finite element meshing.    
\end{verbatim}

\vspace{0.5em}
\textbf{User Prompt:}
\begin{verbatim}
You are given multiple images of a single CAD part along with their 
camera angles marked in (e: elevation, a: azimuth) format. These angles 
symbolize the location of the camera with respect to the objects' 
center in spherical coordinates.
The first {num_perspective_views} images show general 3D views of 
the object (not gridded) with camera situated at various angles 
respectively: {angles_list_perspective_views}.
The remaining {num_gridded_views} images show orthographic views with 
camera positions in the list respectively (e: elevation, a: azimuth): 
{angles_list_gridded_views}.
Each orthographic view has a visible grid with numbered cells.
{intro_loading[loading_case]}
Task:
For EACH gridded orthographic image, identify all grid cells that 
contain inside them stress critical features.
Use the non-gridded {num_perspective_views} 3D views ONLY to understand 
the overall shape. The grid numbers should be marked only for the 
corresponding gridded orthographic images.
Base all cell predictions ONLY on what is visible in the corresponding 
gridded orthographic image.
Do NOT guess or infer cells that are not clearly visible in that image.
Output format (repeat for each gridded image):
***Cells: c1, c2, c3 ***
Rules:
- List only cell numbers.
- If no cells apply, write: NONE.
- DO NOT Add explanations
\end{verbatim}
\end{tcolorbox}

\begin{tcolorbox}[
  title={Loading Case Specific Information},
  colback=gray!5,
  colframe=blue,
  fonttitle=\bfseries,
  breakable
]
\textbf{Loading Introduction:}
\begin{verbatim}
Compression: A compressive displacement load acts on the object in 
the -y direction on its top surface as indicated by the blue arrows.    

Bending: A bending displacement load acts  on the object in the +y/-y 
direction on the top surface as indicated by the blue arrows about the 
horizontal axis indicated via dashed red line.

Torsion: A torsion displacement load acts on the object in the 
anti-clockwise direction along the x-z plane as indicated by the blue 
curved arrows about the vertical axis of loading passing through the 
center of the top surface as indicated by the dashed red line.

Bending-Compression: A bending displacement load acts on the object in 
the +y/-y direction on the top surface as indicated by the blue arrows 
about the horizontal axis indicated via dashed red line. Also a 
compressive load acts on the object in the +/- x direction as shown by
the blue arrows.

Torsion-Compression: A torsion displacement load acts on the object in 
the anti-clockwise direction along the x-z plane as indicated by the 
blue curved arrows about the vertical axis of loading passing through 
the center of the top surface as indicated by the dashed red line. Also 
the compressive load acts on the object in the +/- x direction as shown 
by the blue arrows.
\end{verbatim}

\textbf{Additional Information for Spotting Stress Critical Features:}
\begin{verbatim}
Compression: For the top and bottom views, only consider the BH 
features. All edges or surfaces along the loading direction are not 
stress critical as they are not perpendicular to the load and do not 
create stress concentrations. Neglect the features too far away 
if the load acts on a narrow part of the top surface passing through 
the center of the object as they are not stress critical.

Bending: For the top and bottom views, only consider the BH features. 
All edges or surfaces along the loading direction are not stress 
critical as they are not perpendicular to the load and do not create 
stress concentrations. Geometric features ICE, BH, TH too close to 
the axis about which load is applied are usually not stress critical, in
the case of bending loads features away from the axis are more likely to 
be stress critical.

Torsion: Geometric features ICE, BH, TH too close to the axis about 
which load is applied are usually not stress critical, in the case of 
torsion loads features away from the axis are more likely to be stress 
critical.

Bending-Compression: For each load (bending or compression), only 
consider the BH features when viewed parallel to the loading direction. 
All edges or surfaces along the loading direction for each load are 
not stress critical but perpendicular to the respective loads are.
Geometric features ICE, BH, TH too close to the axis about which 
bending load is applied are usually not stress critical, in the case 
of bending loads features away from the axis are more likely to be 
stress critical. Consider stress critical areas taking care of both the 
loads.   

Torsion-compression: For the torsion load geometric features ICE, BH, TH 
too close to the axis about which load is applied are usually not stress 
critical, in the case of torsion loads features away from the axis are 
more likely to be stress critical. Also, features along the loading 
direction of the compressive load are not stress critical as they 
do not create stress concentrations. Consider stress critical areas 
taking care of both the loads.
    
\end{verbatim}    
\end{tcolorbox}

Using the prompts described above, we query each of the MLLMs considered in the previous section and report their responses. For each query, we provide a perspective views of the geometry and the top 5 orthographic views selected by \textit{orthoViews}.

\begin{figure}[h]
\centering

\subfloat[Perspective View]{
  \includegraphics[width=0.3\textwidth]{figures/Load_Cases_Figures/00210097_view_e18_a30_compression.png}
}
\hfill
\subfloat[Gridded view 1]{
  \includegraphics[width=0.3\textwidth]{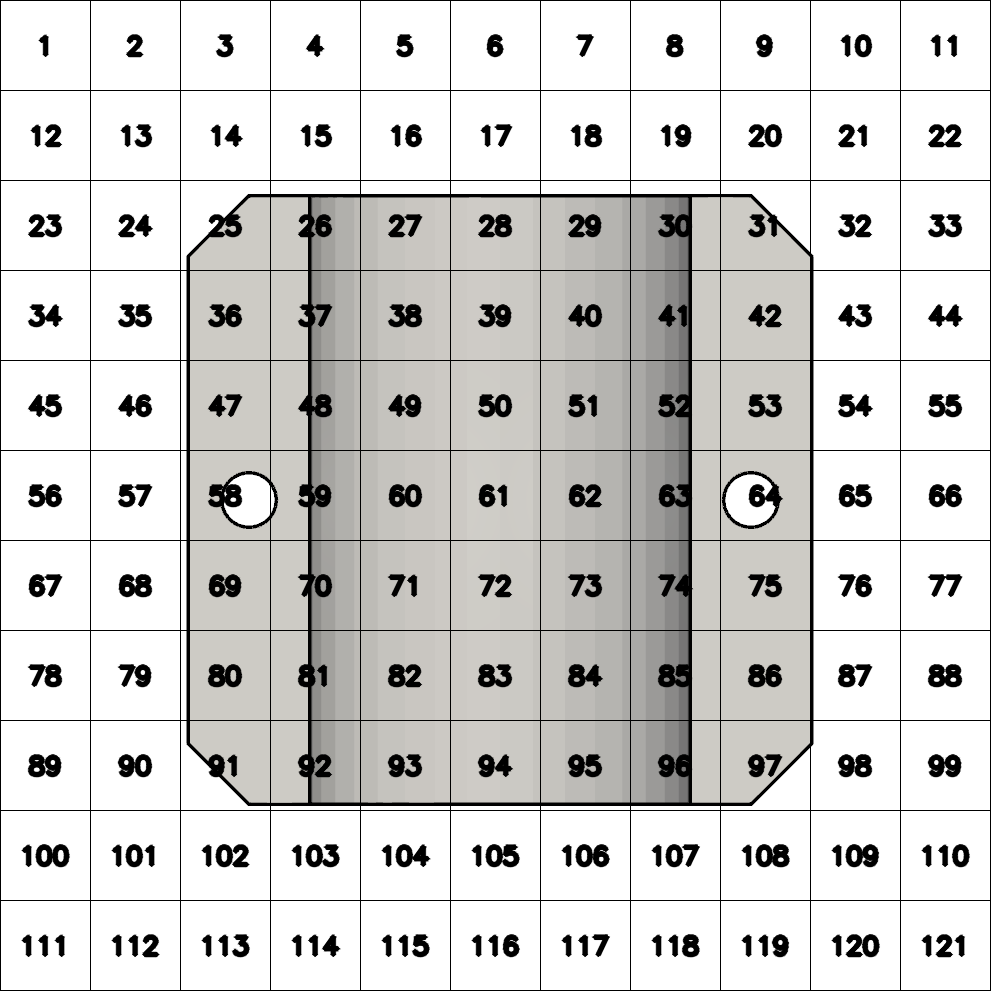}
}
\hfill
\subfloat[Gridded view 2]{
  \includegraphics[width=0.3\textwidth]{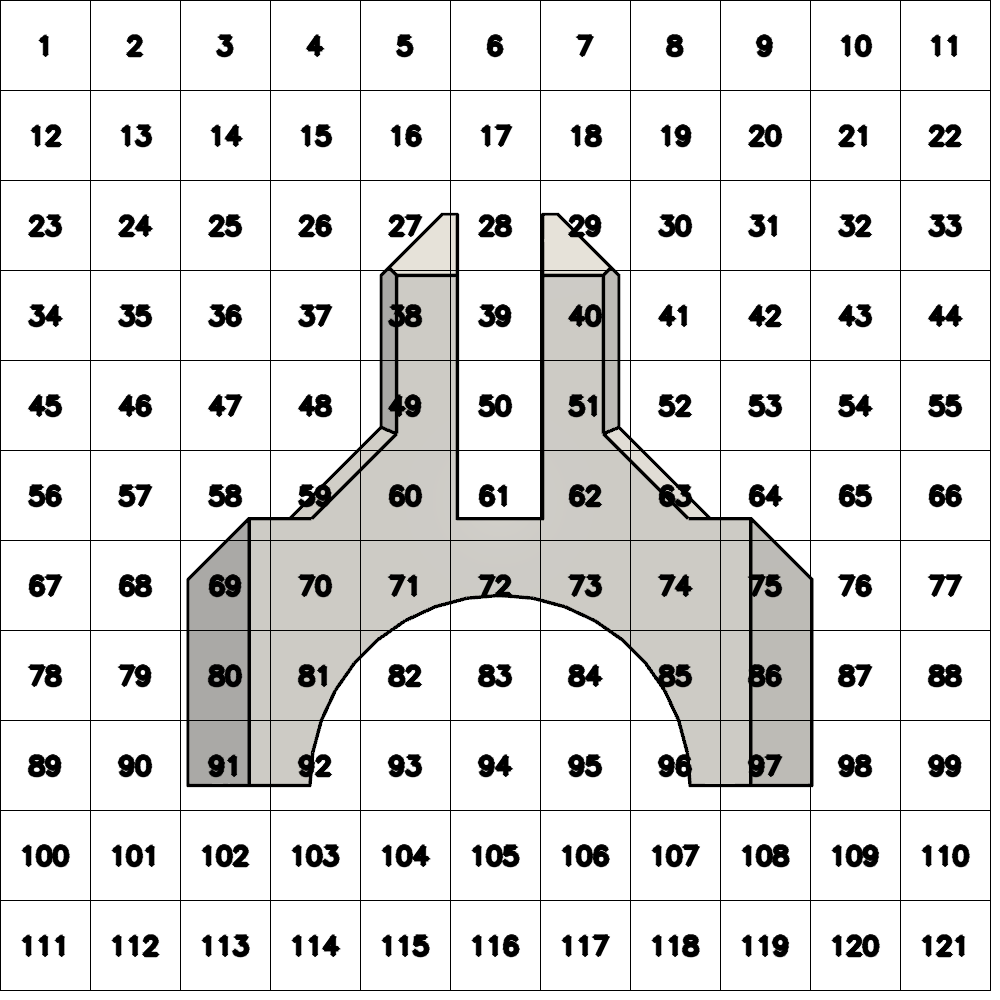}
}

\vspace{1mm}

\subfloat[Gridded view 3]{
  \includegraphics[width=0.3\textwidth]{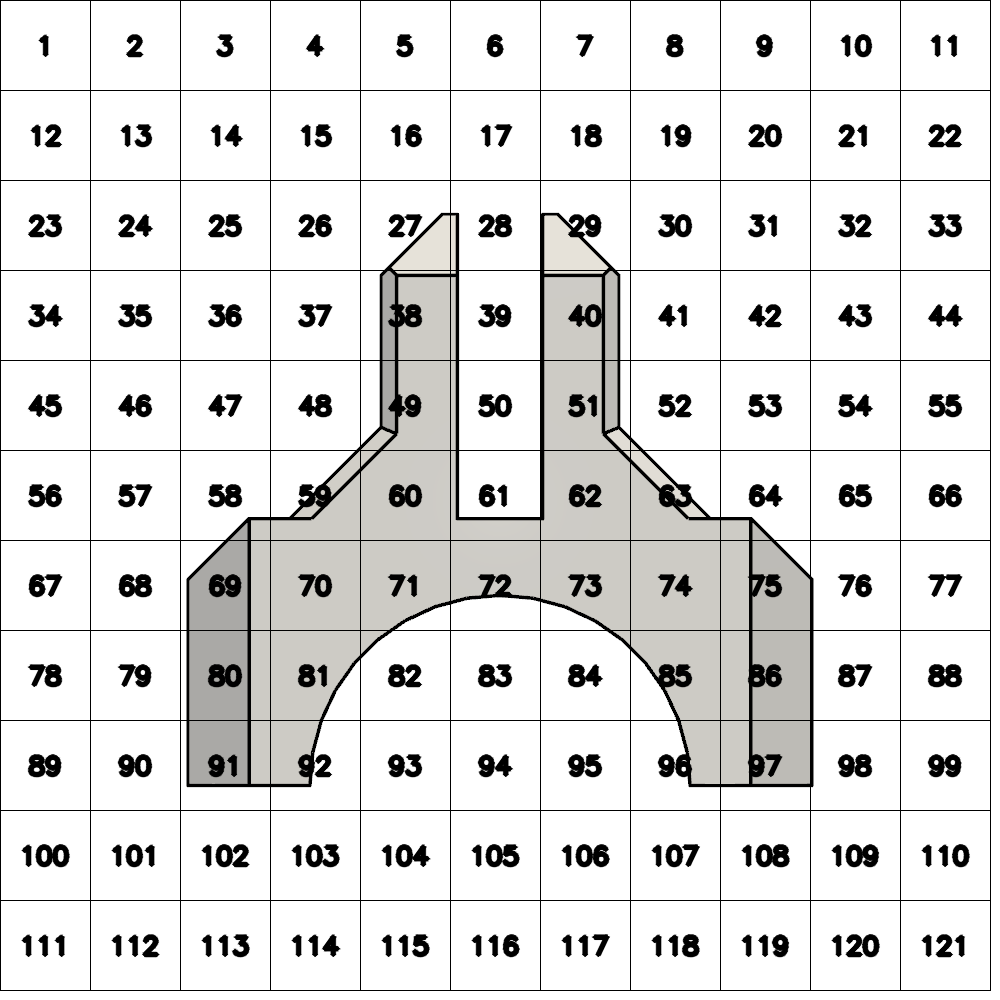}
}
\hfill
\subfloat[Gridded view 4]{
  \includegraphics[width=0.3\textwidth]{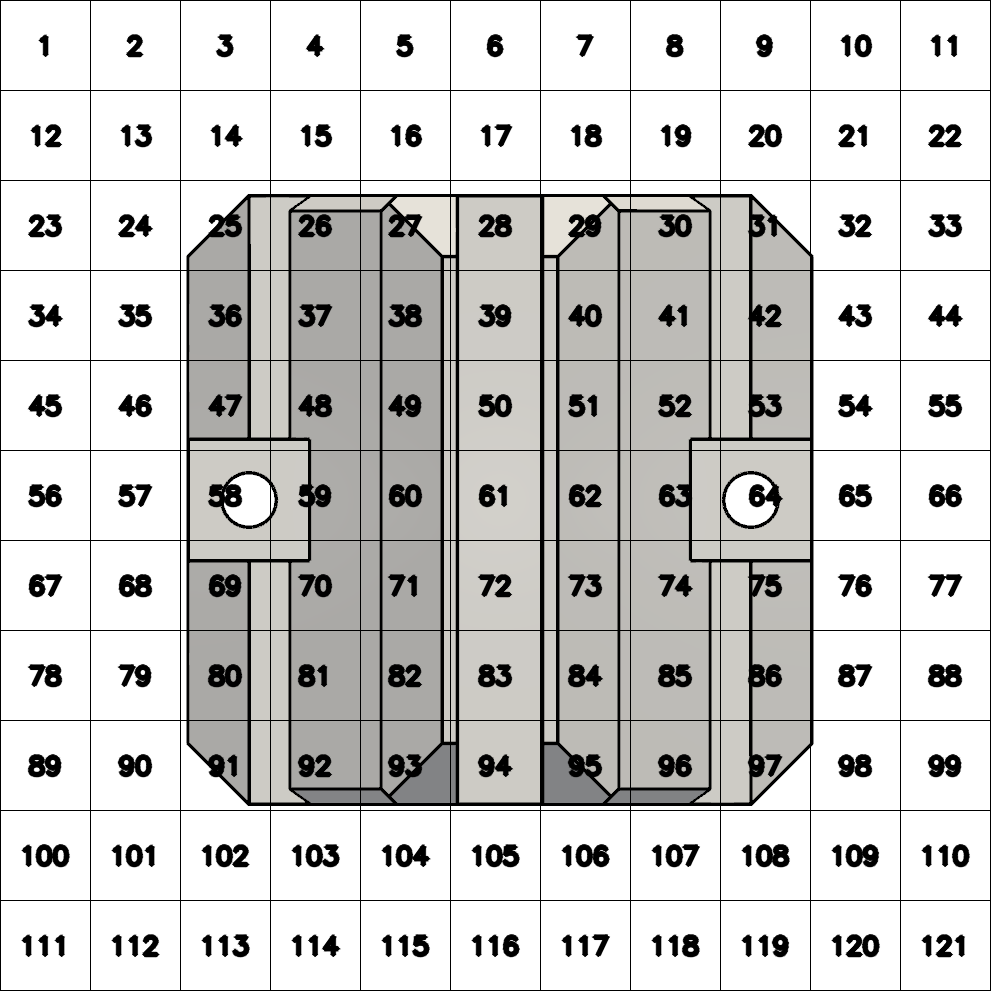}
}
\hfill
\subfloat[Gridded view 5]{
  \includegraphics[width=0.3\textwidth]{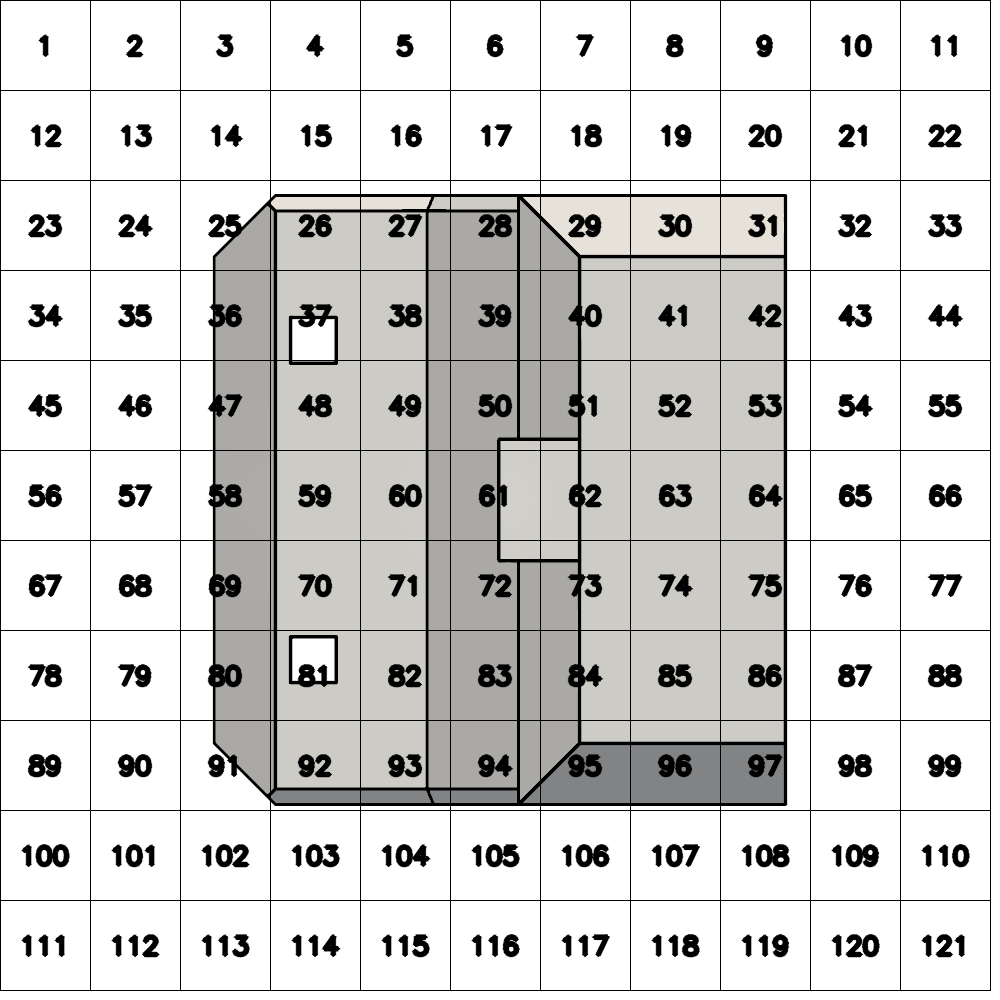}
}
\caption{
Five images provided jointly with example prompt 1.
Top row: non-gridded perspective views for global 3D context.
Bottom row: gridded orthographic views used for cell-level region localization.
}
\label{fig:prompt_images_1}
\end{figure}

\subsubsection{\textbf{Example Prompt 1: Compression (\textit{Load+Features+Heuristics}, views=5, Figure \ref{fig:prompt_images_1})}}

\begin{tcolorbox}[
  title={Response: Gemini 3 Flash Preview},
  colback=gray!5,
  colframe=blue,
  fonttitle=\bfseries,
  breakable
]
\begin{verbatim}
 Gridded Image ('e0', 'a0'):
***I.C.E: 26, 30, 37, 41, 48, 52, 59, 63, 70, 74, 81, 85, 92, 96***
***E.C/P.C: NONE***
***T.H: 58, 64***
***F: NONE***

Gridded Image ('e90', 'a0'):
***I.C.E: NONE***
***E.C/P.C: NONE***
***T.H: NONE***
***F: NONE***

Gridded Image ('e-90', 'a0'):
***I.C.E: NONE***
***E.C/P.C: NONE***
***T.H: NONE***
***F: 71, 72, 73, 82, 83, 84***

Gridded Image ('e0', 'a180'):
***I.C.E: 28, 39, 50, 61, 72, 83, 94***
***E.C/P.C: 58, 59, 63, 64***
***T.H: 58, 64***
***F: NONE***

Gridded Image ('e0', 'a270'):
***I.C.E: 28, 39, 50, 61, 72, 83, 94***
***E.C/P.C: 61, 62, 72, 73***
***T.H: 37, 81***
***F: NONE***
\end{verbatim}
\end{tcolorbox}

\begin{tcolorbox}[
  title={Response: Claude Haiku 4.5},
  colback=gray!5,
  colframe=blue,
  fonttitle=\bfseries,
  breakable
]
\begin{verbatim}
Response: Front View (e0, a0)

***I.C.E: 26, 27, 28, 29, 30, 31, 37, 38, 39, 40, 41, 42, 48, 49, 50, 
51, 52, 53, 59, 60, 61, 62, 63, 64***

***E.C/P.C: 37, 38, 39, 40, 41, 42, 59, 60, 61, 62, 63, 64***

***T.H: NONE***

***F: 82, 83, 84, 85***

## Image 2: Top View (e90, a0)

***I.C.E: NONE***

***E.C/P.C: NONE***

***T.H: NONE***

***F: NONE***

## Image 3: Bottom View (e-90, a0)

***I.C.E: NONE***

***E.C/P.C: NONE***

***T.H: NONE***

***F: NONE***

## Image 4: Back View (e0, a180)

***I.C.E: 26, 27, 28, 29, 30, 31, 37, 38, 39, 40, 41, 42, 48, 49, 50, 
51, 52, 53, 59, 60, 61, 62, 63, 64***

***E.C/P.C: 37, 38, 39, 40, 41, 42, 59, 60, 61, 62, 63, 64***

***T.H: NONE***

***F: 82, 83, 84, 85***

## Image 5: Right View (e0, a270)

***I.C.E: 26, 27, 28, 29, 30, 31, 37, 38, 39, 40, 41, 42, 48, 49, 50, 
51, 52, 53, 59, 60, 61, 62, 63, 64***

***E.C/P.C: 37, 38, 39, 40, 41, 42, 59, 60, 61, 62, 63, 64***

***T.H: 37, 38, 39, 40, 41, 42, 59, 60, 61, 62, 63, 64***

***F: 82, 83, 84, 85***
\end{verbatim}
\end{tcolorbox}

\begin{tcolorbox}[
  title={Response: GPT 5.4 mini},
  colback=gray!5,
  colframe=blue,
  fonttitle=\bfseries,
  breakable
]
\begin{verbatim}
***I.C.E: 27, 28, 29, 38, 39, 40, 49, 50, 51, 59, 60, 61, 62, 63, 71, 
72, 73, 74, 75, 81, 82, 83, 84, 85, 86, 91, 92, 93, 94, 95, 96, 
97 ***
***E.C/P.C: 26, 30, 37, 41, 48, 52, 58, 64, 70, 74, 80, 86, 92, 96 ***
***T.H: NONE ***
***F: 27, 29, 31, 38, 40, 42, 49, 51, 53, 59, 63, 69, 75, 81, 87, 91,
97 ***

***I.C.E: 27, 28, 29, 38, 39, 40, 49, 50, 51, 59, 60, 61, 62, 63, 71, 
72, 73, 74, 75, 81, 82, 83, 84, 85, 86, 91, 92, 93, 94, 95, 96, 97 ***
***E.C/P.C: 26, 30, 37, 41, 48, 52, 58, 64, 70, 74, 80, 86, 92, 96 ***
***T.H: NONE ***
***F: 27, 29, 31, 38, 40, 42, 49, 51, 53, 59, 63, 69, 75, 81, 87, 91, 
97 ***

***I.C.E: 26, 27, 28, 29, 38, 39, 40, 41, 49, 50, 51, 52, 60, 61, 62, 
63, 71, 72, 73, 74, 82, 83, 84, 85, 92, 93, 94, 95 ***
***E.C/P.C: 25, 30, 36, 42, 47, 53, 58, 64, 69, 75, 91, 97 ***
***T.H: 58, 64 ***
***F: 25, 27, 29, 38, 40, 49, 51, 59, 63, 69, 75, 91, 97 ***

***I.C.E: 25, 26, 27, 28, 29, 37, 38, 39, 40, 41, 47, 48, 49, 50, 51, 
58, 59, 60, 61, 62, 69, 70, 71, 72, 73, 80, 81, 82, 83, 84, 91, 92, 93,
94, 95 ***
***E.C/P.C: 25, 30, 36, 42, 47, 53, 58, 64, 69, 75, 91, 97 ***
***T.H: 58, 64 ***
***F: 25, 27, 29, 37, 40, 42, 47, 49, 51, 58, 62, 69, 73, 80, 84, 91, 
95 ***

***I.C.E: 25, 27, 29, 36, 38, 40, 47, 49, 51, 58, 60, 62, 69, 71, 73, 
80, 82, 84, 91, 93, 95 ***
***E.C/P.C: 26, 30, 37, 41, 48, 52, 59, 63, 70, 74, 92, 96 ***
***T.H: 37, 81 ***
***F: 25, 29, 40, 47, 51, 58, 62, 69, 73, 80, 84, 91, 95 ***

\end{verbatim}
\end{tcolorbox}

\begin{tcolorbox}[
  title={Response: Qwen3 VL 235B A22B Instruct},
  colback=gray!5,
  colframe=blue,
  fonttitle=\bfseries,
  breakable
]
\begin{verbatim}
***I.C.E: 58, 64 ***
***E.C/P.C: 25, 31, 91, 97 ***
***T.H: NONE ***
***F: NONE ***

***I.C.E: 49, 63 ***
***E.C/P.C: 27, 29, 69, 75, 91, 97 ***
***T.H: NONE ***
***F: 82, 83, 84, 85, 92, 93, 94, 95, 96 ***

***I.C.E: 49, 63 ***
***E.C/P.C: 27, 29, 69, 75, 91, 97 ***
***T.H: NONE ***
***F: 82, 83, 84, 85, 92, 93, 94, 95, 96 ***

***I.C.E: 58, 64 ***
***E.C/P.C: 25, 31, 91, 97 ***
***T.H: NONE ***
***F: NONE ***

***I.C.E: 37, 81 ***
***E.C/P.C: 26, 28, 36, 38, 47, 49, 50, 51, 52, 61, 62, 63, 64, 72, 73, 
74, 75, 80, 82, 83, 84, 85, 86, 91, 92, 93, 94, 95, 96, 97 ***
***T.H: NONE ***
***F: NONE ***
    
\end{verbatim}
\end{tcolorbox}

\begin{figure}[h]
\centering

\subfloat[Perspective view]{
  \includegraphics[width=0.3\textwidth]{figures/Load_Cases_Figures/00510024_view_e18_a30_bending.png}
}
\hfill
\subfloat[Gridded view 1]{
  \includegraphics[width=0.3\textwidth]{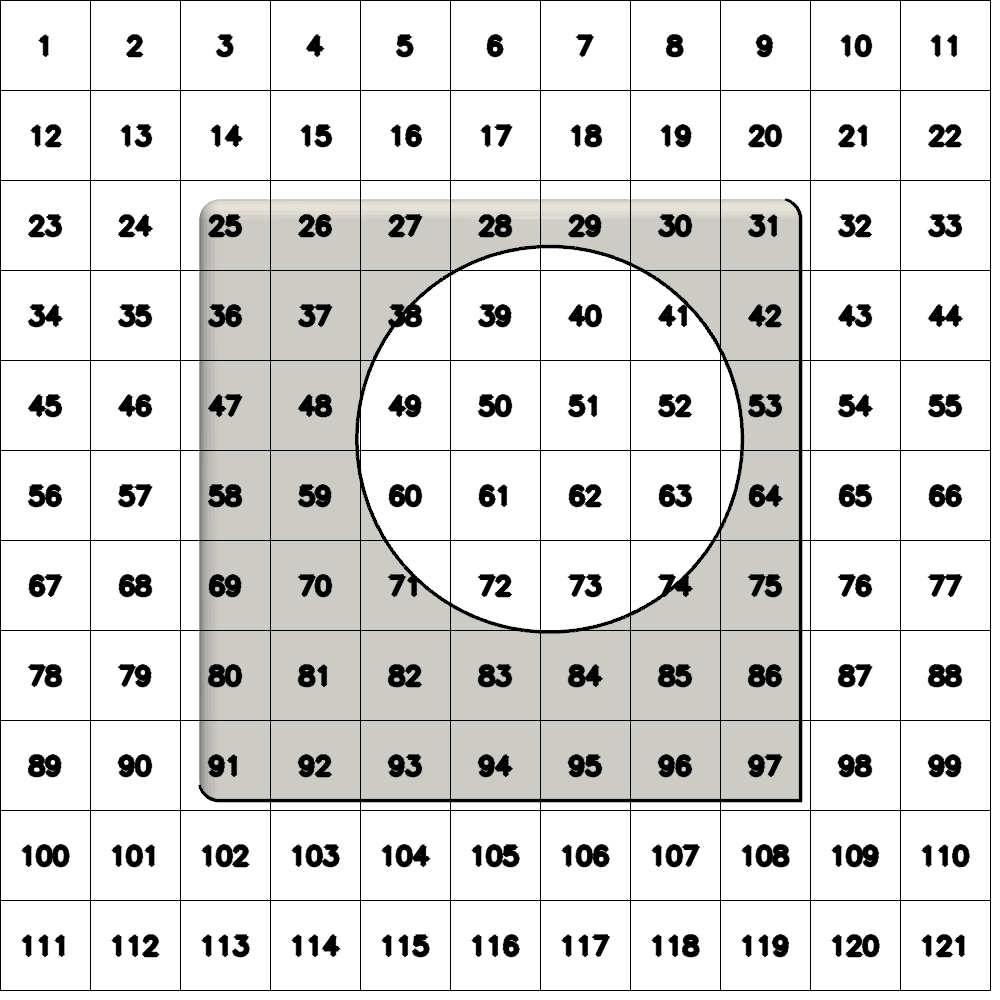}
}
\hfill
\subfloat[Gridded view 2]{
  \includegraphics[width=0.3\textwidth]{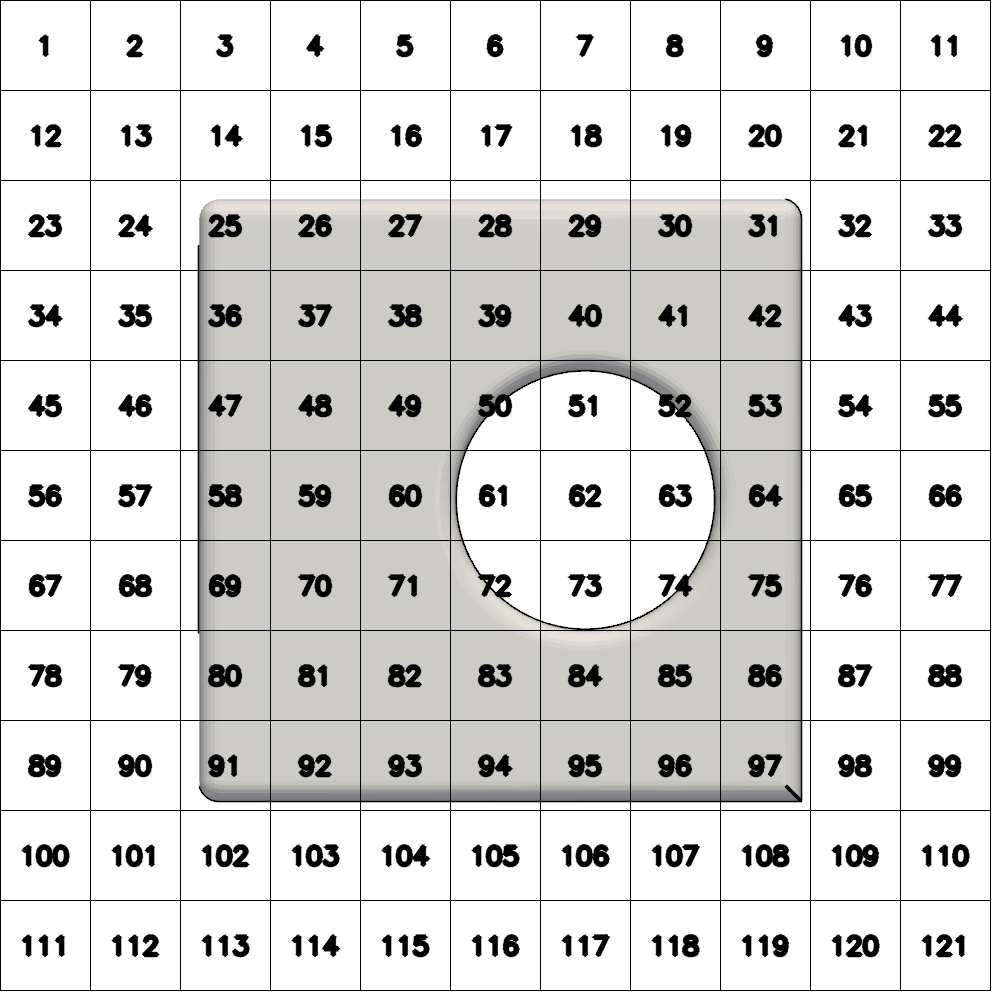}
}

\vspace{1mm}

\subfloat[Gridded view 3]{
  \includegraphics[width=0.3\textwidth]{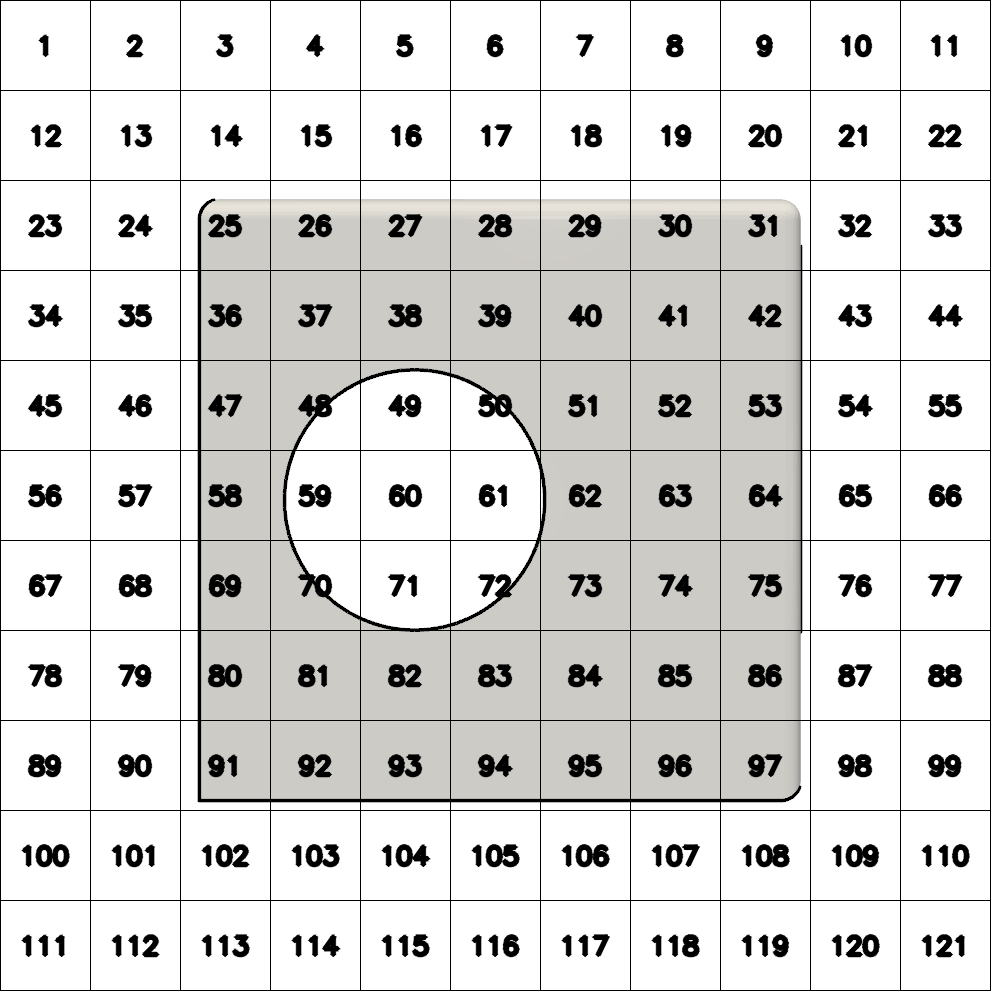}
}
\hfill
\subfloat[Gridded view 4]{
  \includegraphics[width=0.3\textwidth]{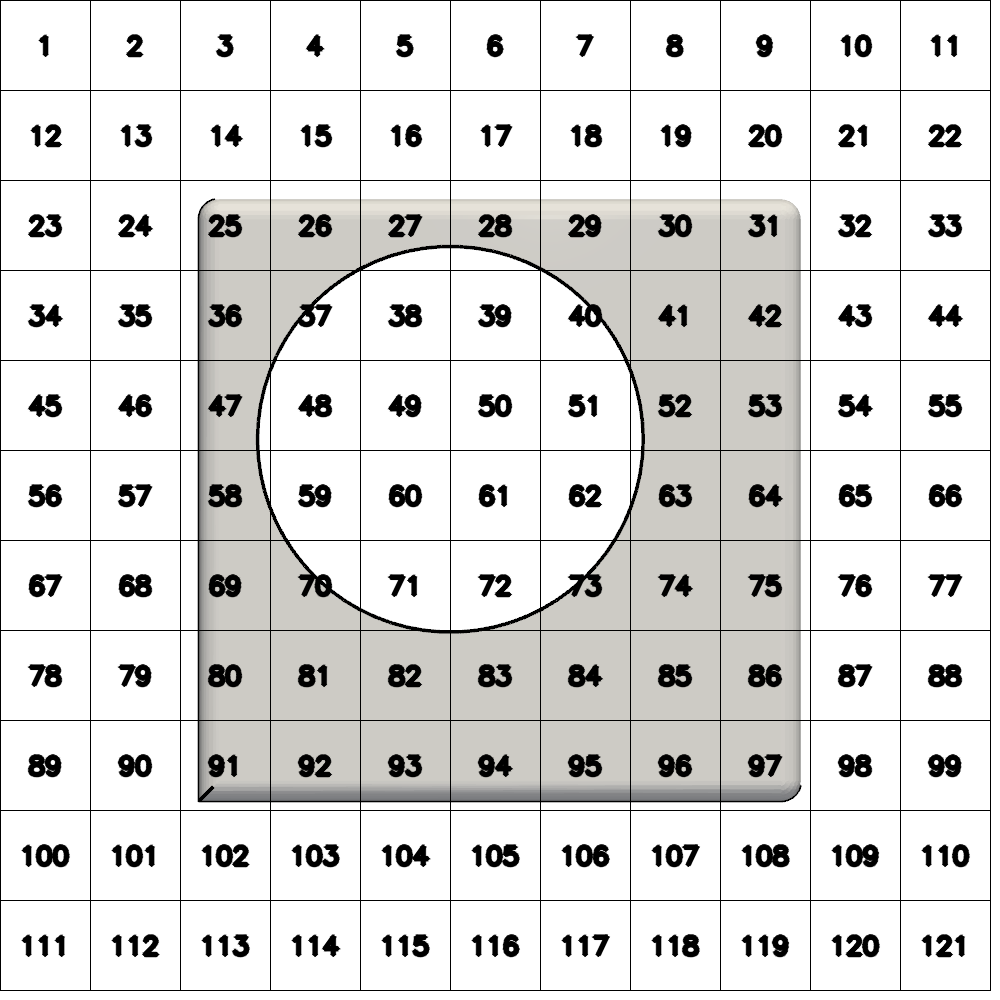}
}
\hfill
\subfloat[Gridded view 5]{
  \includegraphics[width=0.3\textwidth]{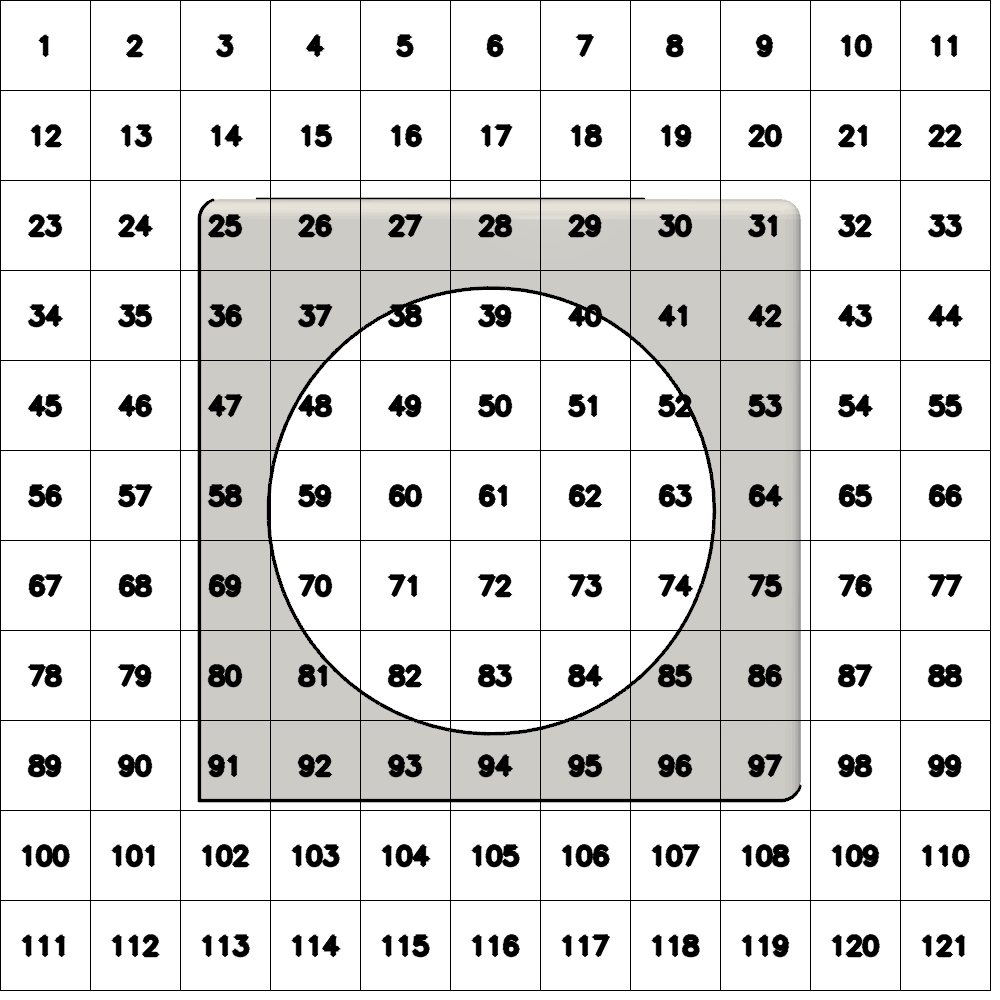}
}
\caption{
Five images provided jointly with example prompt 2.
Top row: non-gridded perspective views for global 3D context.
Bottom row: gridded orthographic views used for cell-level region localization.
}
\label{fig:prompt_images_2}
\end{figure}

\subsubsection{\textbf{Example Prompt 2: Bending  (\textit{Load+Features}, views=5, Figure \ref{fig:prompt_images_2})}}

\begin{tcolorbox}[
  title={Response: Gemini 3 Flash Preview},
  colback=gray!5,
  colframe=blue,
  fonttitle=\bfseries,
  breakable
]
\begin{verbatim}
 Image 3: (e0, a0)
***I.C.E: NONE***
***E.C/P.C: NONE***
***T.H: 28, 29, 38, 39, 40, 41, 49, 50, 51, 52, 60, 61, 62, 63, 71, 72, 
73, 74, 83, 84***
***F: NONE***

Image 4: (e0, a270)
***I.C.E: NONE***
***E.C/P.C: NONE***
***T.H: 50, 51, 52, 61, 62, 63, 72, 73, 74***
***F: NONE***

Image 5: (e0, a90)
***I.C.E: NONE***
***E.C/P.C: NONE***
***T.H: 48, 49, 50, 59, 60, 61, 70, 71, 72***
***F: NONE***

Image 6: (e0, a180)
***I.C.E: NONE***
***E.C/P.C: NONE***
***T.H: 27, 28, 29, 30, 37, 38, 39, 40, 41, 48, 49, 50, 51, 52, 59, 60, 
61, 62, 63, 70, 71, 72, 73, 74, 82, 83, 84, 85***
***F: NONE***

Image 7: (e-90, a0)
***I.C.E: NONE***
***E.C/P.C: NONE***
***T.H: 38, 39, 40, 41, 48, 49, 50, 51, 52, 59, 60, 61, 62, 63, 70, 71, 
72, 73, 74, 82, 83, 84, 85***
***F: NONE***
\end{verbatim}
\end{tcolorbox}

\begin{tcolorbox}[
  title={Response: Claude Haiku 4.5},
  colback=gray!5,
  colframe=blue,
  fonttitle=\bfseries,
  breakable
]
\begin{verbatim}
Response: (e0, a0) - Front View

***I.C.E: 48, 49, 50, 51, 52, 59, 60, 61, 62, 63***
***E.C/P.C: NONE***
***T.H: 50, 51, 52, 61, 62, 63***
***F: 48, 49, 50, 51, 52, 59, 60, 61, 62, 63***

## Image 2: (e0, a270) - Left View

***I.C.E: 25, 26, 31, 32, 36, 37, 42, 43, 47, 48, 52, 53, 58, 59, 63, 
64, 69, 70, 74, 75, 80, 81, 85, 86, 91, 92, 96, 97***
***E.C/P.C: NONE***
***T.H: NONE***
***F: NONE***

## Image 3: (e0, a90) - Right View

***I.C.E: 25, 26, 31, 32, 36, 37, 42, 43, 47, 48, 52, 53, 58, 59, 63, 
64, 69, 70, 74, 75, 80, 81, 85, 86, 91, 92, 96, 97***
***E.C/P.C: NONE***
***T.H: NONE***
***F: NONE***

## Image 4: (e0, a180) - Back View

***I.C.E: 48, 49, 50, 51, 52, 59, 60, 61, 62, 63***
***E.C/P.C: NONE***
***T.H: 50, 51, 52, 61, 62, 63***
***F: 48, 49, 50, 51, 52, 59, 60, 61, 62, 63***

## Image 5: (e-90, a0) - Bottom View

***I.C.E: 25, 26, 31, 32, 36, 37, 42, 43, 47, 48, 52, 53, 58, 59, 63, 
64, 69, 70, 74, 75, 80, 81, 85, 86, 91, 92, 96, 97***
***E.C/P.C: NONE***
***T.H: 50, 51, 52, 61, 62, 63***
***F: NONE***
\end{verbatim}
\end{tcolorbox}

\begin{tcolorbox}[
  title={Response: GPT 5.4 mini},
  colback=gray!5,
  colframe=blue,
  fonttitle=\bfseries,
  breakable
]
\begin{verbatim}
***I.C.E: 25, 31, 91, 97 ***
***E.C/P.C: NONE ***
***T.H: 36, 37, 38, 39, 40, 41, 42, 47, 48, 49, 50, 51, 52, 53, 58, 59, 
60, 61, 62, 63, 64, 69, 70, 71, 72, 73, 74, 75, 80, 81, 82, 83, 84, 85, 
86, 91, 92, 93, 94, 95, 96, 97 ***
***F: 25, 26, 27, 28, 29, 30, 31, 36, 42, 47, 53, 58, 64, 69, 75, 80, 
86, 91, 92, 93, 94, 95, 96, 97 ***

***I.C.E: 25, 31, 91, 97 ***
***E.C/P.C: NONE ***
***T.H: 25, 26, 27, 28, 29, 30, 31, 36, 37, 38, 39, 40, 41, 42, 47, 48, 
49, 50, 51, 52, 53, 58, 59, 60, 61, 62, 63, 64, 69, 70, 71, 72, 73, 74, 
75, 80, 81, 82, 83, 84, 85, 86, 91, 92, 93, 94, 95, 96, 97 ***
***F: 25, 26, 27, 28, 29, 30, 31, 37, 38, 39, 40, 41, 42, 48, 49, 50, 
51, 52, 53, 59, 60, 61, 62, 63, 64, 70, 71, 72, 73, 74, 75, 81, 82, 83, 
84, 85, 86, 91, 92, 93, 94, 95, 96, 97 ***

***I.C.E: 25, 91 ***
***E.C/P.C: NONE ***
***T.H: 38, 39, 40, 41, 48, 49, 50, 51, 52, 59, 60, 61, 62, 63, 70, 71, 
72, 73, 74, 81, 82, 83, 84, 85, 92, 93, 94, 95, 96 ***
***F: 25, 26, 27, 28, 29, 30, 31, 38, 39, 40, 41, 48, 49, 50, 51, 52, 
59, 60, 61, 62, 63, 70, 71, 72, 73, 74, 81, 82, 83, 84, 85, 92, 93, 94, 
95, 96 ***

***I.C.E: 25, 31, 91, 97 ***
***E.C/P.C: NONE ***
***T.H: 38, 39, 40, 41, 48, 49, 50, 51, 52, 59, 60, 61, 62, 63, 70, 71, 
72, 73, 74, 81, 82, 83, 84, 85, 92, 93, 94, 95, 96 ***
***F: 25, 26, 27, 28, 29, 30, 31, 38, 39, 40, 41, 48, 49, 50, 51, 52, 
59, 60, 61, 62, 63, 70, 71, 72, 73, 74, 81, 82, 83, 84, 85, 92, 93, 94, 
95, 96 ***

***I.C.E: 25, 31, 91, 97 ***
***E.C/P.C: NONE ***
***T.H: 25, 26, 27, 28, 29, 30, 31, 36, 37, 38, 39, 40, 41, 42, 47, 48, 
49, 50, 51, 52, 53, 58, 59, 60, 61, 62, 63, 64, 69, 70, 71, 72, 73, 74, 
75, 80, 81, 82, 83, 84, 85, 86, 91, 92, 93, 94, 95, 96, 97 ***
***F: 25, 26, 27, 28, 29, 30, 31, 36, 42, 47, 53, 58, 64, 69, 75, 80, 
86, 91, 92, 93, 94, 95, 96, 97 ***
\end{verbatim}
\end{tcolorbox}

\begin{tcolorbox}[
  title={Response: Qwen3 VL 235B A22B Instruct},
  colback=gray!5,
  colframe=blue,
  fonttitle=\bfseries,
  breakable
]
\begin{verbatim}
***I.C.E: NONE ***
***E.C/P.C: NONE ***
***T.H: 38,39,40,41,42,48,49,50,51,52,58,59,60,61,62,68,69,70,71,72,
78,79,80,81,82 ***
***F: NONE ***

***I.C.E: NONE ***
***E.C/P.C: NONE ***
***T.H: 50,51,52,53,61,62,63,64,72,73,74,75,83,84,85,86,94,95,96,97 ***
***F: NONE ***

***I.C.E: NONE ***
***E.C.P.C: NONE ***
***T.H: 48,49,50,51,52,58,59,60,61,62,68,69,70,71,72,78,79,80,81,82 ***
***F: NONE ***

***I.C.E: NONE ***
***E.C/P.C: NONE ***
***T.H: 37,38,39,40,41,47,48,49,50,51,57,58,59,60,61,67,68,69,70,71,77,
78,79,80,81 ***
***F: NONE ***

***I.C.E: NONE ***
***E.C/P.C: NONE ***
***T.H: 38,39,40,41,42,48,49,50,51,52,58,59,60,61,62,68,69,70,71,72,
78,79,80,81,82 ***
***F: NONE ***

    
\end{verbatim}
\end{tcolorbox}

\begin{figure}[h]
\centering

\subfloat[Perspective view]{
  \includegraphics[width=0.3\textwidth]{figures/Load_Cases_Figures/00890079_view_e18_a30_torsion.png}
}
\hfill
\subfloat[Gridded view 1]{
  \includegraphics[width=0.3\textwidth]{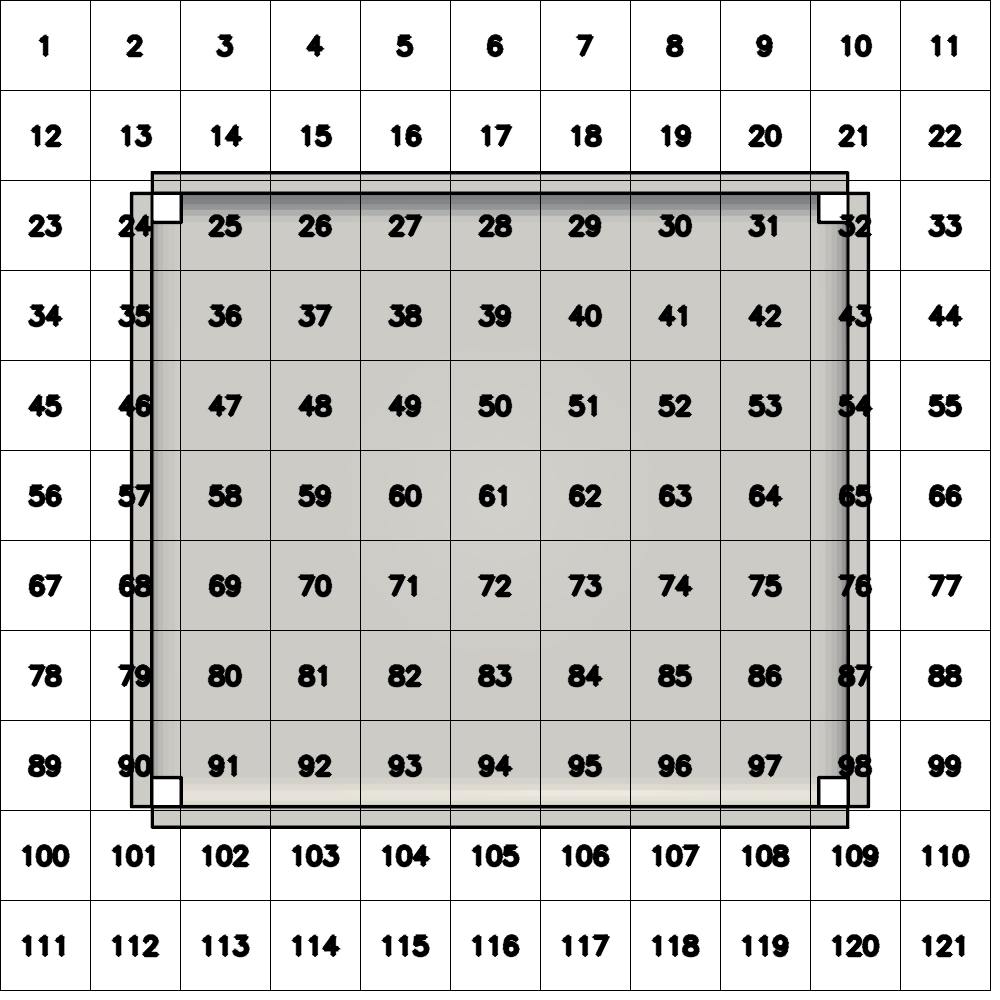}
}
\hfill
\subfloat[Gridded view 2]{
  \includegraphics[width=0.3\textwidth]{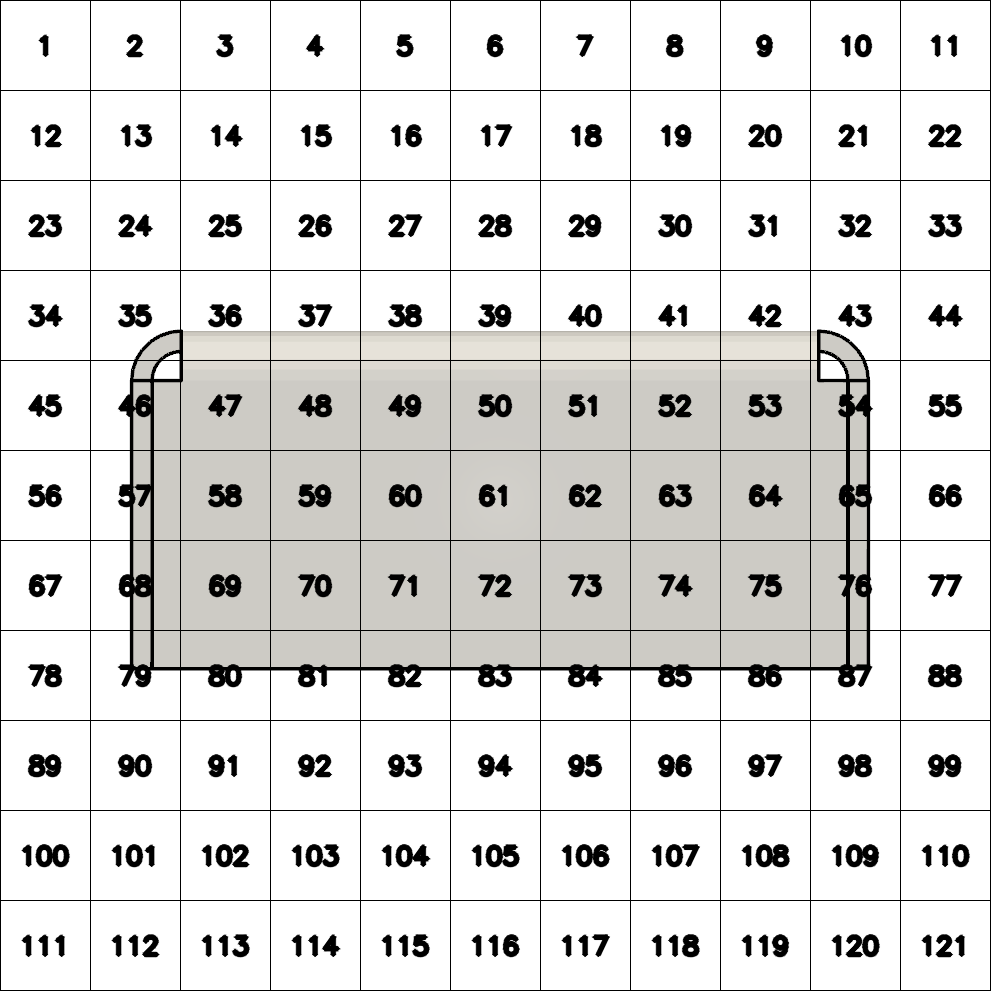}
}

\vspace{1mm}

\subfloat[Gridded view 3]{
  \includegraphics[width=0.3\textwidth]{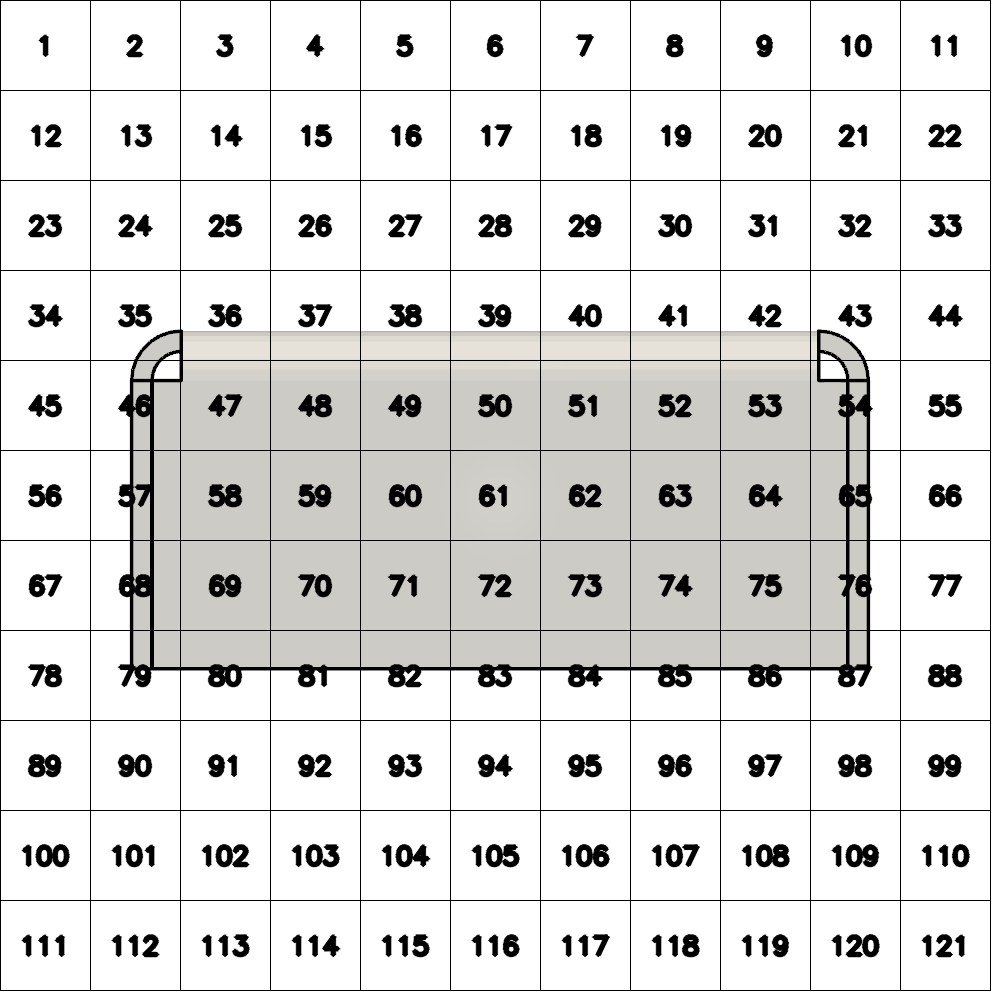}
}
\hfill
\subfloat[Gridded view 4]{
  \includegraphics[width=0.3\textwidth]{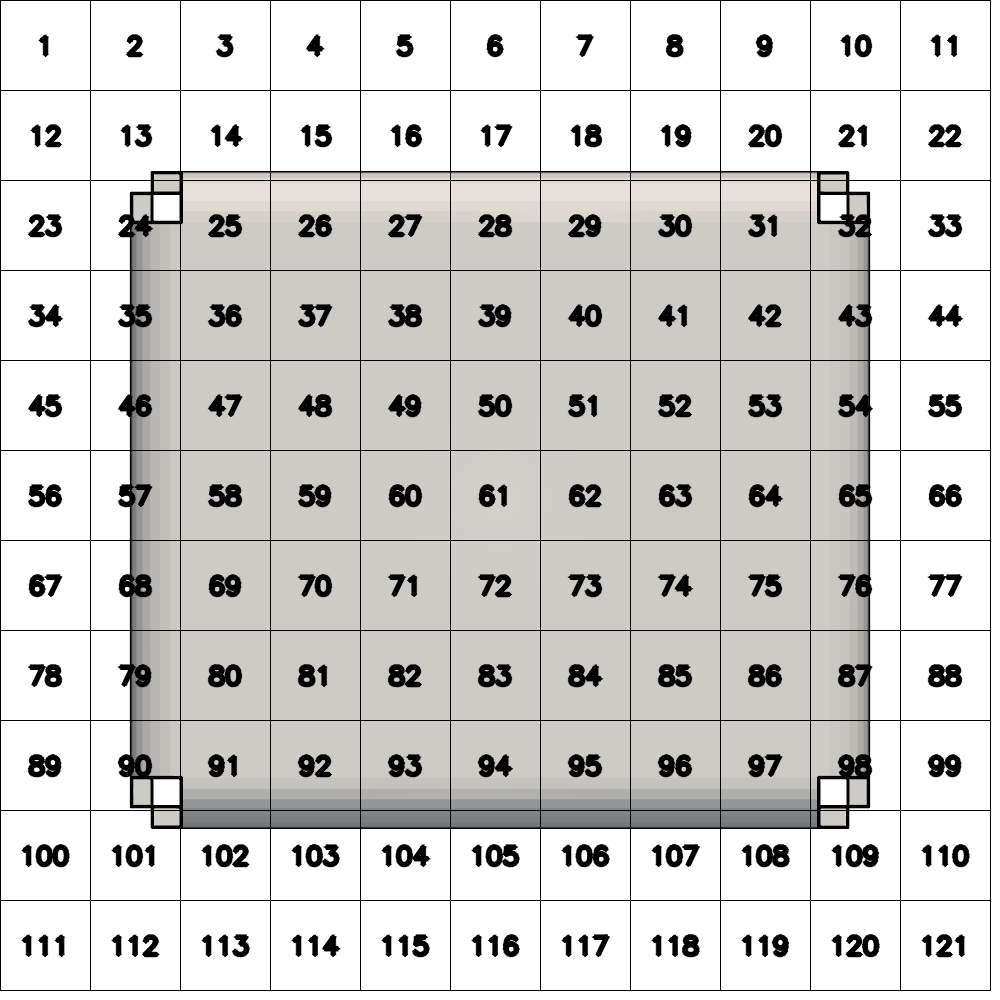}
}
\hfill
\subfloat[Gridded view 5]{
  \includegraphics[width=0.3\textwidth]{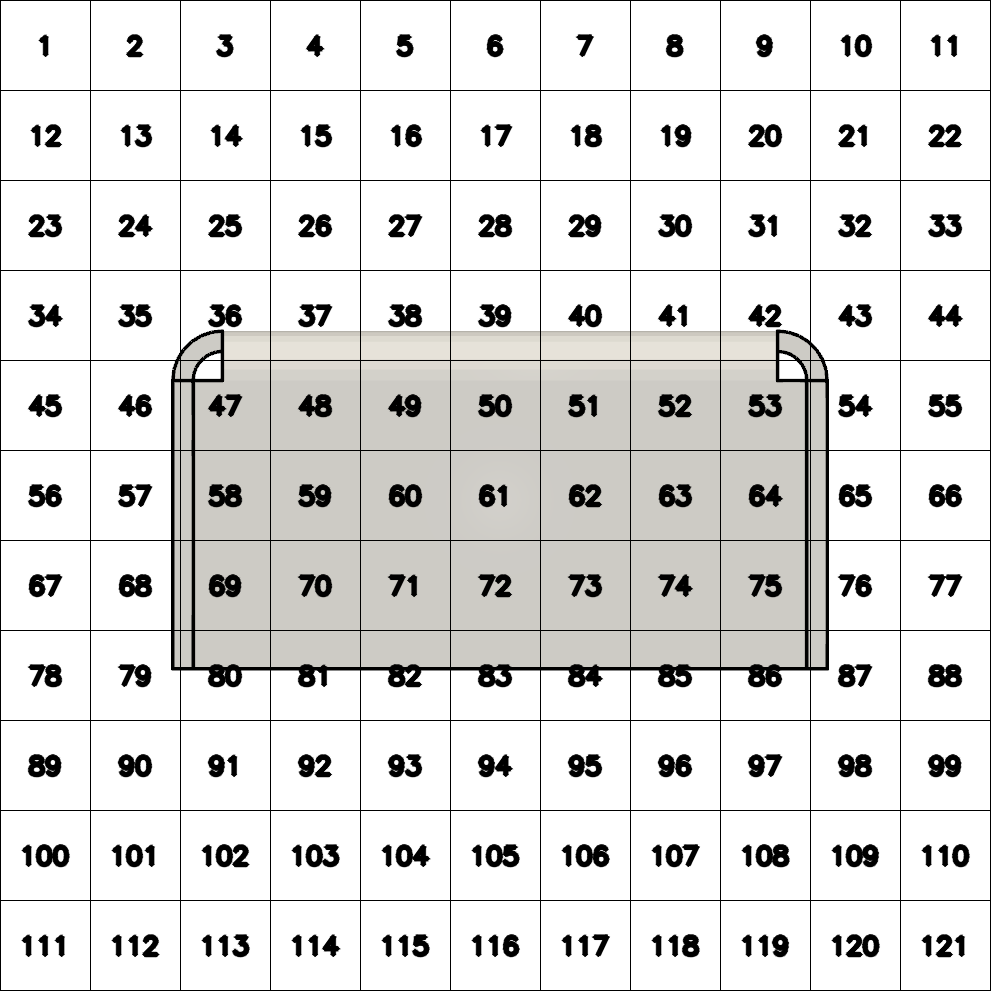}
}
\caption{
Five images provided jointly with example prompt 3.
Top row: non-gridded perspective views for global 3D context.
Bottom row: gridded orthographic views used for cell-level region localization.
}
\label{fig:prompt_images_3}
\end{figure}

\subsubsection{\textbf{Example Prompt 3: Torsion (Load-Only, views=5, Figure \ref{fig:prompt_images_3})}}

\begin{tcolorbox}[
  title={Response: Gemini 3 Flash Preview},
  colback=gray!5,
  colframe=blue,
  fonttitle=\bfseries,
  breakable
]
\begin{verbatim}
Image 1 [('e-90', 'a0')]:
***Cells: 24, 32, 90, 98***

Image 2 [('e0', 'a0')]:
***Cells: 46, 54***

Image 3 [('e0', 'a180')]:
***Cells: 46, 54***

Image 4 [('e90', 'a0')]:
***Cells: 24, 32, 90, 98***

Image 5 [('e0', 'a270')]:
***Cells: 47, 53***
\end{verbatim}
\end{tcolorbox}

\begin{tcolorbox}[
  title={Response: Claude Haiku 4.5},
  colback=gray!5,
  colframe=blue,
  fonttitle=\bfseries,
  breakable
]
\begin{verbatim}
Response: I'll analyze each gridded orthographic image to identify cells
containing stress critical features under torsional loading about the
vertical axis.

**Image 1 (e-90, a0) - Bottom view:**
The visible stress critical features are the corner cutouts/notches at
the four corners of the rectangular part.

***Cells: 24, 32, 101, 109***

**Image 2 (e0, a0) - Front view:**
The visible stress critical features are the corner cutouts/notches on
the left and right sides.

***Cells: 24, 32, 101, 109***

**Image 3 (e0, a180) - Back view:**
The visible stress critical features are the corner cutouts/notches on
the left and right sides (same as front view, mirrored).

***Cells: 24, 32, 101, 109***

**Image 4 (e90, a0) - Top view:**
The visible stress critical features are the corner cutouts/notches at
the four corners and the slot features along the edges.

***Cells: 24, 25, 32, 101, 102, 109***

**Image 5 (e0, a270) - Right side view:**
The visible stress critical features are the corner cutouts/notches on
the top and bottom.

***Cells: 36, 37, 46, 47, 80, 81, 90, 91***
\end{verbatim}
\end{tcolorbox}

\begin{tcolorbox}[
  title={Response: GPT 5.4 mini},
  colback=gray!5,
  colframe=blue,
  fonttitle=\bfseries,
  breakable
]
\begin{verbatim}
***Cells: 24, 25, 26, 27, 28, 29, 30, 31, 32, 35, 36, 37, 38, 39, 40,
41, 42, 43, 46, 47, 48, 49, 50, 51, 52, 53, 54, 57, 58, 59, 60, 61, 62,
63, 64, 65, 68, 69, 70, 71, 72, 73, 74, 75, 76, 79, 80, 81, 82, 83, 84,
85, 86, 87 ***
***Cells: 46, 47, 48, 49, 50, 51, 52, 53, 54, 57, 58, 59, 60, 61, 62,
63, 64, 65, 68, 69, 70, 71, 72, 73, 74, 75, 76, 79, 80, 81, 82, 83, 84,
85, 86, 87 ***
***Cells: 46, 47, 48, 49, 50, 51, 52, 53, 54, 57, 58, 59, 60, 61, 62,
63, 64, 65, 68, 69, 70, 71, 72, 73, 74, 75, 76, 79, 80, 81, 82, 83, 84,
85, 86, 87 ***
***Cells: 24, 25, 26, 27, 28, 29, 30, 31, 32, 35, 36, 37, 38, 39, 40,
41, 42, 43, 46, 47, 48, 49, 50, 51, 52, 53, 54, 57, 58, 59, 60, 61, 62,
63, 64, 65, 68, 69, 70, 71, 72, 73, 74, 75, 76, 79, 80, 81, 82, 83, 84,
85, 86, 87 ***
***Cells: 47, 48, 49, 50, 51, 52, 53, 58, 59, 60, 61, 62, 63, 64, 69,
70, 71, 72, 73, 74, 75, 80, 81, 82, 83, 84, 85, 86 ***
\end{verbatim}
\end{tcolorbox}

\begin{tcolorbox}[
  title={Response: Qwen3 VL 235B A22B Instruct},
  colback=gray!5,
  colframe=blue,
  fonttitle=\bfseries,
  breakable
]
\begin{verbatim}
***Cells: 24, 32, 90, 98 ***
***Cells: 35, 43, 79, 87 ***
***Cells: 35, 43, 79, 87 ***
***Cells: 24, 32, 90, 98 ***
***Cells: 35, 43, 79, 87 ***
\end{verbatim}
\end{tcolorbox}

\begin{figure}[h]
\centering

\subfloat[Perspective View]{
  \includegraphics[width=0.3\textwidth]{figures/Load_Cases_Figures/00520006_view_e18_a30_bending_compression.png}
}
\hfill
\subfloat[Gridded view 1]{
  \includegraphics[width=0.3\textwidth]{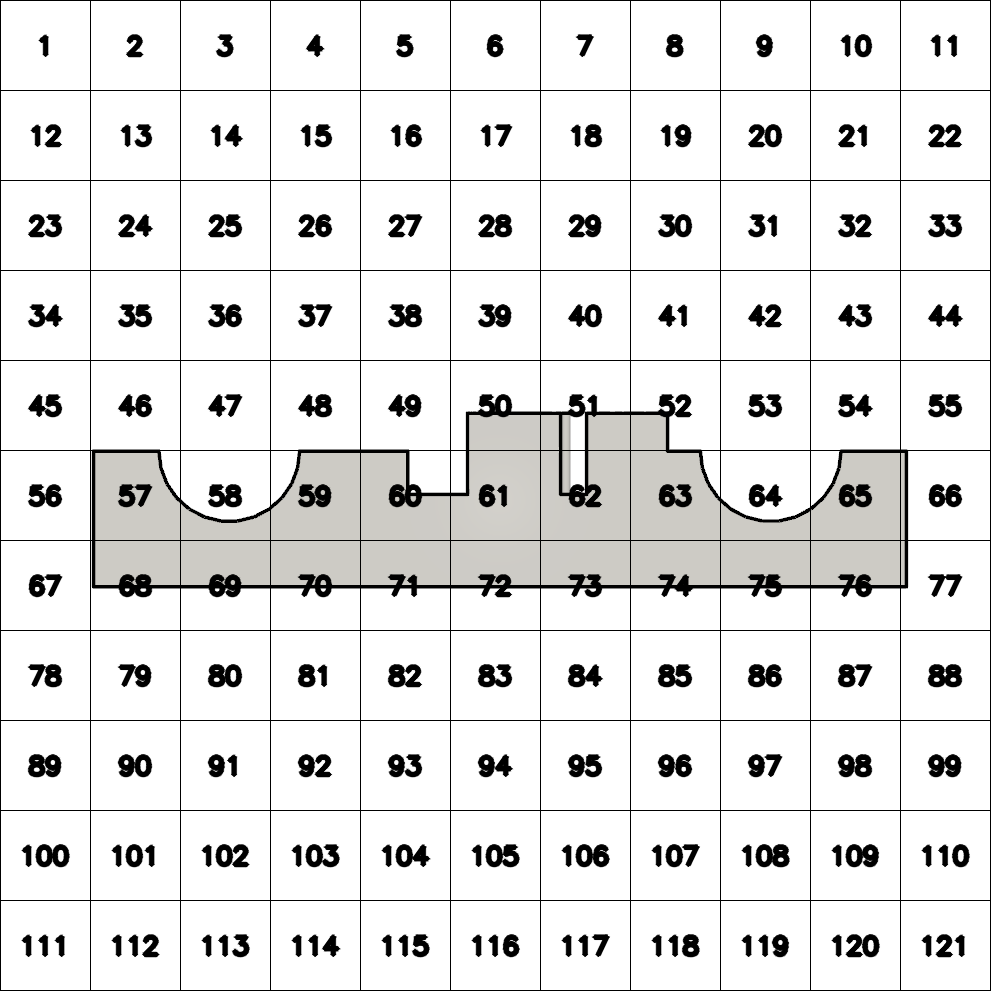}
}
\hfill
\subfloat[Gridded view 2]{
  \includegraphics[width=0.3\textwidth]{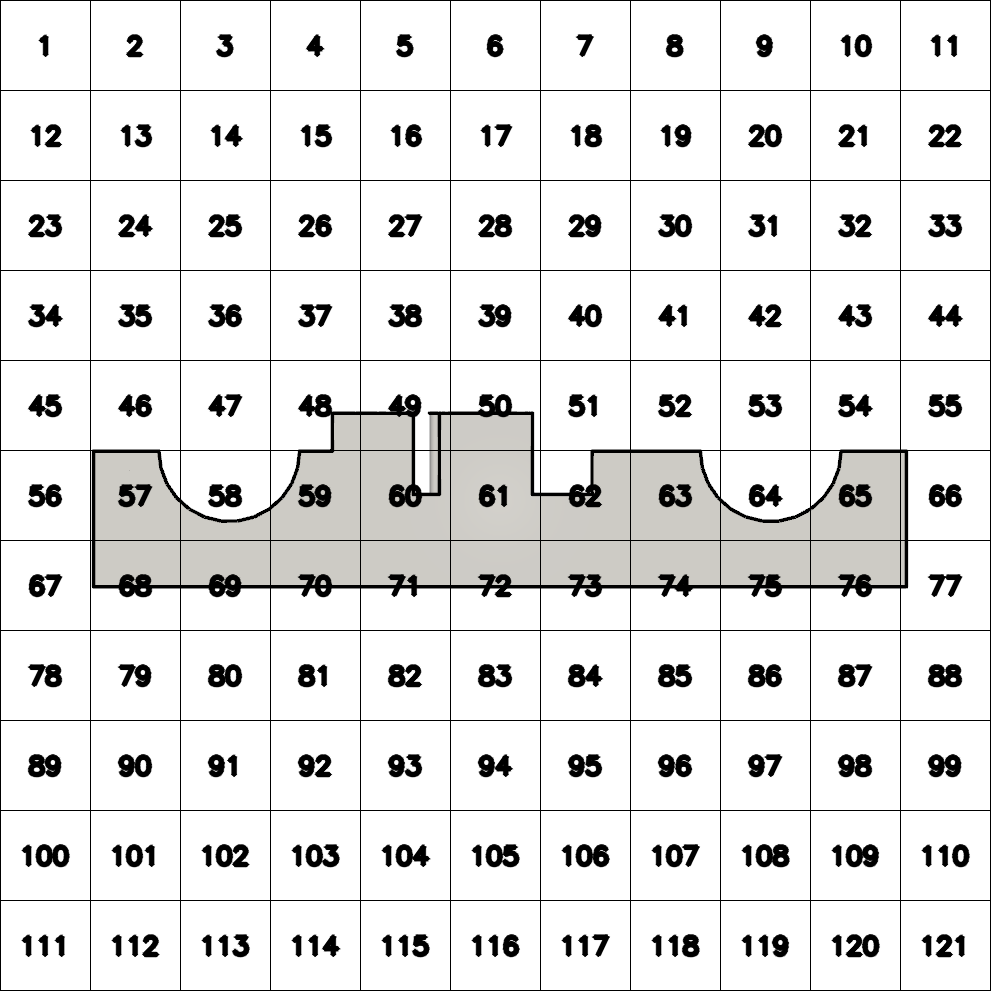}
}

\vspace{1mm}

\subfloat[Gridded view 3]{
  \includegraphics[width=0.3\textwidth]{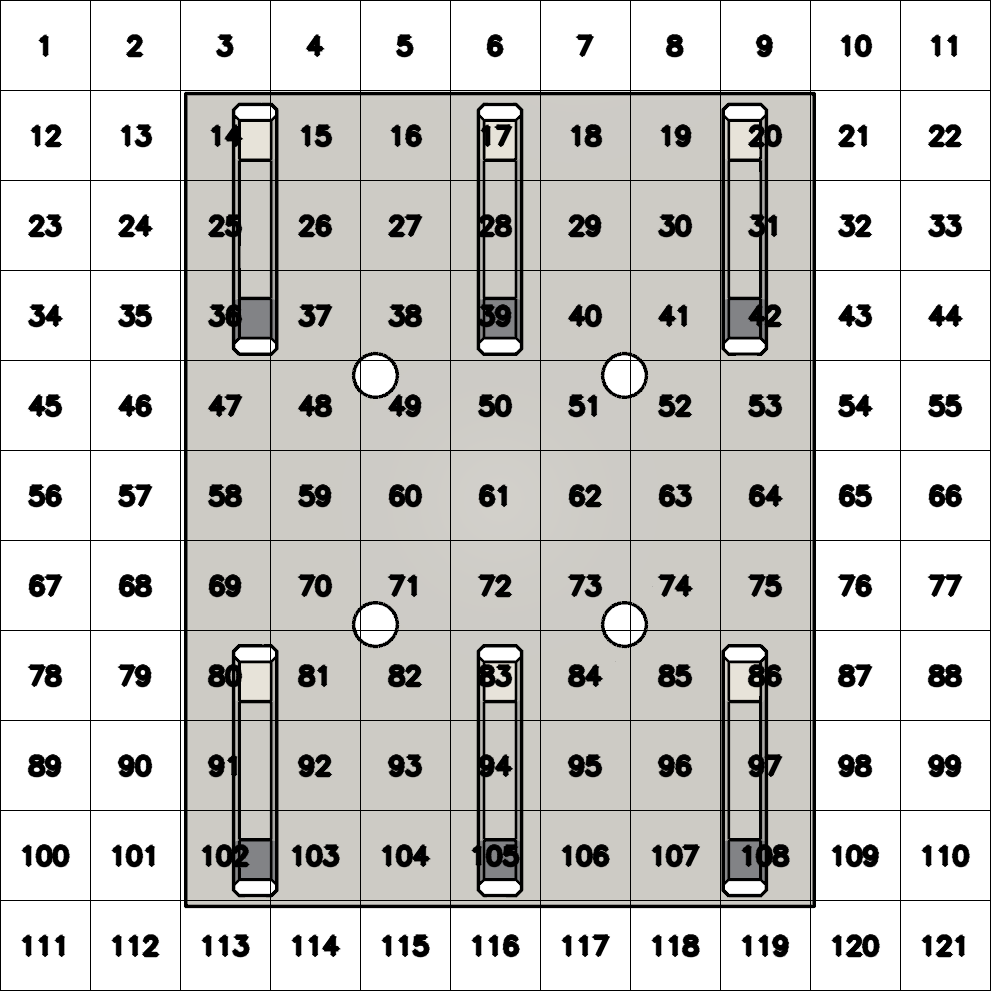}
}
\hfill
\subfloat[Gridded view 4]{
  \includegraphics[width=0.3\textwidth]{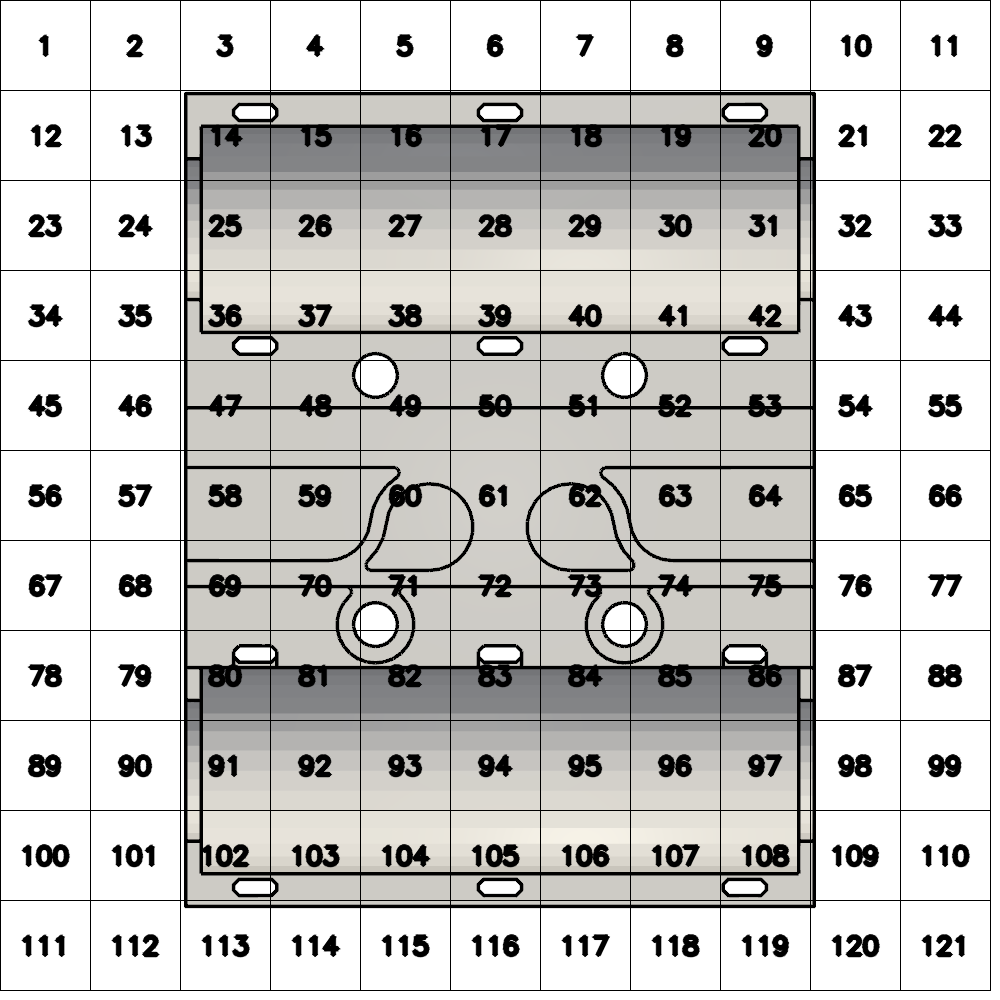}
}
\hfill
\subfloat[Gridded view 5]{
  \includegraphics[width=0.3\textwidth]{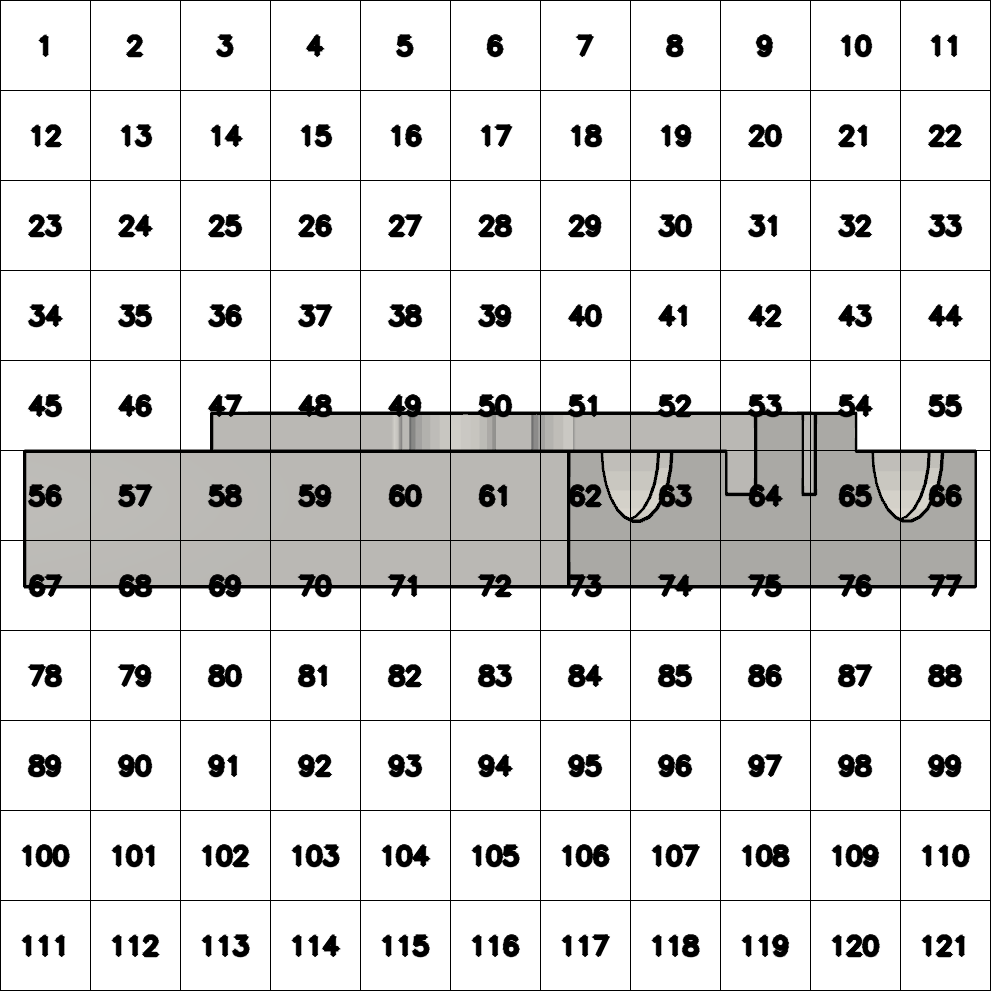}
}
\caption{
Five images provided jointly with example prompt 4.
Top row: non-gridded perspective views for global 3D context.
Bottom row: gridded orthographic views used for cell-level region localization.
}
\label{fig:prompt_images_4}
\end{figure}

\subsubsection{\textbf{Example Prompt 4: Bending-Compression (\textit{Load+Features+Heuristics}, views=5, Figure \ref{fig:prompt_images_4})}}

\begin{tcolorbox}[
  title={Response: Gemini 3 Flash Preview},
  colback=gray!5,
  colframe=blue,
  fonttitle=\bfseries,
  breakable
]
\begin{verbatim}
 **Image: ('e0', 'a270')**
***I.C.E: 50, 52, 60, 62***
***E.C/P.C: 50, 51, 52, 61, 62***
***T.H: NONE***
***F: 57, 58, 59, 64, 65***

**Image: ('e0', 'a90')**
***I.C.E: 48, 51, 60, 62***
***E.C/P.C: 49, 50, 60, 61***
***T.H: NONE***
***F: 57, 58, 59, 64, 65***

**Image: ('e-90', 'a0')**
***I.C.E: NONE***
***E.C/P.C: NONE***
***T.H: 14, 25, 36, 17, 28, 39, 20, 31, 42, 49, 52, 71, 74, 80, 91, 102, 
83, 94, 105, 86, 97, 108***
***F: NONE***

**Image: ('e90', 'a0')**
***I.C.E: 60, 61, 62, 63***
***E.C/P.C: 60, 61, 62, 63***
***T.H: 4, 6, 8, 41, 43, 49, 52, 71, 74, 78, 80, 82, 113, 116, 119***
***F: 14, 15, 16, 17, 18, 19, 20, 25, 26, 27, 28, 29, 30, 31, 36, 37, 38, 
39, 40, 41, 42, 80, 81, 82, 83, 84, 85, 86, 91, 92, 93, 94, 95, 96, 97, 
102, 103, 104, 105, 106, 107, 108***

**Image: ('e0', 'a210')**
***I.C.E: 53, 64***
***E.C/P.C: 47, 48, 49, 50, 51, 52, 53, 54***
***T.H: NONE***
***F: 62, 63, 65, 66***
\end{verbatim}
\end{tcolorbox}

\begin{tcolorbox}[
  title={Response: Claude Haiku 4.5},
  colback=gray!5,
  colframe=blue,
  fonttitle=\bfseries,
  breakable
]
\begin{verbatim}
# Analysis of Stress-Critical Features

## Image 1: (e0, a270) - Left View
***I.C.E: 57, 58, 59, 64, 65***
***E.C/P.C: 50, 51, 52, 60, 61, 62, 63***
***T.H: 49, 52***
***F: 57, 58, 59, 64, 65***

## Image 2: (e0, a90) - Right View
***I.C.E: 48, 49, 50, 51, 52***
***E.C/P.C: 48, 49, 50, 51, 52***
***T.H: 48, 52***
***F: NONE***

## Image 3: (e-90, a0) - Bottom View
***I.C.E: NONE***
***E.C/P.C: NONE***
***T.H: 49, 52, 71, 74***
***F: 60, 62, 71, 74***

## Image 4: (e90, a0) - Top View
***I.C.E: 14, 17, 20, 36, 39, 42, 80, 83, 86***
***E.C/P.C: 14, 17, 20, 36, 39, 42, 80, 83, 86***
***T.H: 49, 52, 71, 74***
***F: 60, 62, 71, 74***

## Image 5: (e0, a210) - Oblique View
***I.C.E: 57, 58, 59, 63, 64, 65***
***E.C/P.C: 50, 51, 52, 60, 61, 62, 63***
***T.H: 49, 52***
***F: 57, 58, 59, 63, 64, 65***
\end{verbatim}
\end{tcolorbox}

\begin{tcolorbox}[
  title={Response: GPT 5.4 mini},
  colback=gray!5,
  colframe=blue,
  fonttitle=\bfseries,
  breakable
]
\begin{verbatim}
***I.C.E: 62, 73, 74, 75 ***
***E.C/P.C: 57, 59, 60, 61, 63, 64, 65, 68, 70, 71, 72, 76, 77 ***
***T.H: 49, 52, 60, 62, 71, 74, 82, 85 ***
***F: 58, 61, 64, 69, 72, 73, 81, 84, 91, 94, 102, 105 ***

***I.C.E: 60, 61, 62, 63, 71, 72, 73, 74, 75, 76, 77 ***
***E.C/P.C: 57, 59, 48, 49, 50, 62, 63, 64, 65, 68, 70, 71, 72, 76, 77, 
60, 61, 62, 63, 64, 65, 66 ***
***T.H: 49, 52, 71, 74, 82, 85 ***
***F: 58, 69, 72, 81, 84, 91, 94, 102, 105 ***

***I.C.E: 14, 17, 20, 25, 28, 31, 36, 39, 42, 47, 50, 53, 58, 61, 64, 
69, 72, 75, 80, 83, 86, 91, 94, 97, 102, 105, 108 ***
***E.C/P.C: 14, 17, 20, 25, 28, 31, 36, 39, 42, 47, 50, 53, 58, 61, 64, 
69, 72, 75, 80, 83, 86, 91, 94, 97, 102, 105, 108 ***
***T.H: 49, 52, 71, 74, 82, 85 ***
***F: NONE ***

***I.C.E: 14, 17, 20, 25, 28, 31, 36, 39, 42, 47, 50, 53, 58, 61, 64, 
69, 72, 75, 80, 83, 86, 91, 94, 97, 102, 105, 108 ***
***E.C/P.C: 14, 17, 20, 25, 28, 31, 36, 39, 42, 47, 50, 53, 58, 61, 64, 
69, 72, 75, 80, 83, 86, 91, 94, 97, 102, 105, 108 ***
***T.H: 49, 52, 71, 74, 82, 85 ***
***F: 60, 62, 63, 65, 66, 71, 74, 82, 85 ***

***I.C.E: 62, 73, 74, 75 ***
***E.C/P.C: 47, 48, 49, 50, 51, 52, 53, 54, 56, 57, 58, 59, 60, 61, 62, 
63, 64, 65, 66, 67, 68, 69, 70, 71, 72, 73, 74, 75, 76, 77 ***
***T.H: 63, 66 ***
***F: 62, 64, 65, 73, 74 ***

\end{verbatim}
\end{tcolorbox}

\begin{tcolorbox}[
  title={Response: Qwen3 VL 235B A22B Instruct},
  colback=gray!5,
  colframe=blue,
  fonttitle=\bfseries,
  breakable
]
\begin{verbatim}
 ***I.C.E: 57, 58, 59, 60, 61, 62, 63, 64, 65, 68, 69, 70, 71, 72, 73, 
 74, 75, 76 ***
***E.C/P.C: 50, 51, 52, 61, 62, 63 ***
***T.H: 49, 50, 51, 52, 53, 54, 55, 56, 57, 58, 59, 60, 61, 62, 63, 64, 
65, 66, 67, 68, 69, 70, 71, 72, 73, 74, 75, 76, 77 ***
***F: 57, 58, 59, 60, 61, 62, 63, 64, 65, 68, 69, 70, 71, 72, 73, 74, 
75, 76 ***

***I.C.E: 57, 58, 59, 60, 61, 62, 63, 64, 65, 68, 69, 70, 71, 72, 73, 
74, 75, 76 ***
***E.C/P.C: 50, 51, 52, 61, 62, 63 ***
***T.H: 49, 50, 51, 52, 53, 54, 55, 56, 57, 58, 59, 60, 61, 62, 63, 64, 
65, 66, 67, 68, 69, 70, 71, 72, 73, 74, 75, 76, 77 ***
***F: 57, 58, 59, 60, 61, 62, 63, 64, 65, 68, 69, 70, 71, 72, 73, 74, 
75, 76 ***

***I.C.E: 14, 17, 20, 80, 83, 86, 102, 105, 108 ***
***E.C/P.C: 14, 17, 20, 80, 83, 86, 102, 105, 108 ***
***T.H: 49, 51, 71, 73 ***
***F: NONE ***

***I.C.E: 14, 17, 20, 80, 83, 86, 102, 105, 108 ***
***E.C/P.C: 14, 17, 20, 80, 83, 86, 102, 105, 108 ***
***T.H: 49, 51, 71, 73 ***
***F: NONE ***

***I.C.E: 57, 58, 59, 60, 61, 62, 63, 64, 65, 68, 69, 70, 71, 72, 73, 
74, 75, 76 ***
***E.C/P.C: 50, 51, 52, 61, 62, 63 ***
***T.H: 49, 50, 51, 52, 53, 54, 55, 56, 57, 58, 59, 60, 61, 62, 63, 64, 
65, 66, 67, 68, 69, 70, 71, 72, 73, 74, 75, 76, 77 ***
***F: 57, 58, 59, 60, 61, 62, 63, 64, 65, 68, 69, 70, 71, 72, 73, 74, 
75, 76 ***
    
\end{verbatim}
\end{tcolorbox}

\begin{figure}[h]
\centering

\subfloat[Perspective view]{
  \includegraphics[width=0.3\textwidth]{figures/Load_Cases_Figures/00550014_view_e-18_a30_torsion_compression.png}
}
\hfill
\subfloat[Gridded view 1]{
  \includegraphics[width=0.3\textwidth]{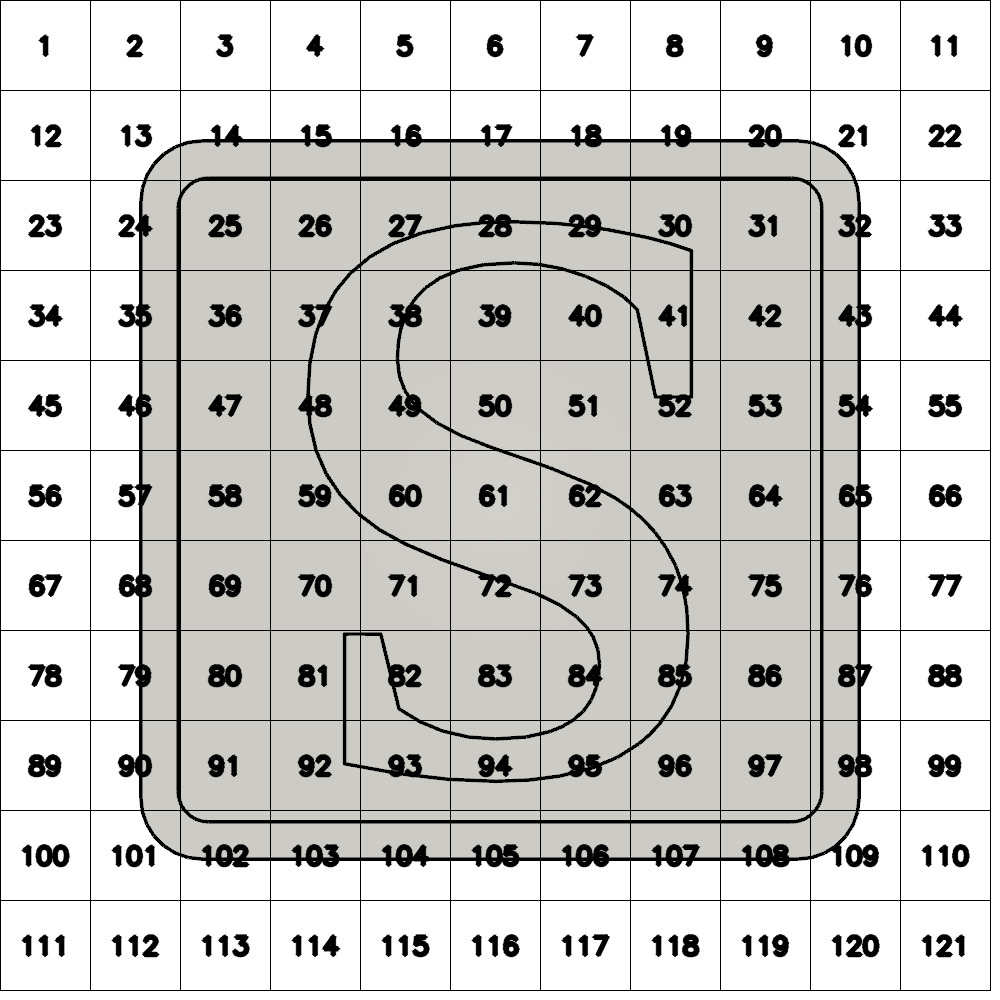}
}
\hfill
\subfloat[Gridded view 2]{
  \includegraphics[width=0.3\textwidth]{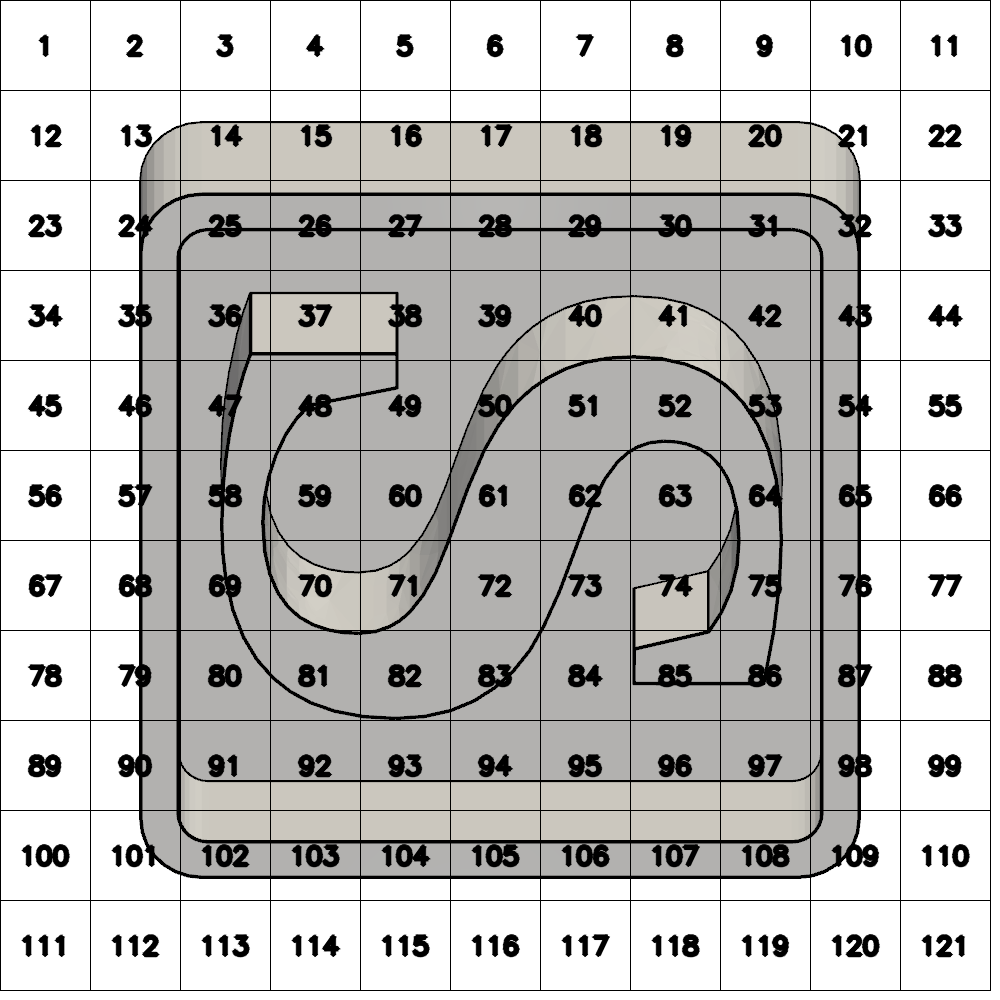}
}

\vspace{1mm}

\subfloat[Gridded view 3]{
  \includegraphics[width=0.3\textwidth]{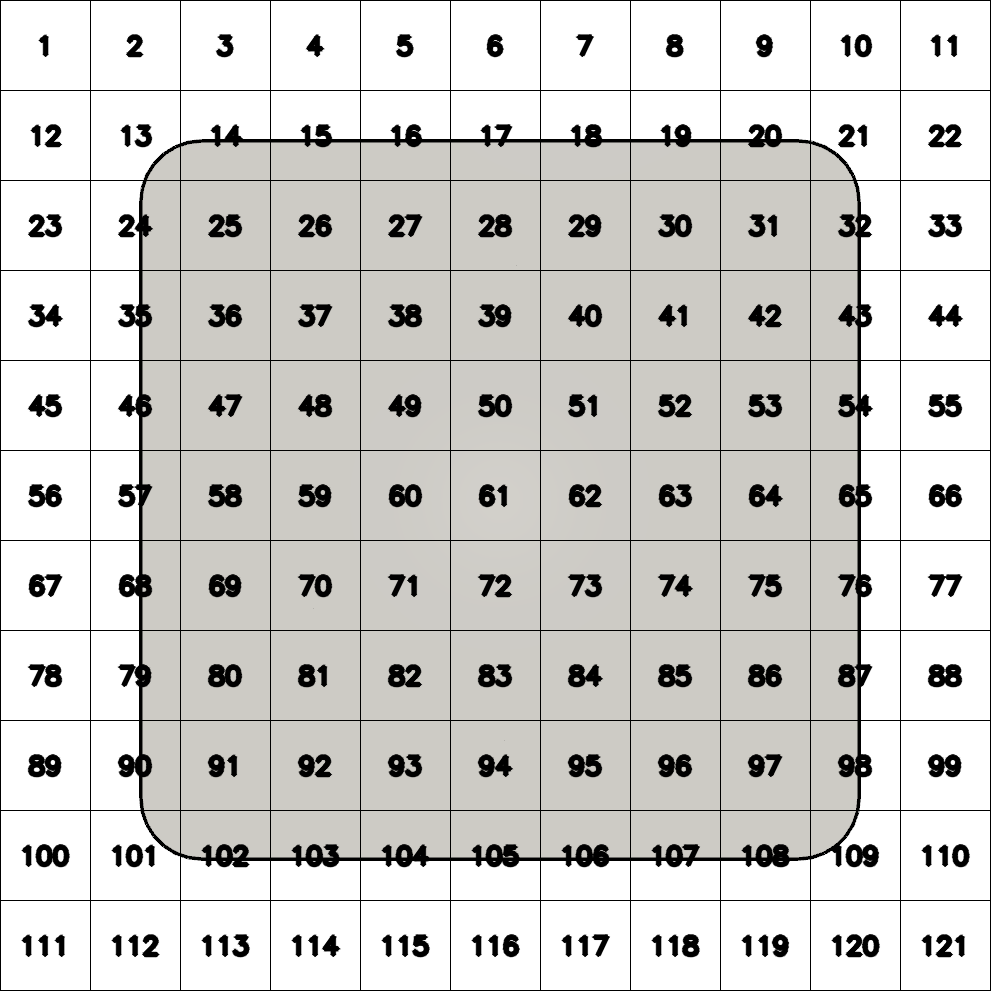}
}
\hfill
\subfloat[Gridded view 4]{
  \includegraphics[width=0.3\textwidth]{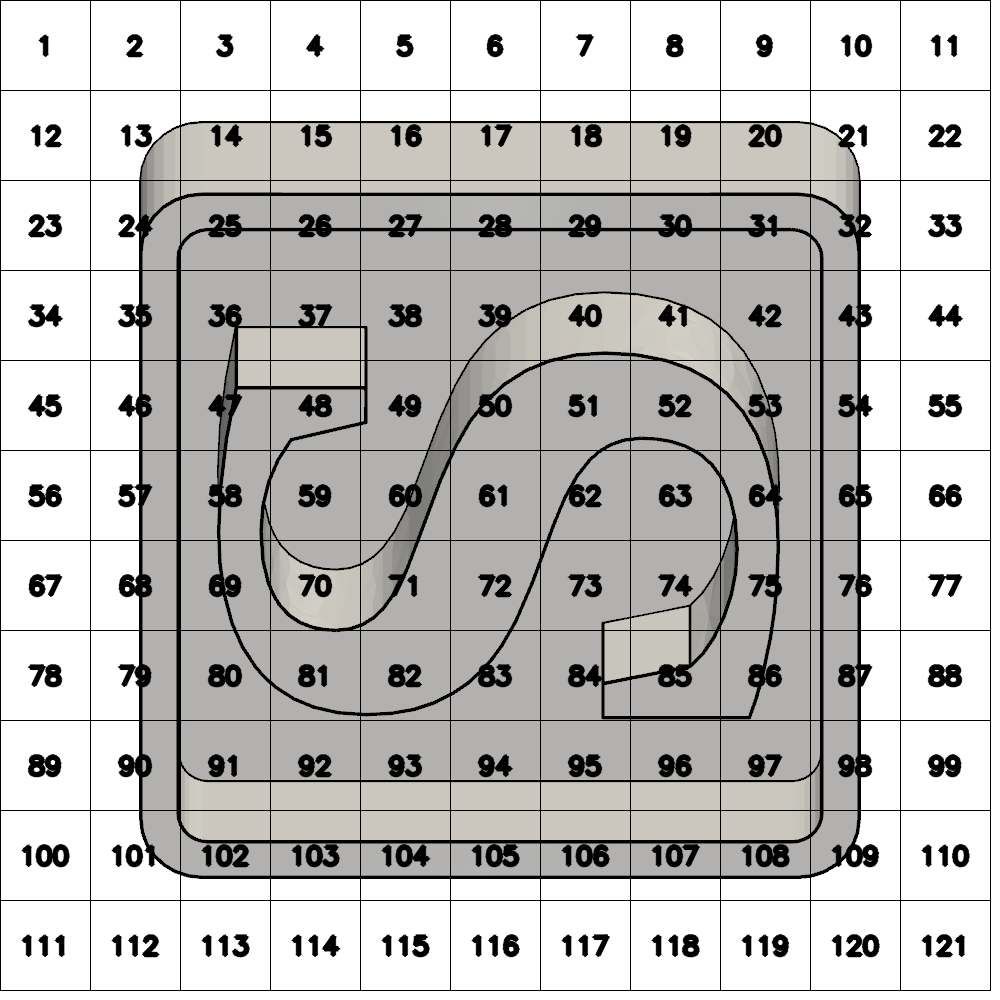}
}
\hfill
\subfloat[Gridded view 5]{
  \includegraphics[width=0.3\textwidth]{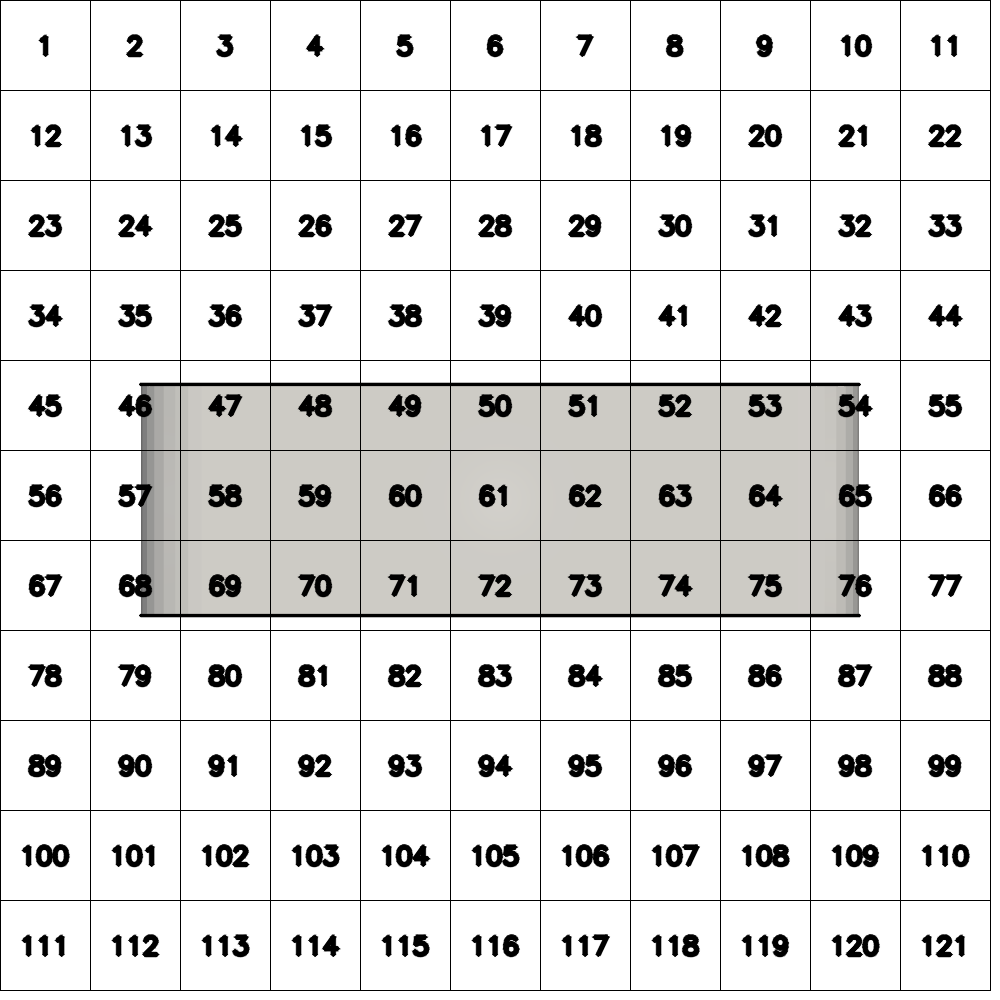}
}
\caption{
Five images provided jointly with example prompt 5.
Top row: non-gridded perspective views for global 3D context.
Bottom row: gridded orthographic views used for cell-level region localization.
}
\label{fig:prompt_images_5}
\end{figure}

\subsubsection{\textbf{Example Prompt 5: Torsion-Compression  (\textit{Load+Features}, views=5, Figure \ref{fig:prompt_images_5})}}

\begin{tcolorbox}[
  title={Response: Gemini 3 Flash Preview},
  colback=gray!5,
  colframe=blue,
  fonttitle=\bfseries,
  breakable
]
\begin{verbatim}
**Image 1 (e-90, a0)**
***I.C.E: NONE***
***E.C/P.C: 27, 28, 29, 30, 31, 38, 39, 40, 41, 42, 49, 50, 51, 52, 53, 
60, 61, 62, 63, 64, 71, 72, 73, 74, 75, 82, 83, 84, 85, 86, 93, 94, 95
***
***T.H: NONE***
***F: NONE***

**Image 2 (e-72, a270)**
***I.C.E: 36, 37, 38, 47, 58, 69, 70, 71, 73, 74, 85, 86***
***E.C/P.C: 37, 38, 39, 40, 41, 42, 48, 49, 50, 51, 52, 53, 59, 60, 61, 
62, 63, 64, 71, 72, 73, 74, 75, 81, 82, 83, 84, 85***
***T.H: NONE***
***F: 47, 48, 50, 51, 52, 53, 58, 59, 61, 62, 63, 64, 70, 72, 73, 75, 
81, 82, 83, 84***

**Image 3 (e90, a0)**
***I.C.E: NONE***
***E.C/P.C: NONE***
***T.H: NONE***
***F: NONE***

**Image 4 (e-72, a90)**
***I.C.E: 36, 37, 38, 47, 58, 69, 70, 71, 73, 74, 85, 86***
***E.C/P.C: 37, 38, 39, 40, 41, 42, 48, 49, 50, 51, 52, 53, 59, 60, 61, 
62, 63, 64, 71, 72, 73, 74, 75, 81, 82, 83, 84, 85***
***T.H: NONE***
***F: 47, 48, 50, 51, 52, 53, 58, 59, 61, 62, 63, 64, 70, 72, 73, 75, 
81, 82, 83, 84***

**Image 5 (e0, a0)**
***I.C.E: NONE***
***E.C/P.C: NONE***
***T.H: NONE***
***F: NONE***
\end{verbatim}
\end{tcolorbox}

\begin{tcolorbox}[
  title={Response: Claude Haiku 4.5},
  colback=gray!5,
  colframe=blue,
  fonttitle=\bfseries,
  breakable
]
\begin{verbatim}
# Analysis of Stress-Critical Features

## Image 1: (e-90, a0) - Bottom View

***I.C.E: 24, 32, 35, 43***
***E.C/P.C: 37, 48, 49, 52, 53***
***T.H: 40, 41, 51, 52, 62, 63***
***F: 40, 41, 51, 52, 62, 63***

## Image 2: (e-72, a270) - Side View

***I.C.E: 24, 25, 32, 33***
***E.C/P.C: 36, 37, 42, 43***
***T.H: NONE***
***F: 25, 26, 31, 32***

## Image 3: (e90, a0) - Top View

***I.C.E: 14, 21, 24, 32***
***E.C/P.C: NONE***
***T.H: NONE***
***F: 14, 21, 24, 32***

## Image 4: (e-72, a90) - Side View

***I.C.E: 24, 25, 32, 33***
***E.C/P.C: 36, 37, 42, 43, 46, 47, 53, 54***
***T.H: NONE***
***F: 25, 26, 31, 32, 46, 47, 53, 54***

## Image 5: (e0, a0) - Front View

***I.C.E: 46, 54***
***E.C/P.C: NONE***
***T.H: NONE***
***F: 46, 47, 53, 54***
\end{verbatim}
\end{tcolorbox}

\begin{tcolorbox}[
  title={Response: GPT 5.4 mini},
  colback=gray!5,
  colframe=blue,
  fonttitle=\bfseries,
  breakable
]
\begin{verbatim}
 ***I.C.E: 36, 37, 47, 48, 58, 59, 69, 70, 81, 82, 92, 93 ***
***E.C/P.C: 36, 37, 48, 49, 60, 61, 73, 74, 84, 85, 94, 95 ***
***T.H: NONE ***
***F: 40, 41, 51, 52, 62, 63, 74, 75 ***

***I.C.E: 36, 37, 47, 48, 58, 59, 69, 70, 81, 82, 92, 93 ***
***E.C/P.C: 36, 37, 48, 49, 60, 61, 73, 74, 84, 85, 94, 95 ***
***T.H: NONE ***
***F: 40, 41, 51, 52, 62, 63, 74, 75 ***

***I.C.E: NONE ***
***E.C/P.C: NONE ***
***T.H: NONE ***
***F: NONE ***

***I.C.E: NONE ***
***E.C/P.C: NONE ***
***T.H: NONE ***
***F: NONE ***

***I.C.E: NONE ***
***E.C/P.C: NONE ***
***T.H: NONE ***
***F: NONE ***
\end{verbatim}
\end{tcolorbox}

\begin{tcolorbox}[
  title={Response: Qwen3 VL 235B A22B Instruct},
  colback=gray!5,
  colframe=blue,
  fonttitle=\bfseries,
  breakable
]
\begin{verbatim}
***I.C.E: 30, 41, 82, 93 ***
***E.C/P.C: 36, 37, 38, 84, 85, 86 ***
***T.H: NONE ***
***F: 14, 21, 102, 109 ***

***I.C.E: 30, 41, 82, 93 ***
***E.C/P.C: 36, 37, 38, 84, 85, 86 ***
***T.H: NONE ***
***F: 14, 21, 102, 109 ***

***I.C.E: NONE ***
***E.C/P.C: NONE ***
***T.H: NONE ***
***F: 14, 21, 102, 109 ***

***I.C.E: 30, 41, 82, 93 ***
***E.C/P.C: 36, 37, 38, 84, 85, 86 ***
***T.H: NONE ***
***F: 14, 21, 102, 109 ***

***I.C.E: NONE ***
***E.C/P.C: NONE ***
***T.H: NONE ***
***F: NONE ***
\end{verbatim}
\end{tcolorbox}

\clearpage
\section{CAD Geometry Testset}
\label{app:testset}
The evaluation benchmark comprises 156 CAD parts $\times$ 5 load cases (780 paired cases), drawn from the ABC dataset~\cite{koch2019abc} and FreeCAD~\cite{riegel2016freecad}, disjoint from the \textit{orthoViews} training and validation splits. Figures~\ref{fig:cad_dataset_grid}--\ref{fig:cad_dataset_grid_4} show all 156 parts.

\begin{figure}[h]
    \centering
    \includegraphics[width=0.117\linewidth]{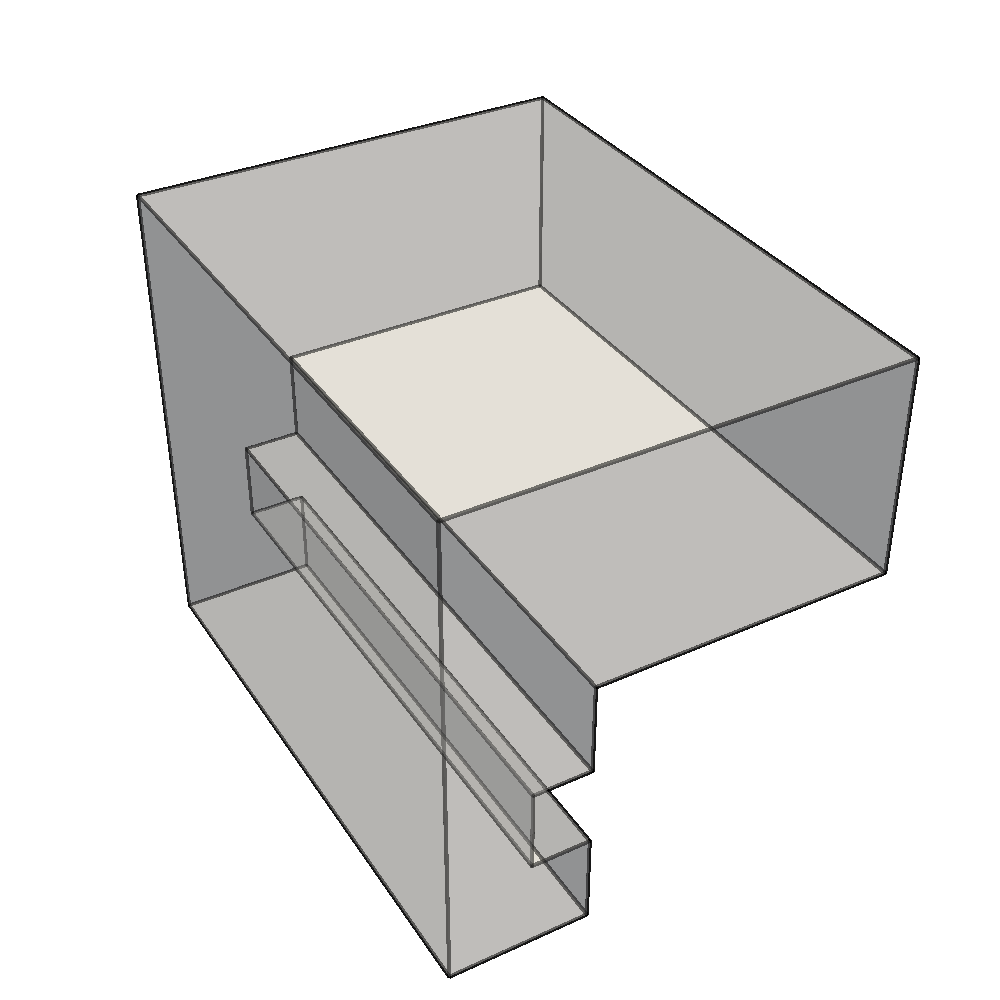}\hfill
    \includegraphics[width=0.117\linewidth]{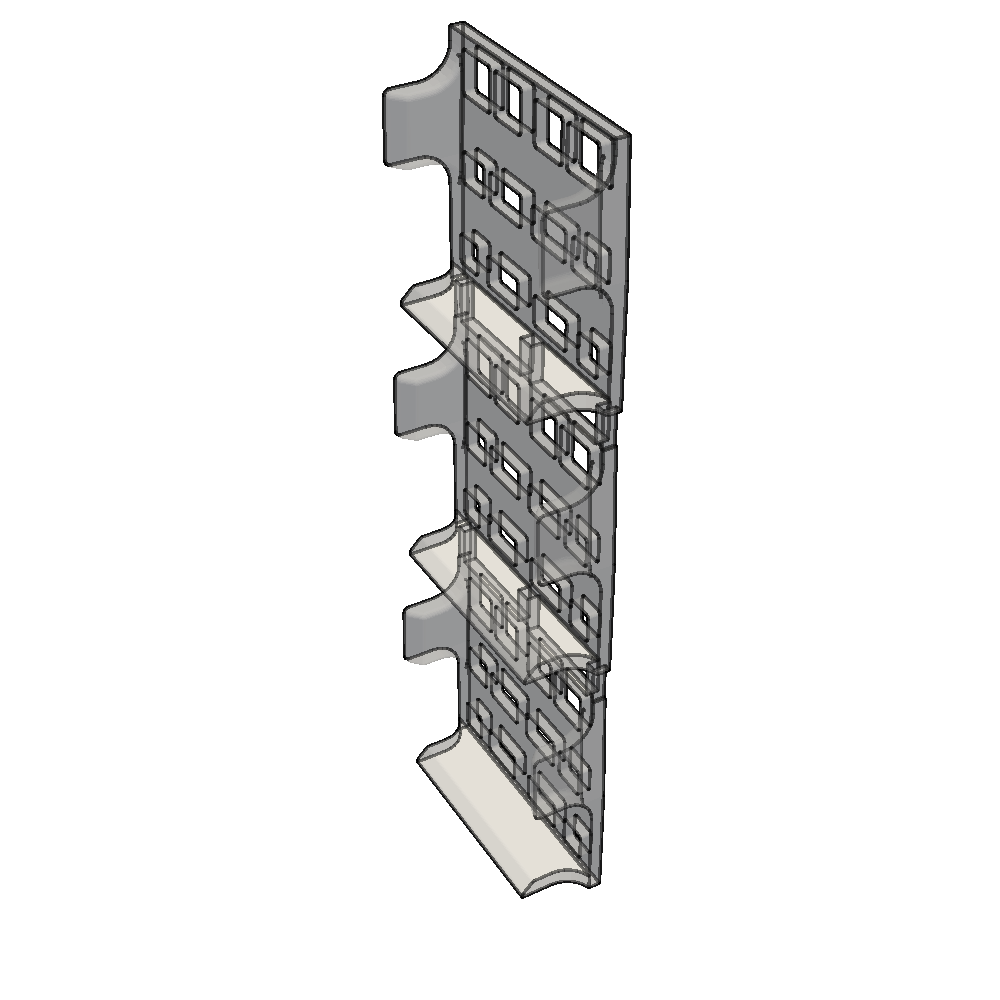}\hfill
    \includegraphics[width=0.117\linewidth]{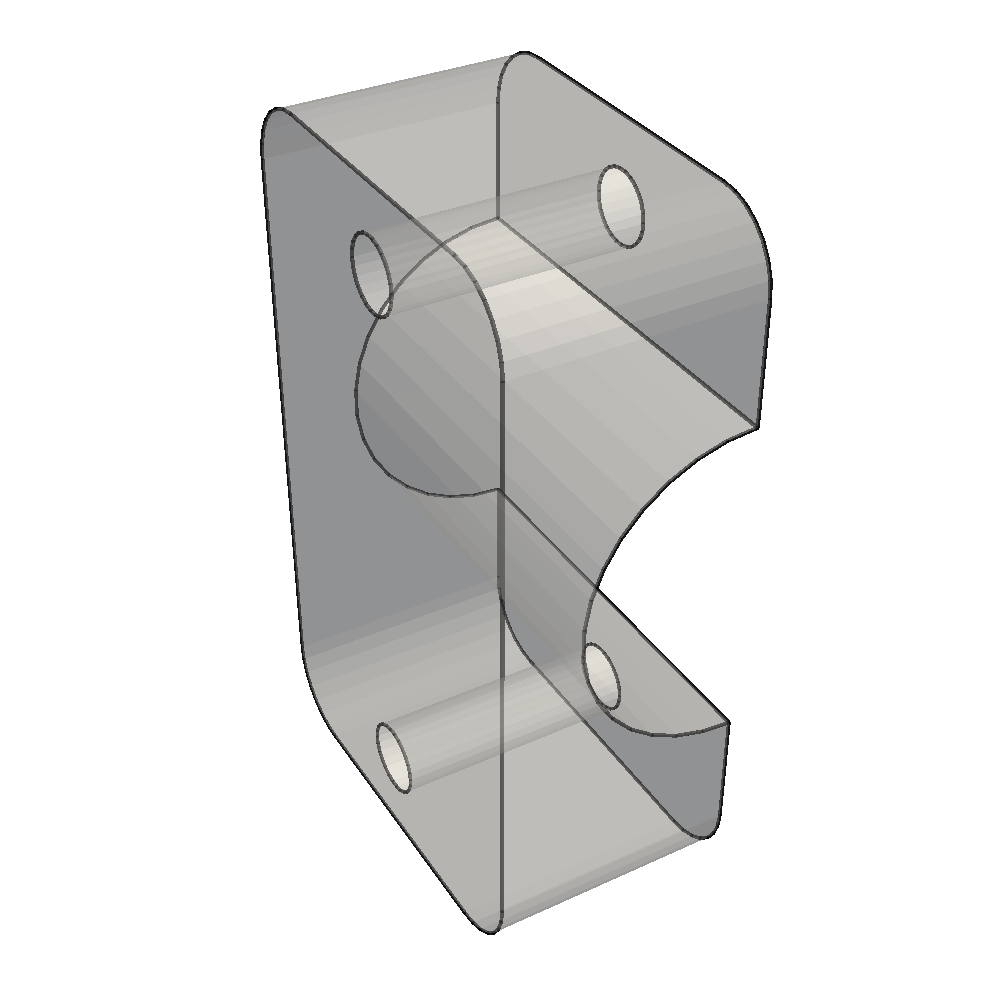}\hfill
    \includegraphics[width=0.117\linewidth]{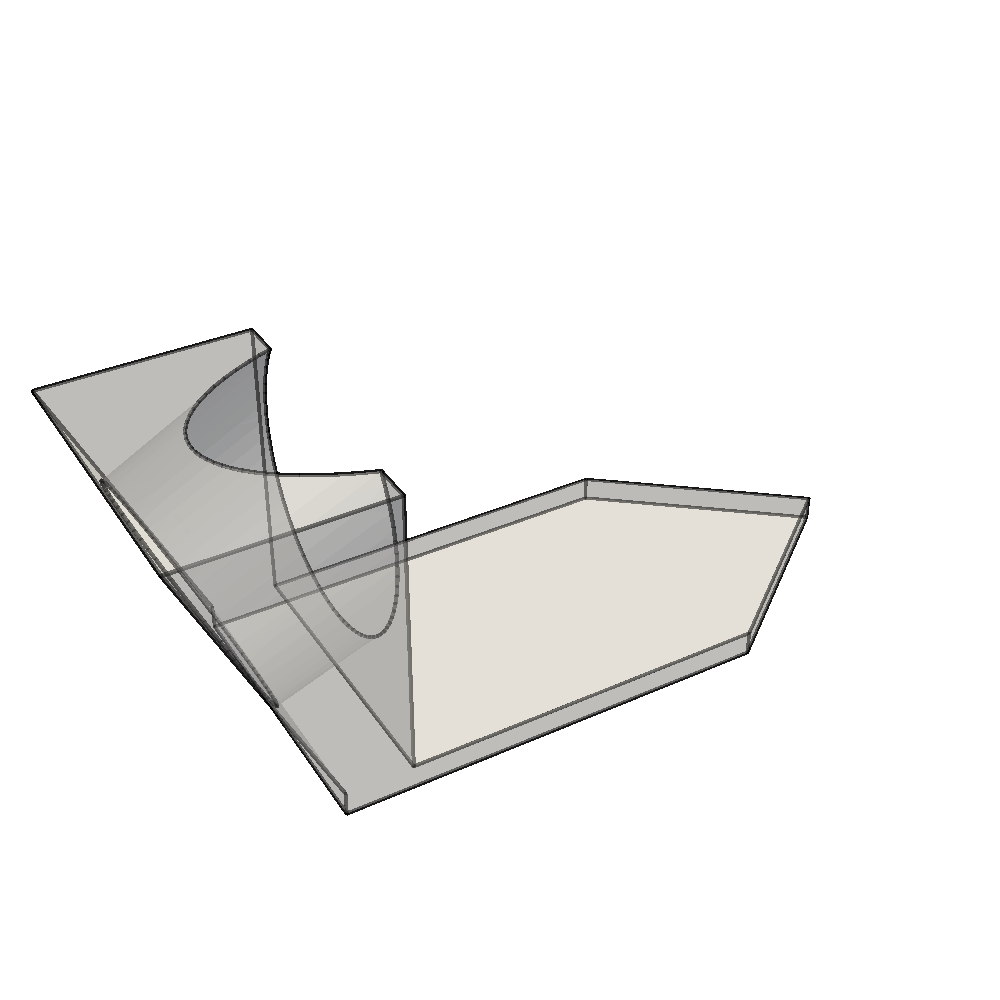}\hfill
    \includegraphics[width=0.117\linewidth]{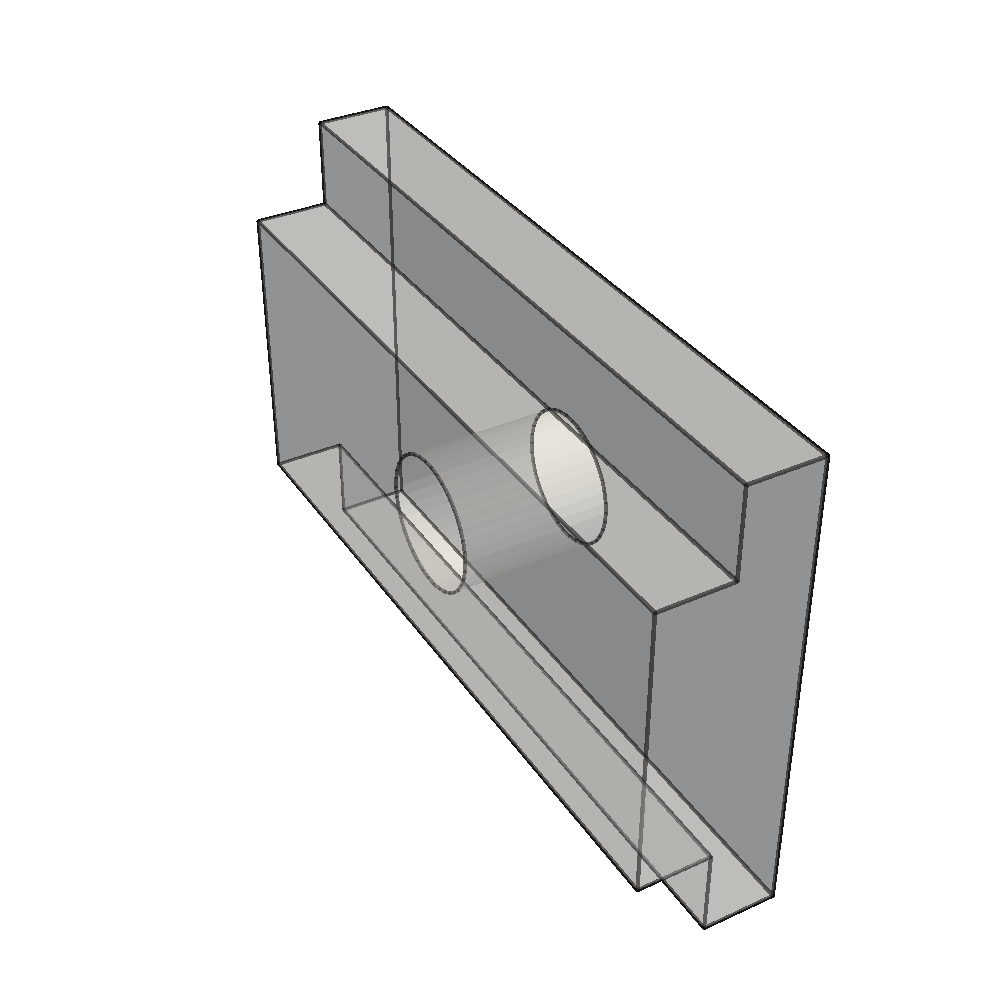}\hfill
    \includegraphics[width=0.117\linewidth]{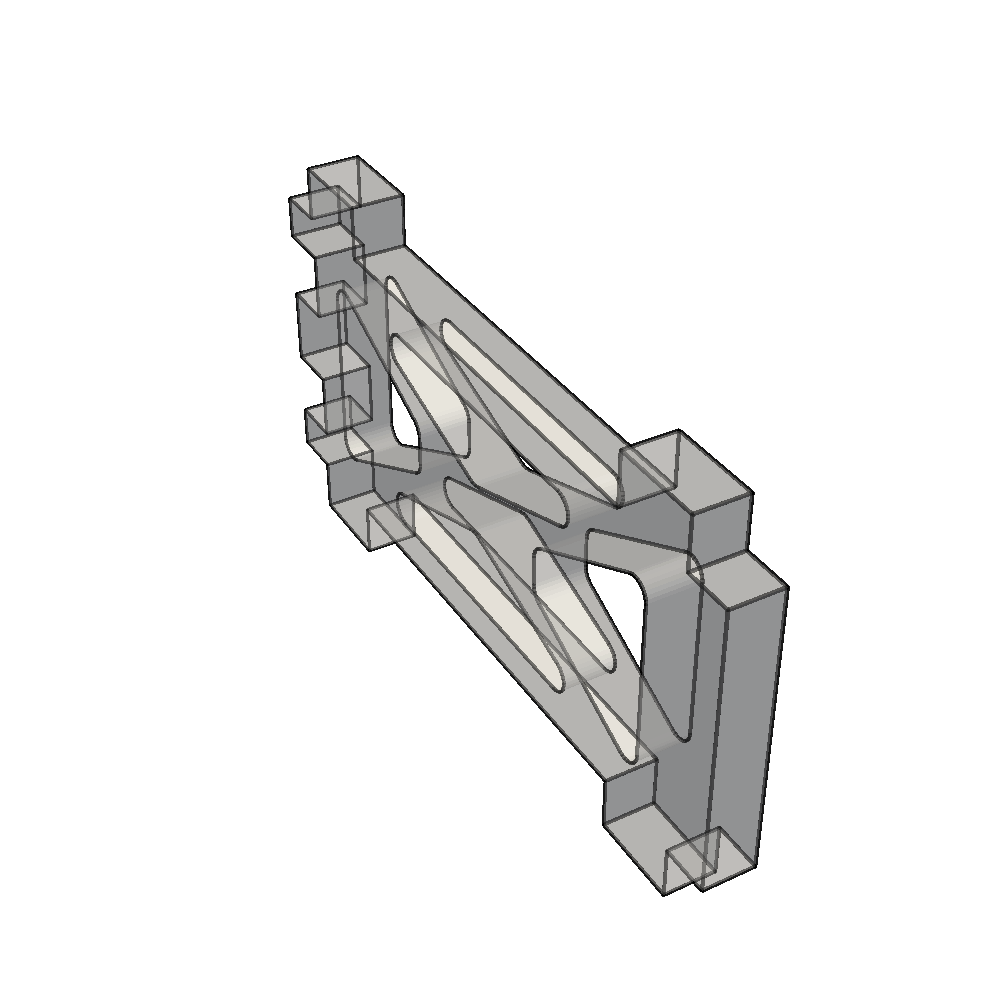}\hfill
    \includegraphics[width=0.117\linewidth]{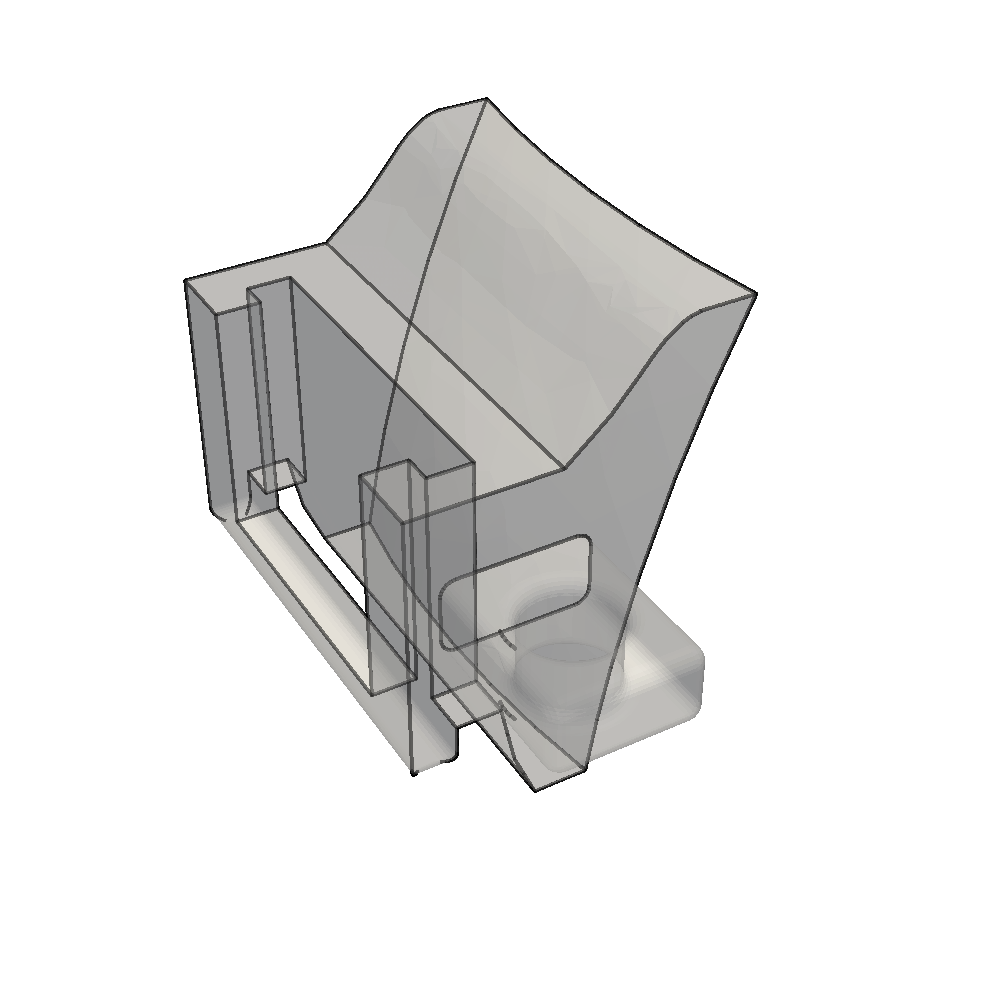}\hfill
    \includegraphics[width=0.117\linewidth]{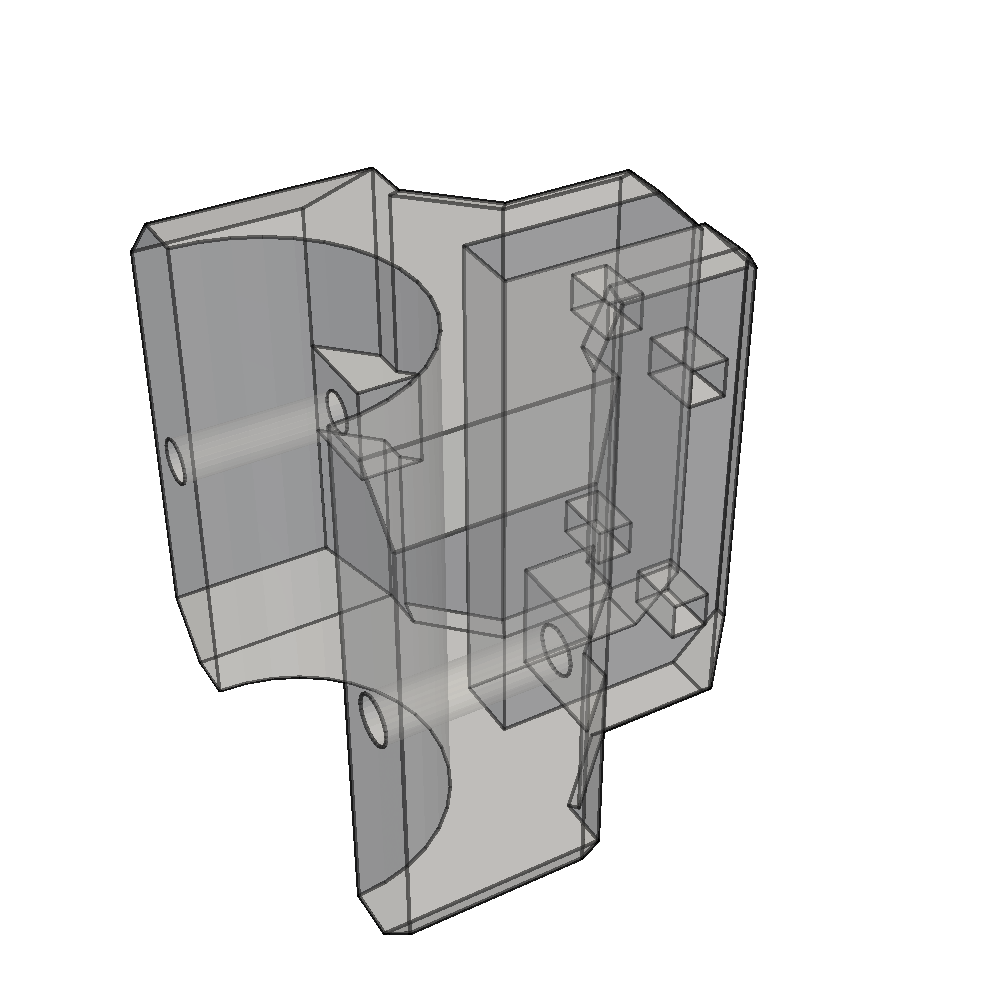}\\[\smallskipamount]
    \includegraphics[width=0.117\linewidth]{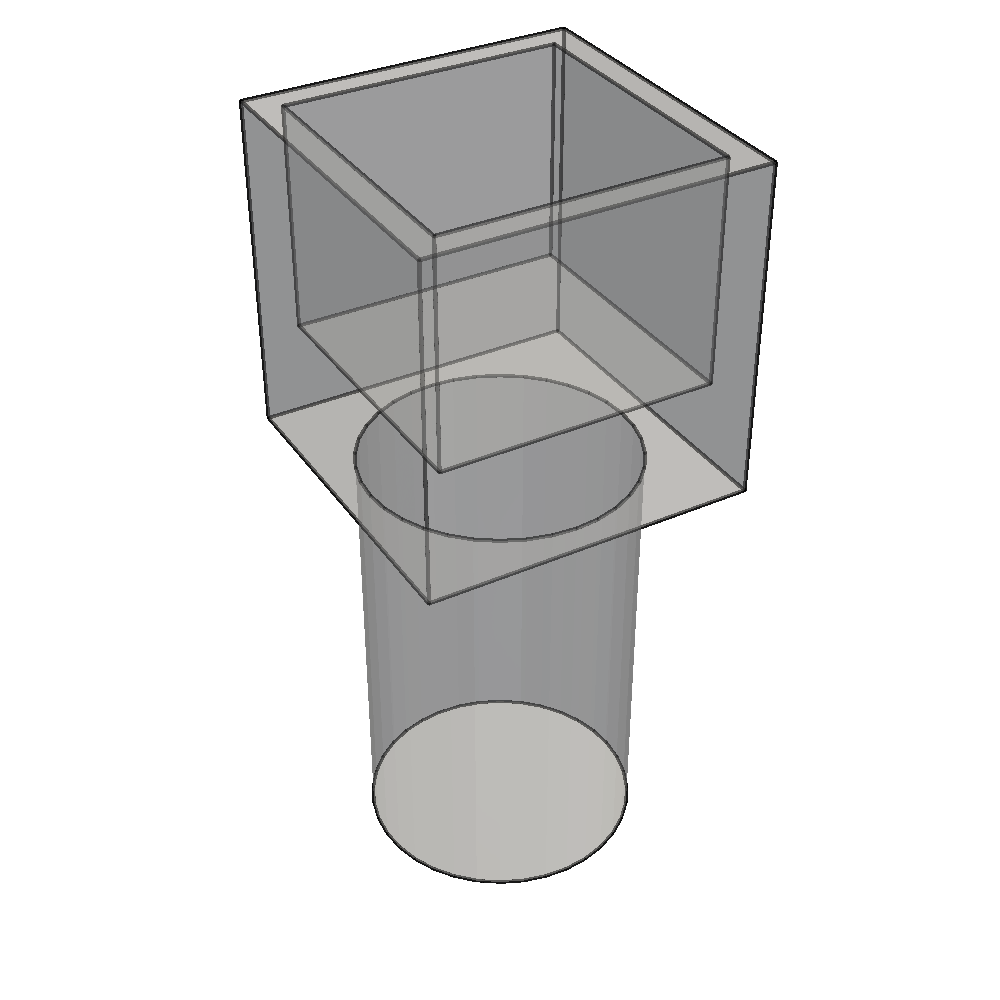}\hfill
    \includegraphics[width=0.117\linewidth]{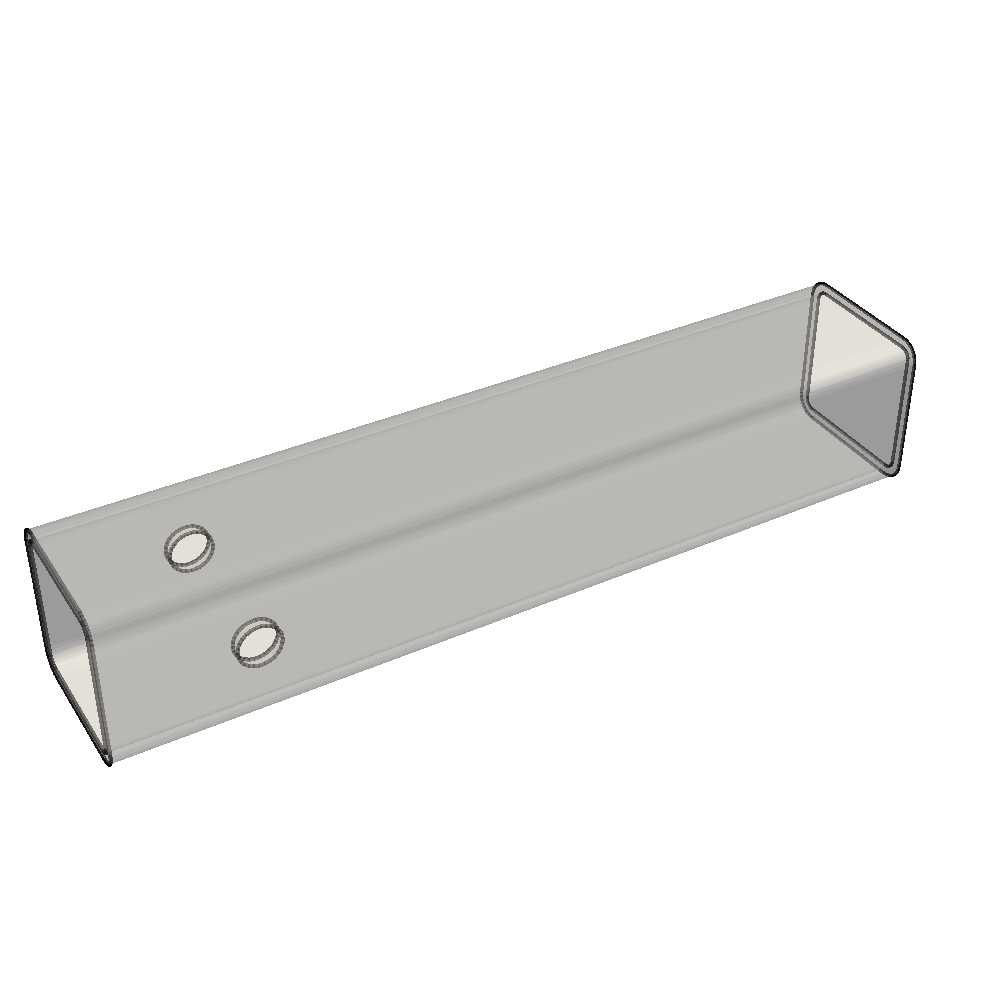}\hfill
    \includegraphics[width=0.117\linewidth]{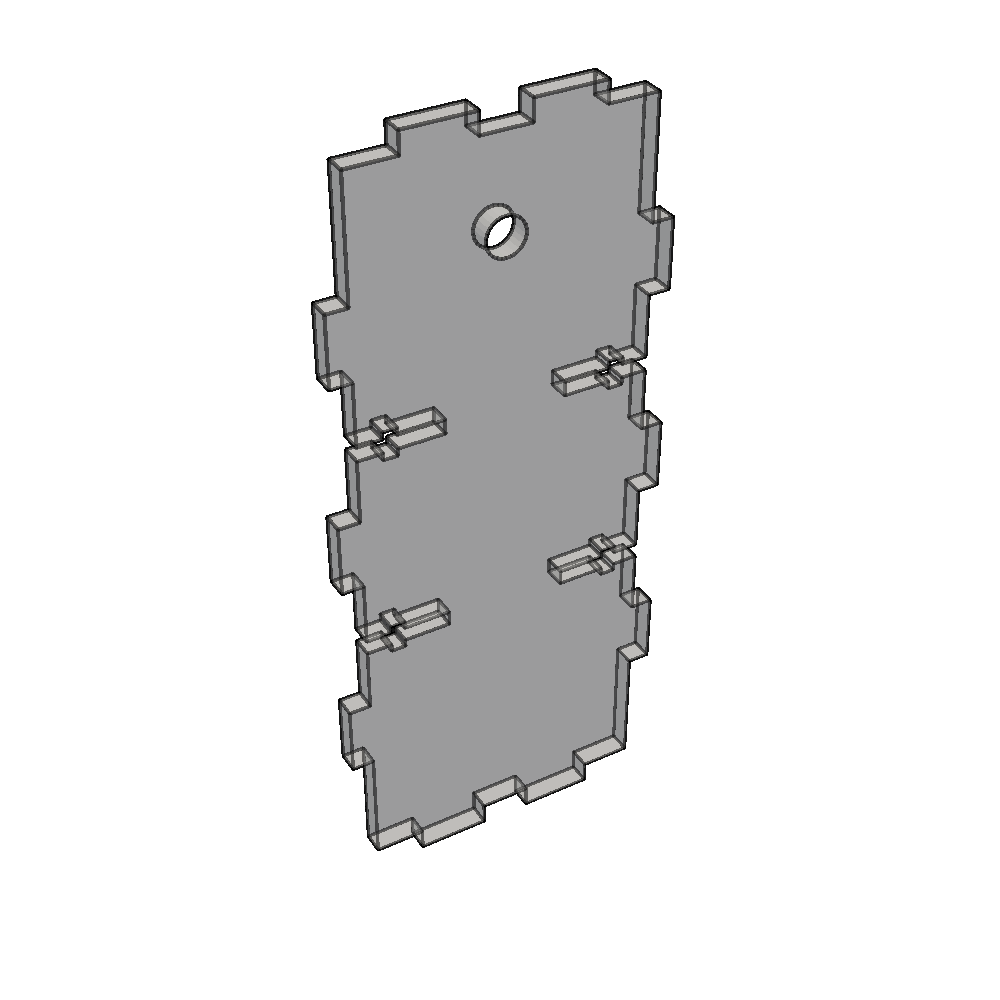}\hfill
    \includegraphics[width=0.117\linewidth]{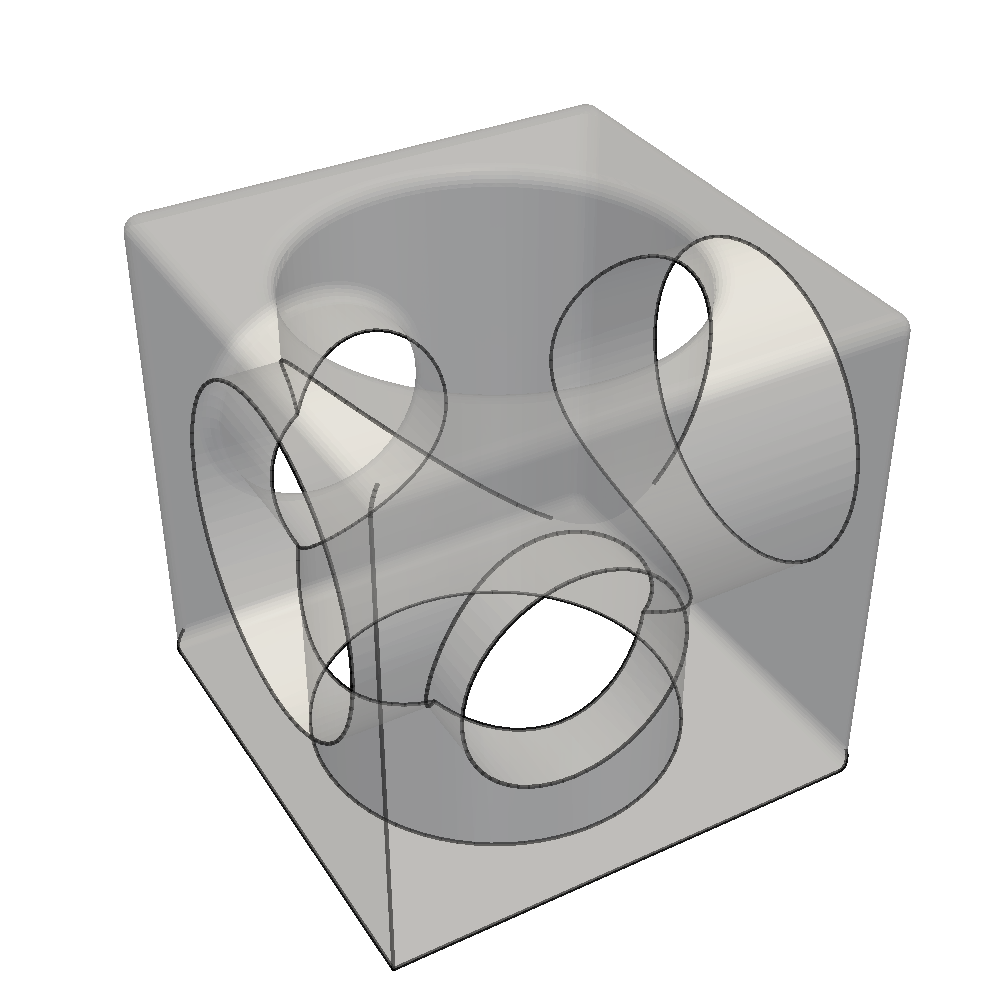}\hfill
    \includegraphics[width=0.117\linewidth]{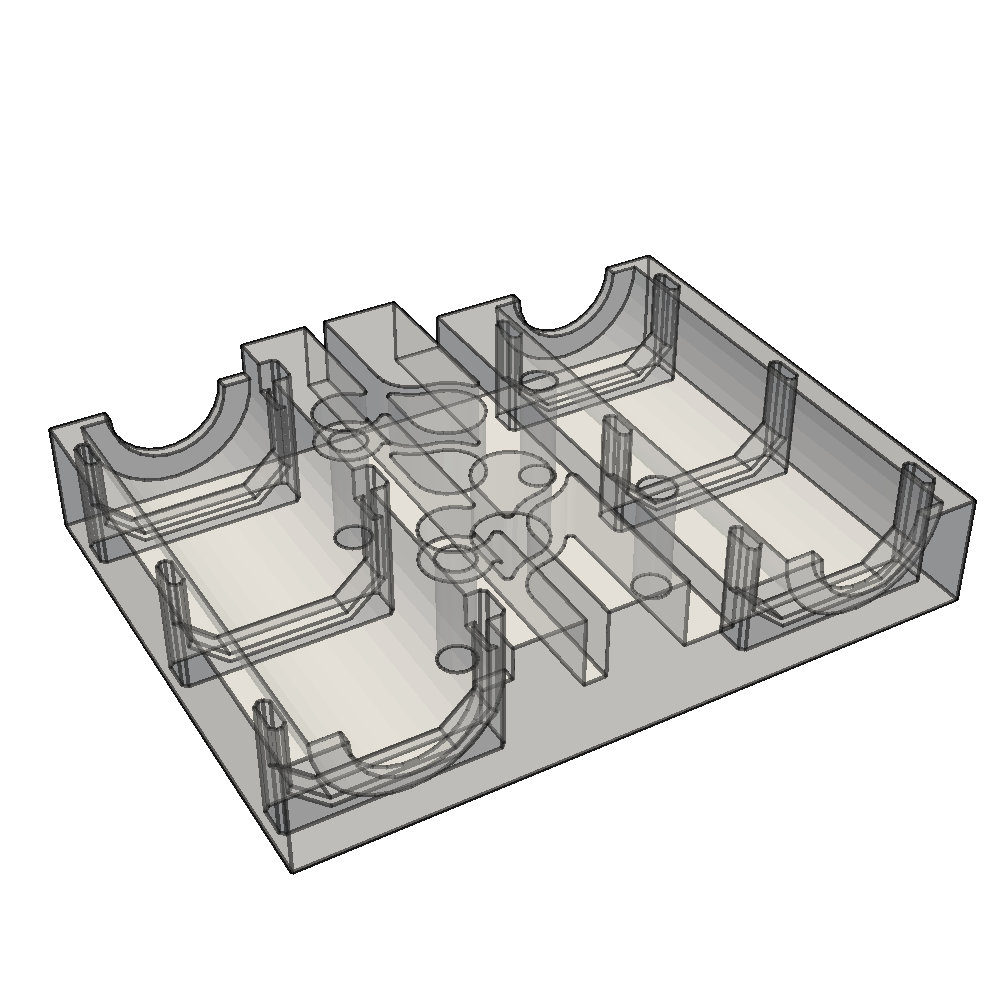}\hfill
    \includegraphics[width=0.117\linewidth]{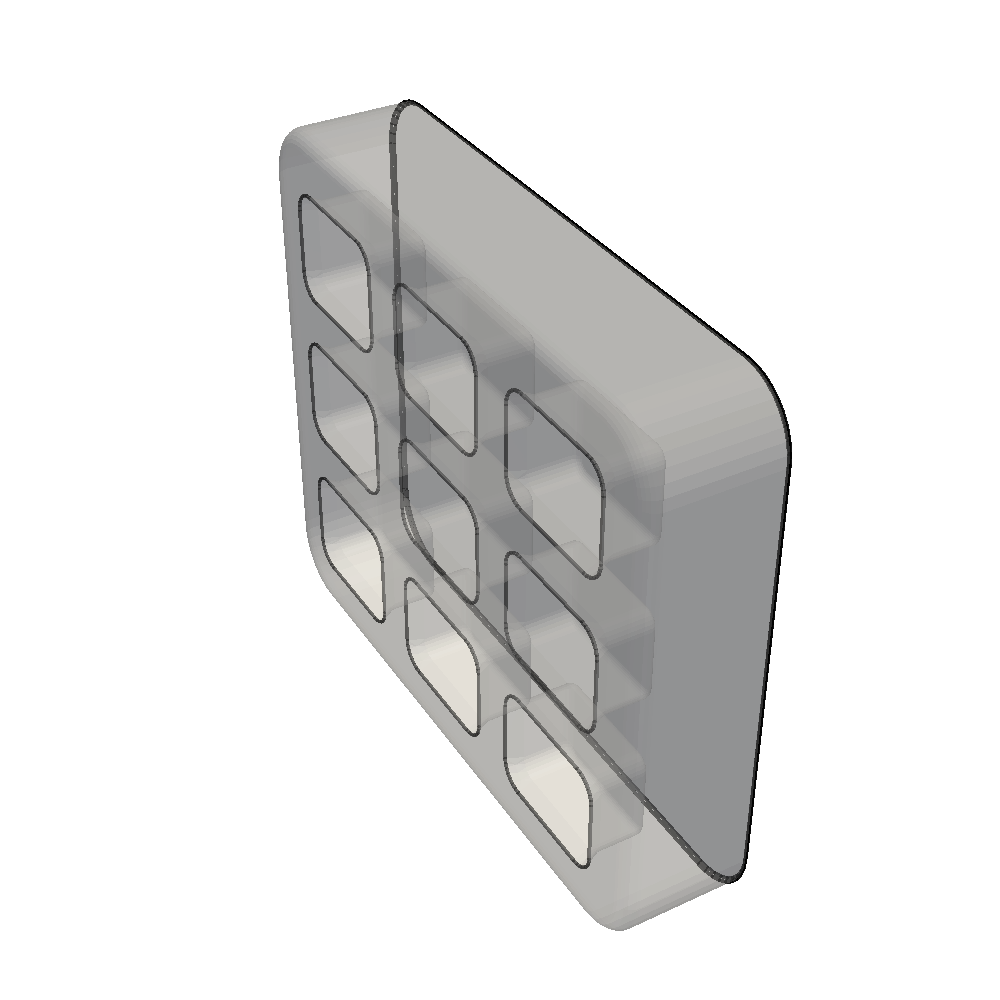}\hfill
    \includegraphics[width=0.117\linewidth]{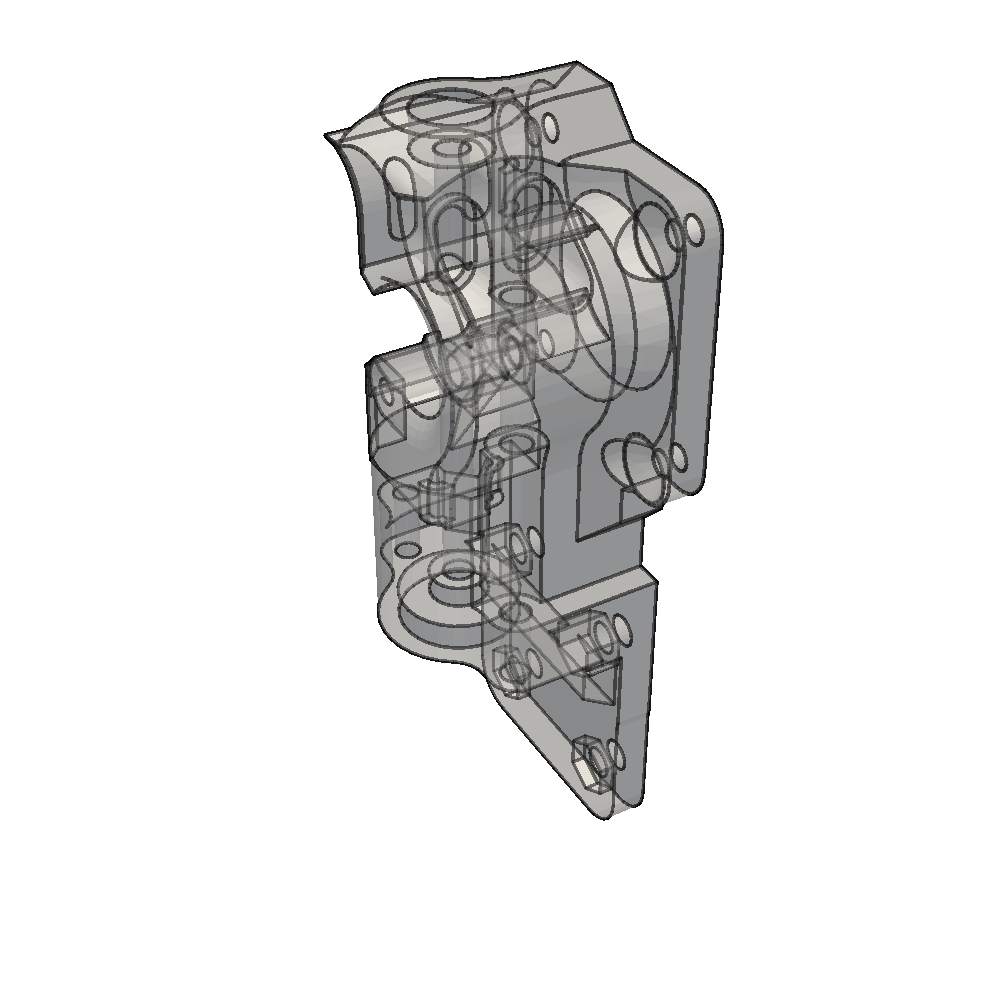}\hfill
    \includegraphics[width=0.117\linewidth]{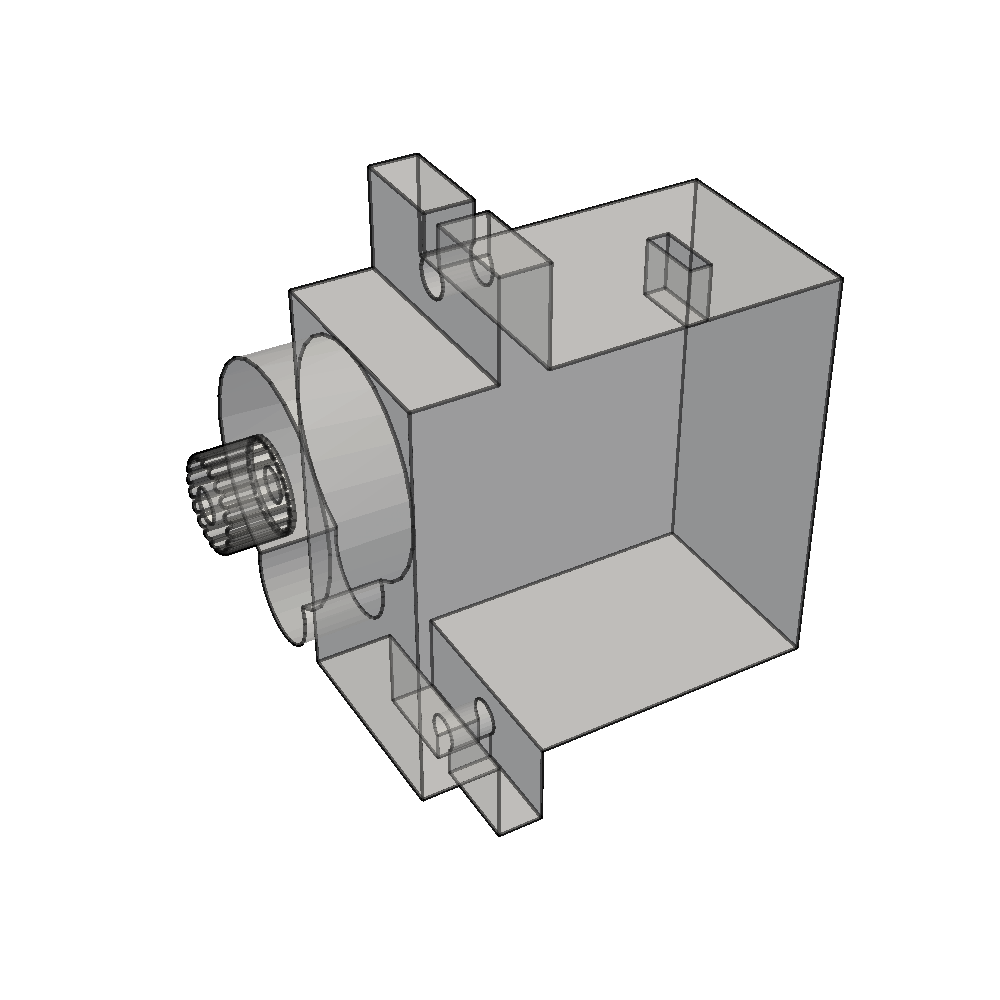}\\[\smallskipamount]
    \includegraphics[width=0.117\linewidth]{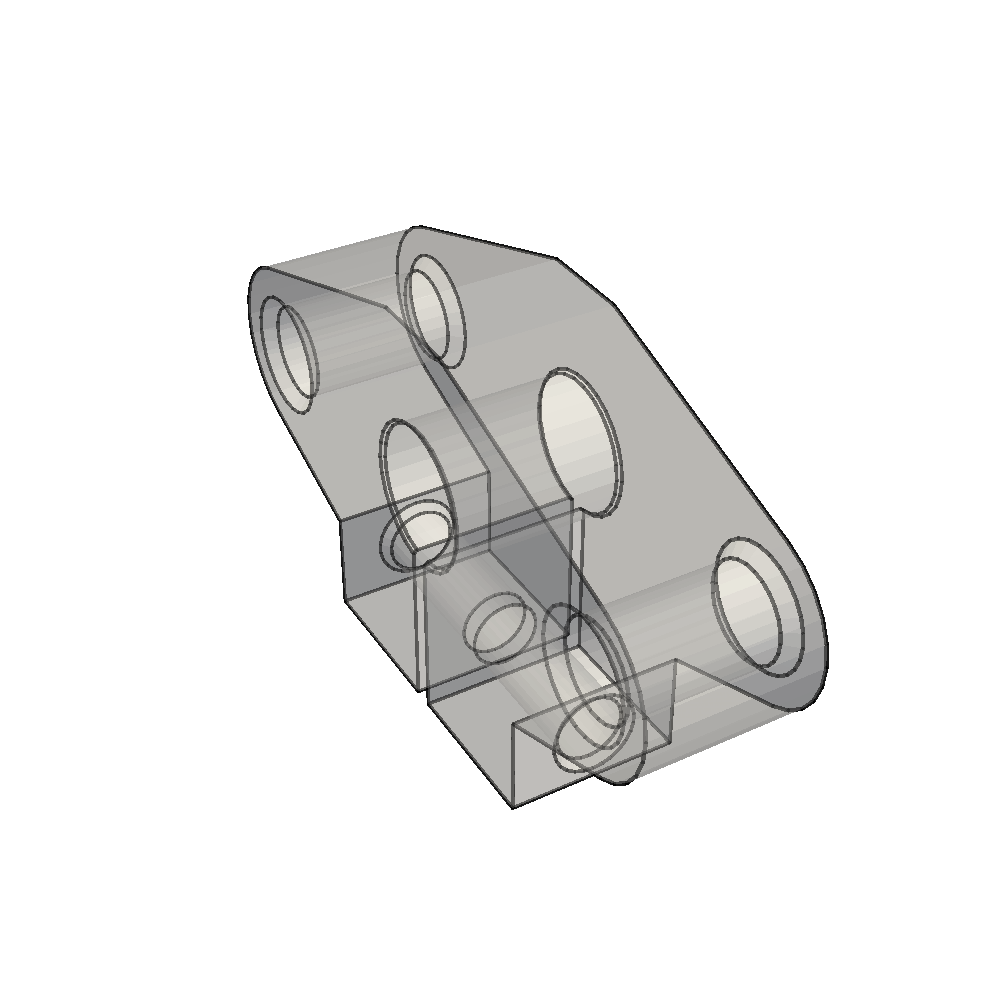}\hfill
    \includegraphics[width=0.117\linewidth]{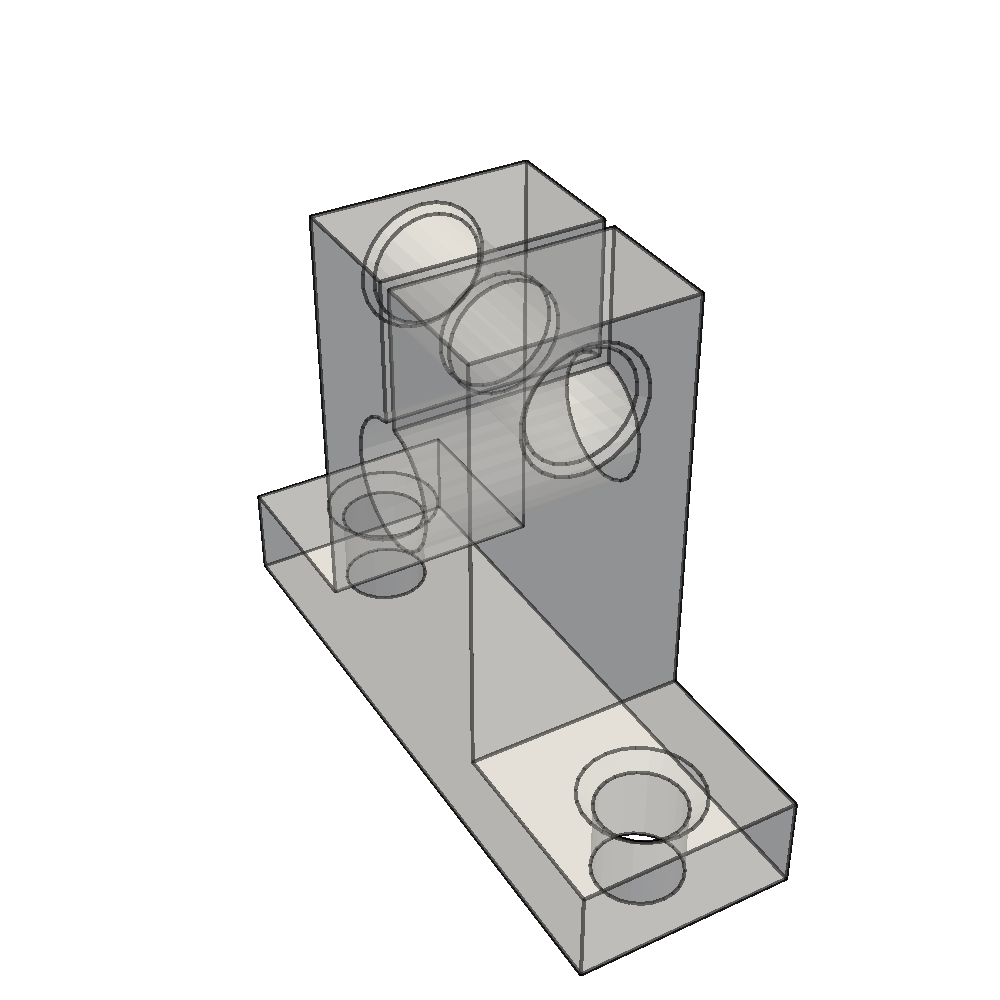}\hfill
    \includegraphics[width=0.117\linewidth]{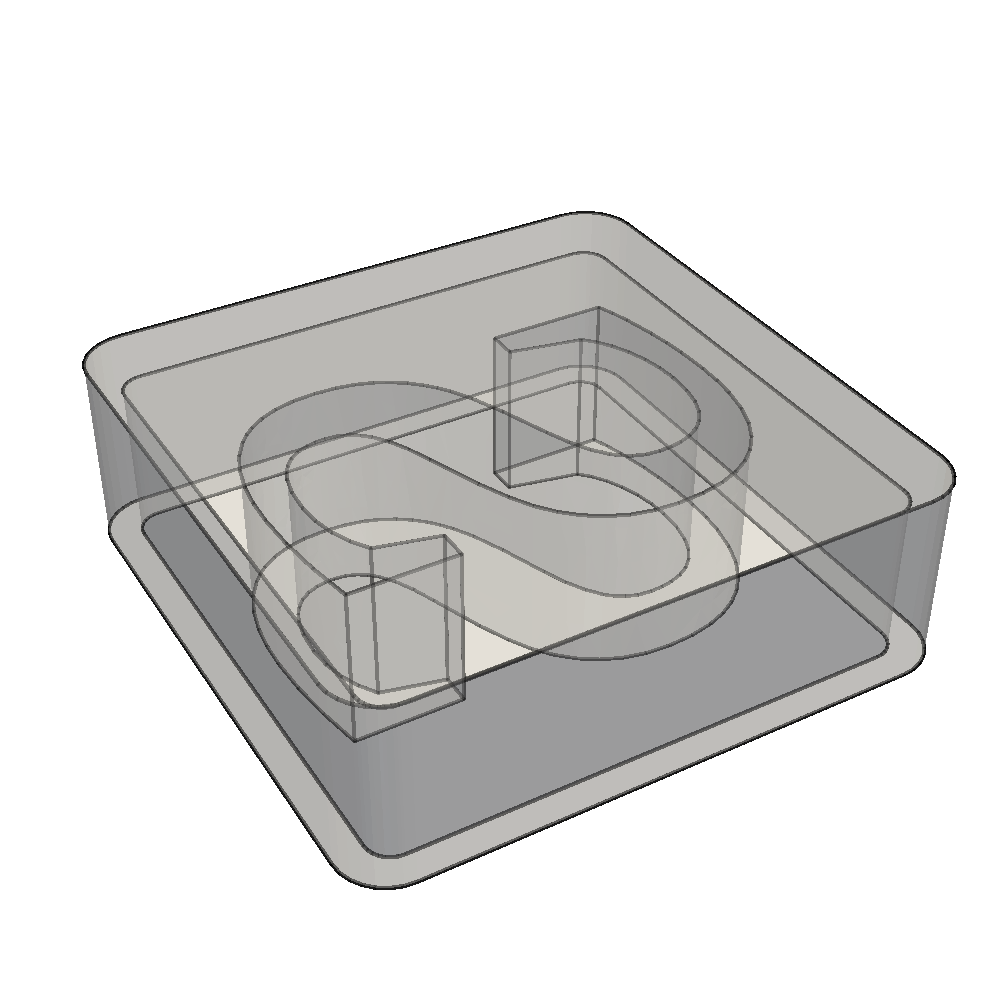}\hfill
    \includegraphics[width=0.117\linewidth]{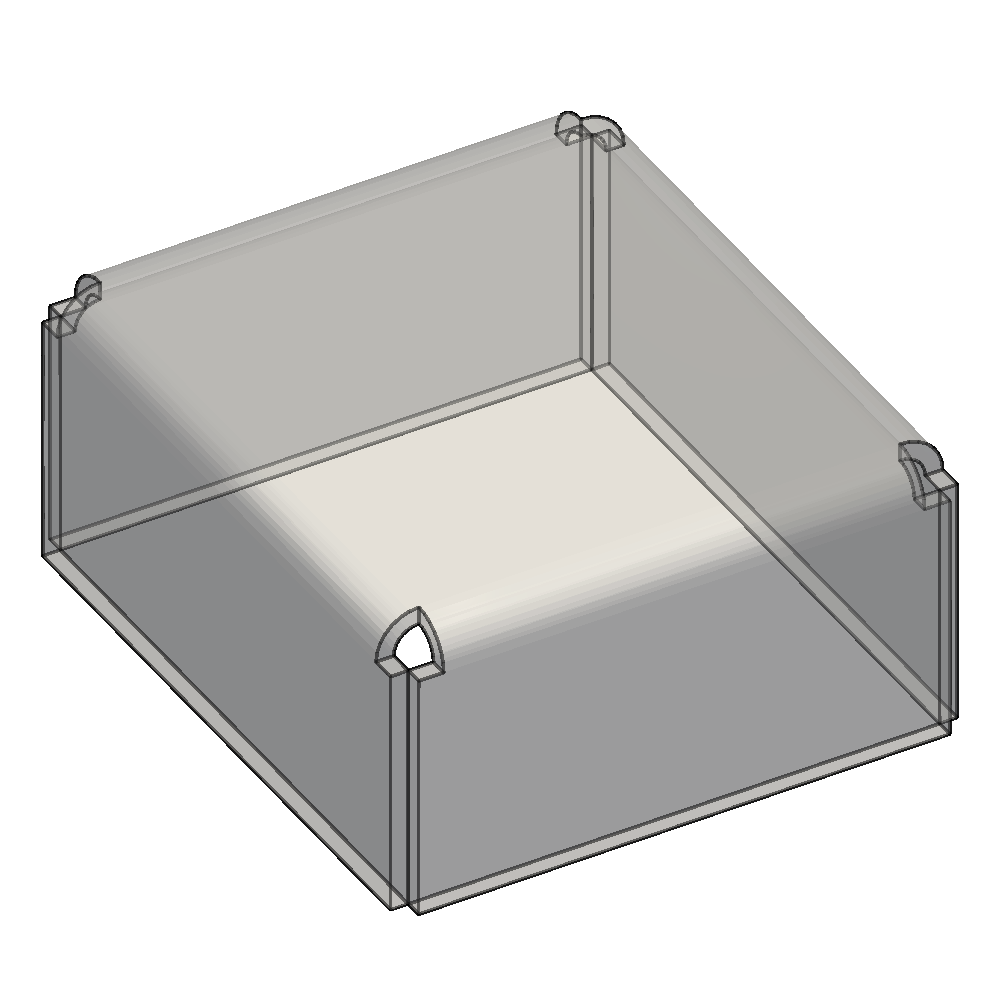}\hfill
    \includegraphics[width=0.117\linewidth]{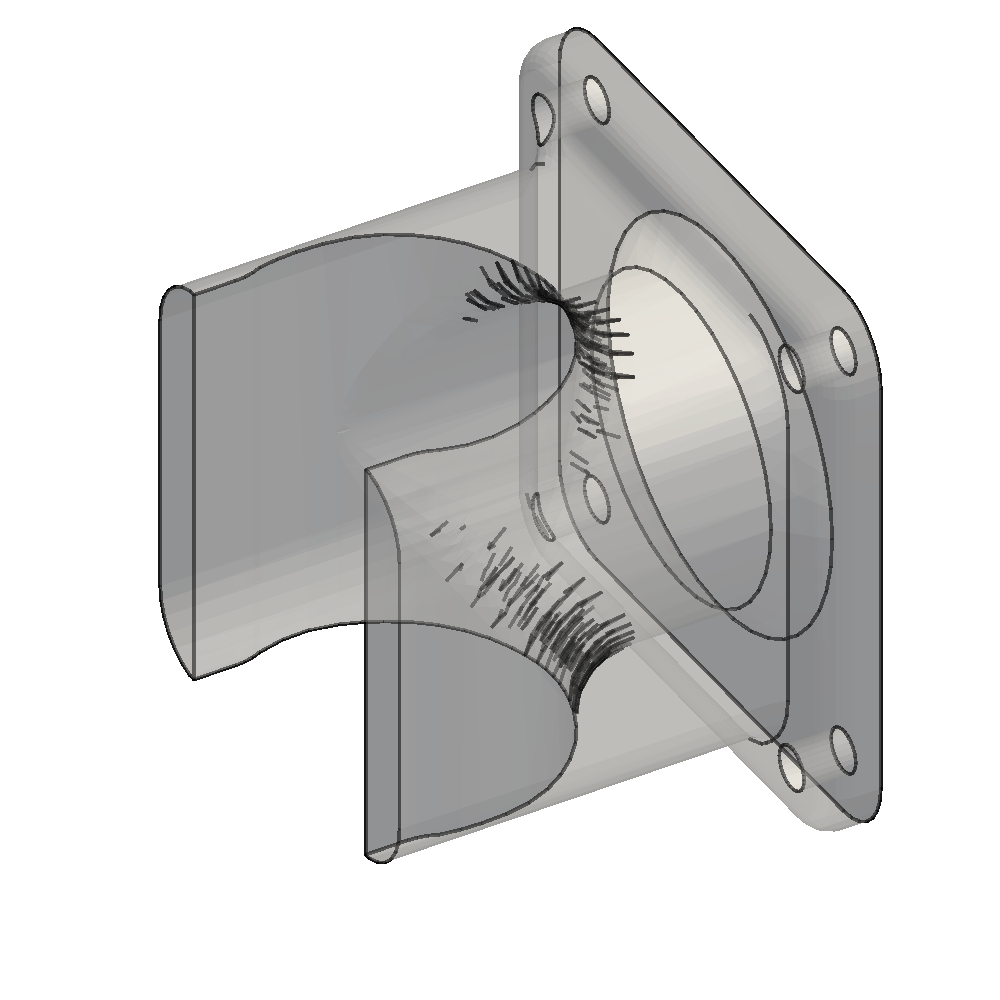}\hfill
    \includegraphics[width=0.117\linewidth]{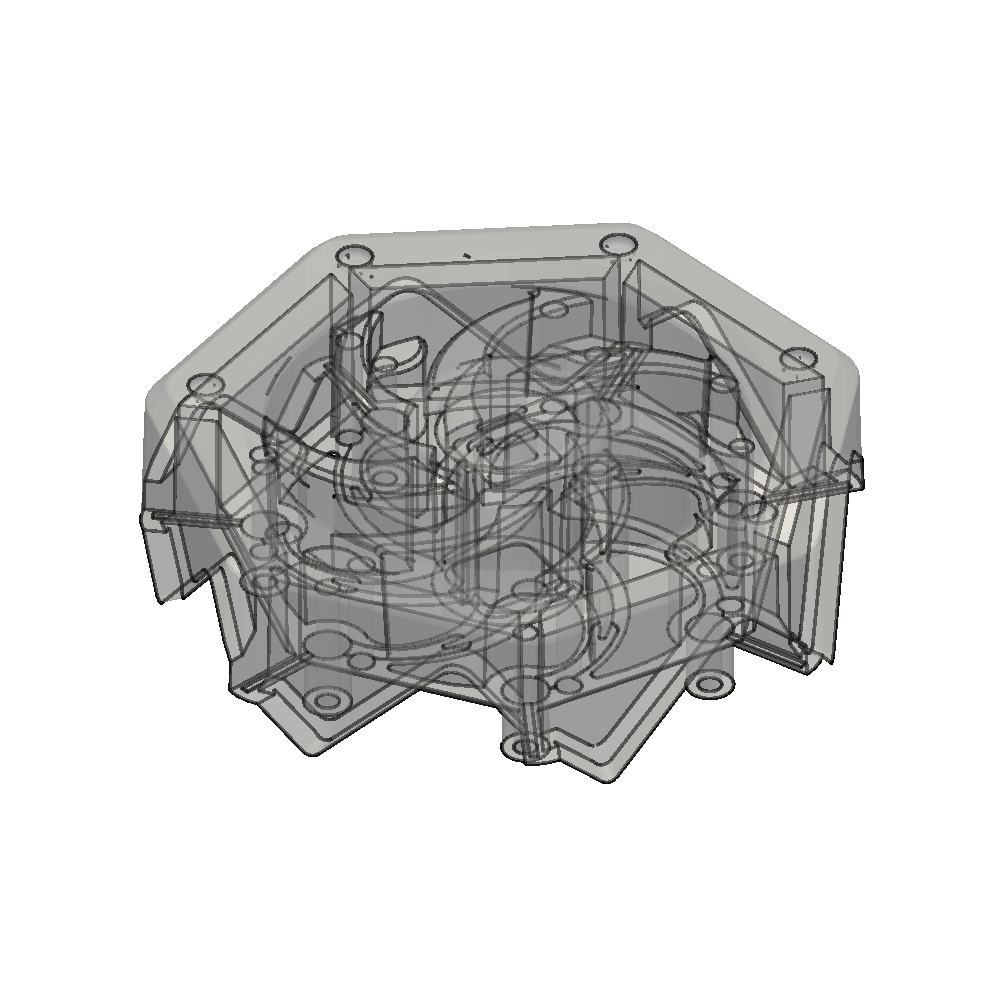}\hfill
    \includegraphics[width=0.117\linewidth]{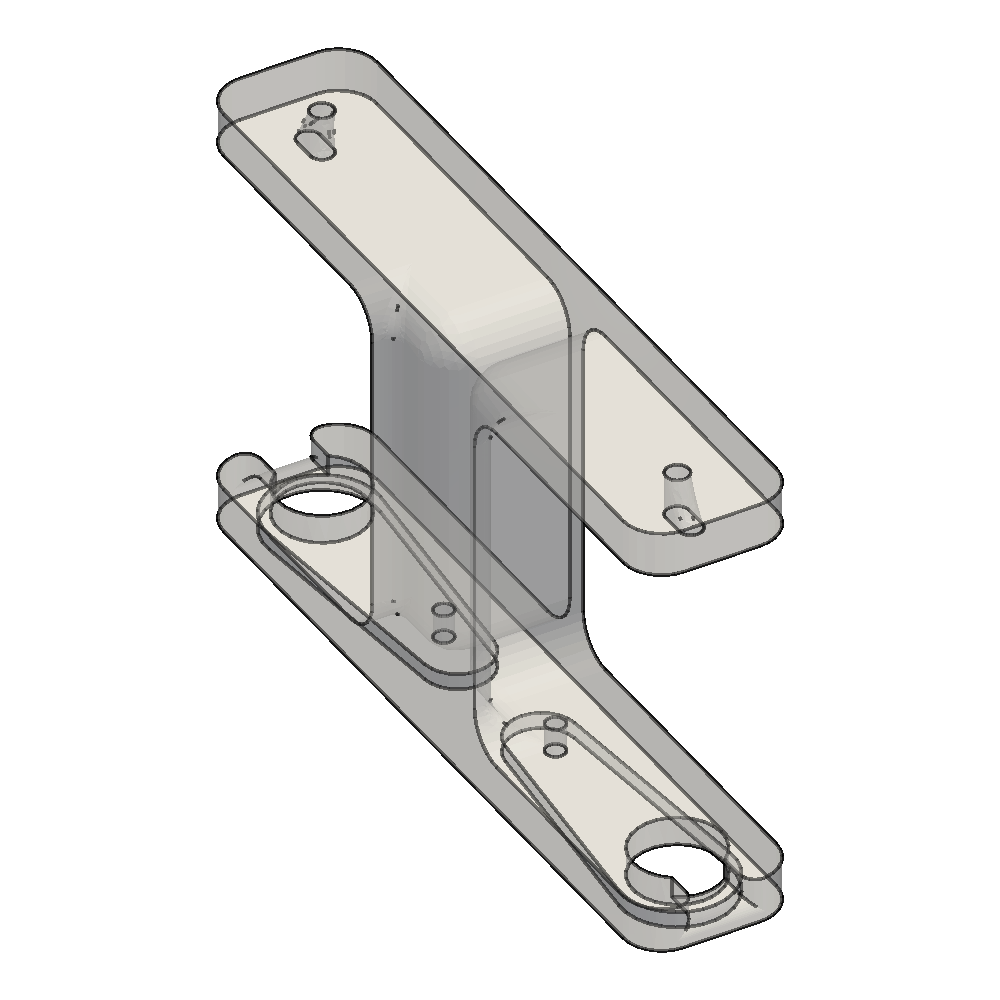}\hfill
    \includegraphics[width=0.117\linewidth]{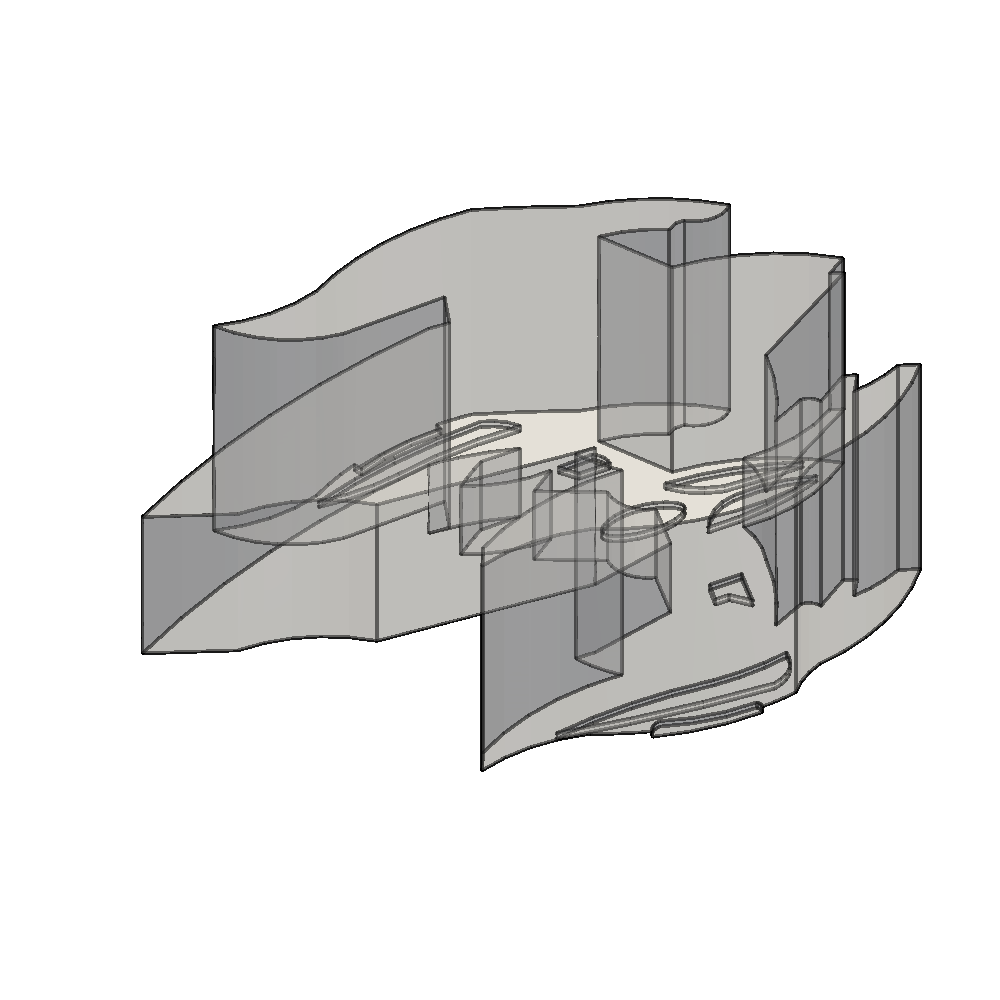}\\[\smallskipamount]
    \includegraphics[width=0.117\linewidth]{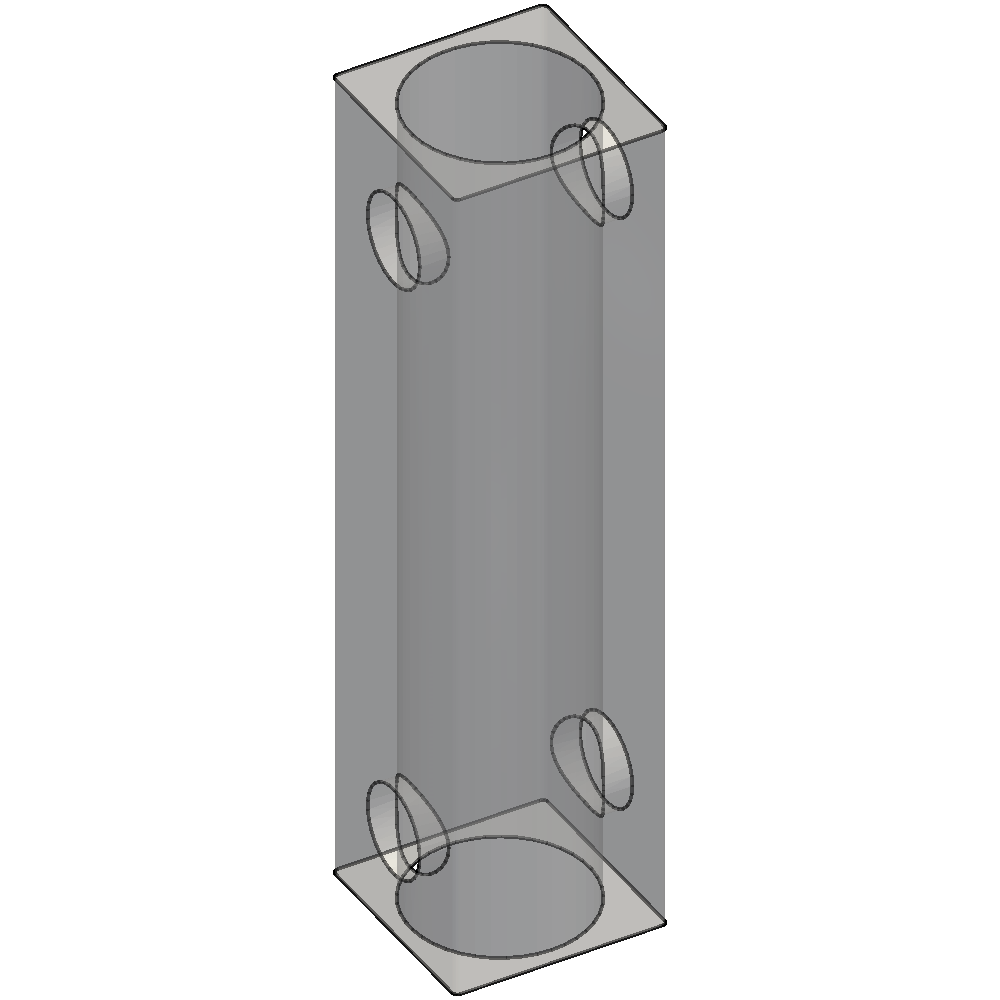}\hfill
    \includegraphics[width=0.117\linewidth]{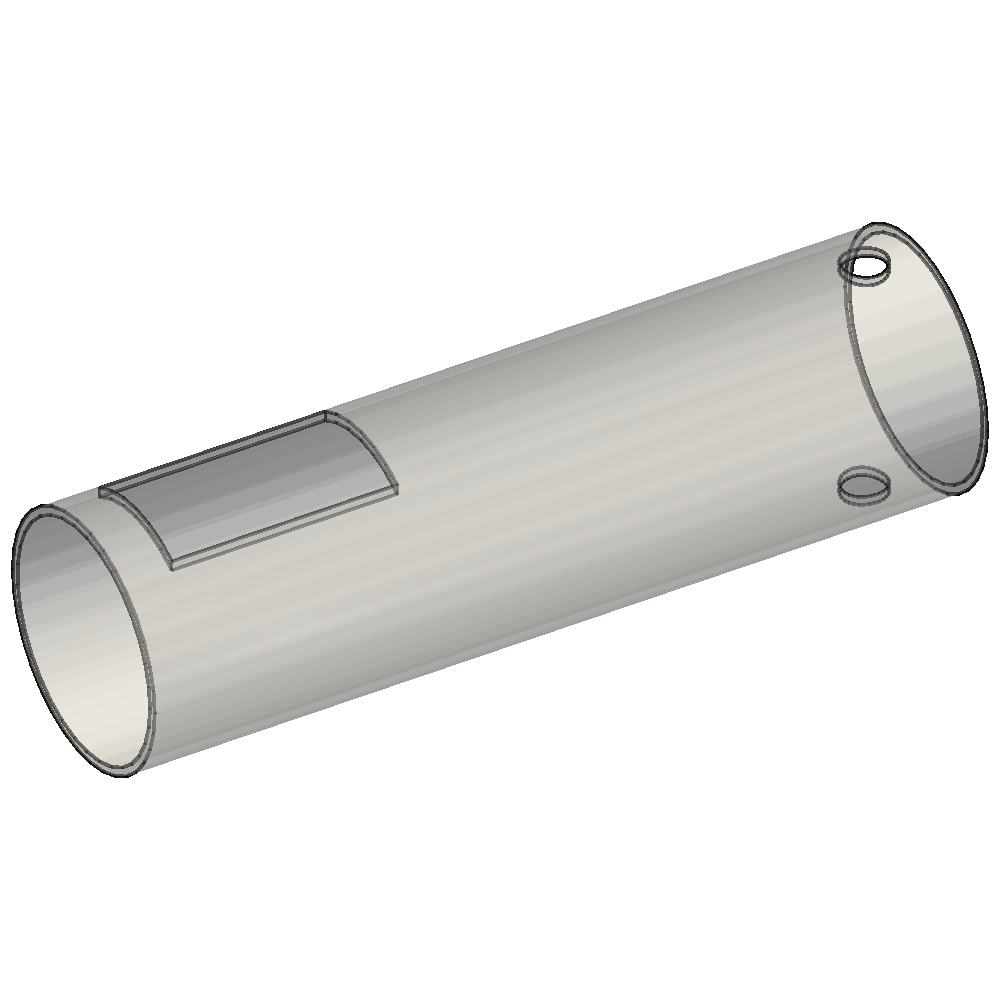}\hfill
    \includegraphics[width=0.117\linewidth]{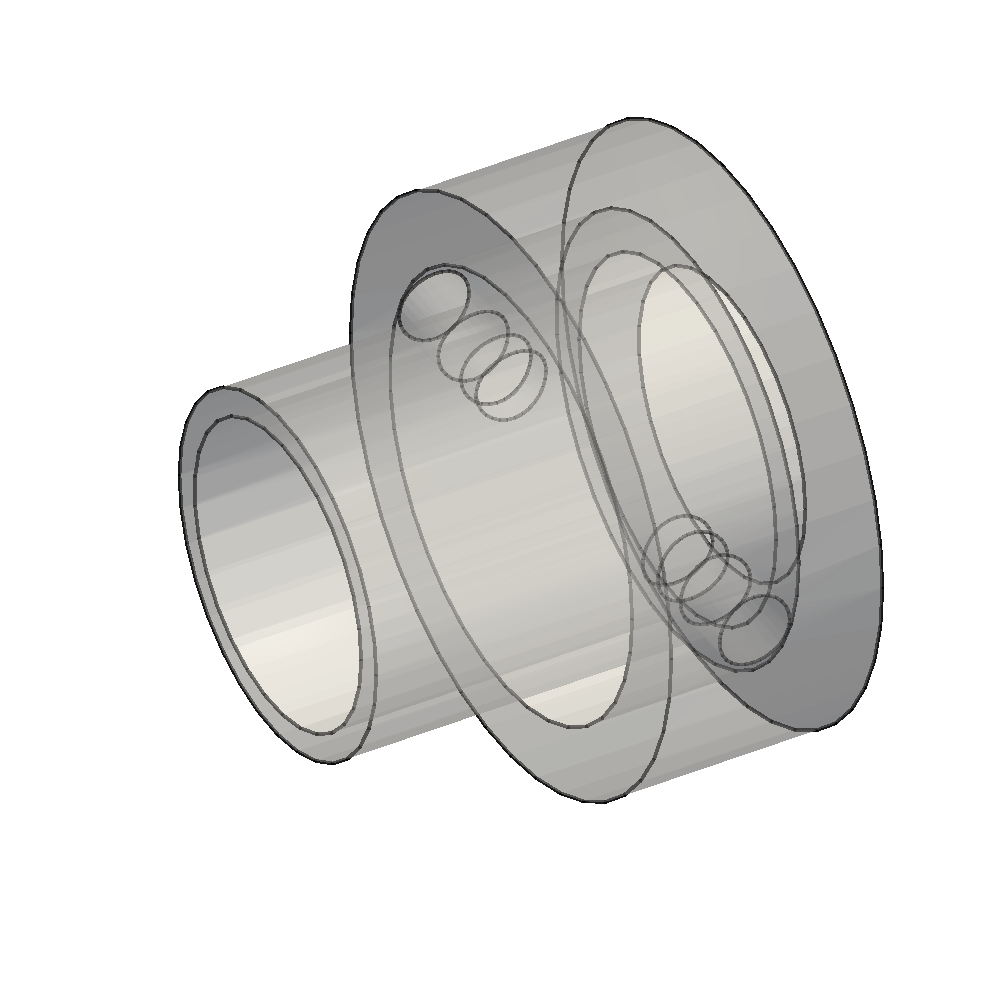}\hfill
    \includegraphics[width=0.117\linewidth]{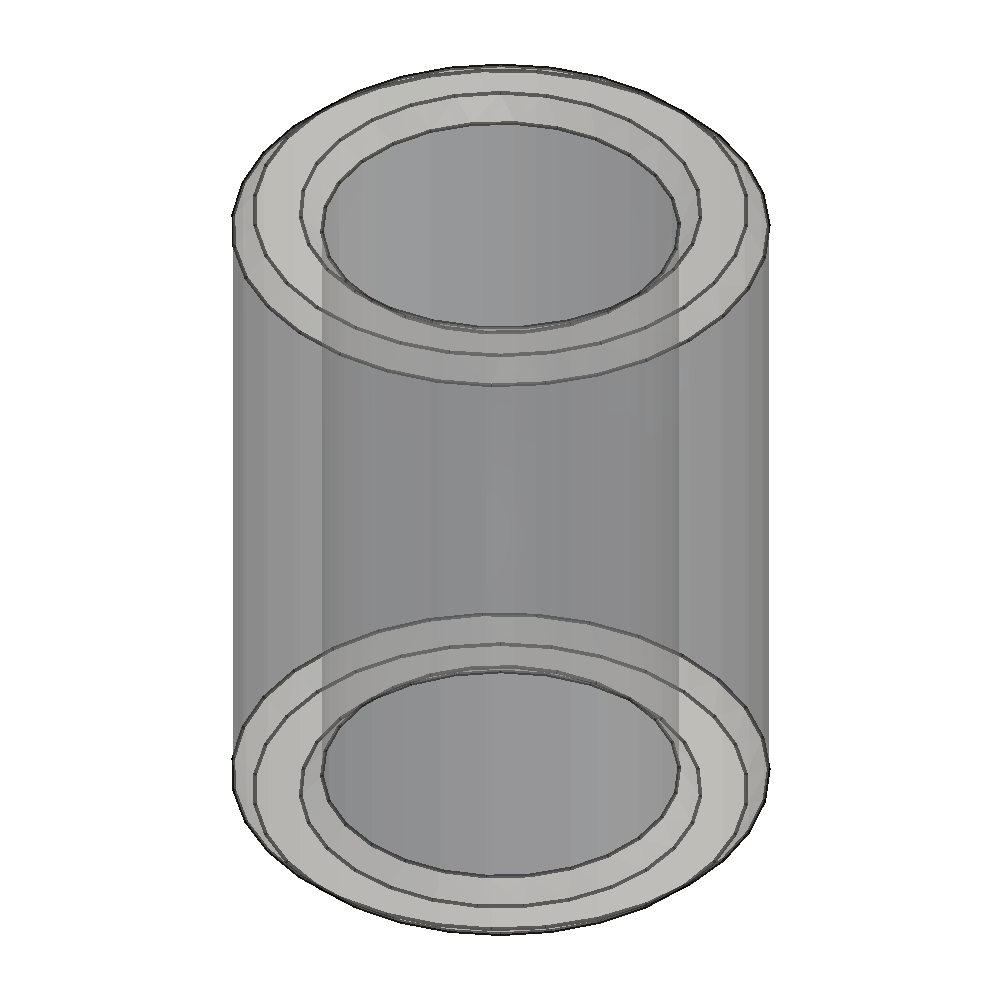}\hfill
    \includegraphics[width=0.117\linewidth]{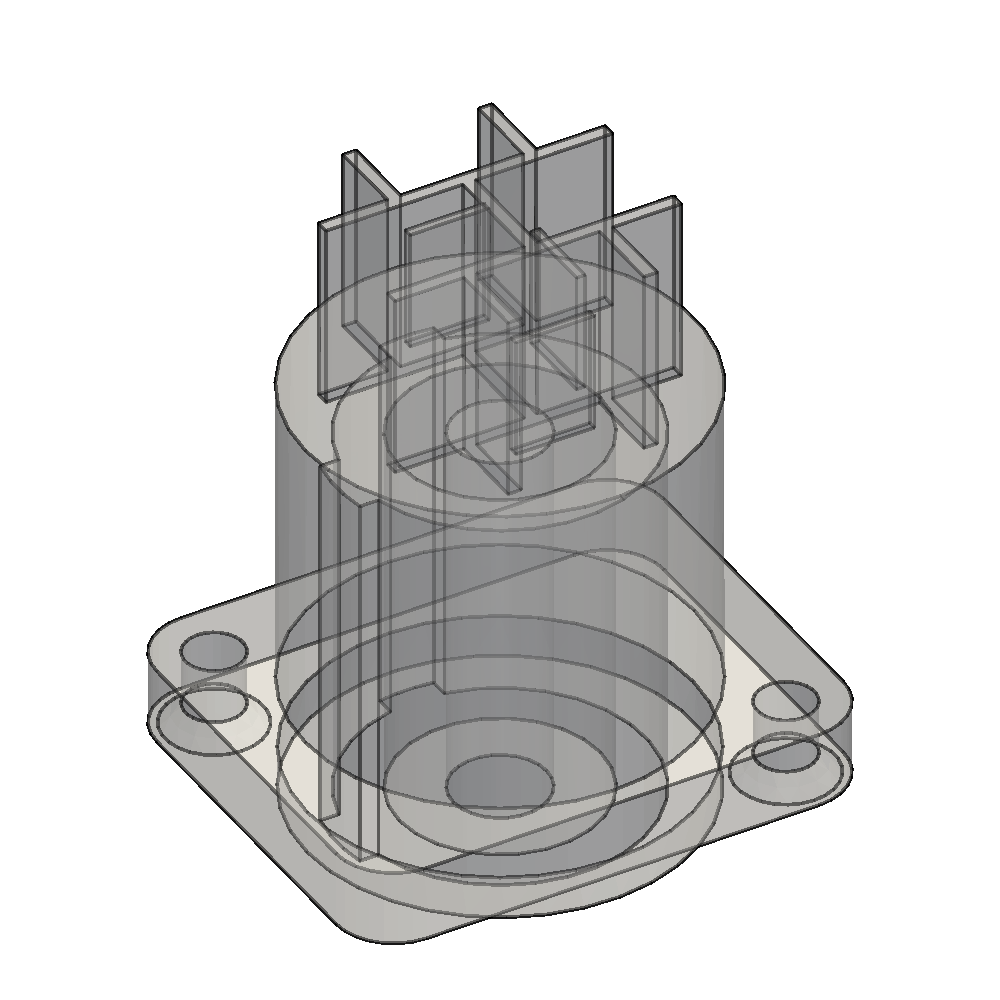}\hfill
    \includegraphics[width=0.117\linewidth]{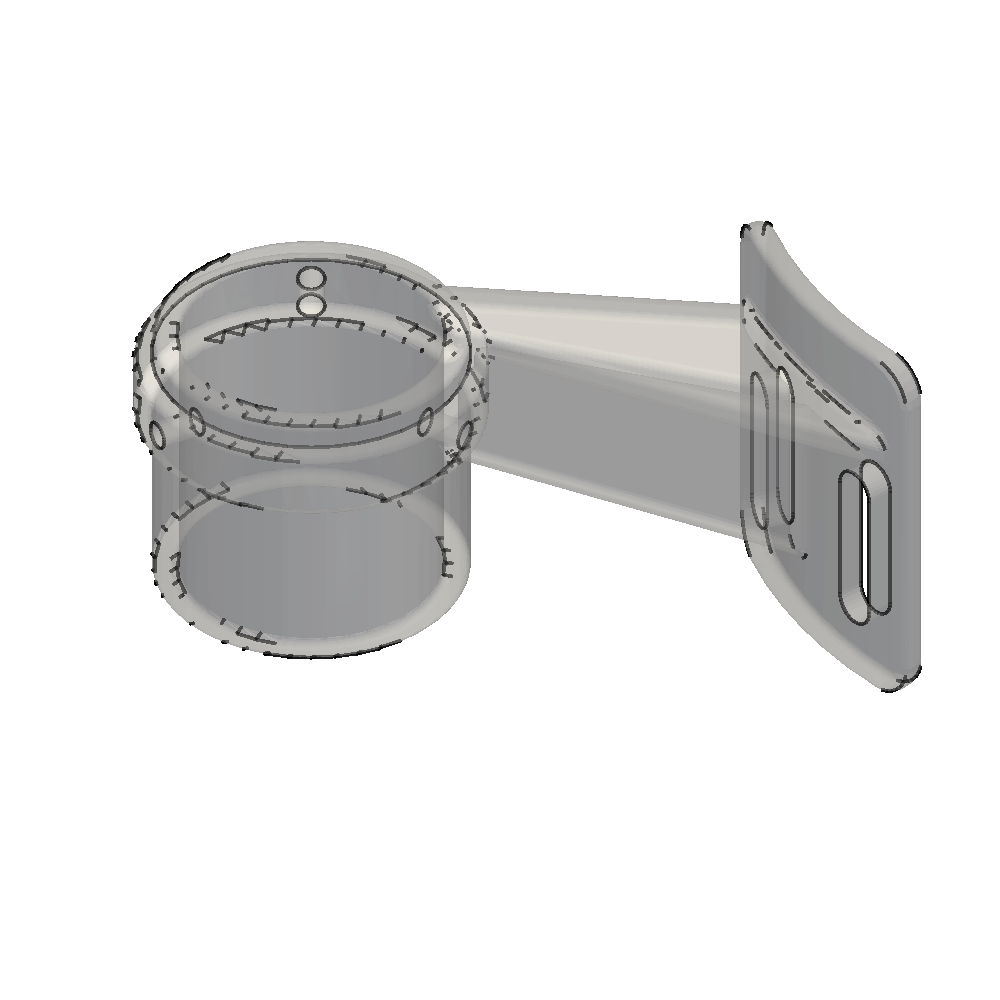}\hfill
    \includegraphics[width=0.117\linewidth]{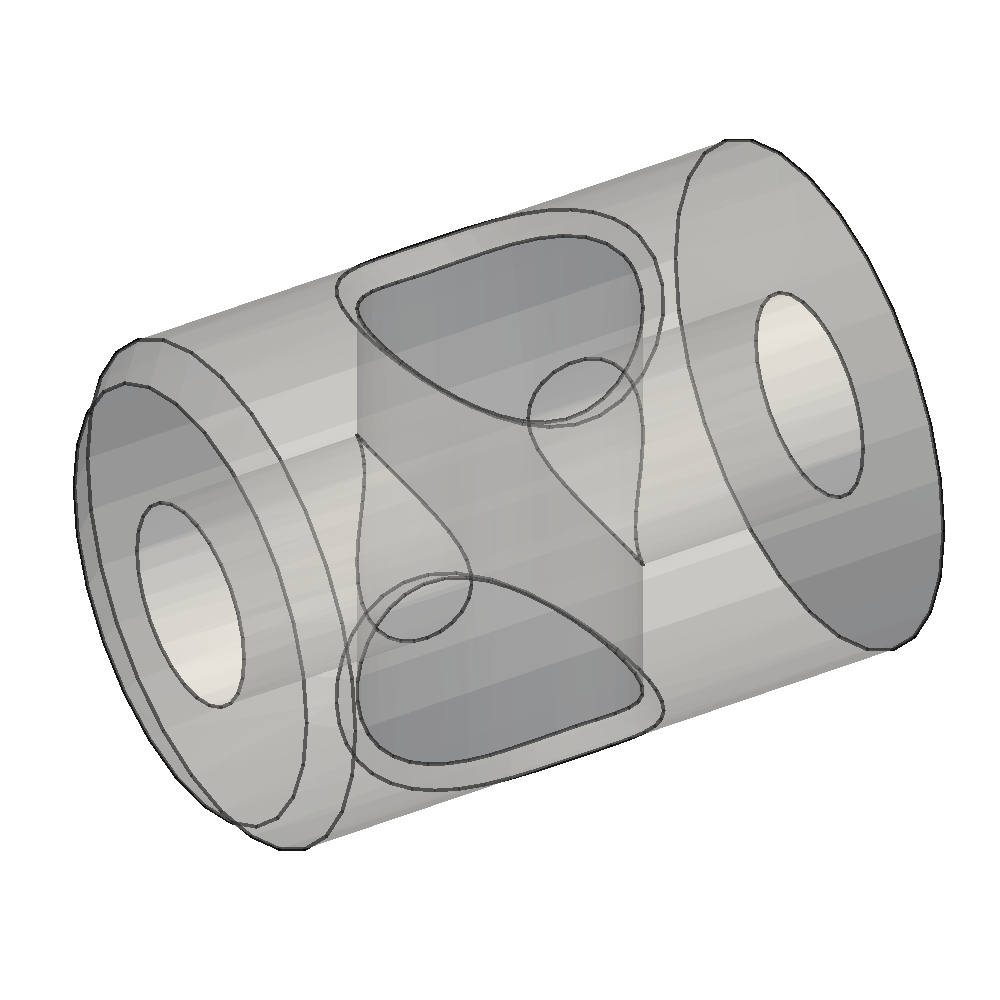}\hfill
    \includegraphics[width=0.117\linewidth]{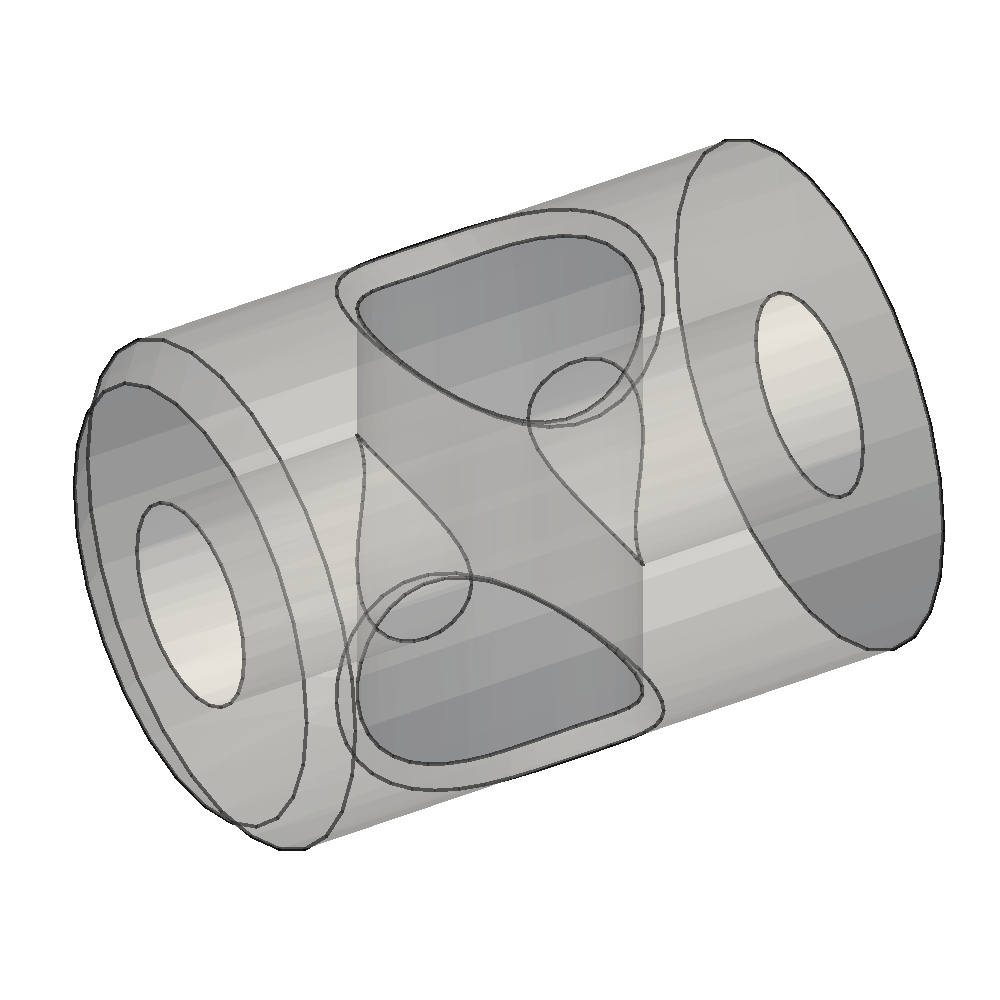}\\[\smallskipamount]
    \includegraphics[width=0.117\linewidth]{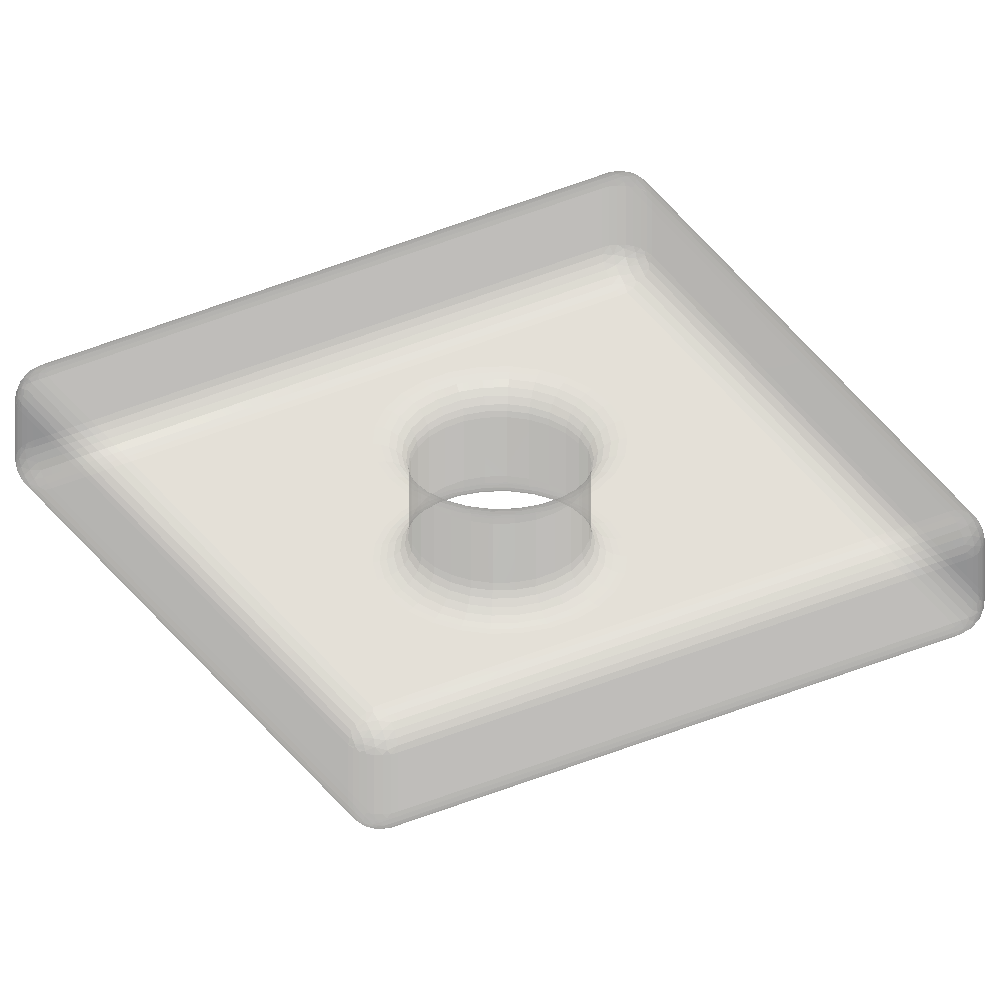}\hfill
    \includegraphics[width=0.117\linewidth]{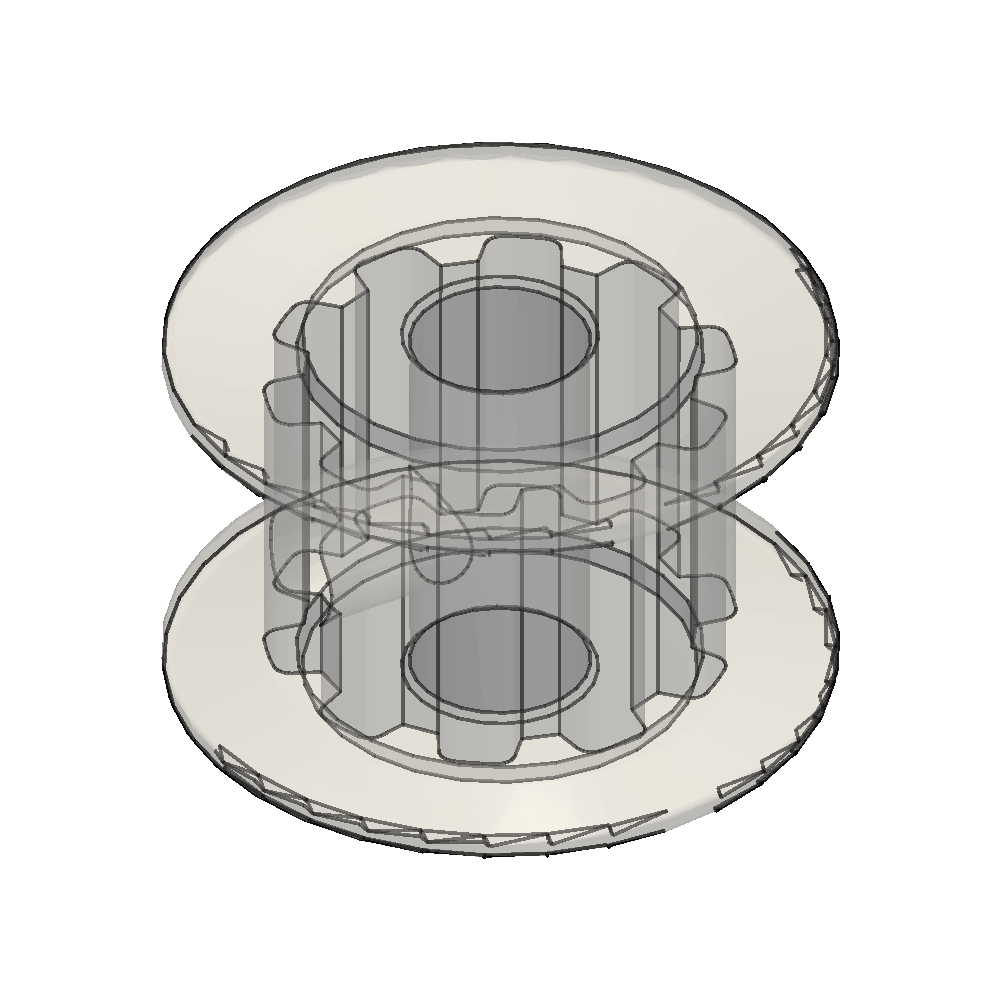}\hfill
    \includegraphics[width=0.117\linewidth]{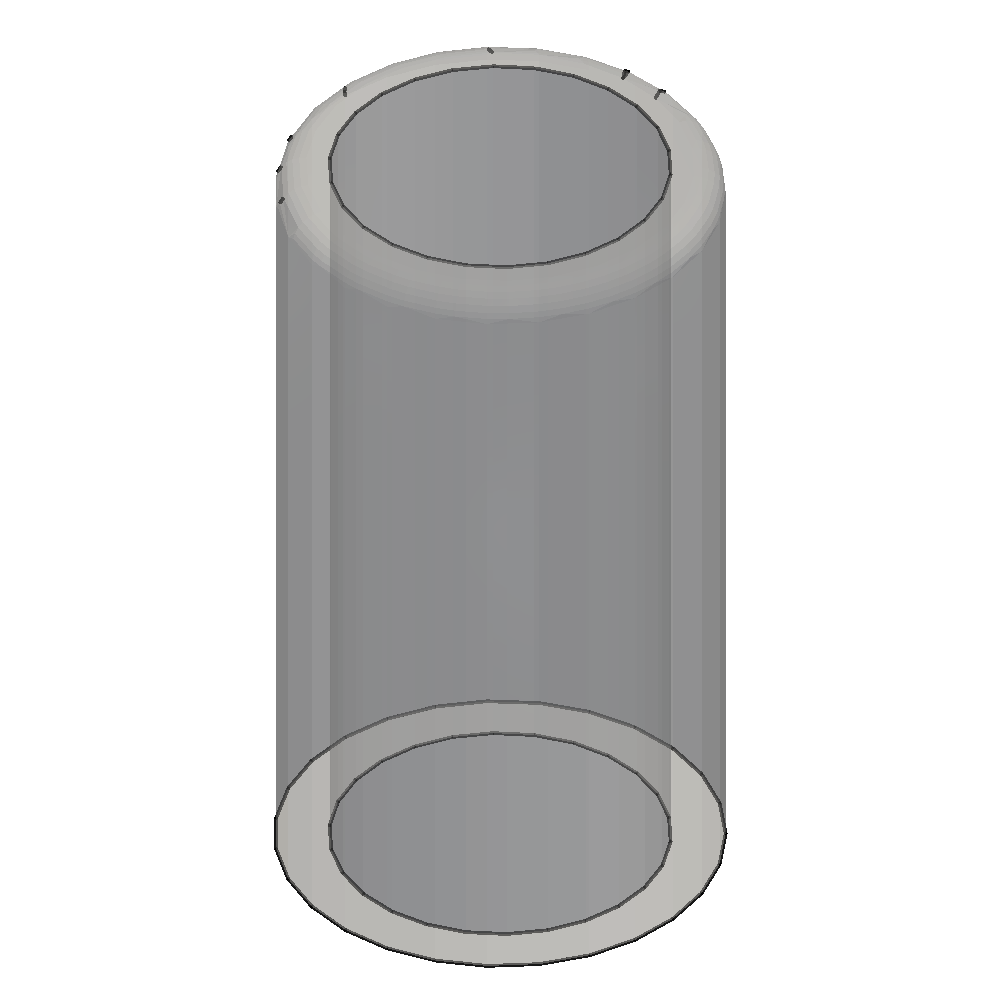}\hfill
    \includegraphics[width=0.117\linewidth]{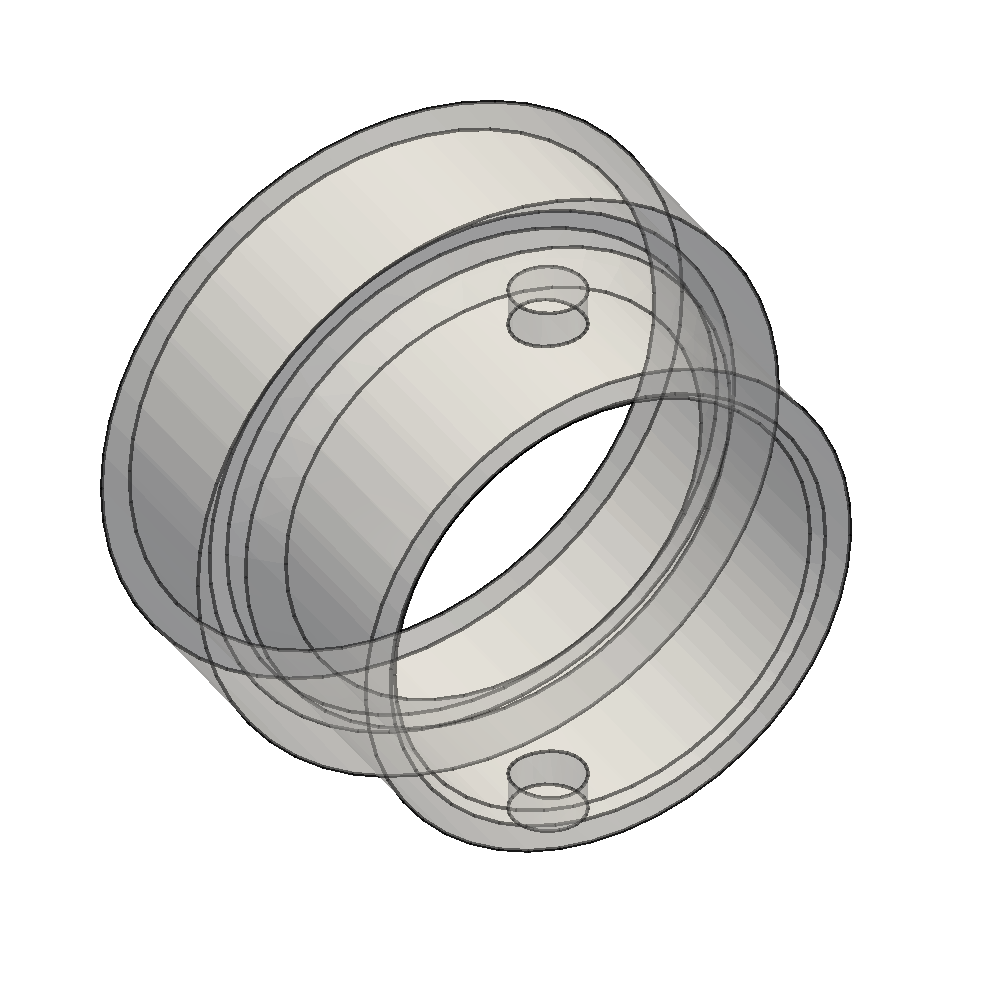}\hfill
    \includegraphics[width=0.117\linewidth]{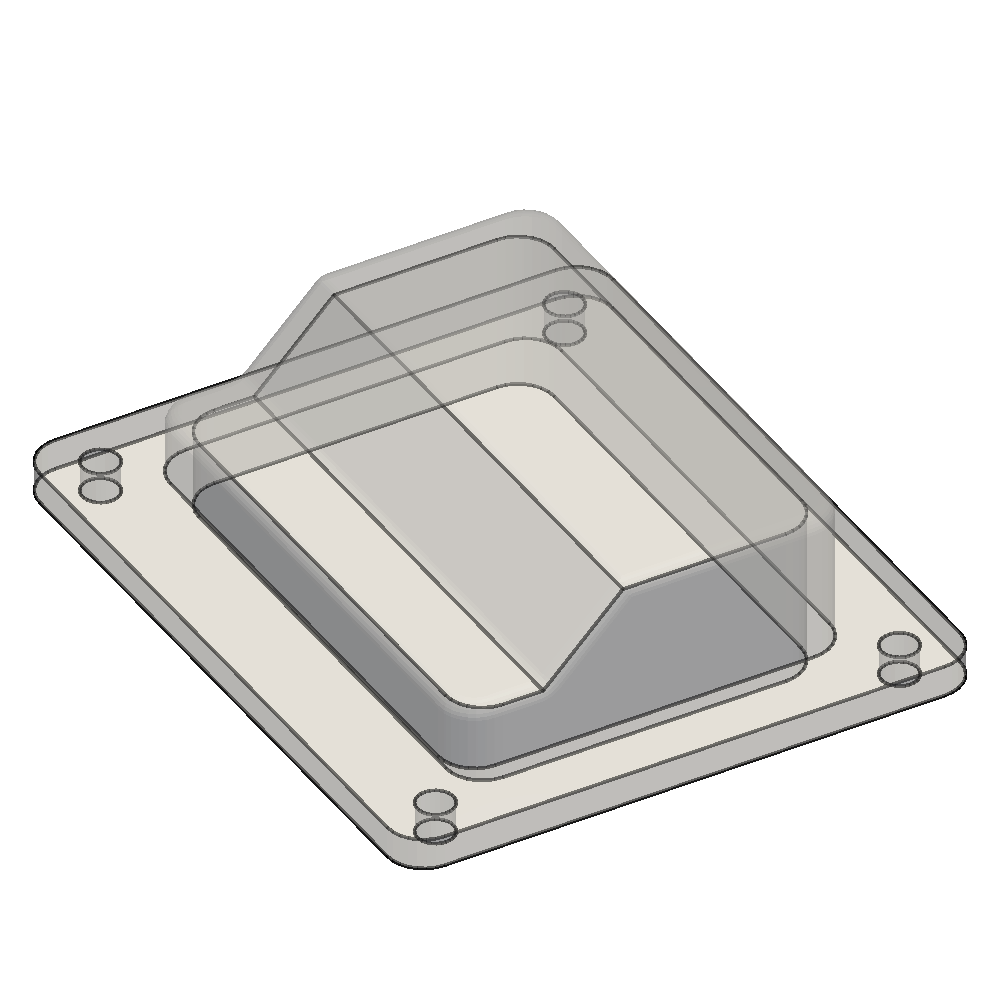}\hfill
    \includegraphics[width=0.117\linewidth]{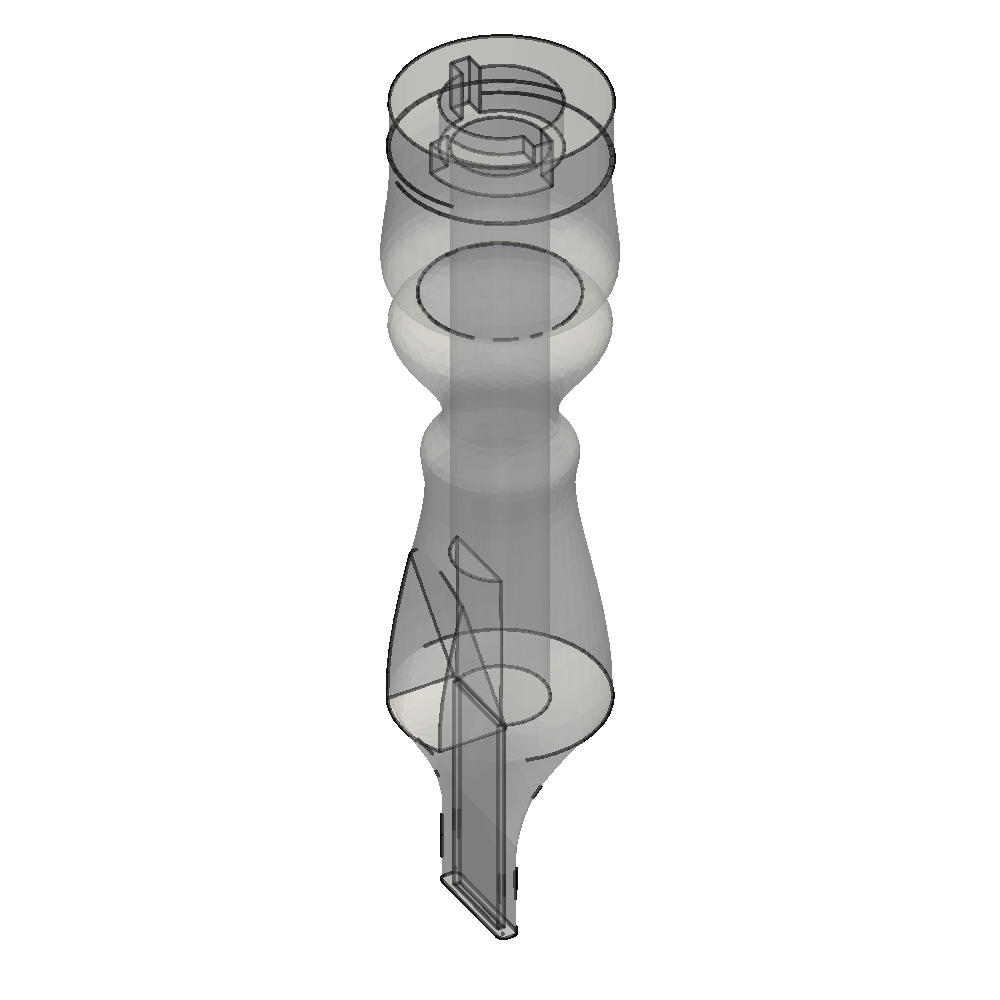}\hfill
    \includegraphics[width=0.117\linewidth]{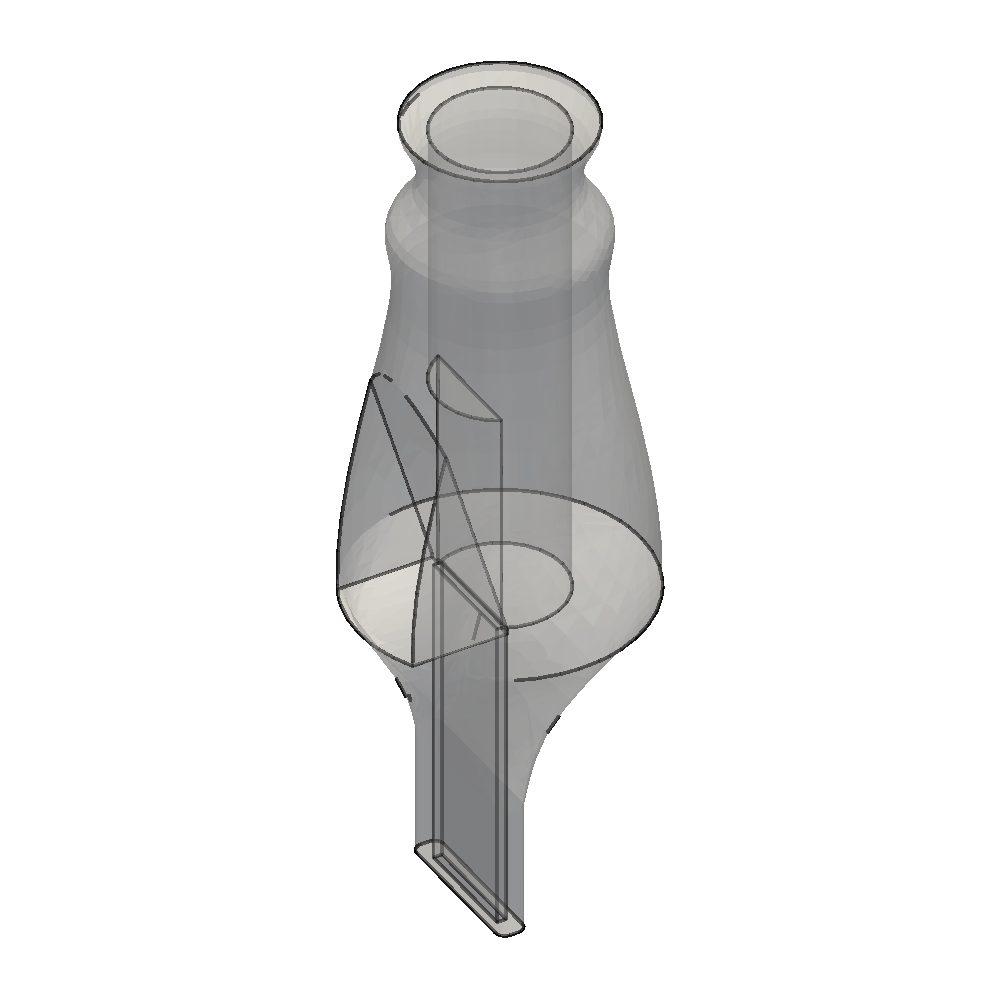}\hfill
    \includegraphics[width=0.117\linewidth]{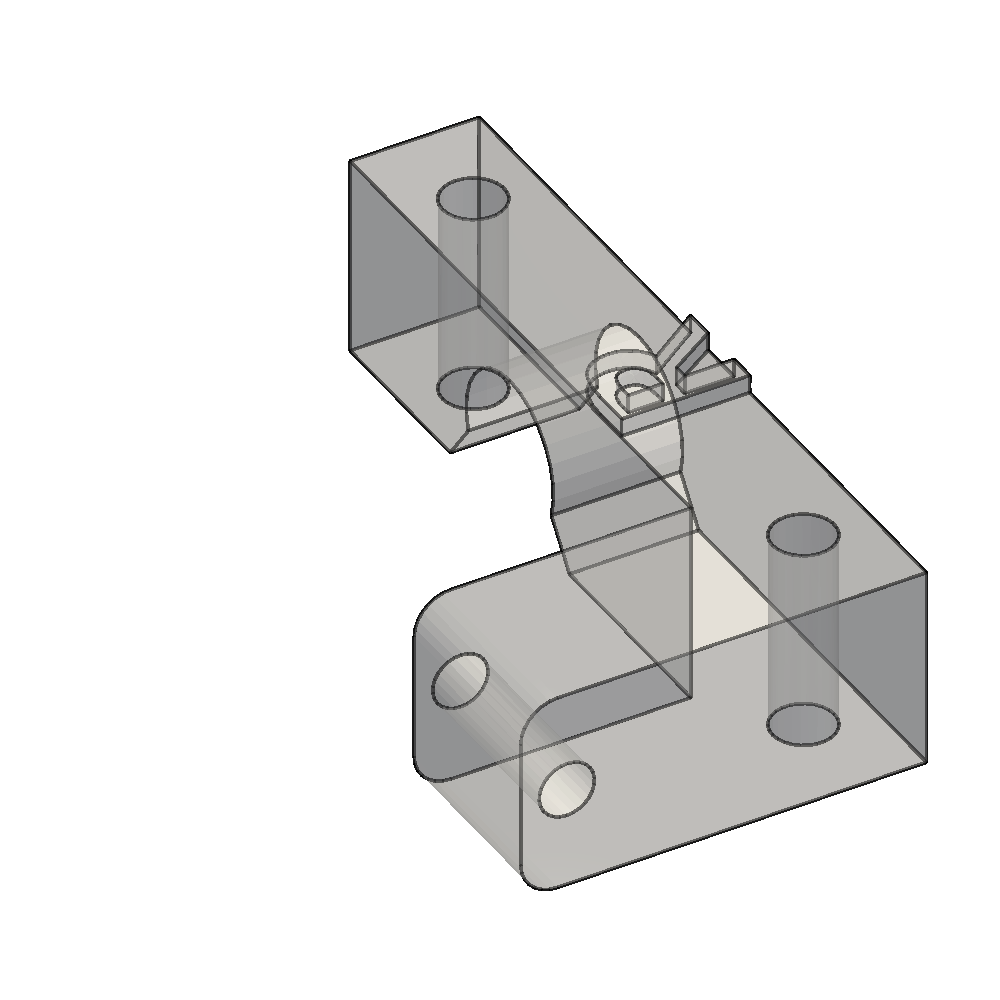}

    \caption{The 156 CAD parts of the evaluation benchmark (part 1 of 4), each rendered under the same semi-transparent orthographic view ($e36$,\,$a60$). All parts are disjoint from the \textit{orthoViews} training and validation splits. The first 20 tiles are the 19 benchmark parts from the original submission set followed by the torsion example of Fig.~\ref{fig:visual_examples_4x5}; the remaining parts follow in order of object identifier. The set spans extrusions, through-holes, spoked and ribbed bosses, brackets, and filleted load-bearing features.}
    \label{fig:cad_dataset_grid}
\end{figure}

\begin{figure}[h]
    \centering
    \includegraphics[width=0.117\linewidth]{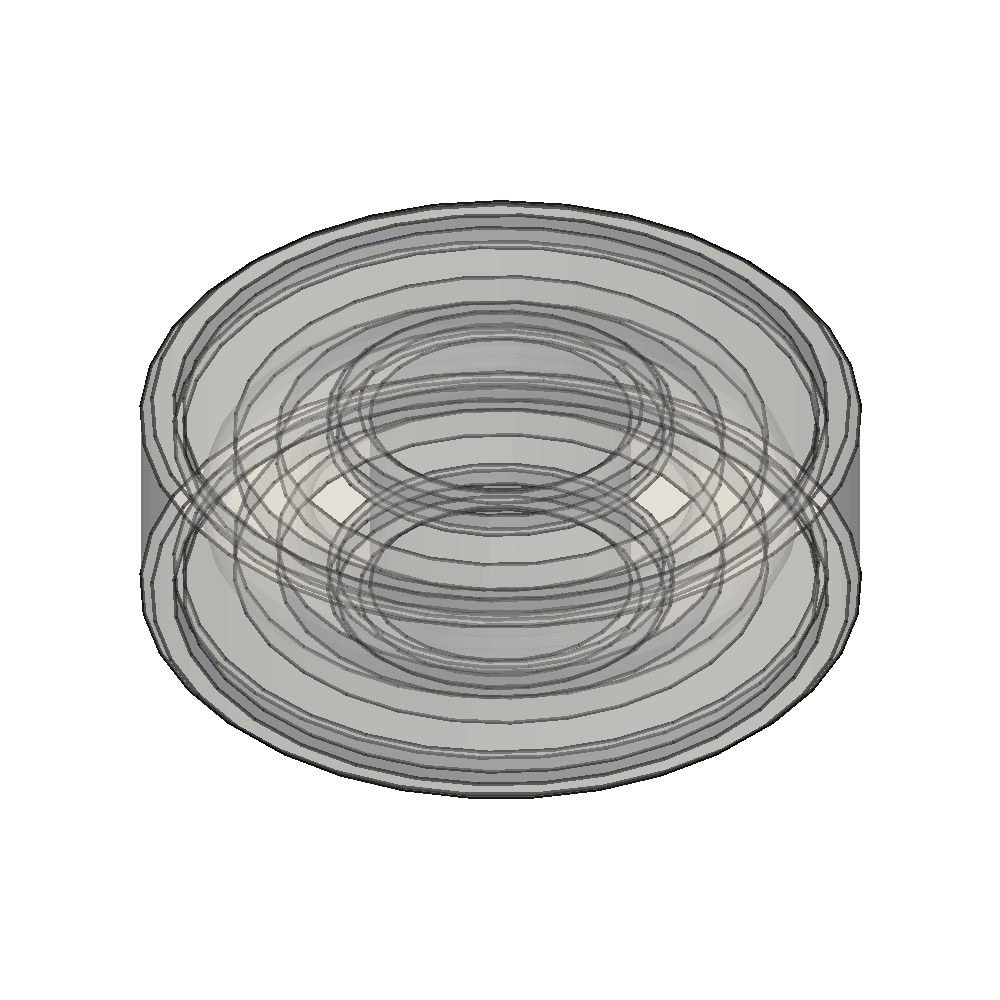}\hfill
    \includegraphics[width=0.117\linewidth]{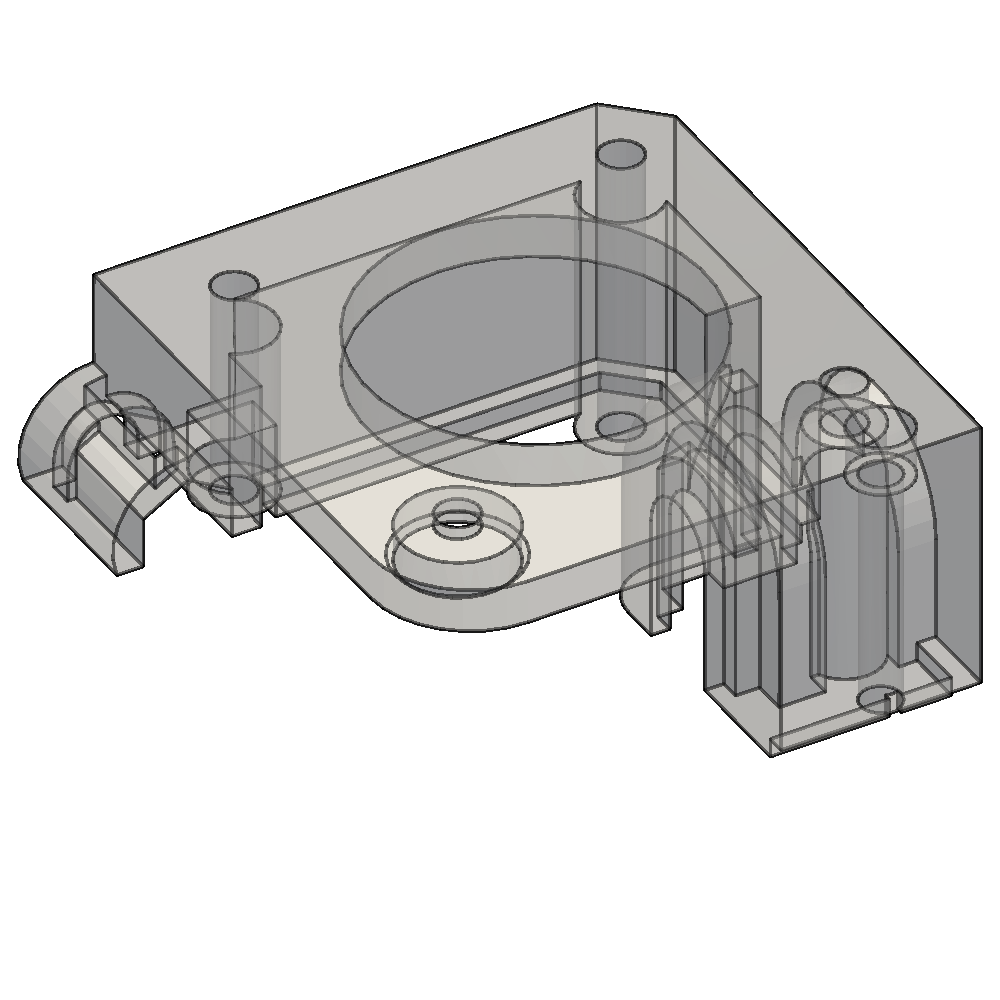}\hfill
    \includegraphics[width=0.117\linewidth]{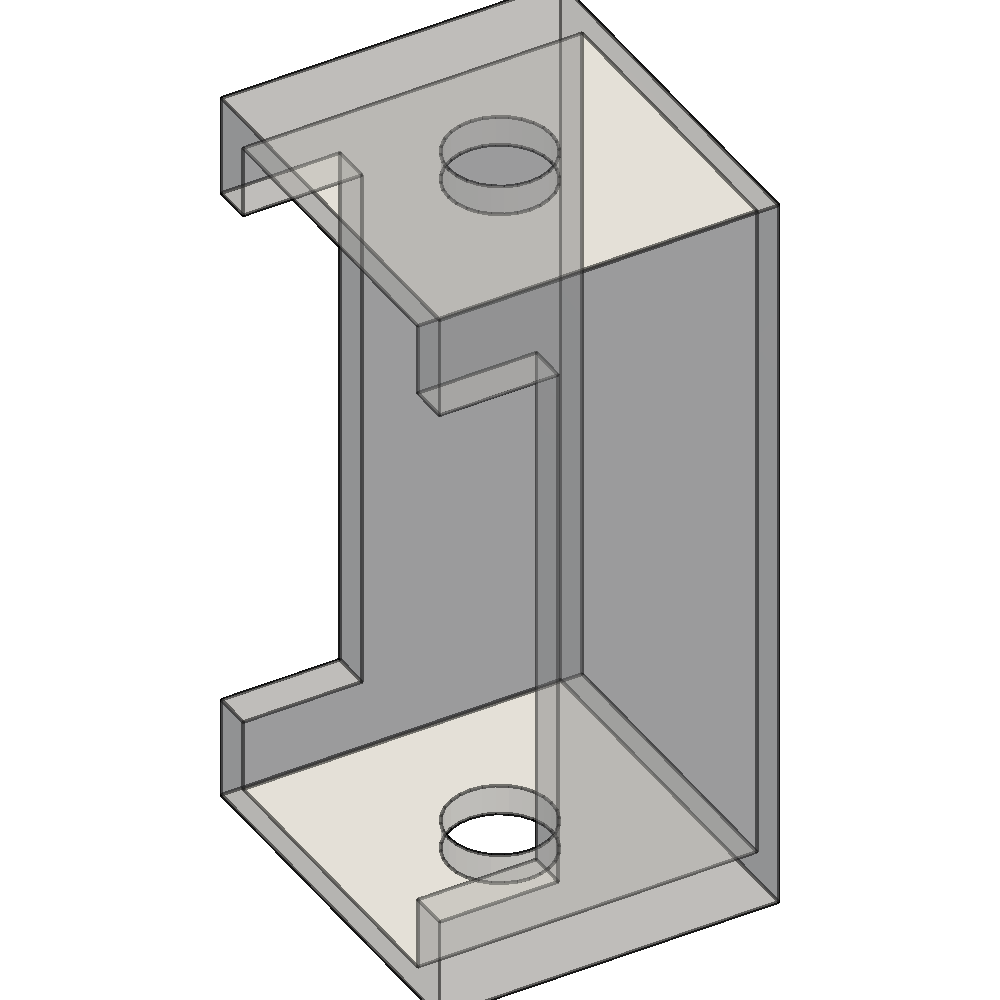}\hfill
    \includegraphics[width=0.117\linewidth]{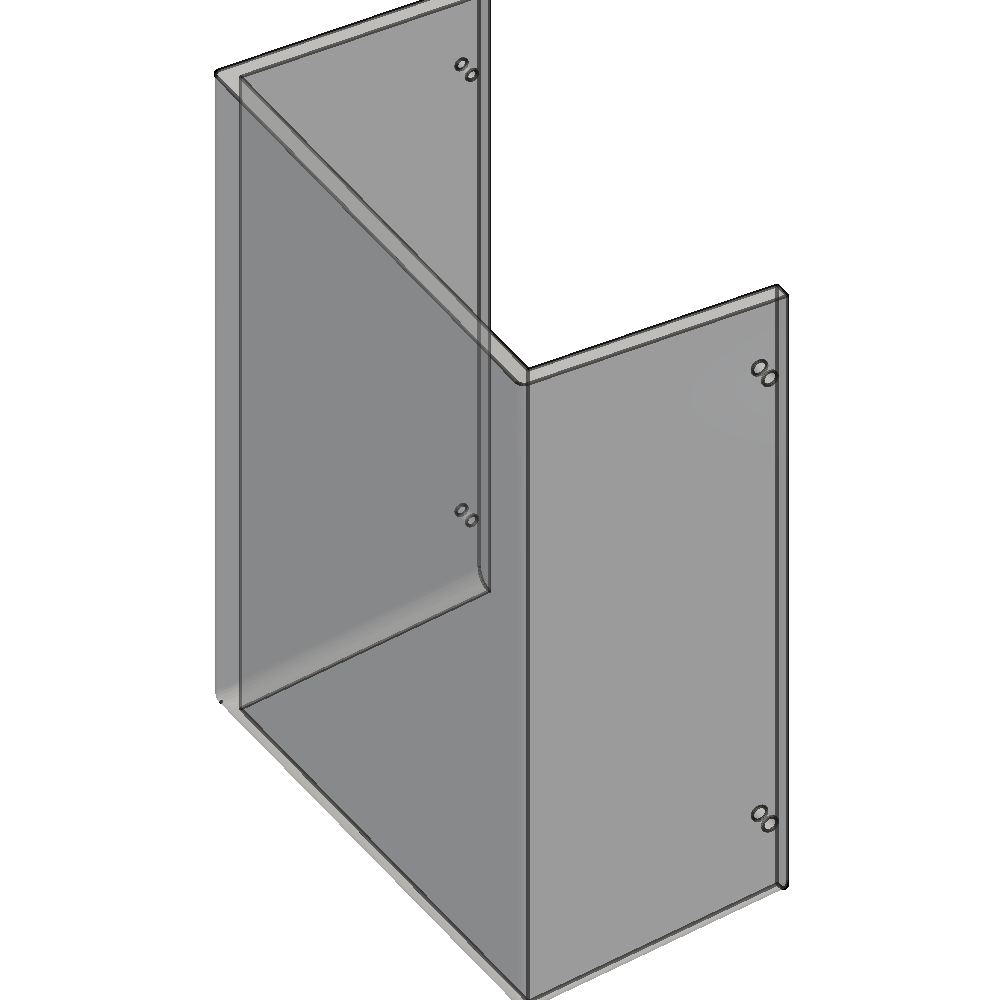}\hfill
    \includegraphics[width=0.117\linewidth]{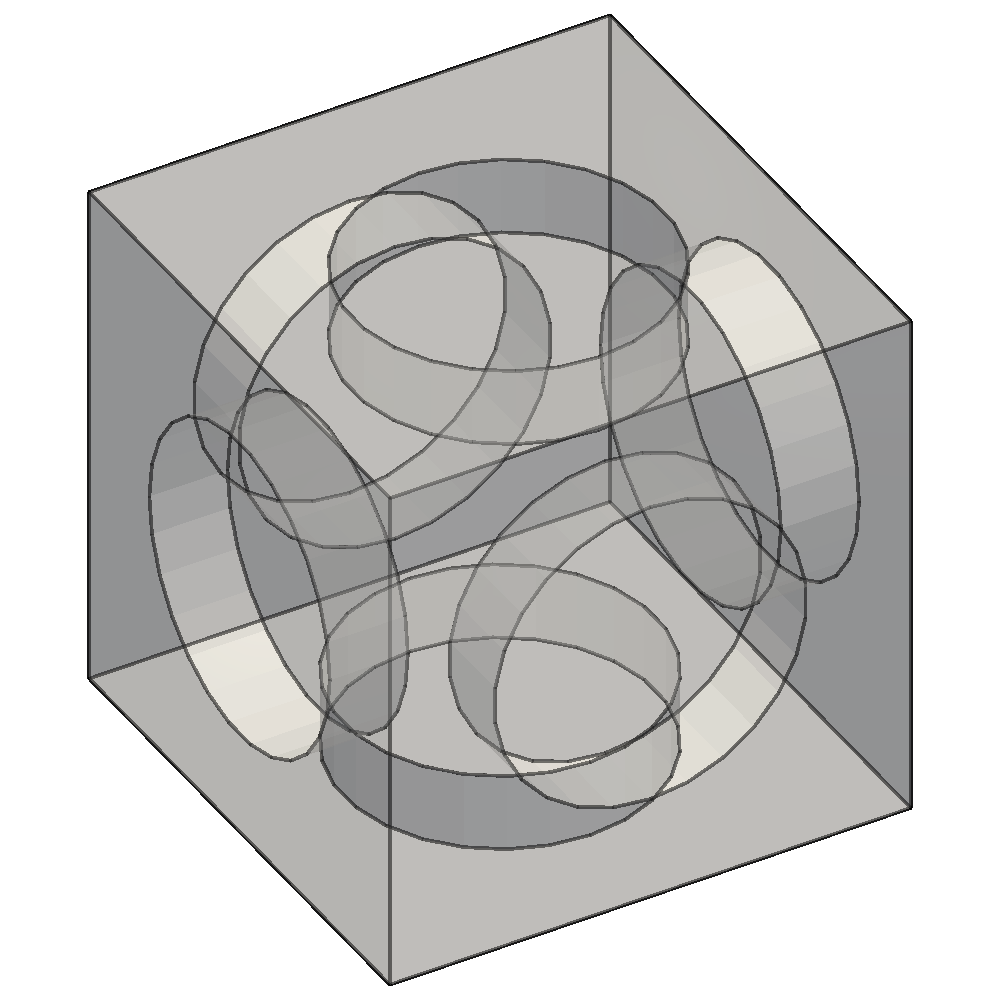}\hfill
    \includegraphics[width=0.117\linewidth]{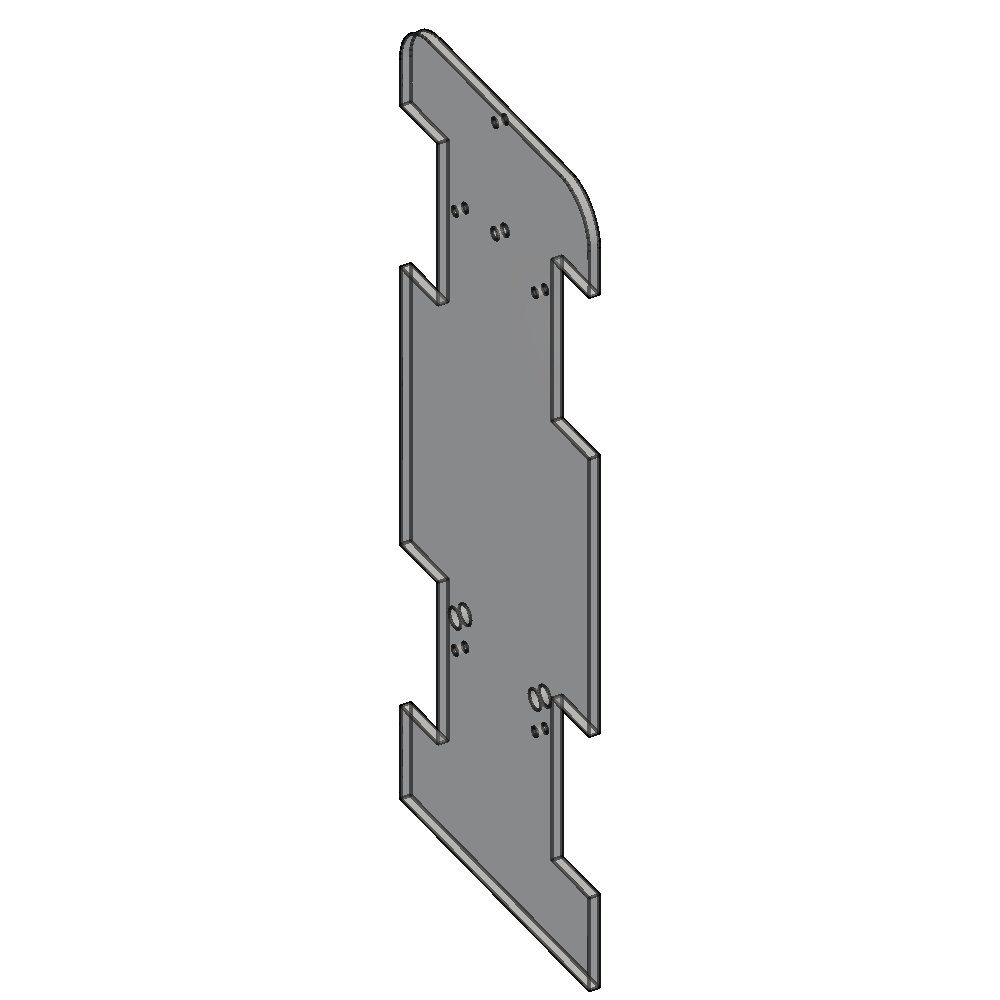}\hfill
    \includegraphics[width=0.117\linewidth]{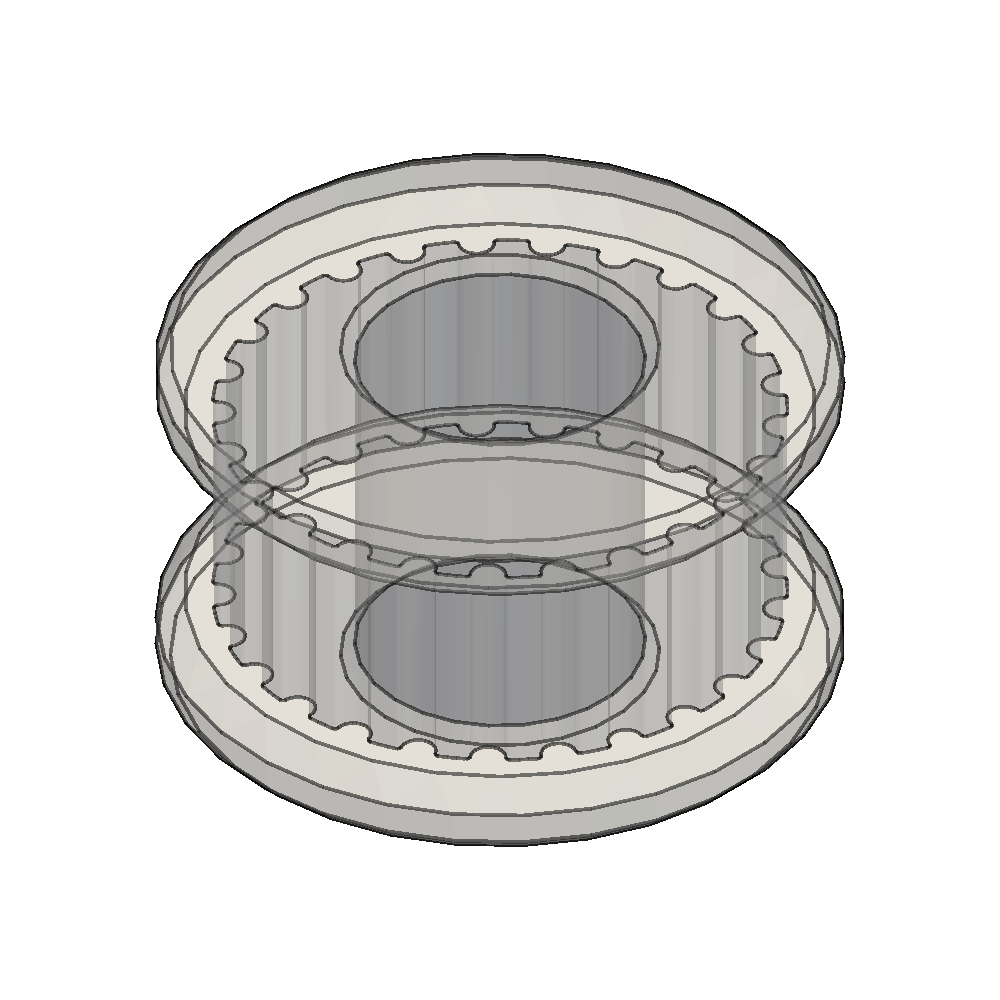}\hfill
    \includegraphics[width=0.117\linewidth]{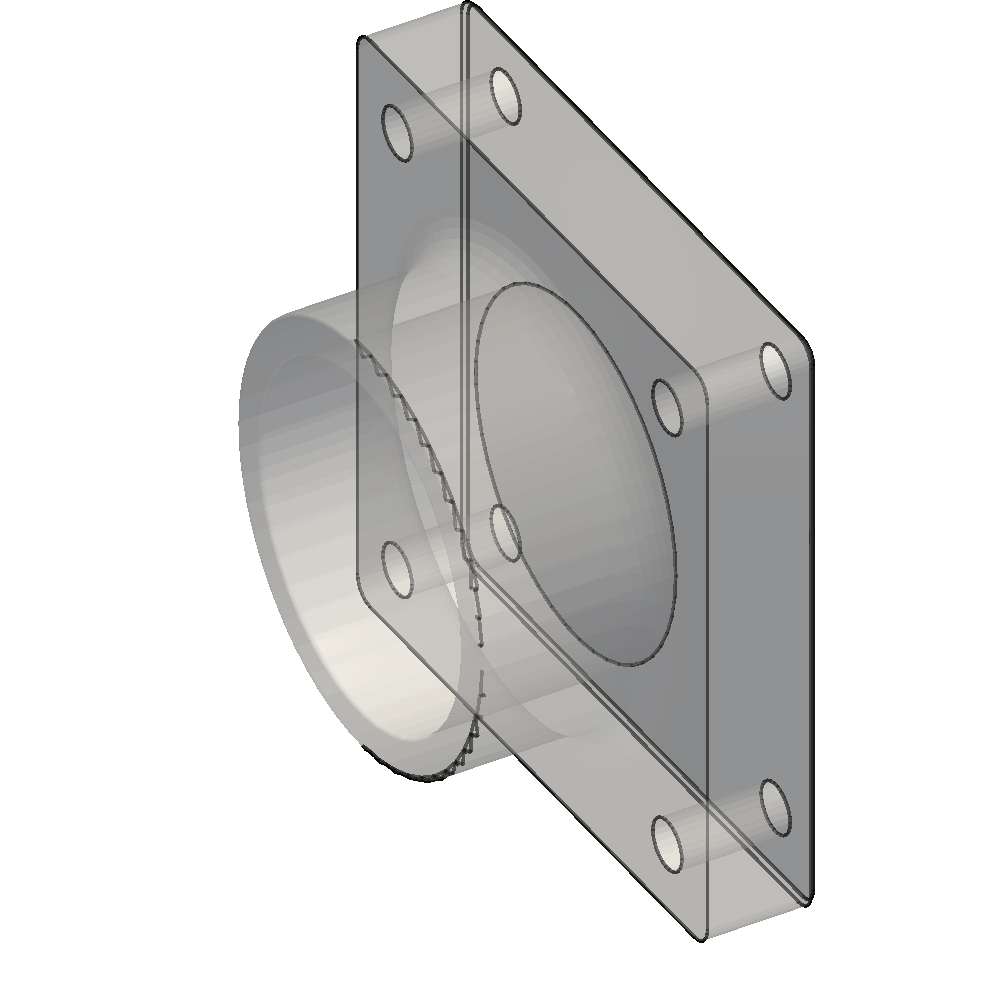}\\[\smallskipamount]
    \includegraphics[width=0.117\linewidth]{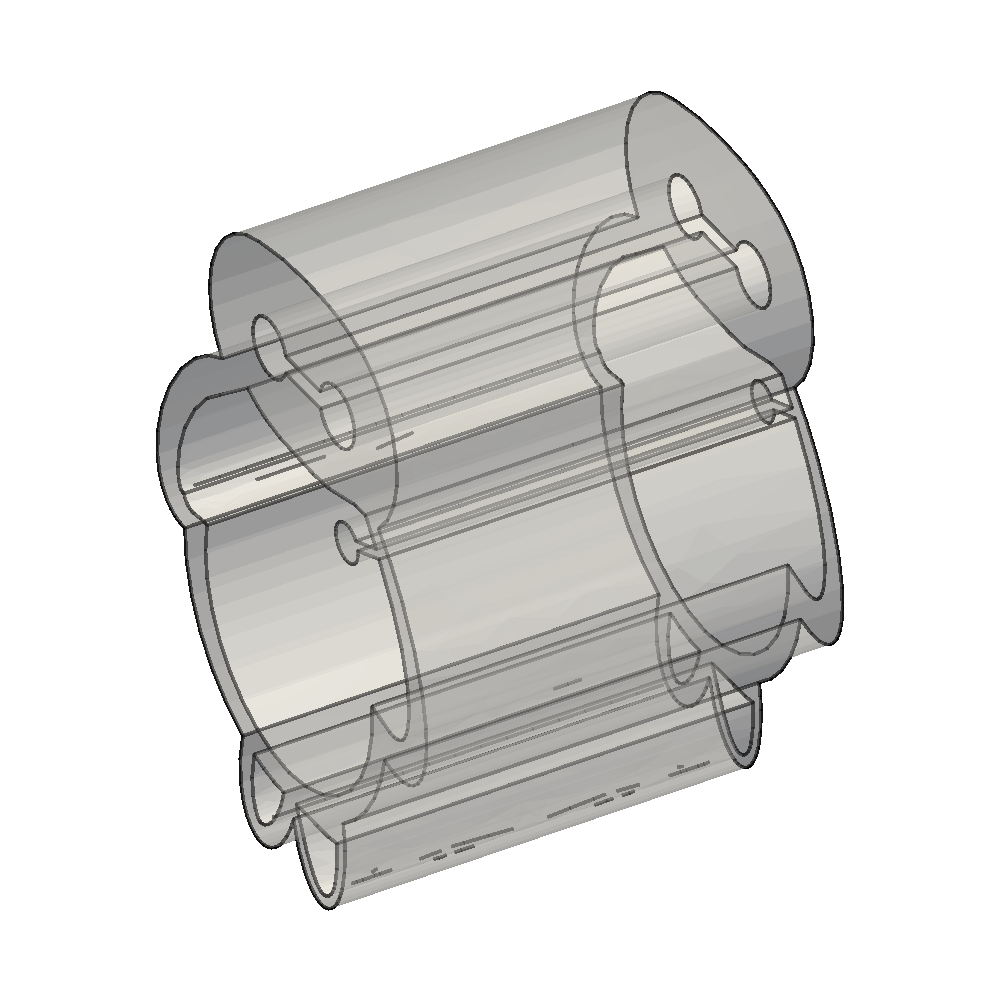}\hfill
    \includegraphics[width=0.117\linewidth]{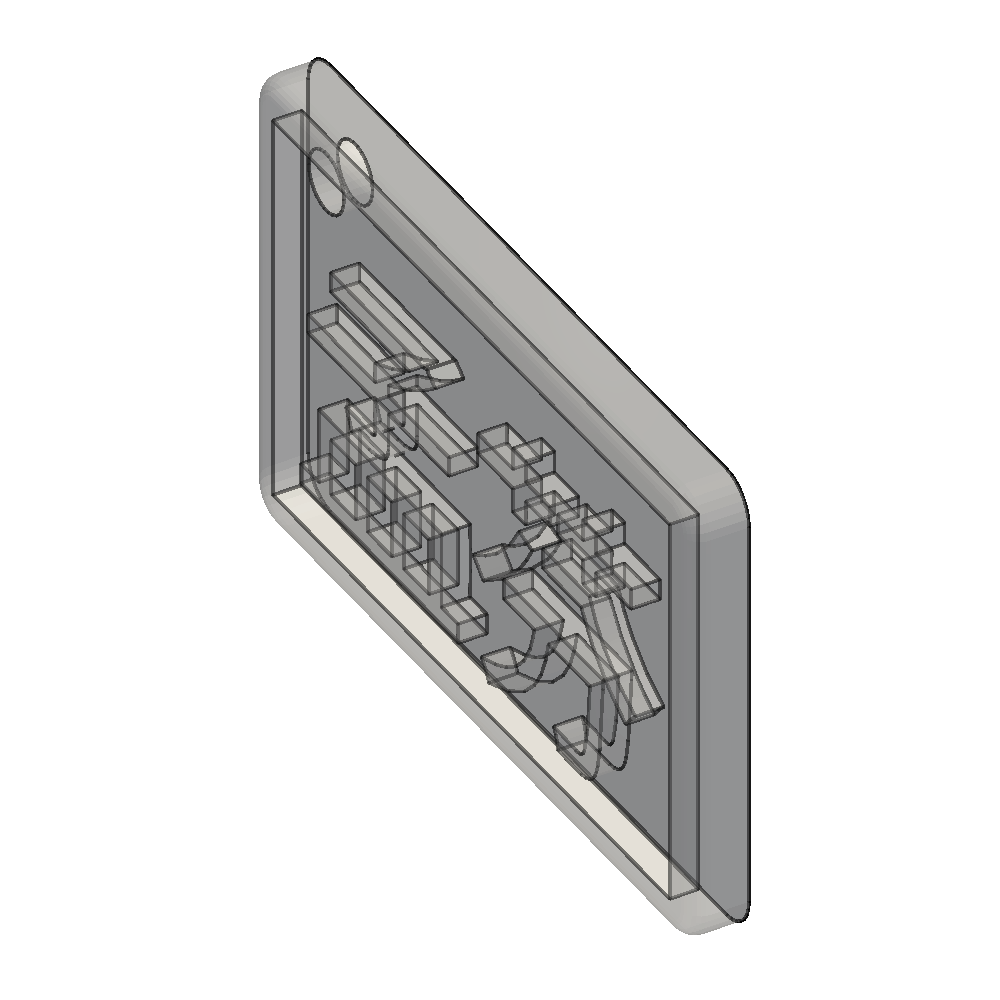}\hfill
    \includegraphics[width=0.117\linewidth]{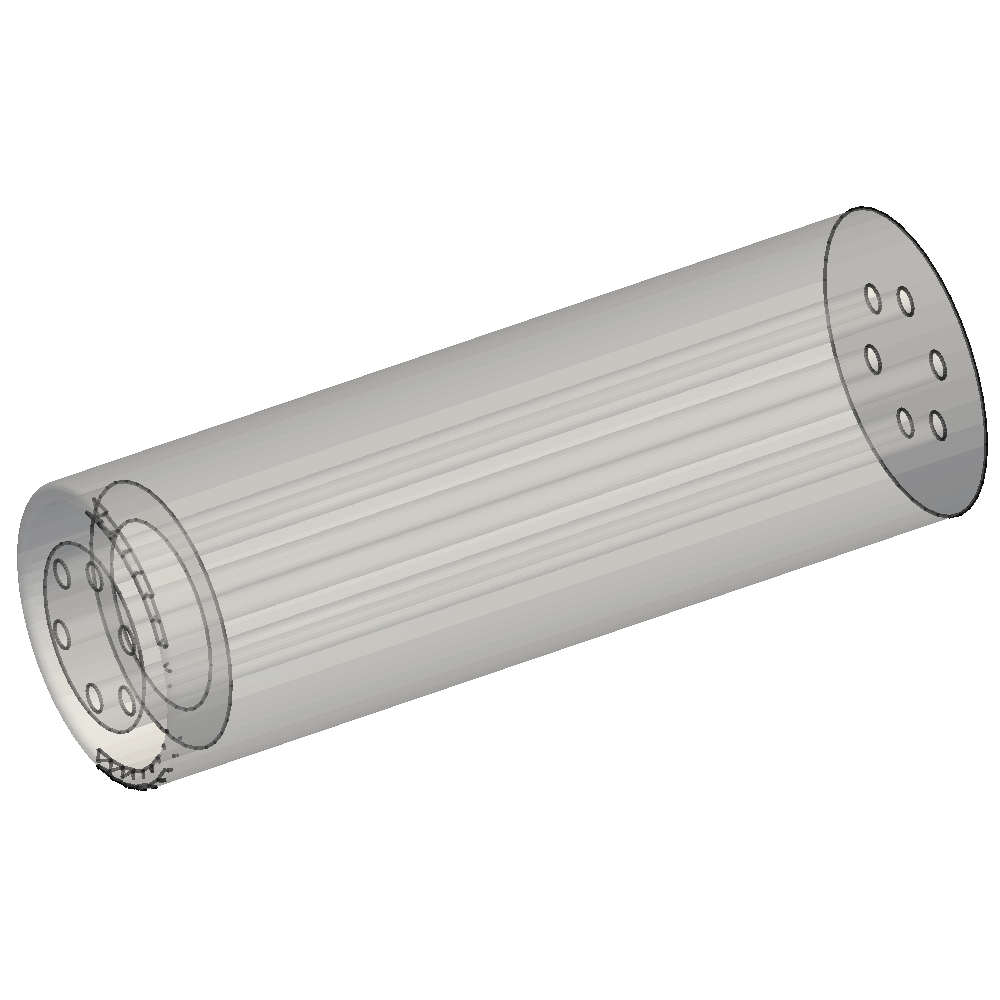}\hfill
    \includegraphics[width=0.117\linewidth]{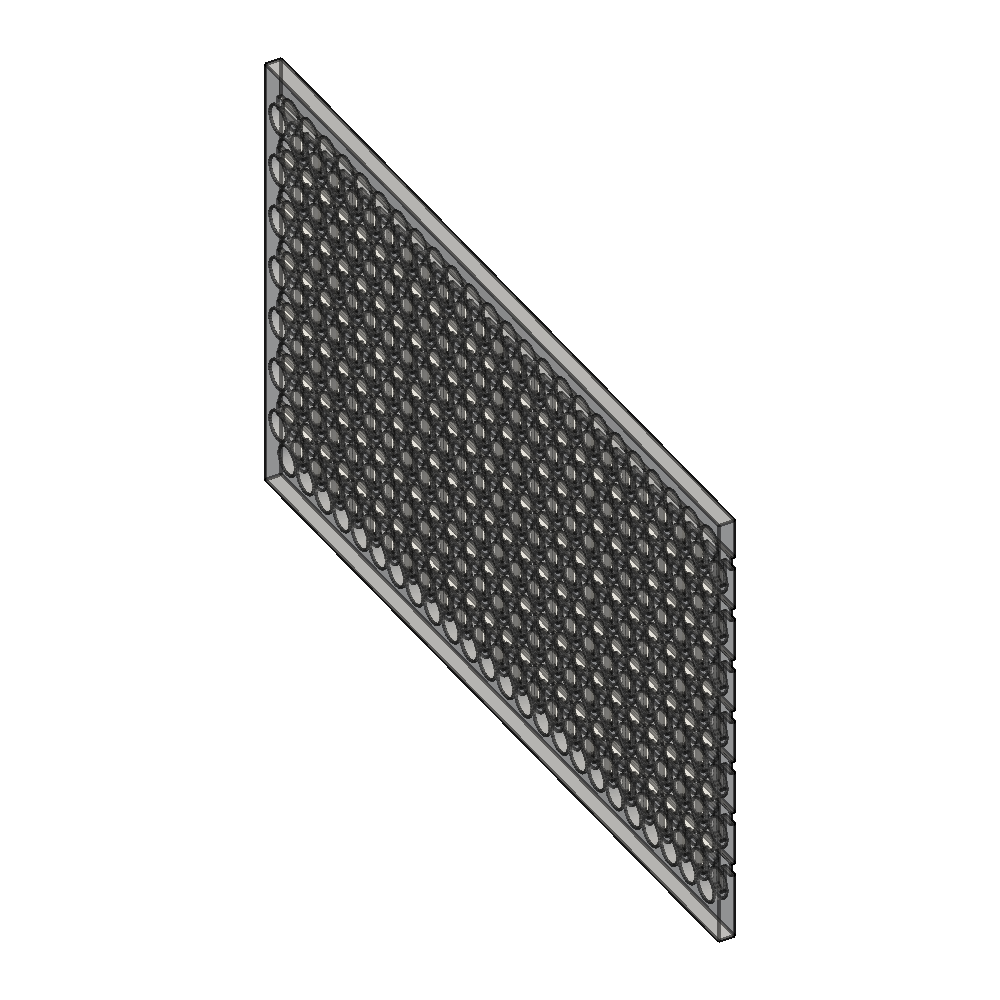}\hfill
    \includegraphics[width=0.117\linewidth]{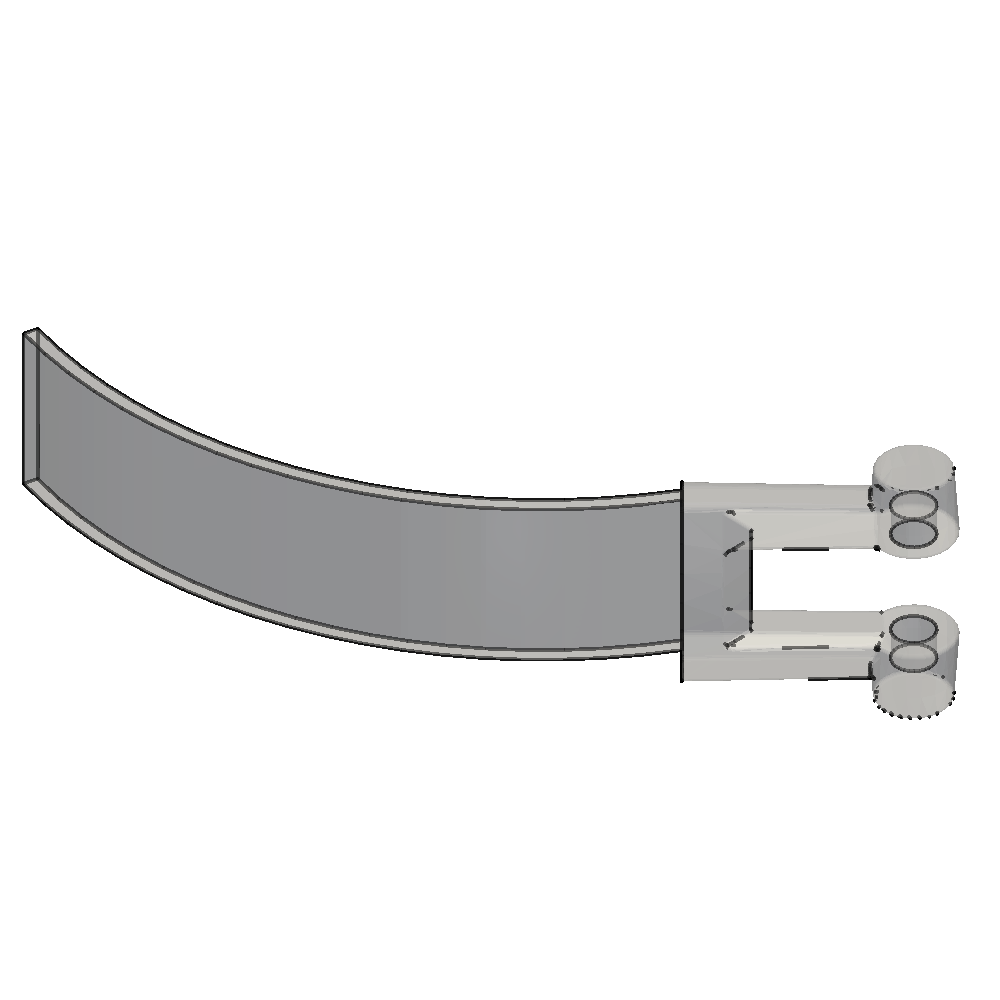}\hfill
    \includegraphics[width=0.117\linewidth]{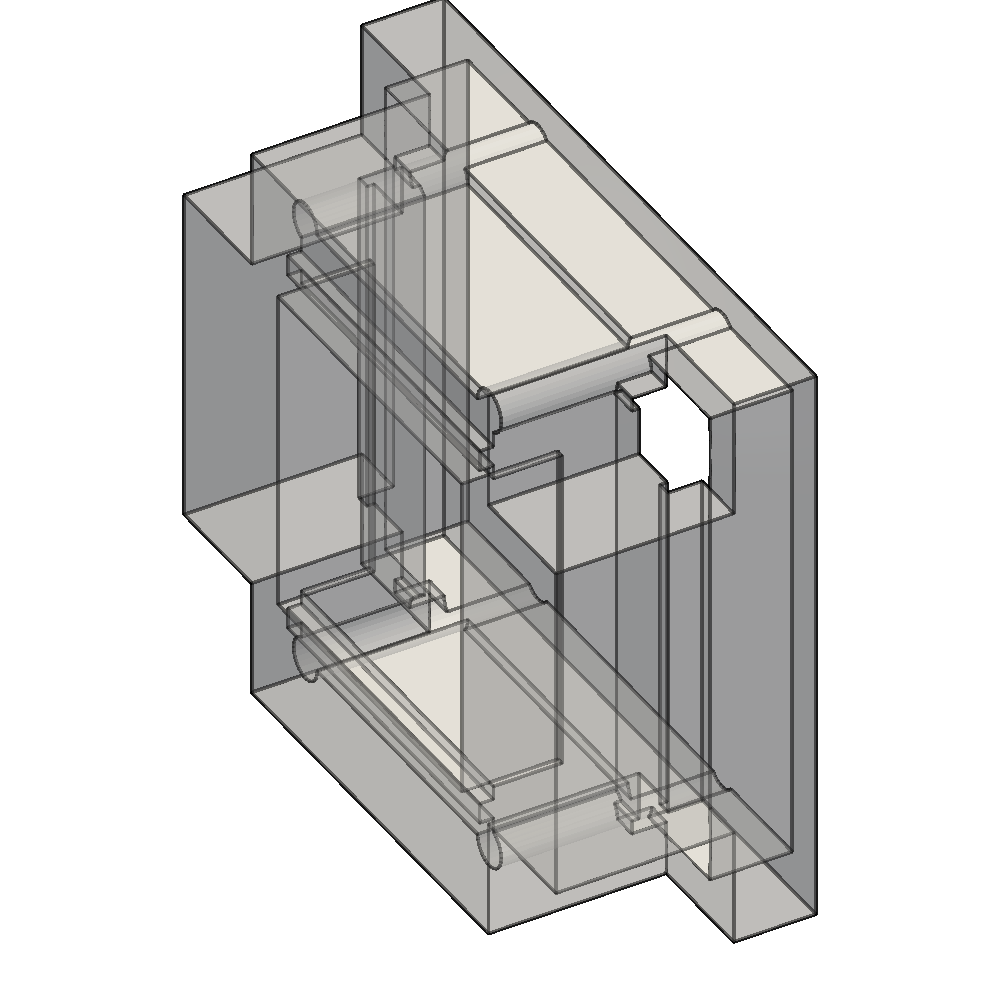}\hfill
    \includegraphics[width=0.117\linewidth]{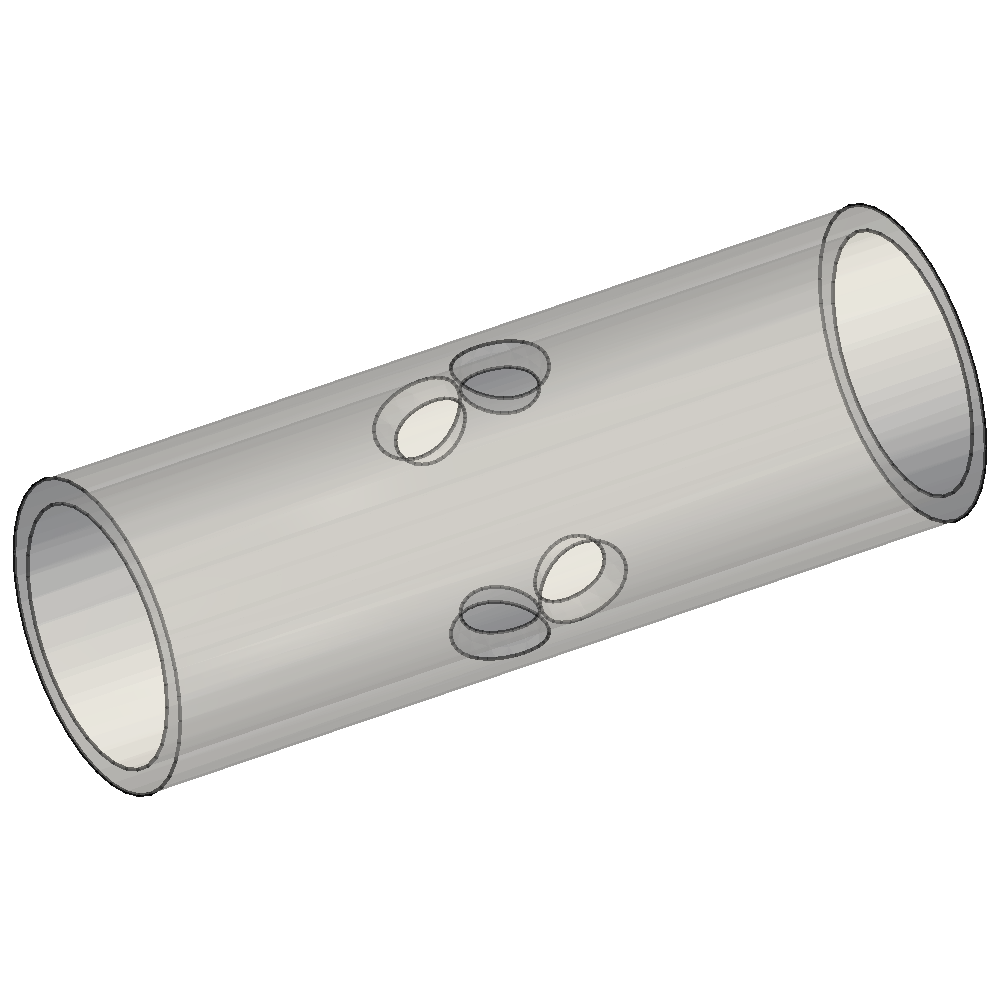}\hfill
    \includegraphics[width=0.117\linewidth]{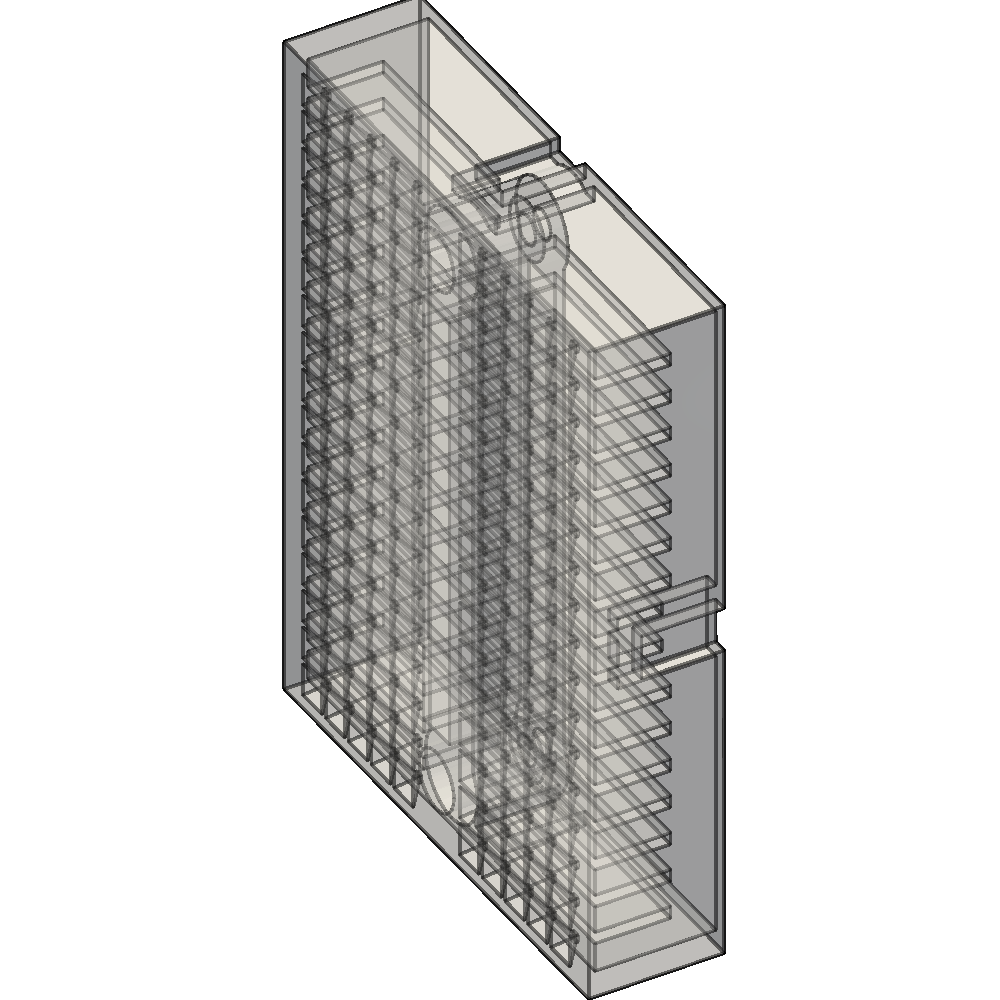}\\[\smallskipamount]
    \includegraphics[width=0.117\linewidth]{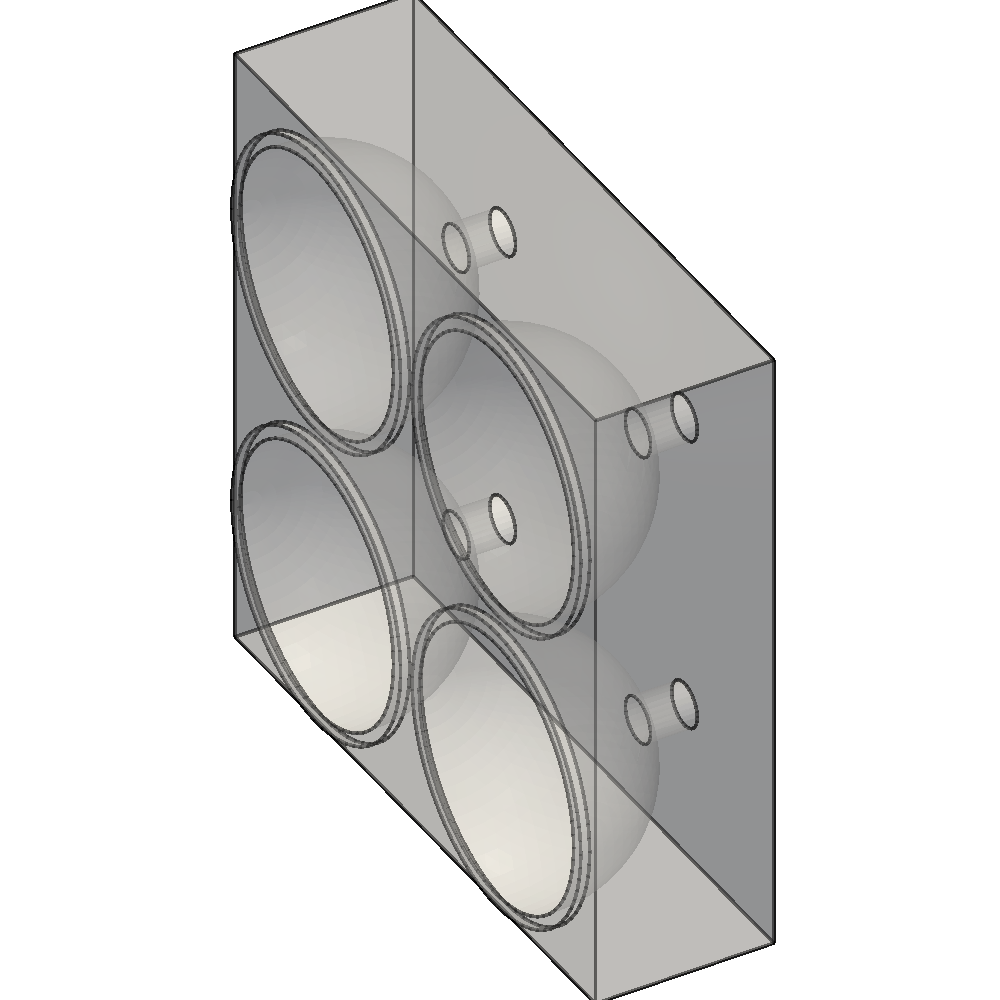}\hfill
    \includegraphics[width=0.117\linewidth]{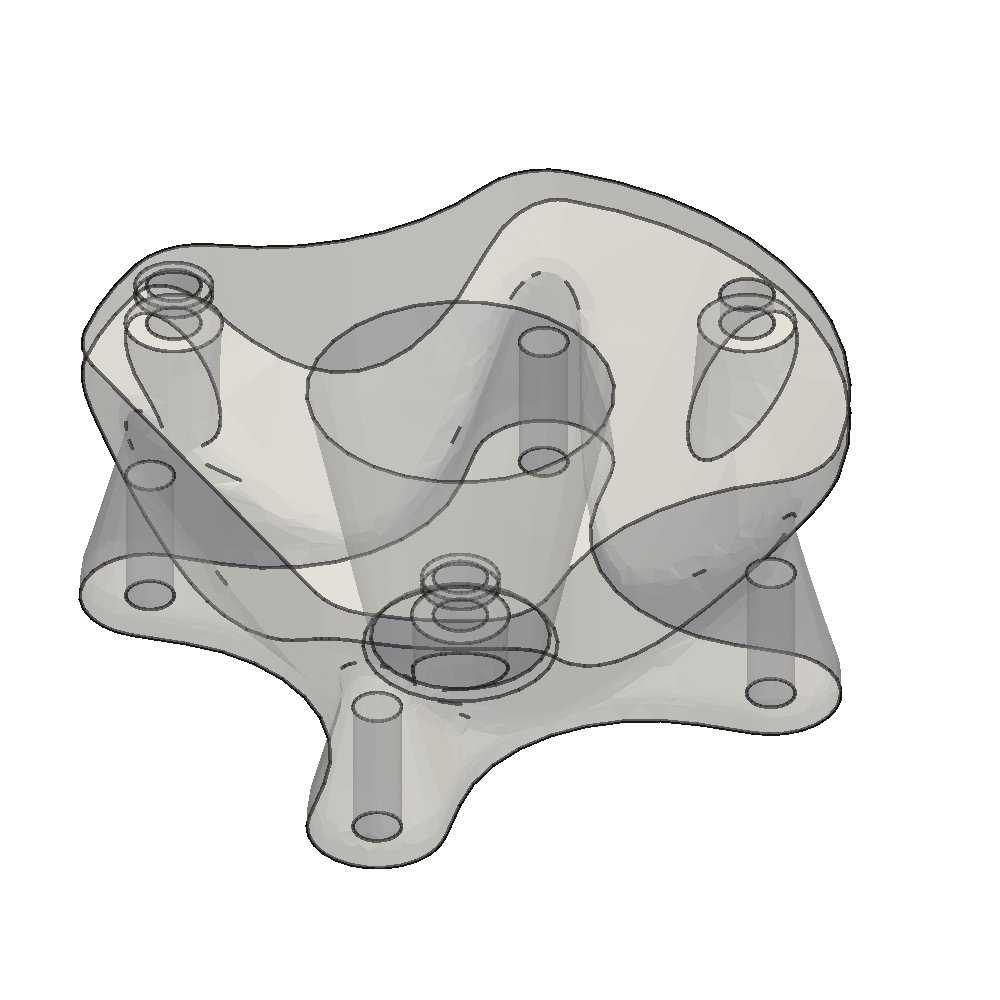}\hfill
    \includegraphics[width=0.117\linewidth]{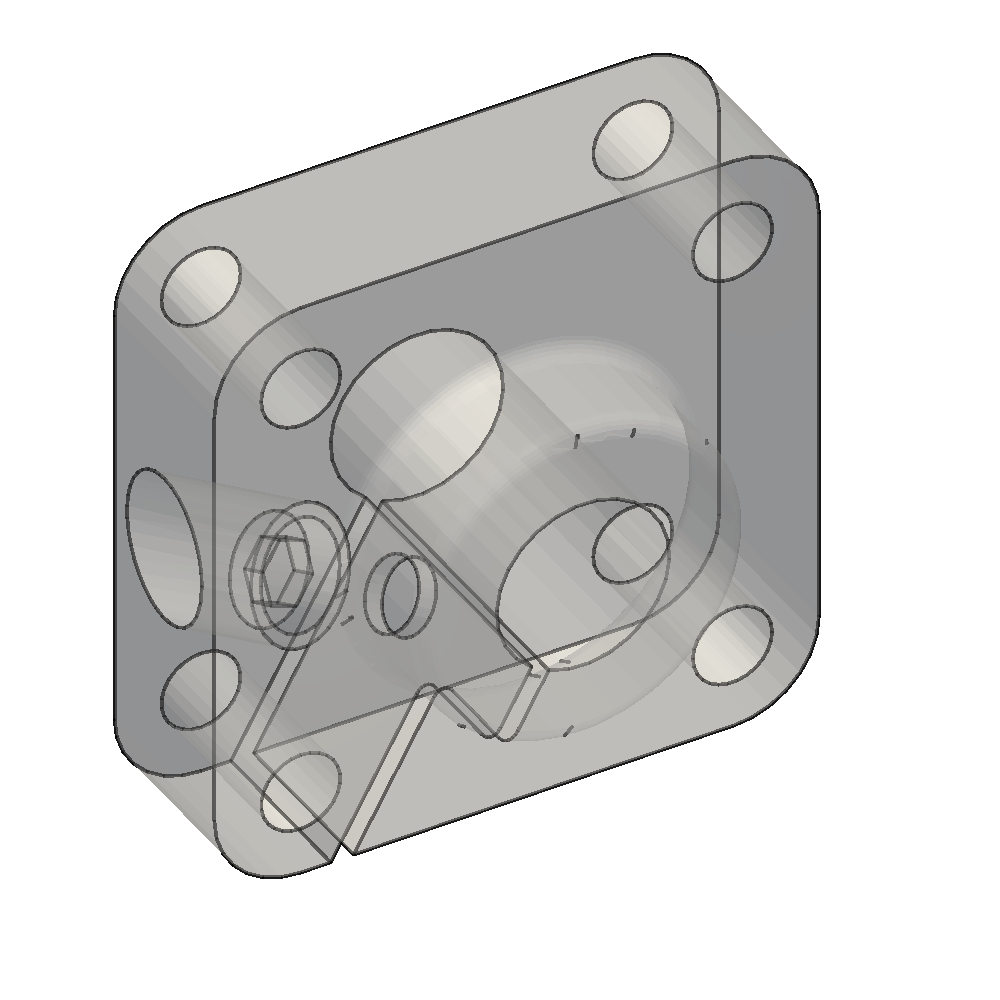}\hfill
    \includegraphics[width=0.117\linewidth]{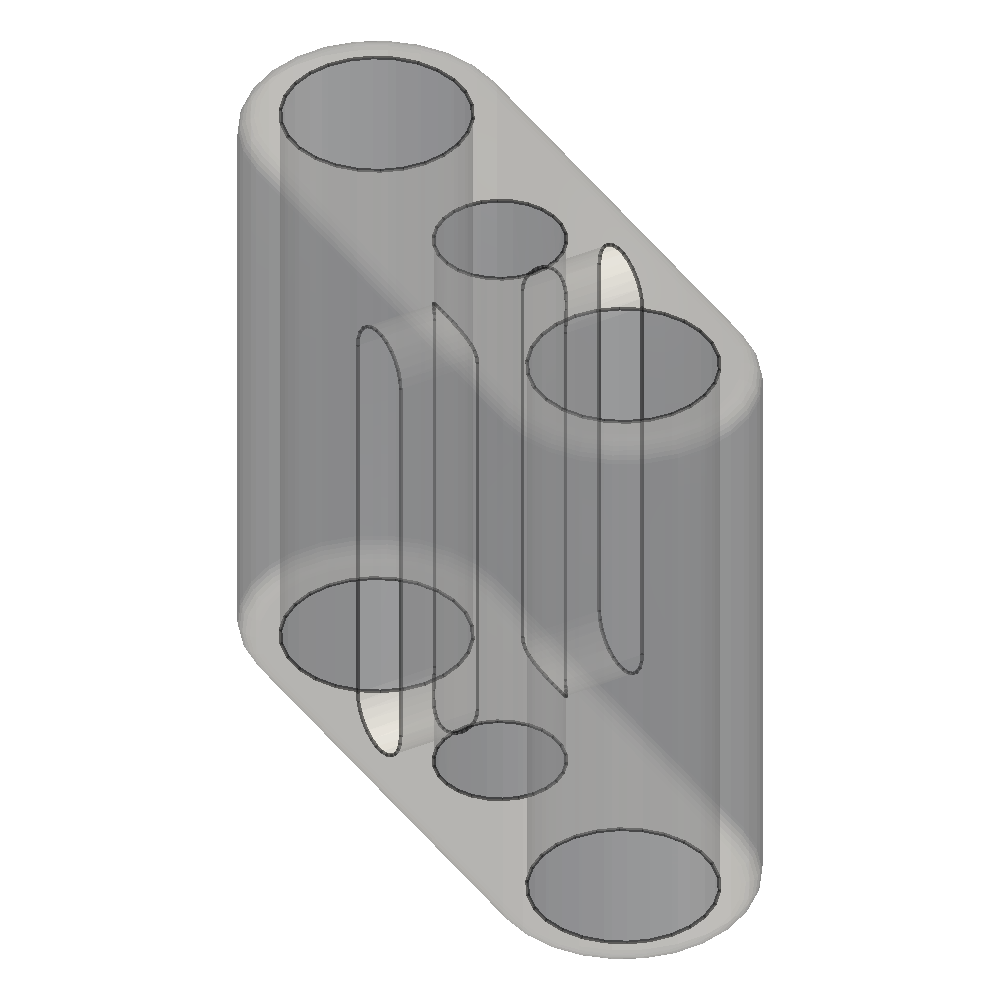}\hfill
    \includegraphics[width=0.117\linewidth]{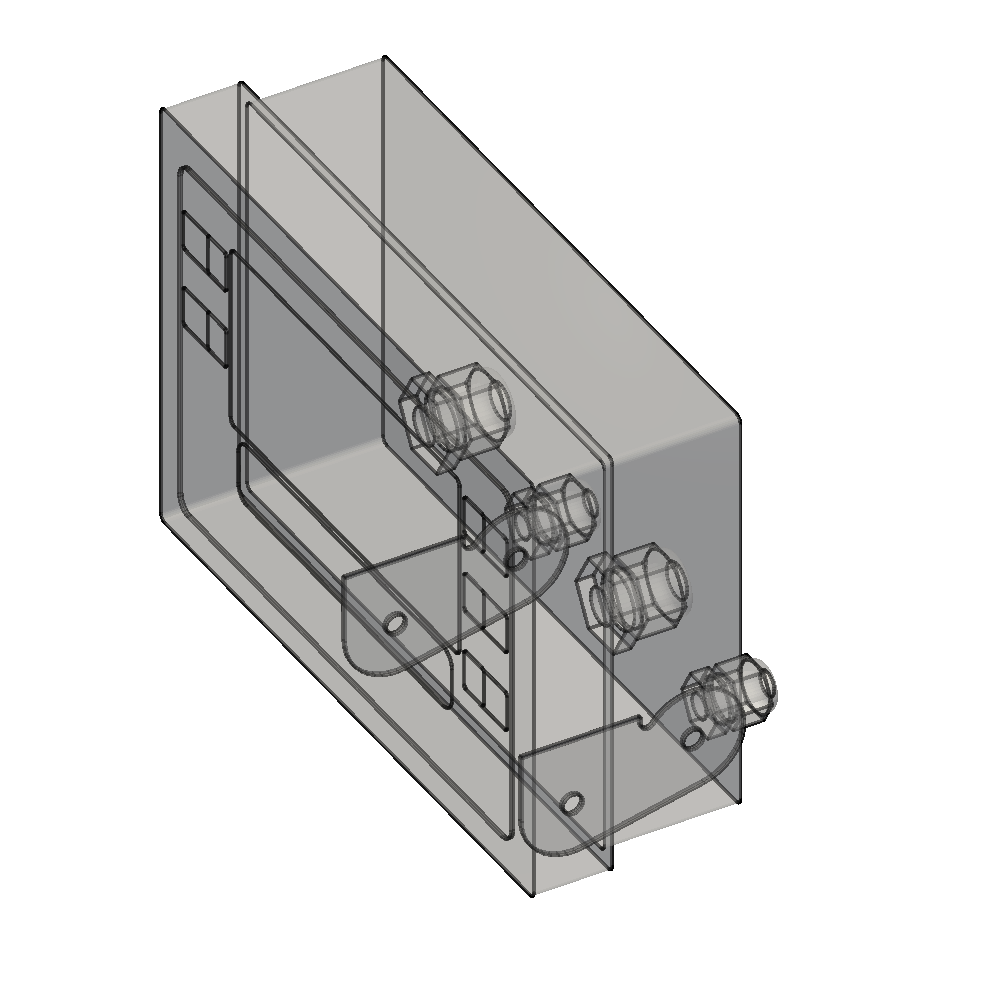}\hfill
    \includegraphics[width=0.117\linewidth]{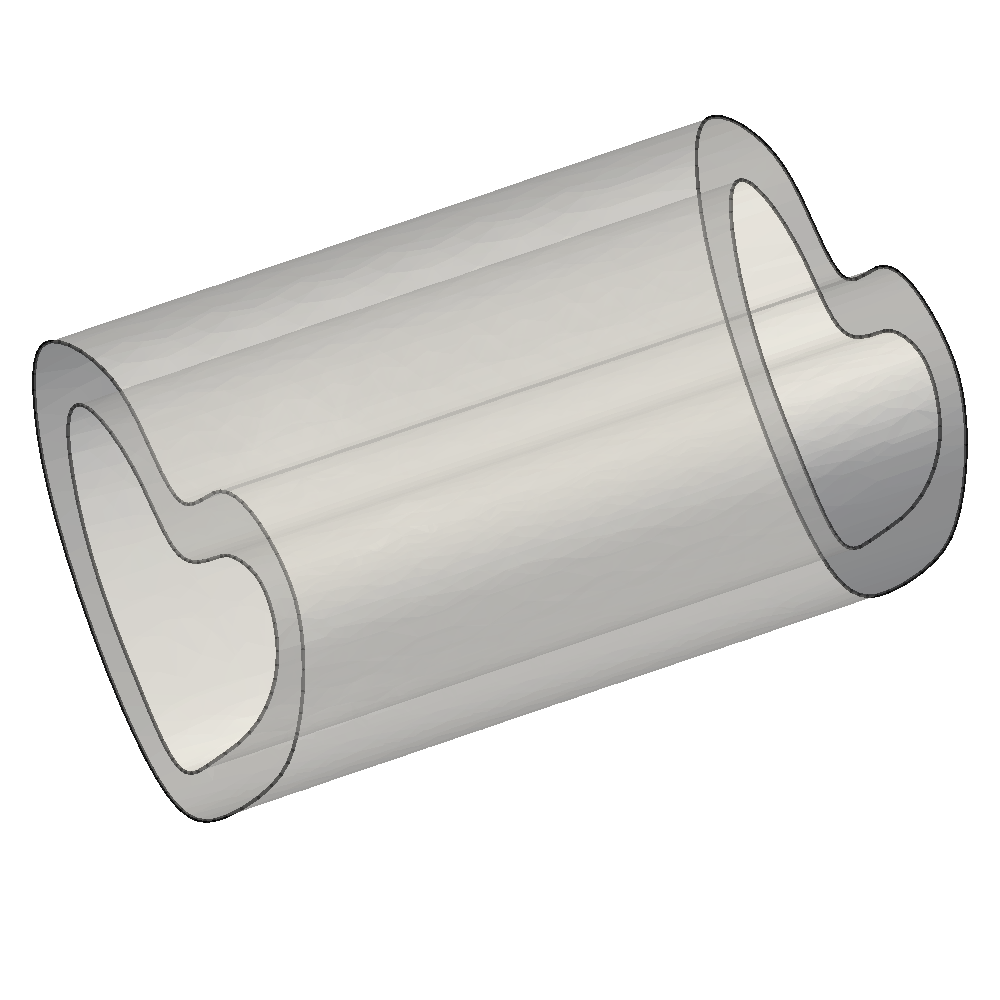}\hfill
    \includegraphics[width=0.117\linewidth]{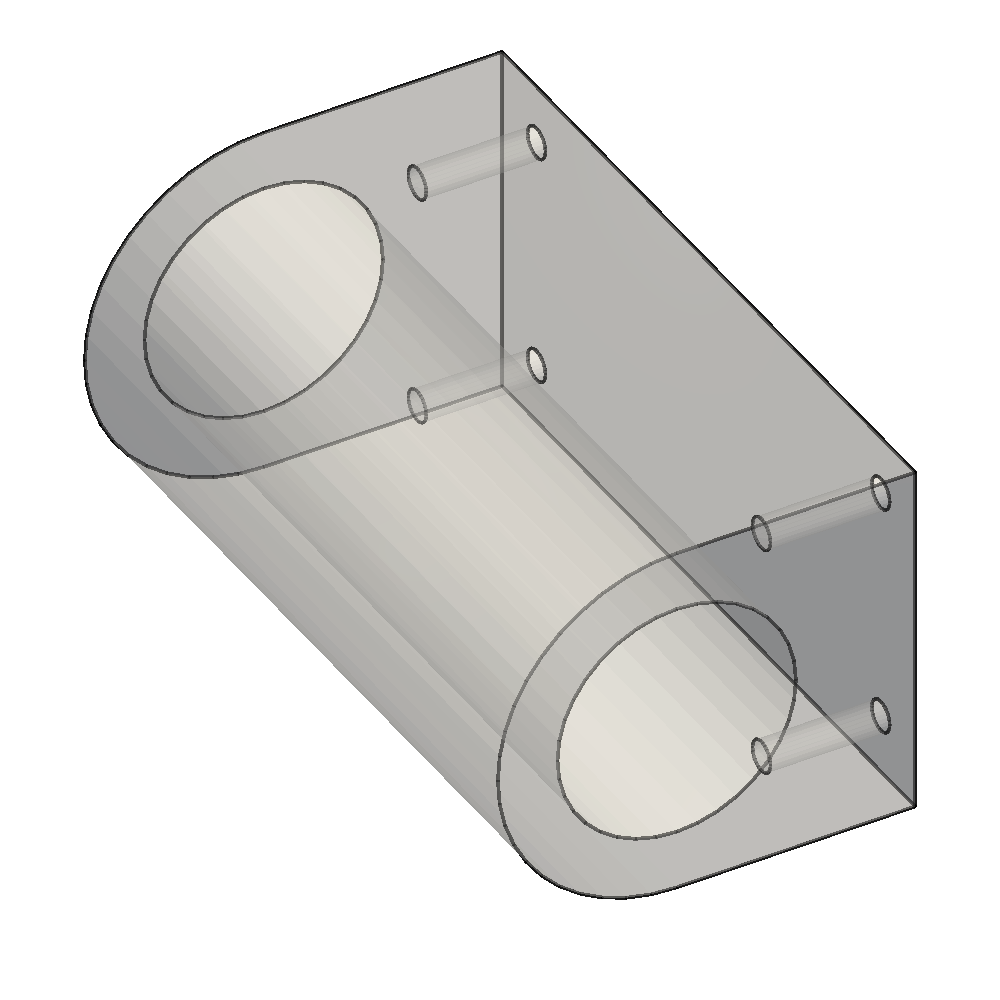}\hfill
    \includegraphics[width=0.117\linewidth]{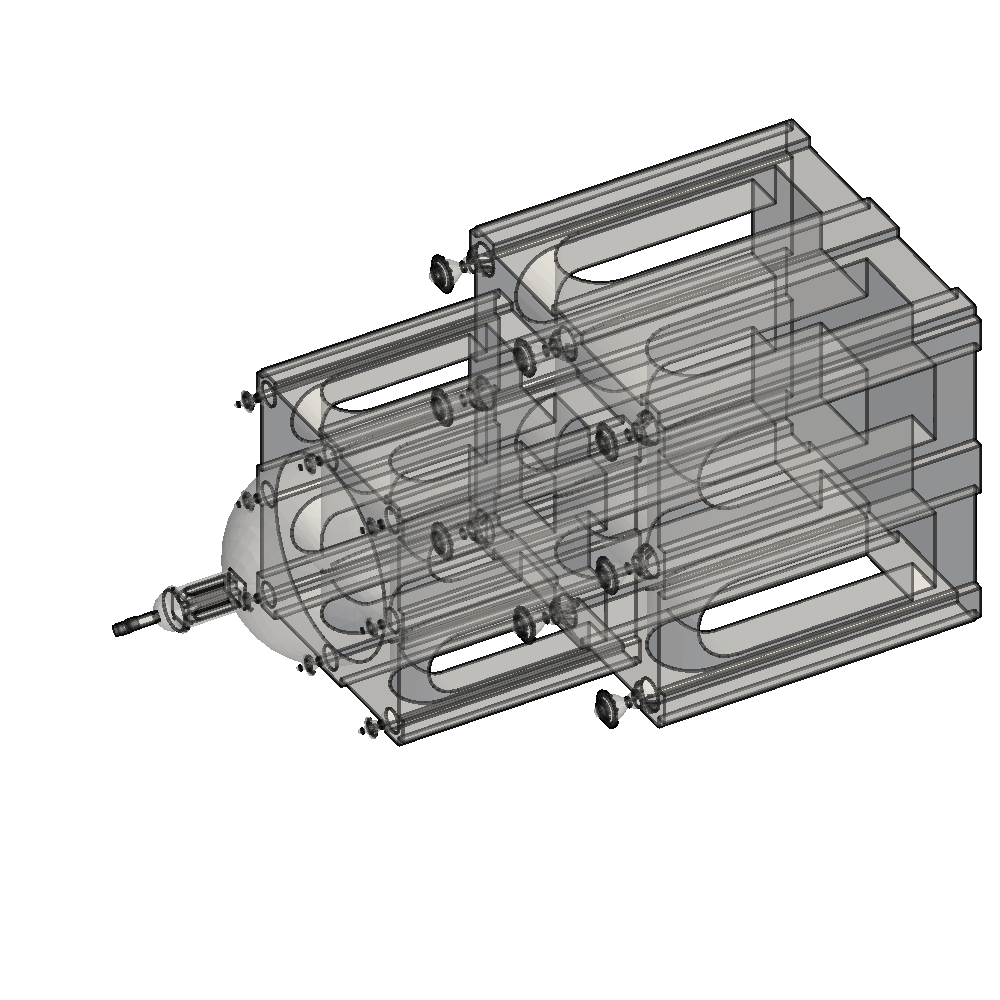}\\[\smallskipamount]
    \includegraphics[width=0.117\linewidth]{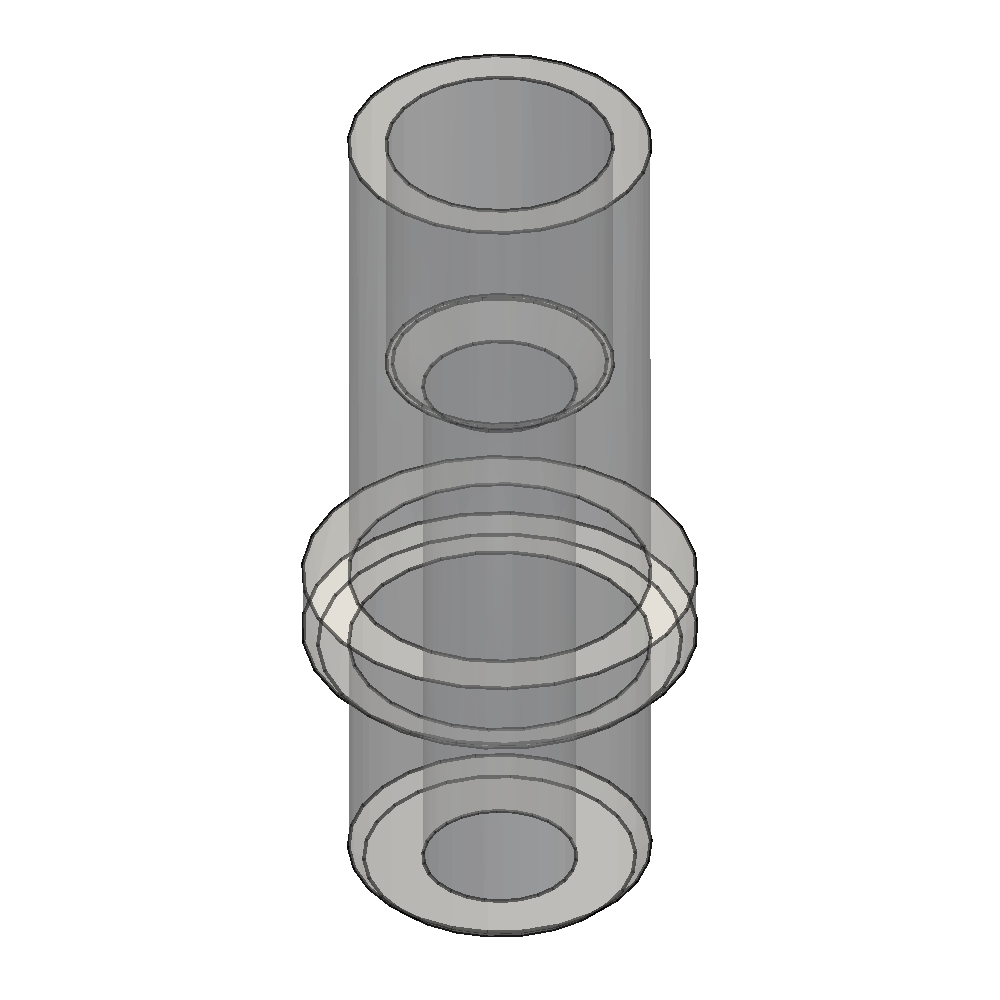}\hfill
    \includegraphics[width=0.117\linewidth]{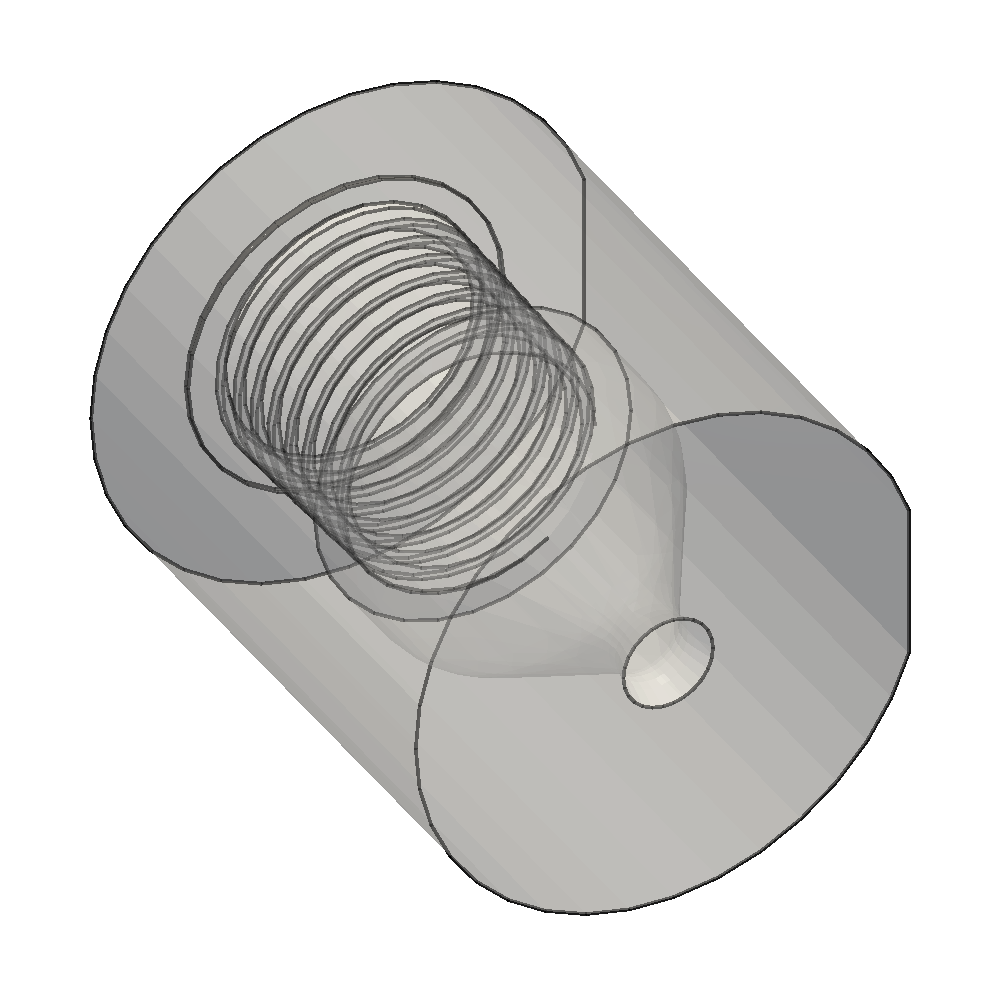}\hfill
    \includegraphics[width=0.117\linewidth]{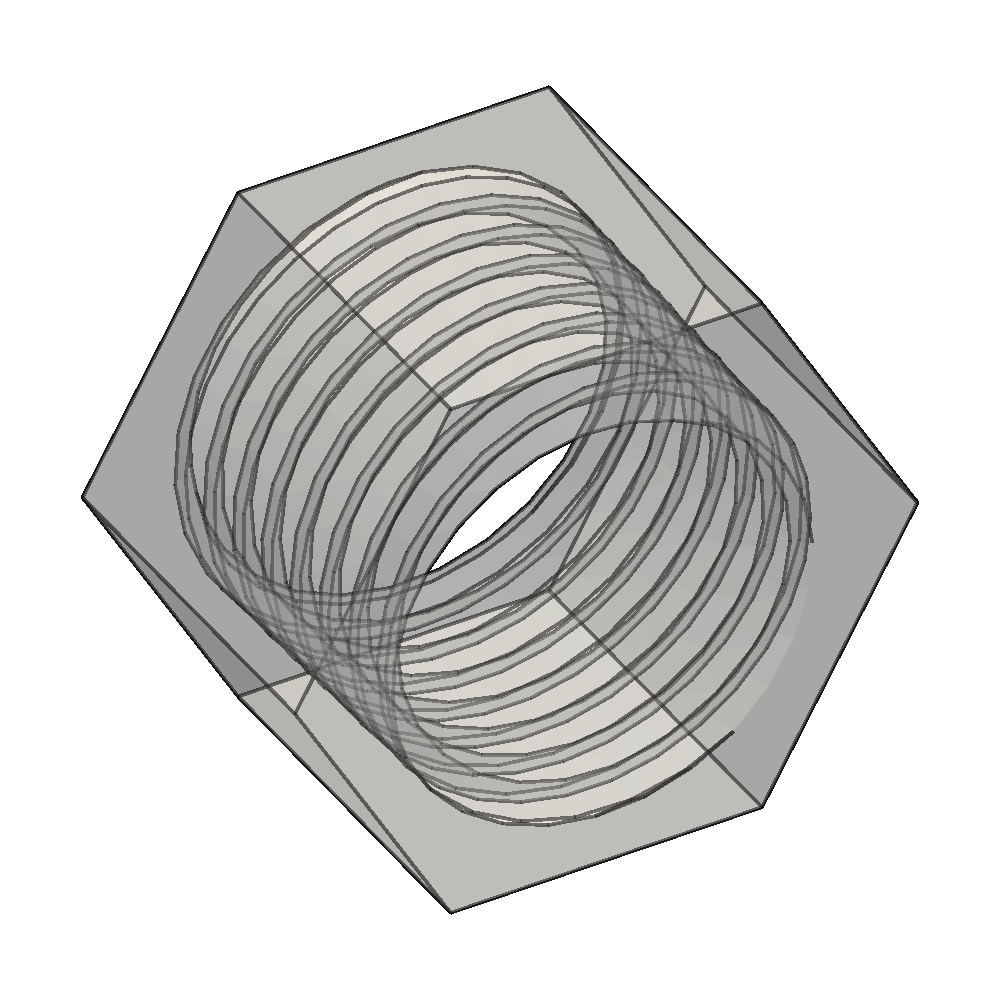}\hfill
    \includegraphics[width=0.117\linewidth]{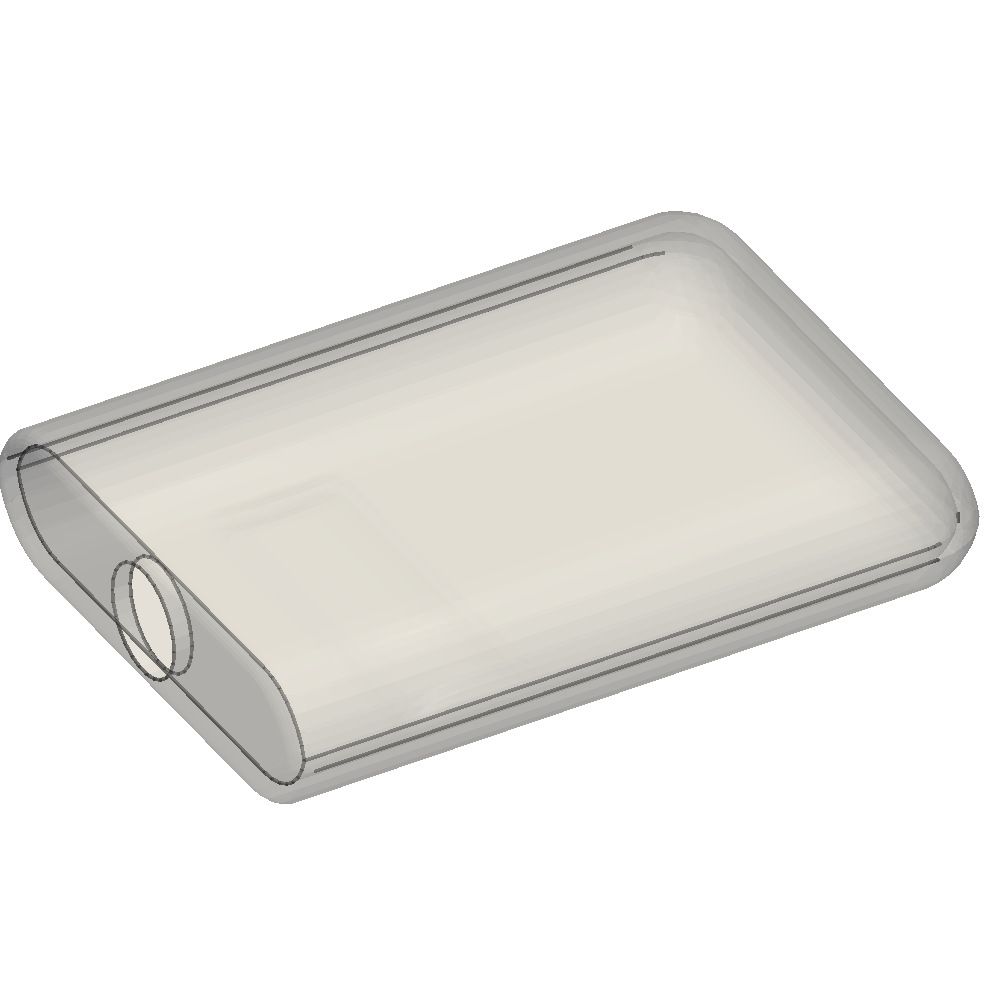}\hfill
    \includegraphics[width=0.117\linewidth]{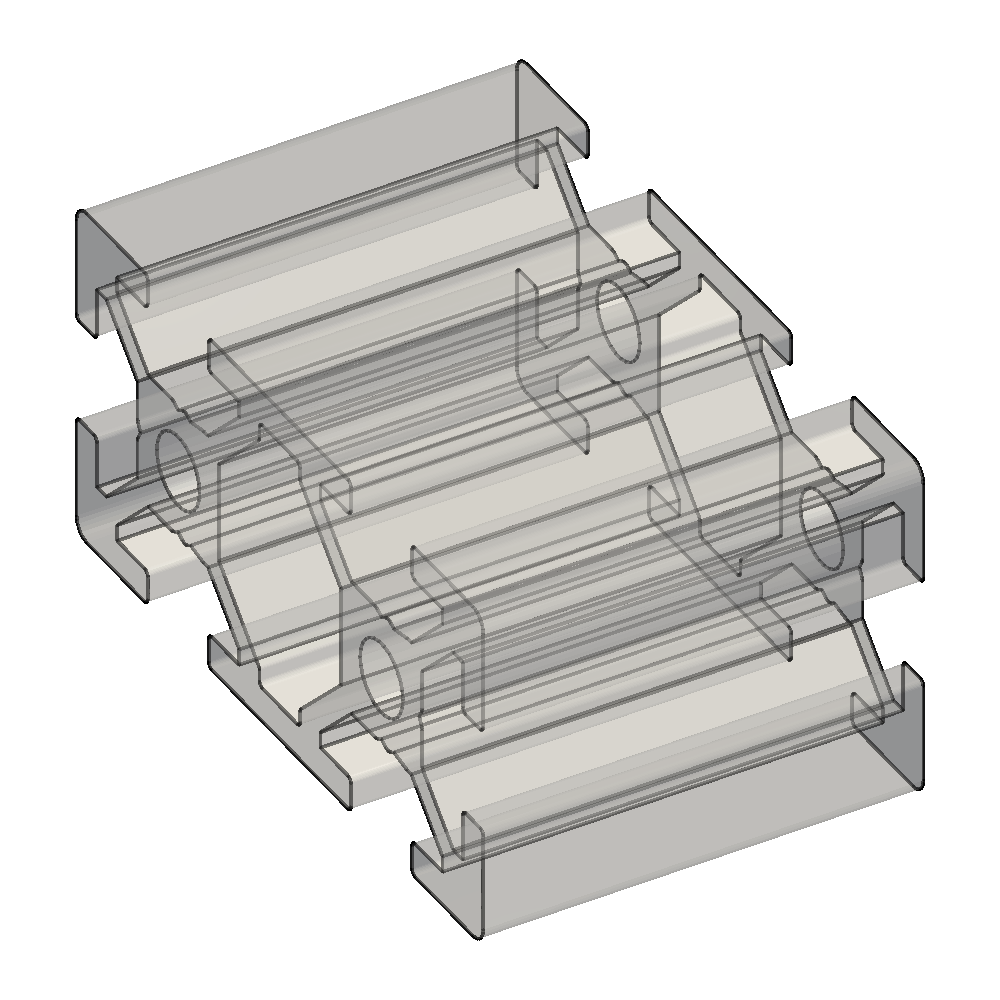}\hfill
    \includegraphics[width=0.117\linewidth]{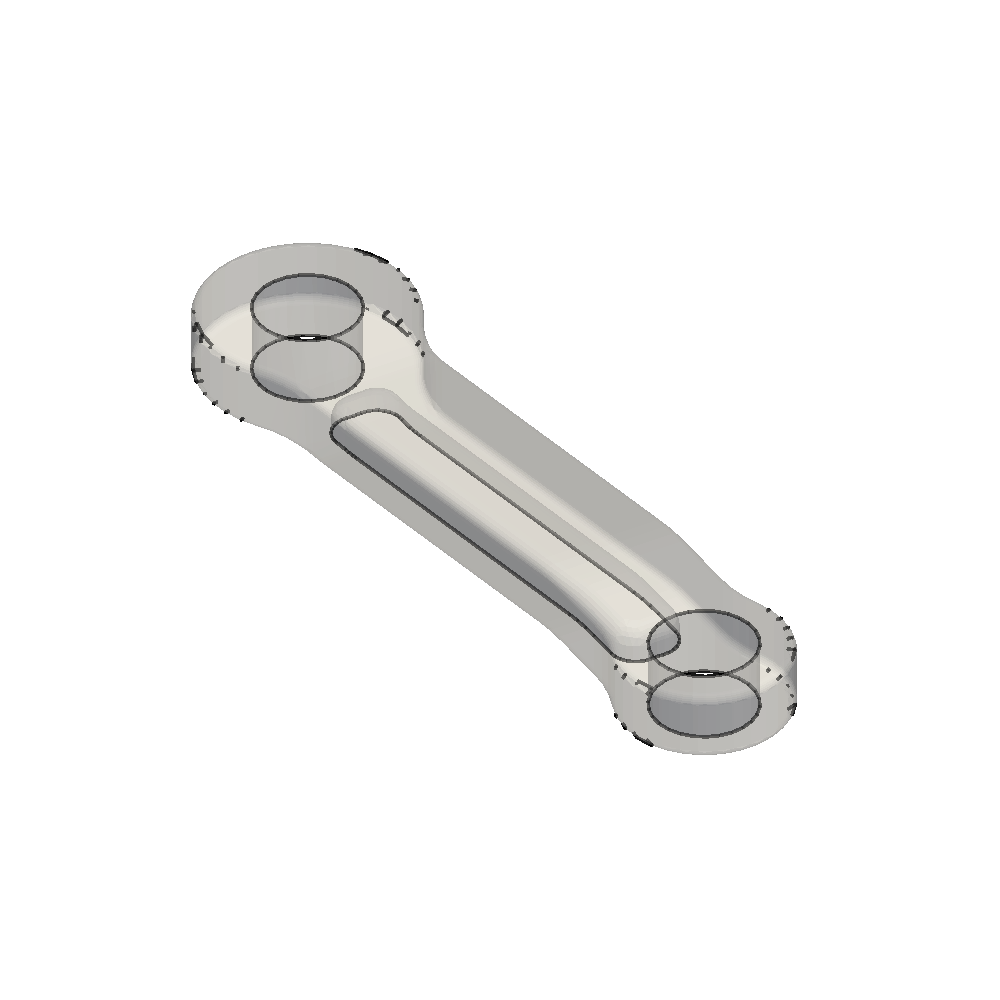}\hfill
    \includegraphics[width=0.117\linewidth]{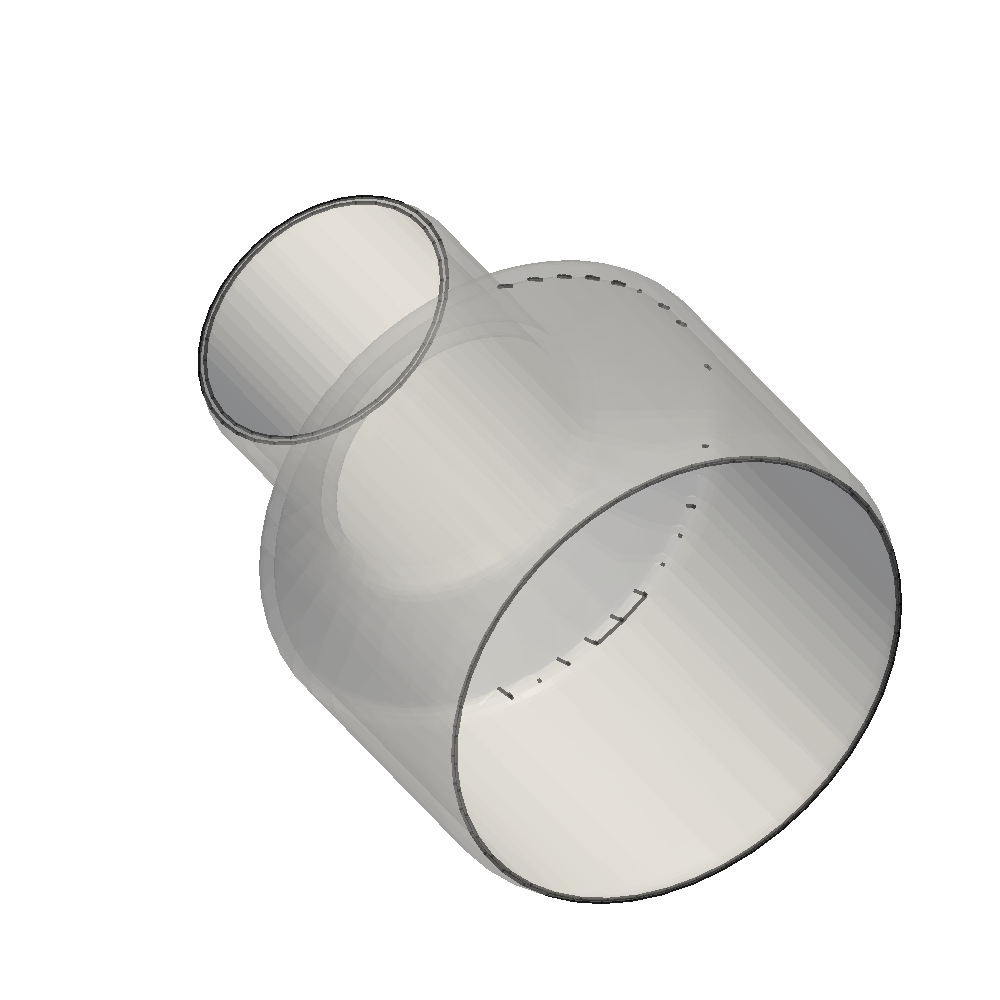}\hfill
    \includegraphics[width=0.117\linewidth]{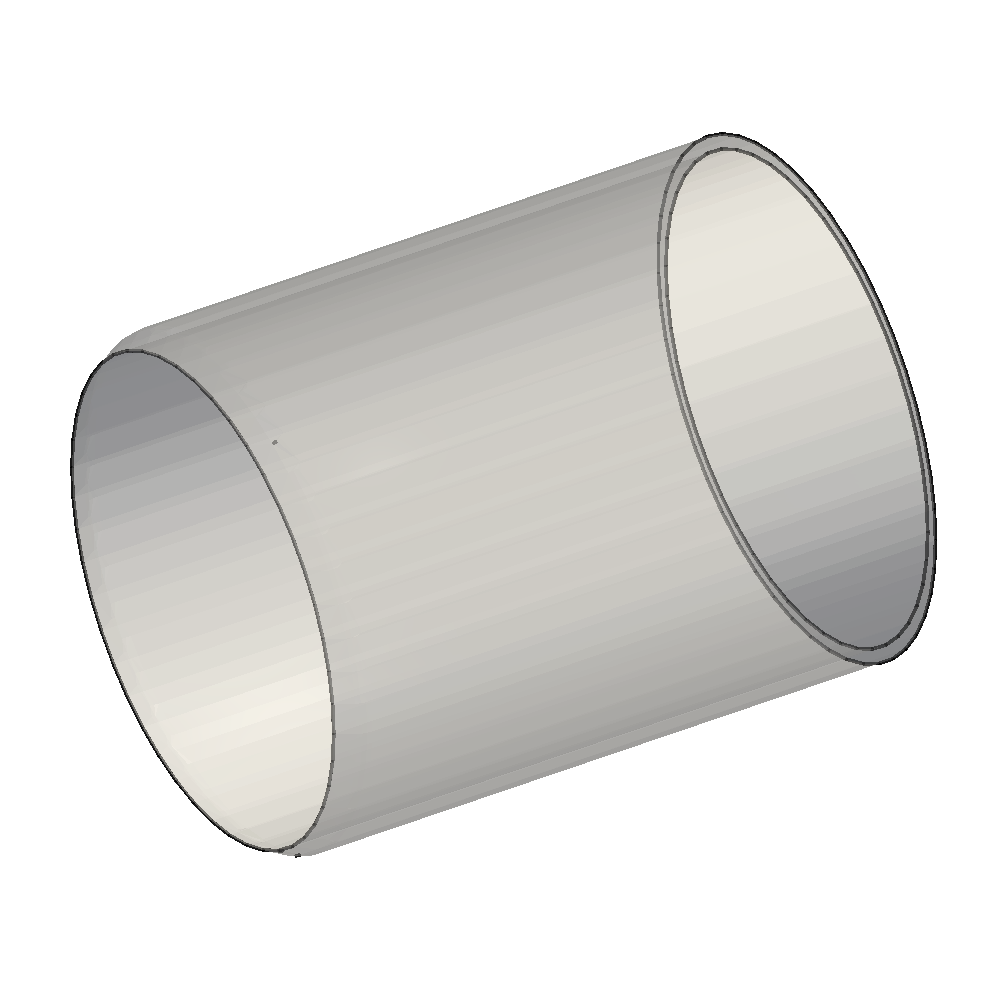}\\[\smallskipamount]
    \includegraphics[width=0.117\linewidth]{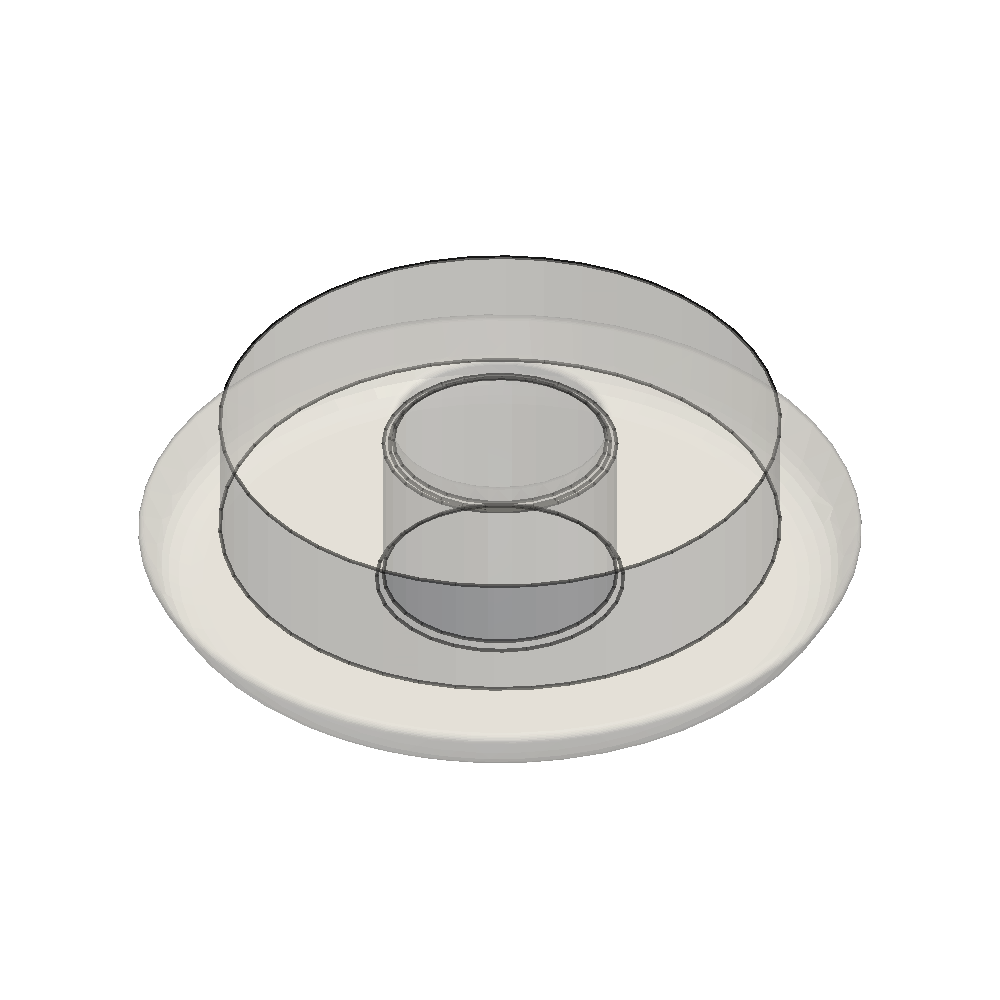}\hfill
    \includegraphics[width=0.117\linewidth]{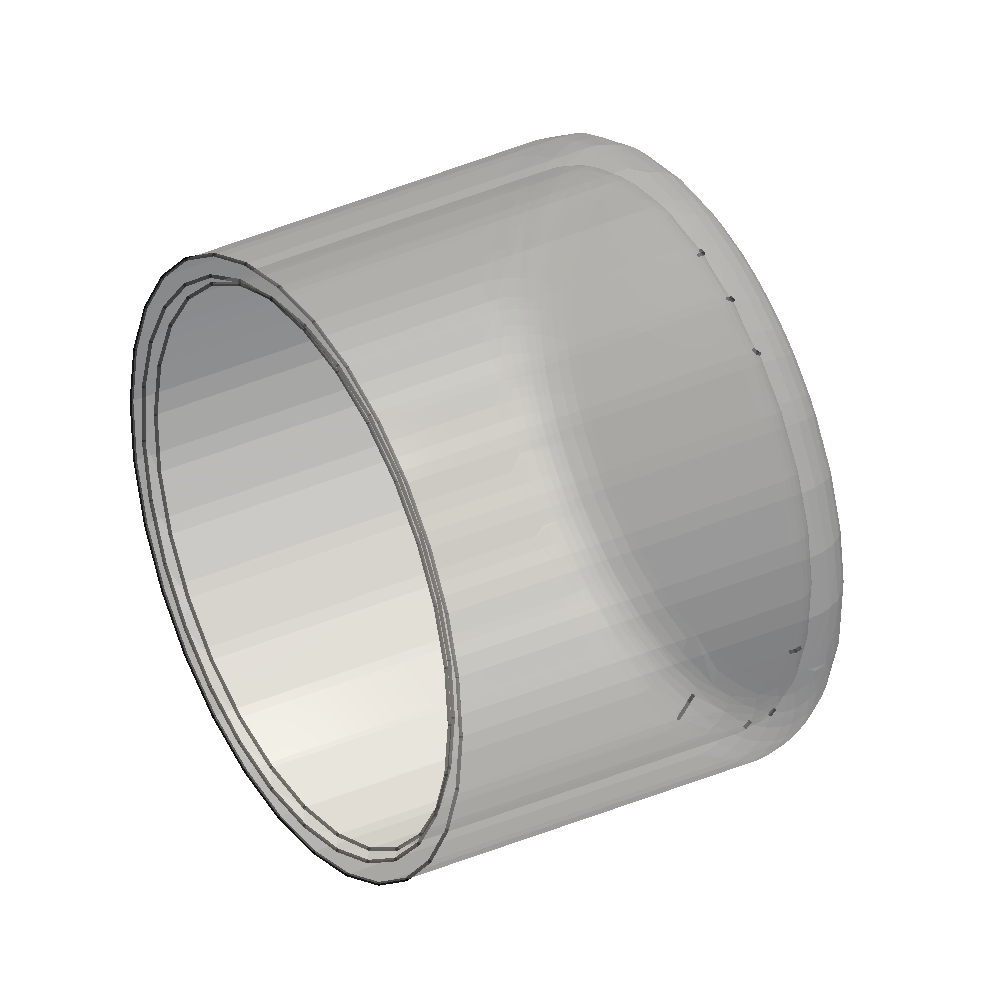}\hfill
    \includegraphics[width=0.117\linewidth]{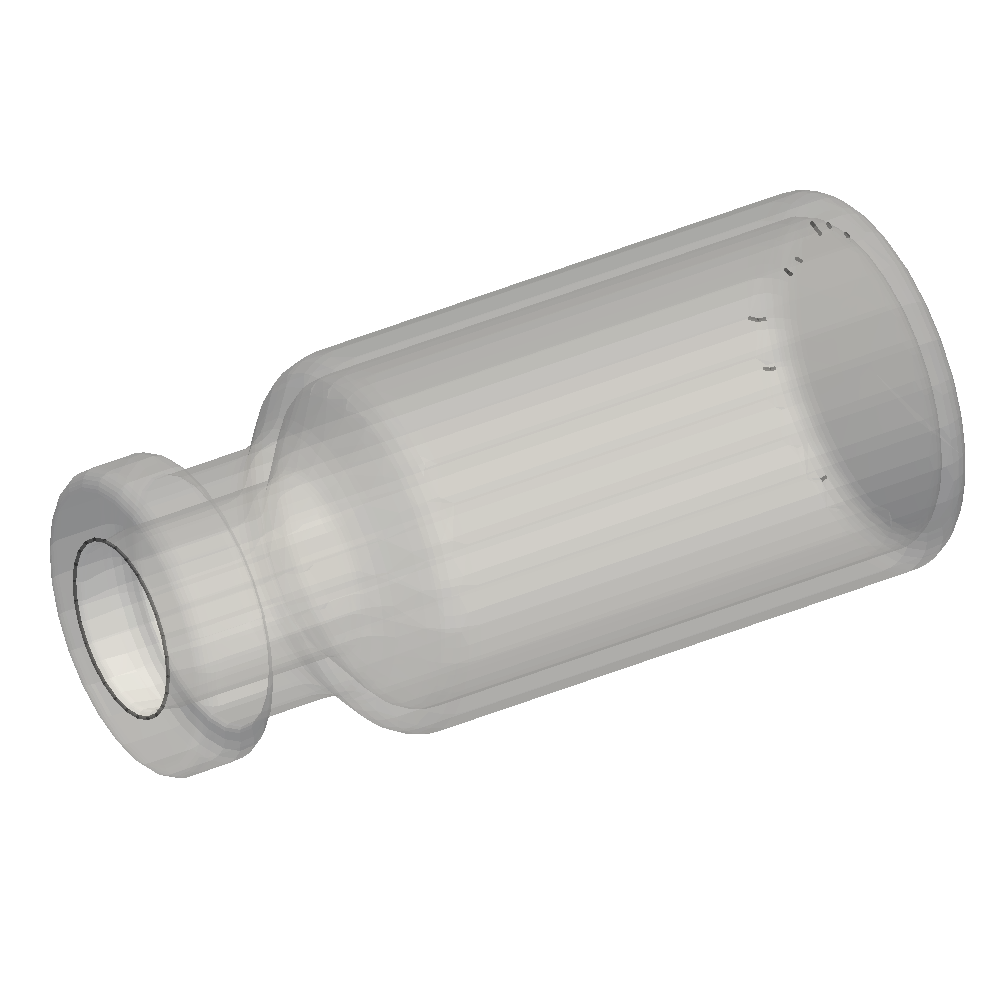}\hfill
    \includegraphics[width=0.117\linewidth]{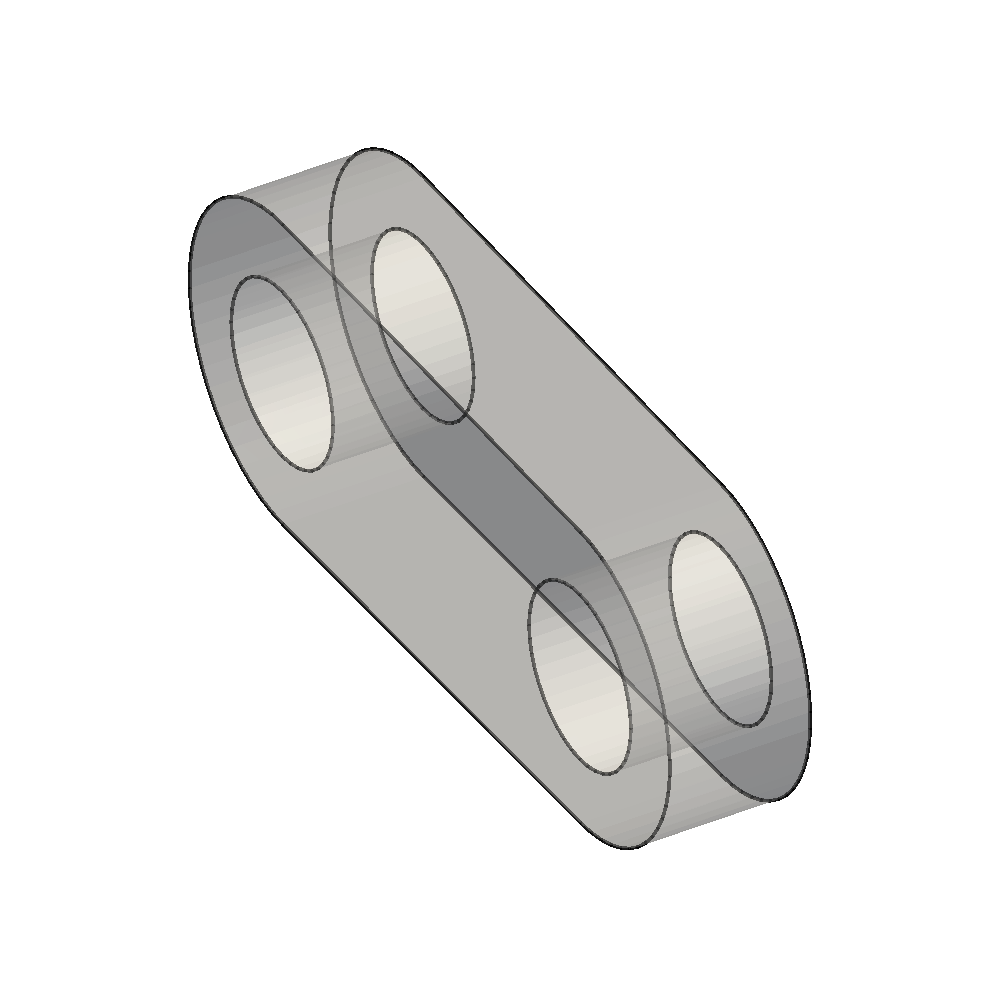}\hfill
    \includegraphics[width=0.117\linewidth]{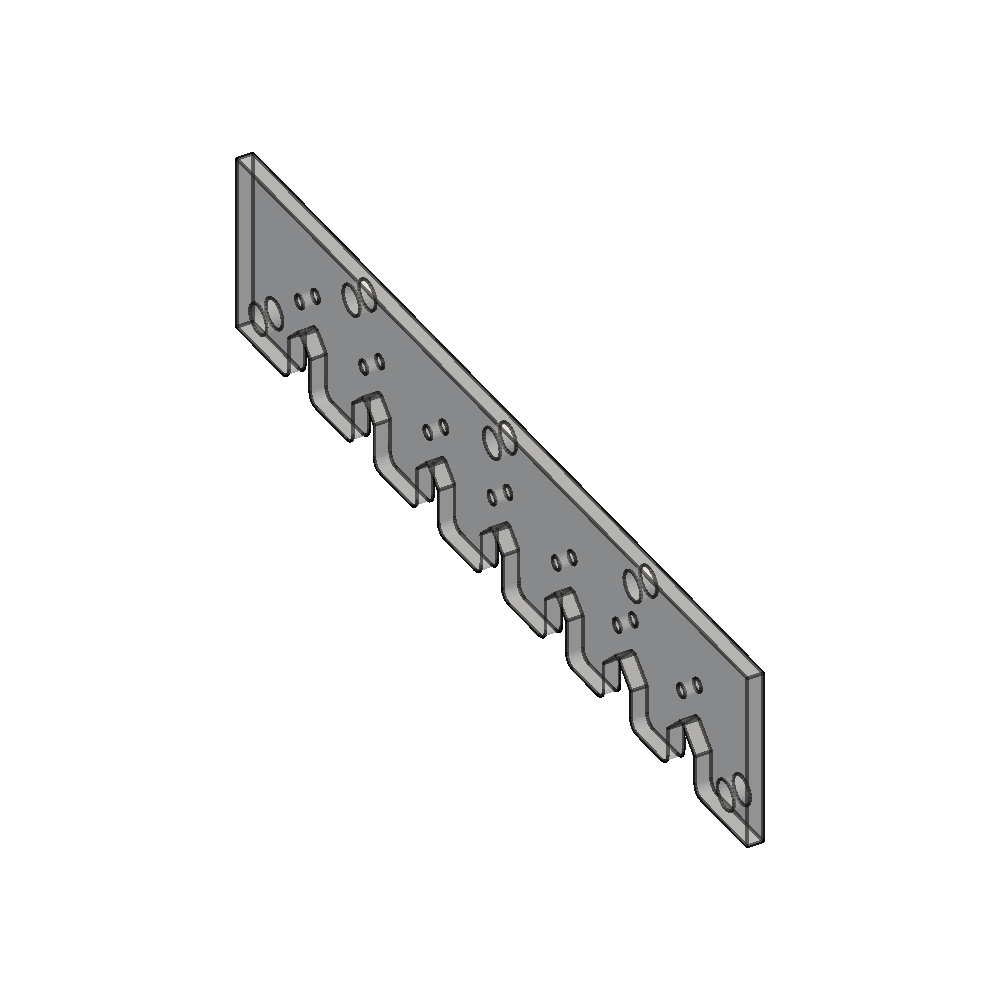}\hfill
    \includegraphics[width=0.117\linewidth]{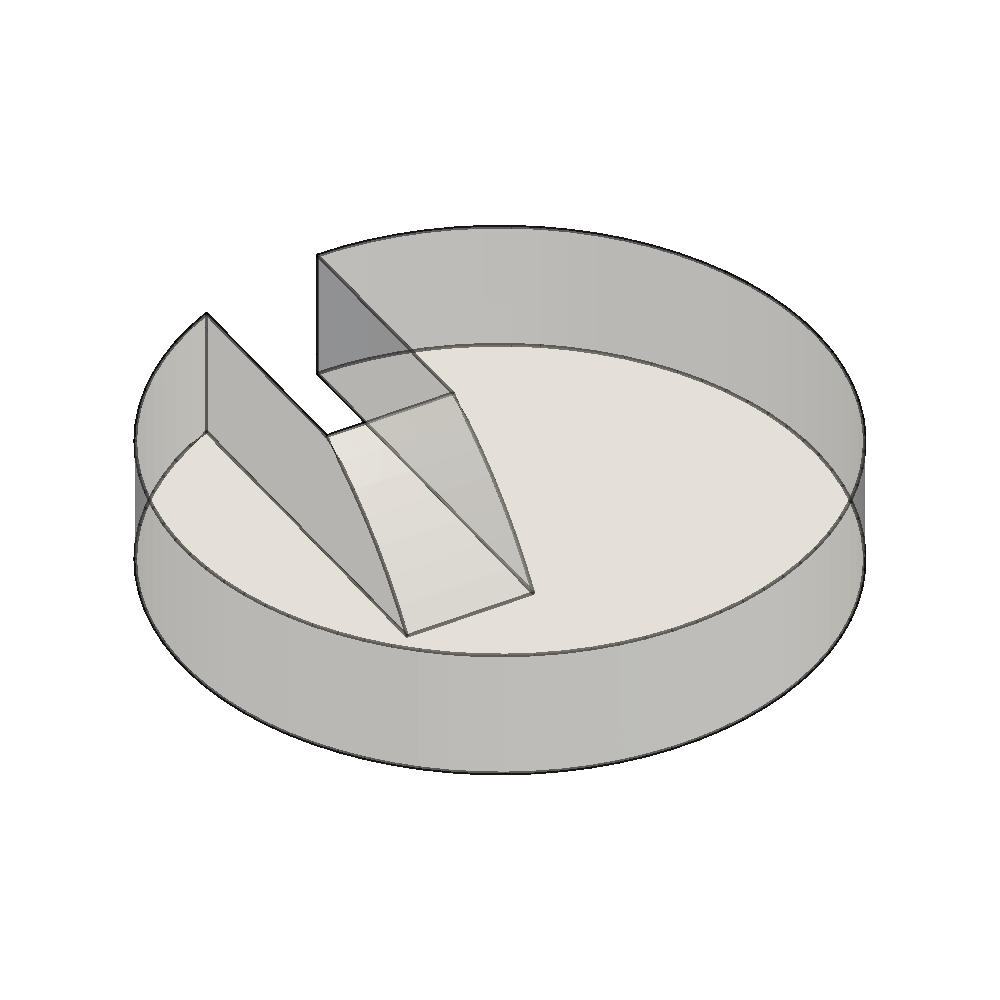}\hfill
    \includegraphics[width=0.117\linewidth]{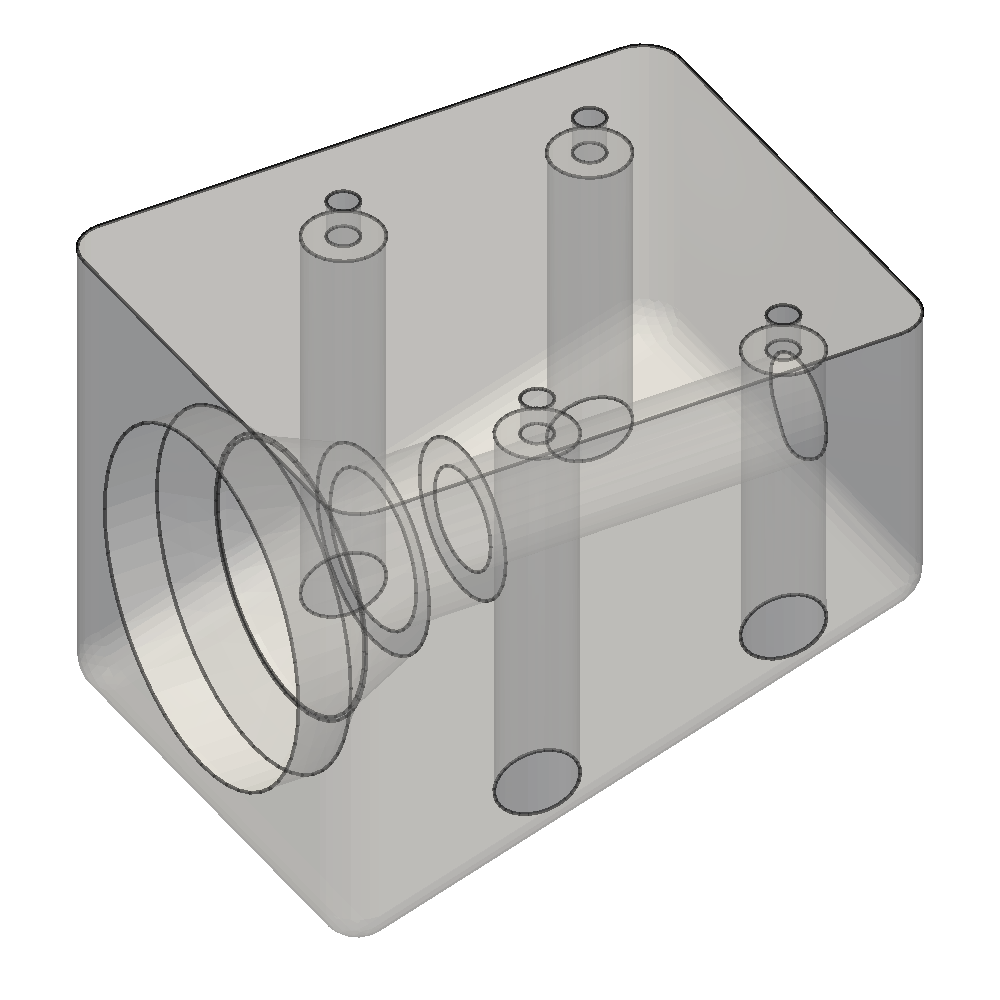}\hfill
    \includegraphics[width=0.117\linewidth]{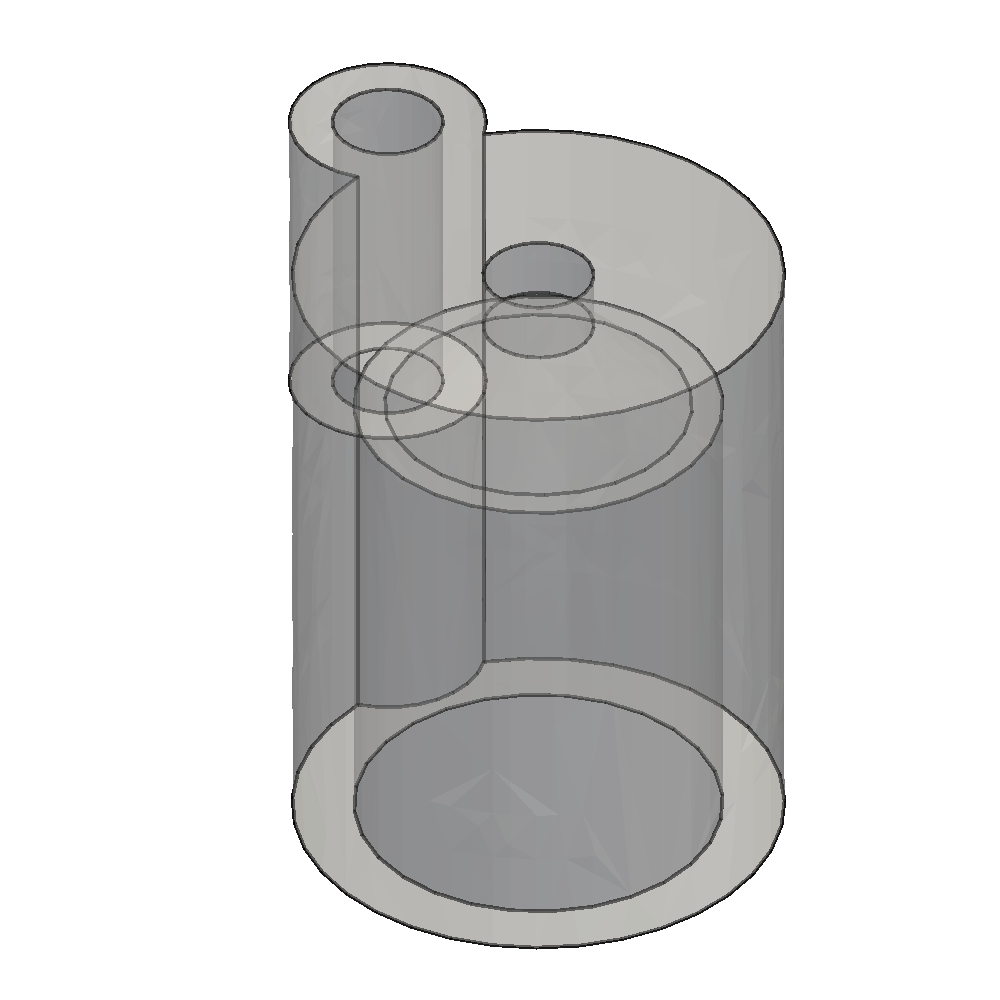}

    \caption{(continued, part 2 of 4) CAD parts of the 156-part evaluation benchmark.}
    \label{fig:cad_dataset_grid_2}
\end{figure}

\begin{figure}[h]
    \centering
    \includegraphics[width=0.117\linewidth]{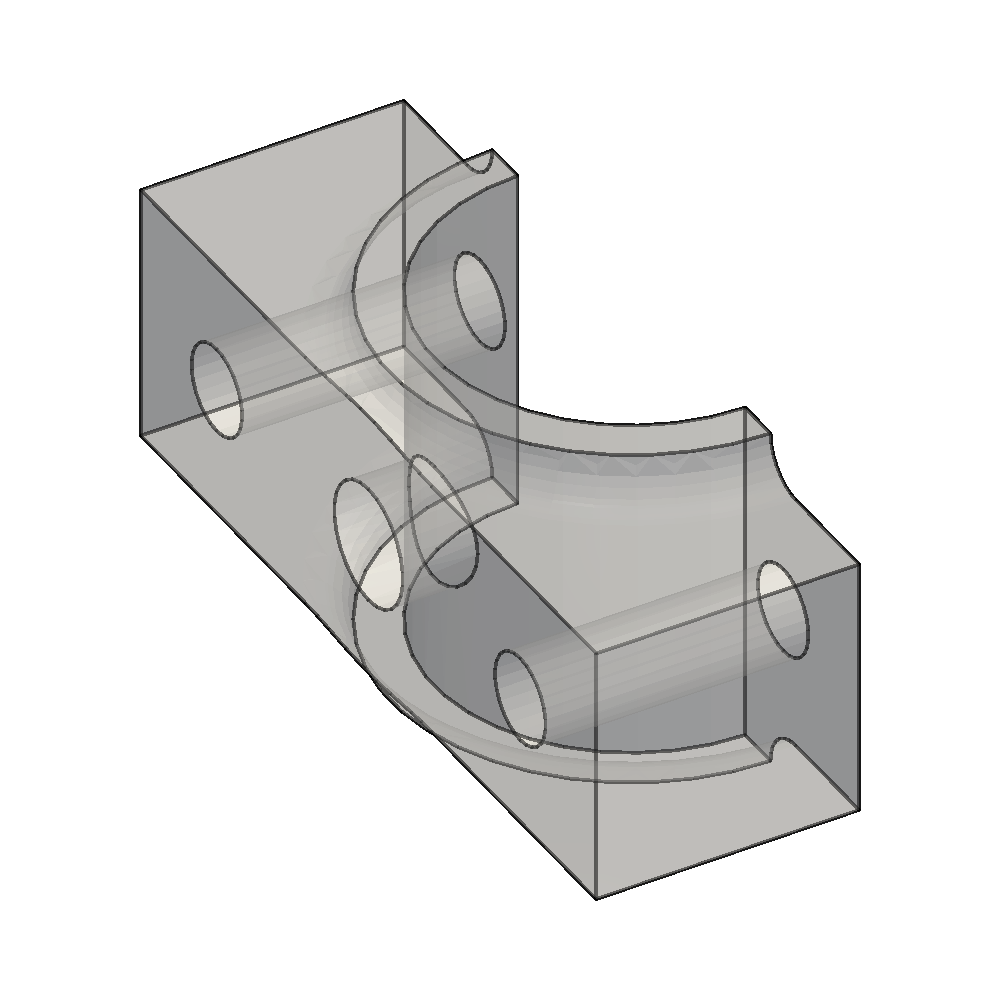}\hfill
    \includegraphics[width=0.117\linewidth]{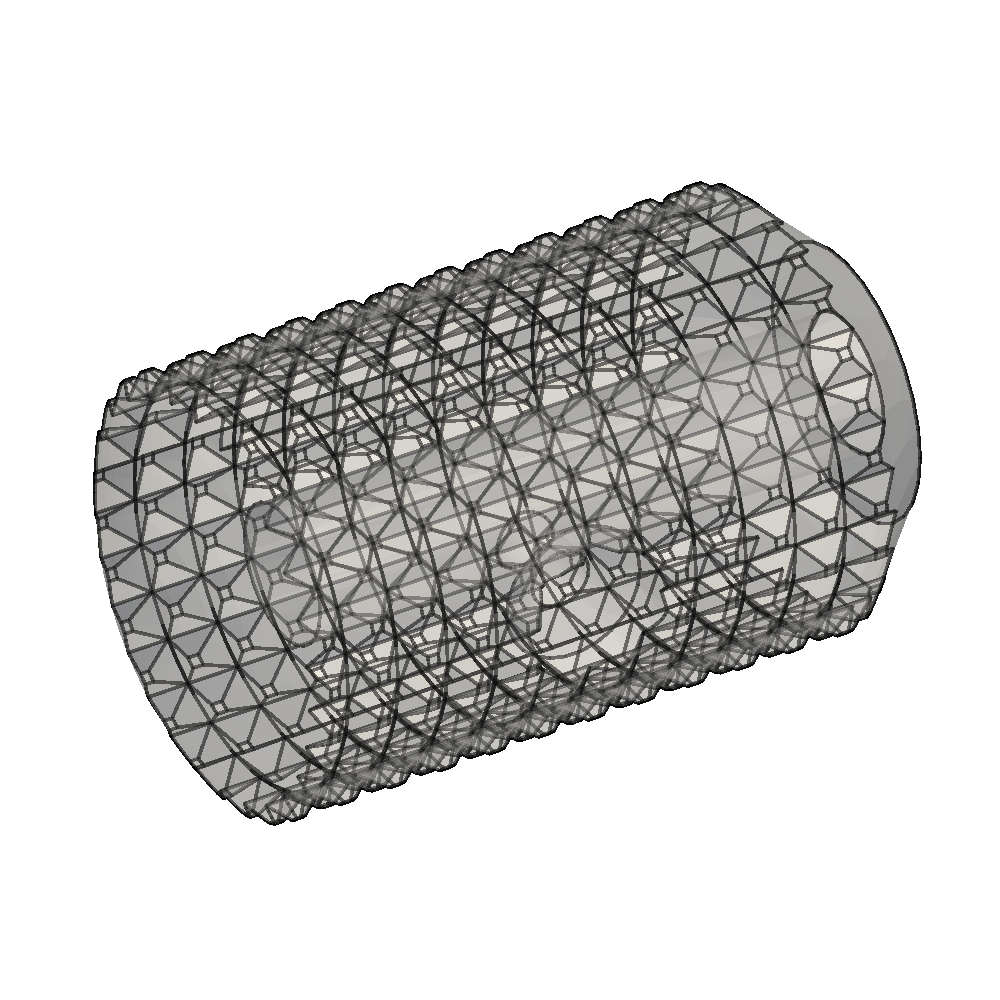}\hfill
    \includegraphics[width=0.117\linewidth]{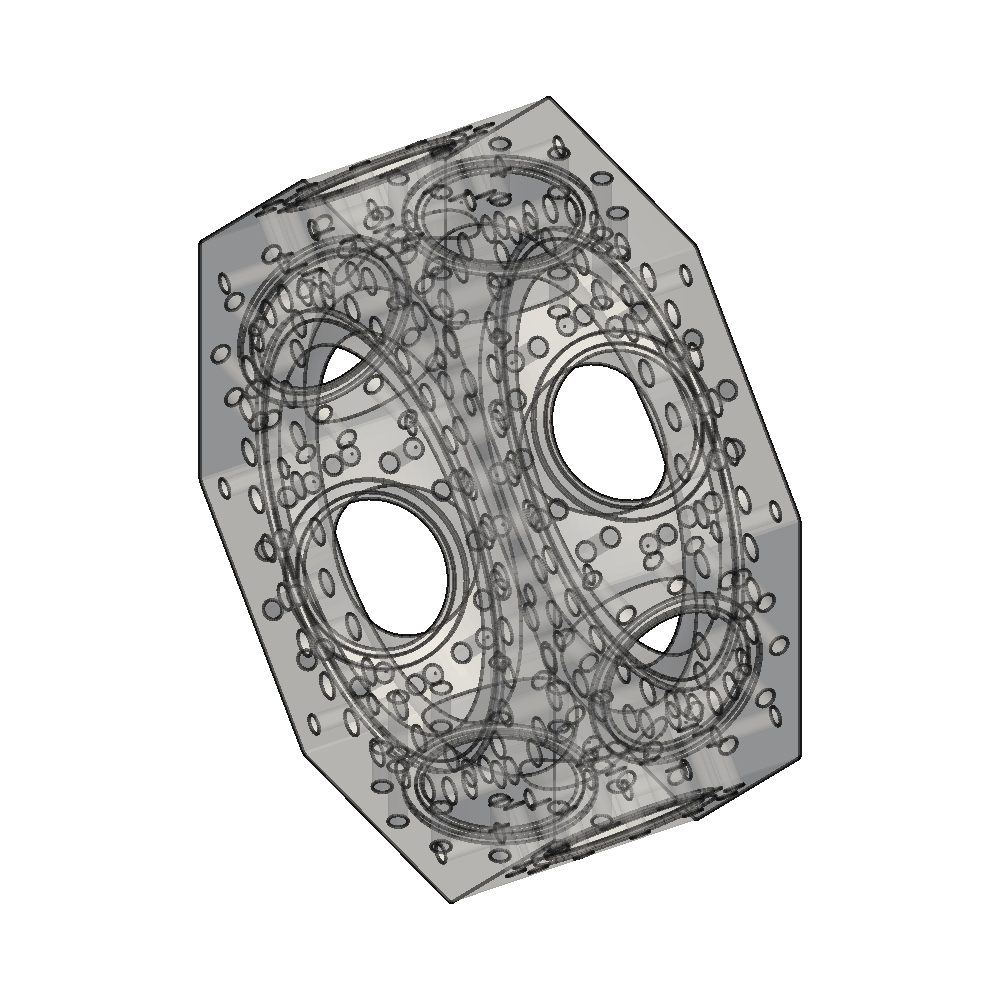}\hfill
    \includegraphics[width=0.117\linewidth]{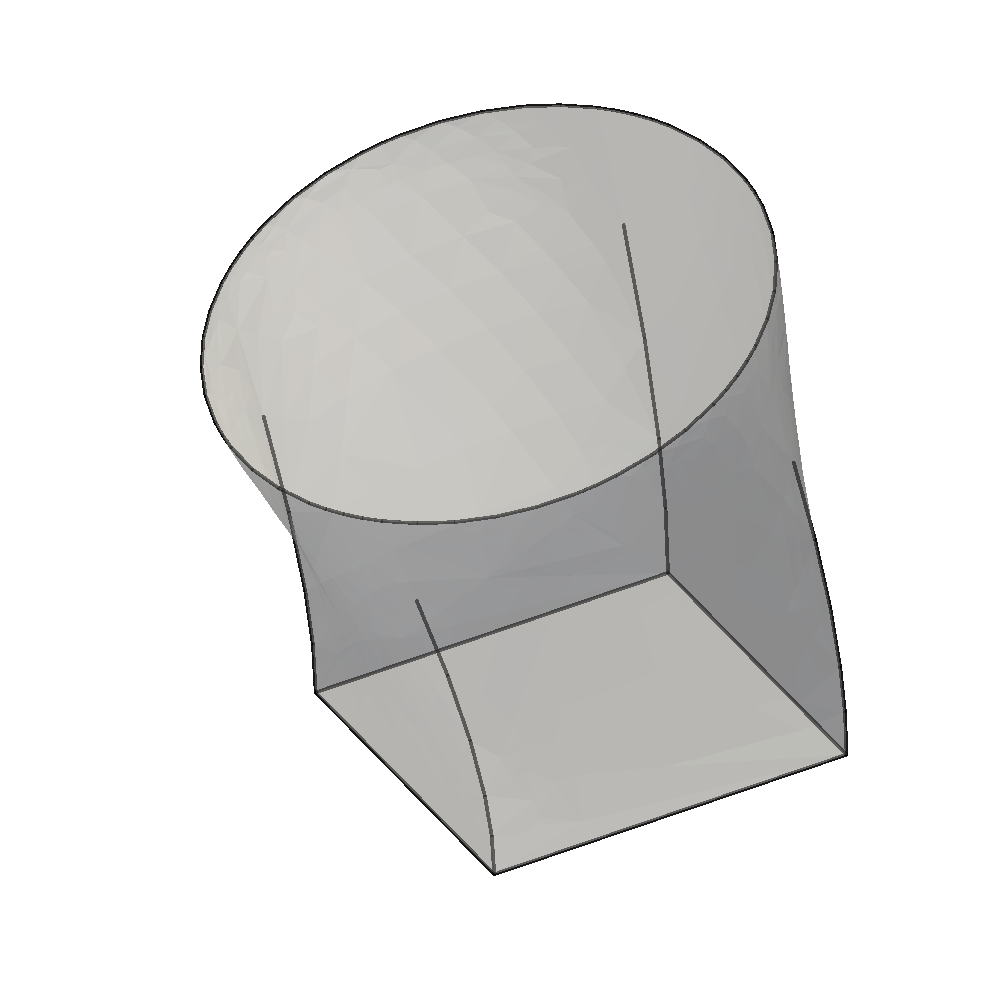}\hfill
    \includegraphics[width=0.117\linewidth]{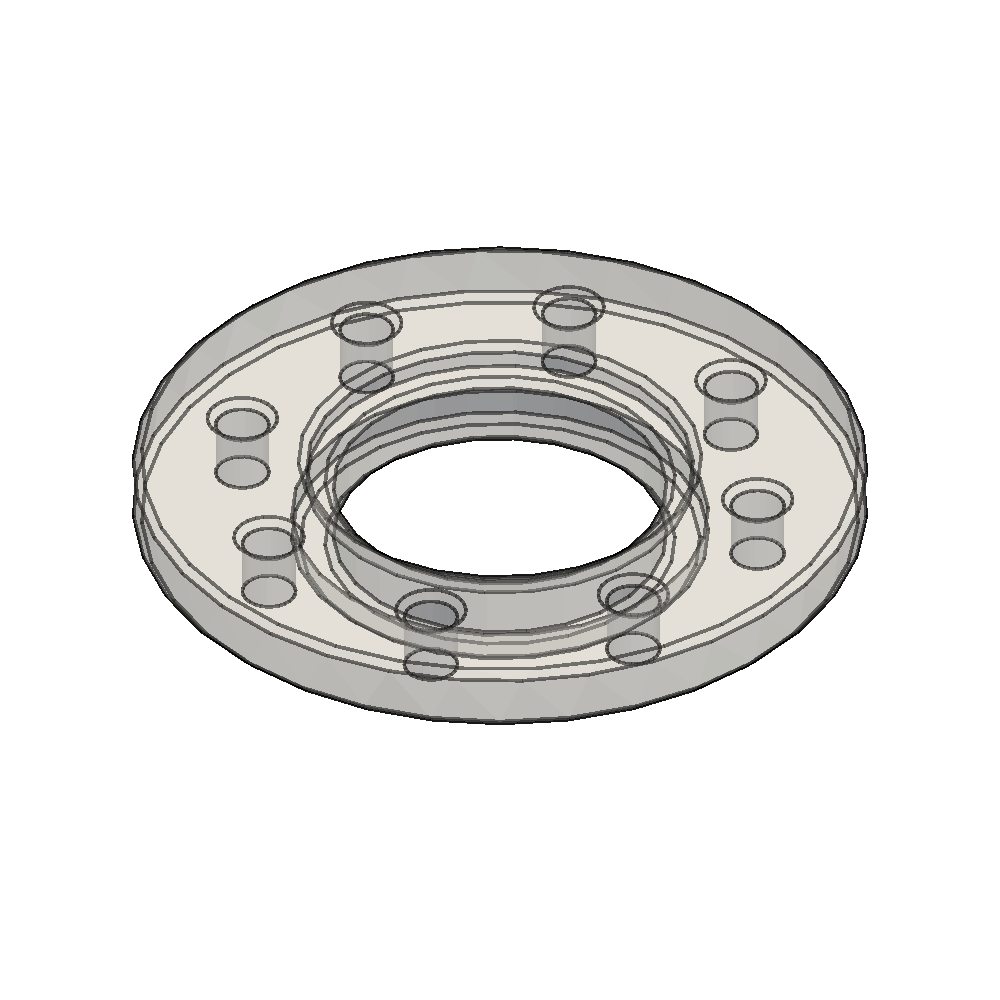}\hfill
    \includegraphics[width=0.117\linewidth]{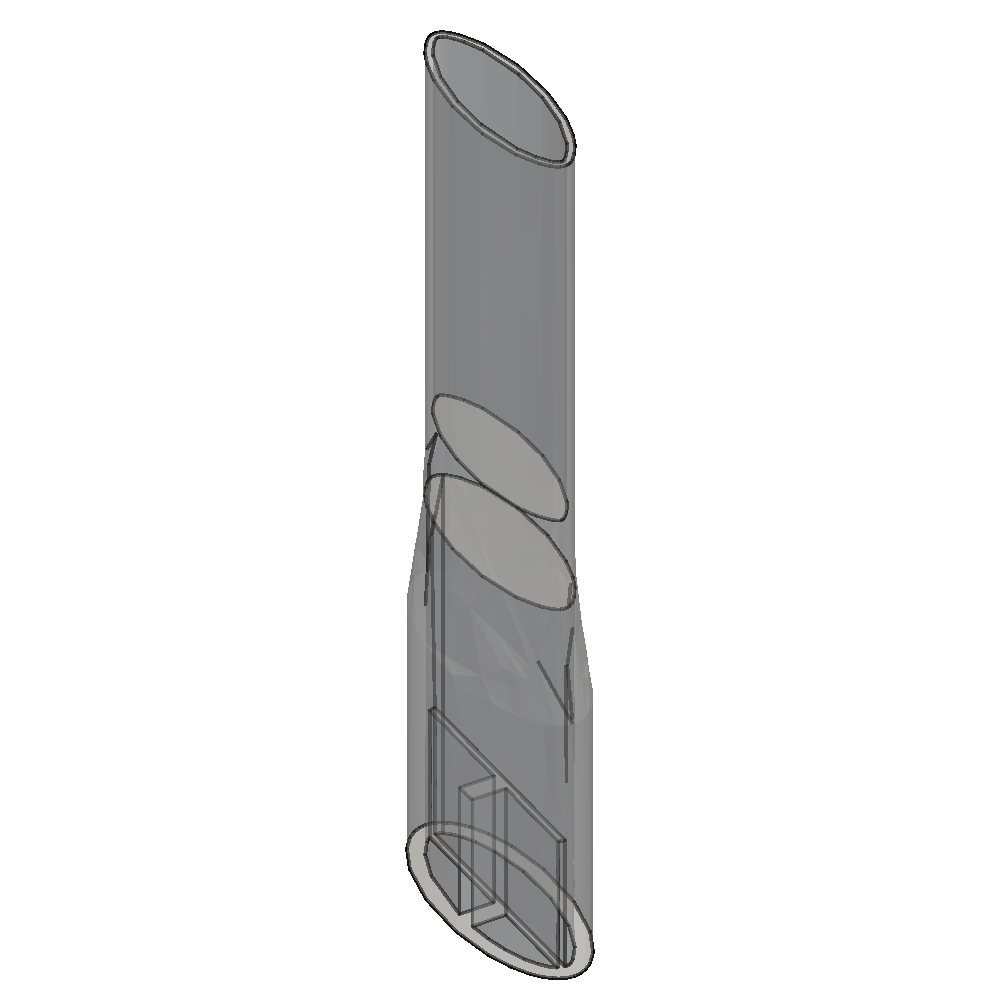}\hfill
    \includegraphics[width=0.117\linewidth]{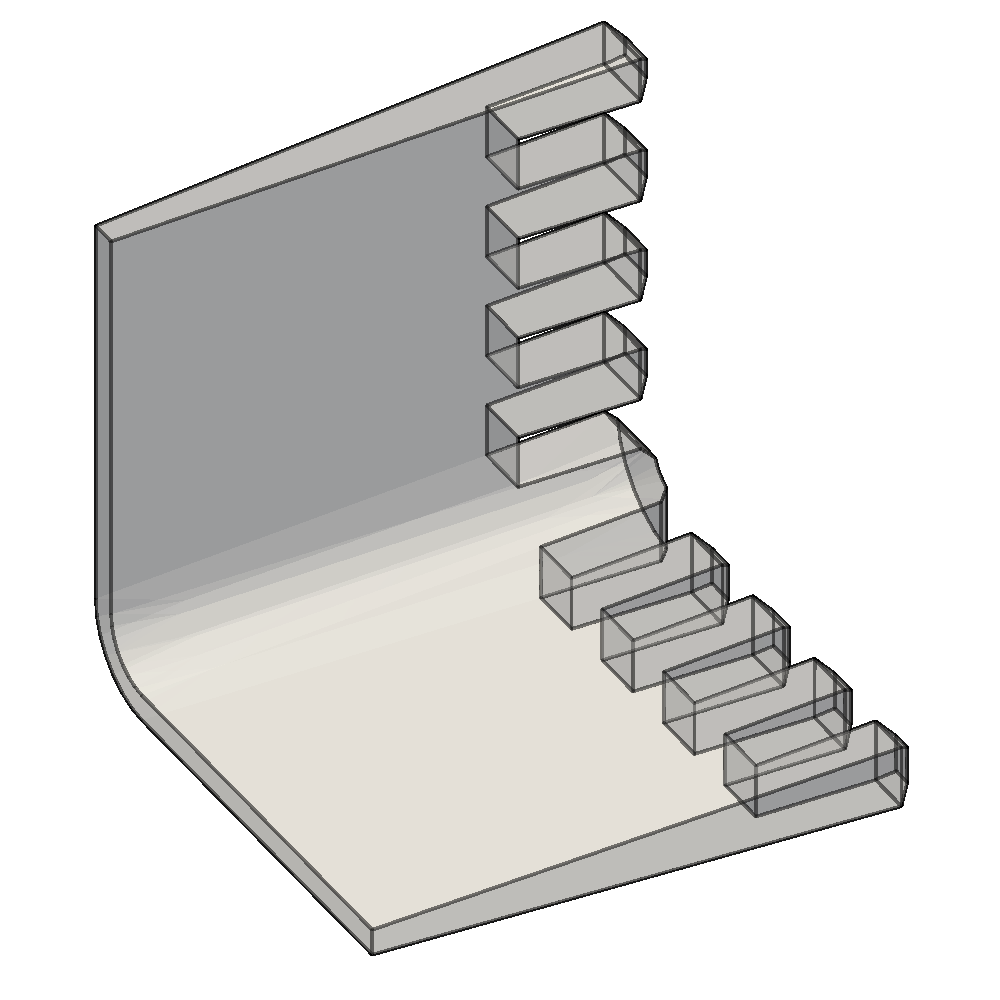}\hfill
    \includegraphics[width=0.117\linewidth]{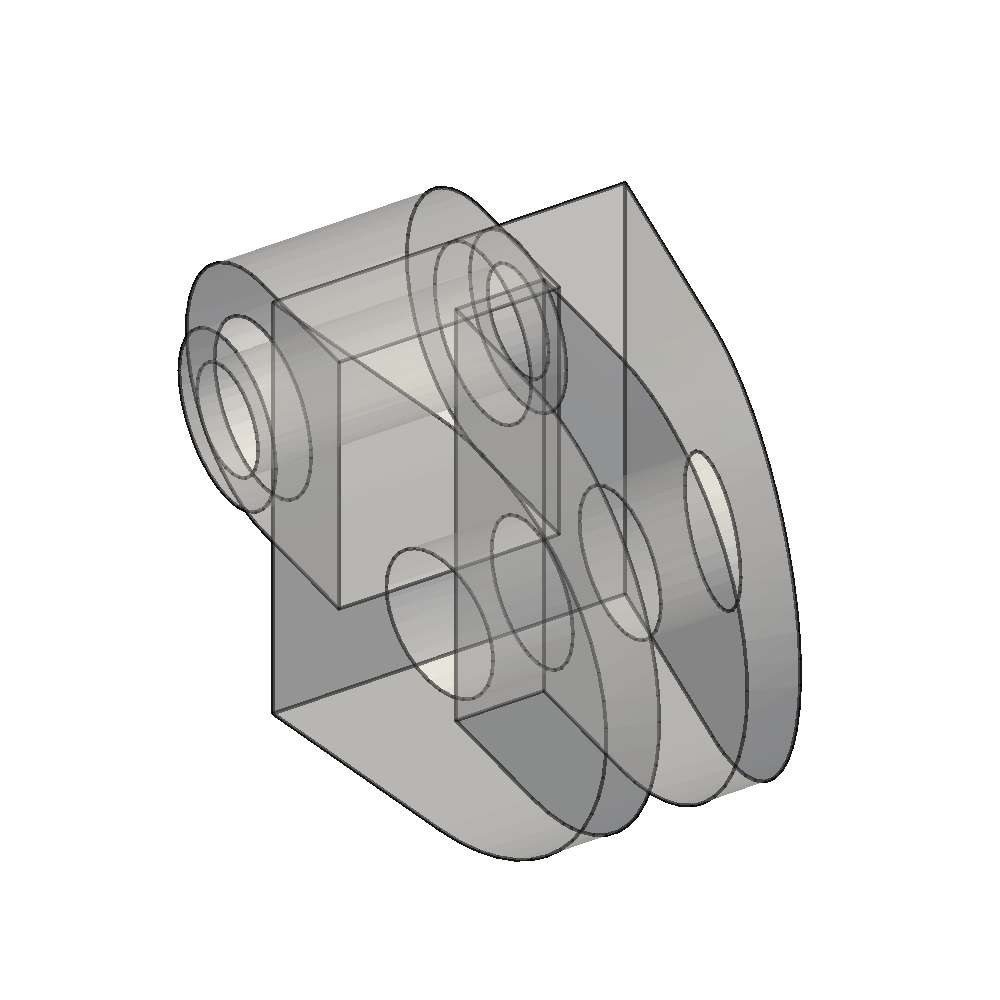}\\[\smallskipamount]
    \includegraphics[width=0.117\linewidth]{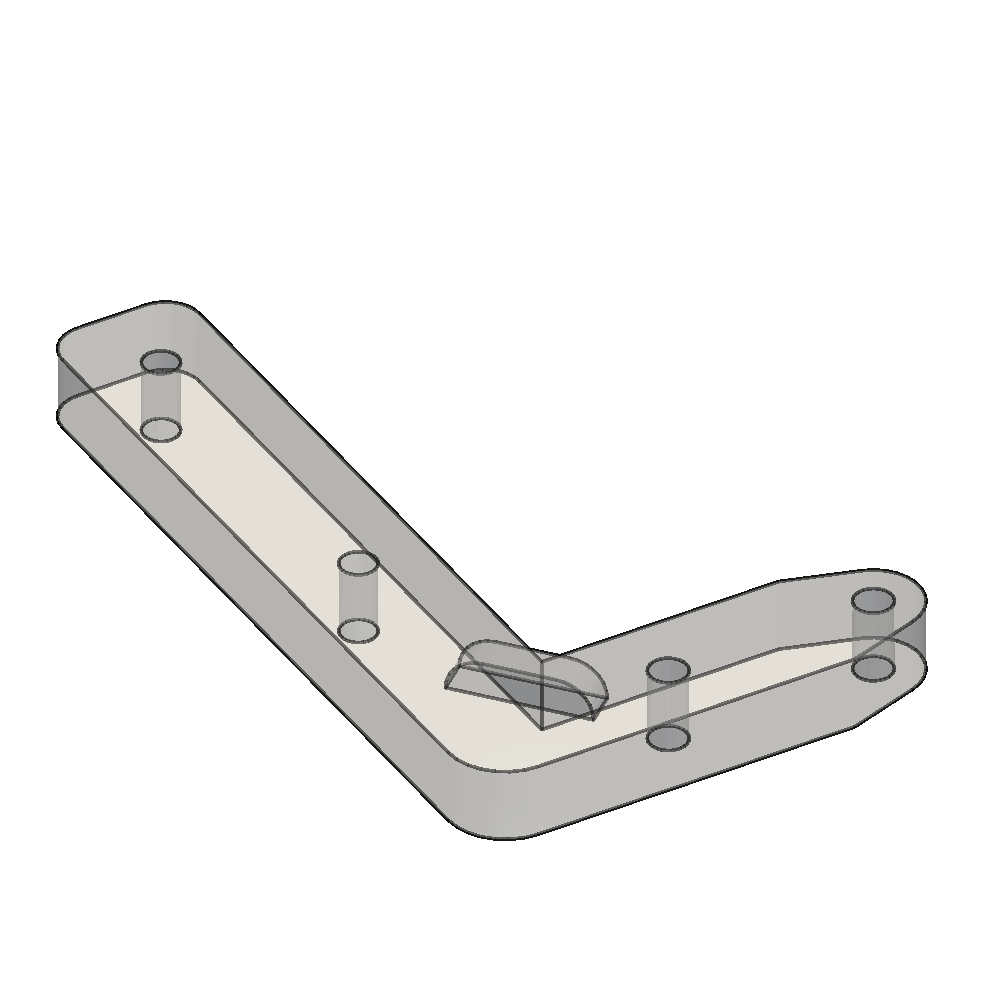}\hfill
    \includegraphics[width=0.117\linewidth]{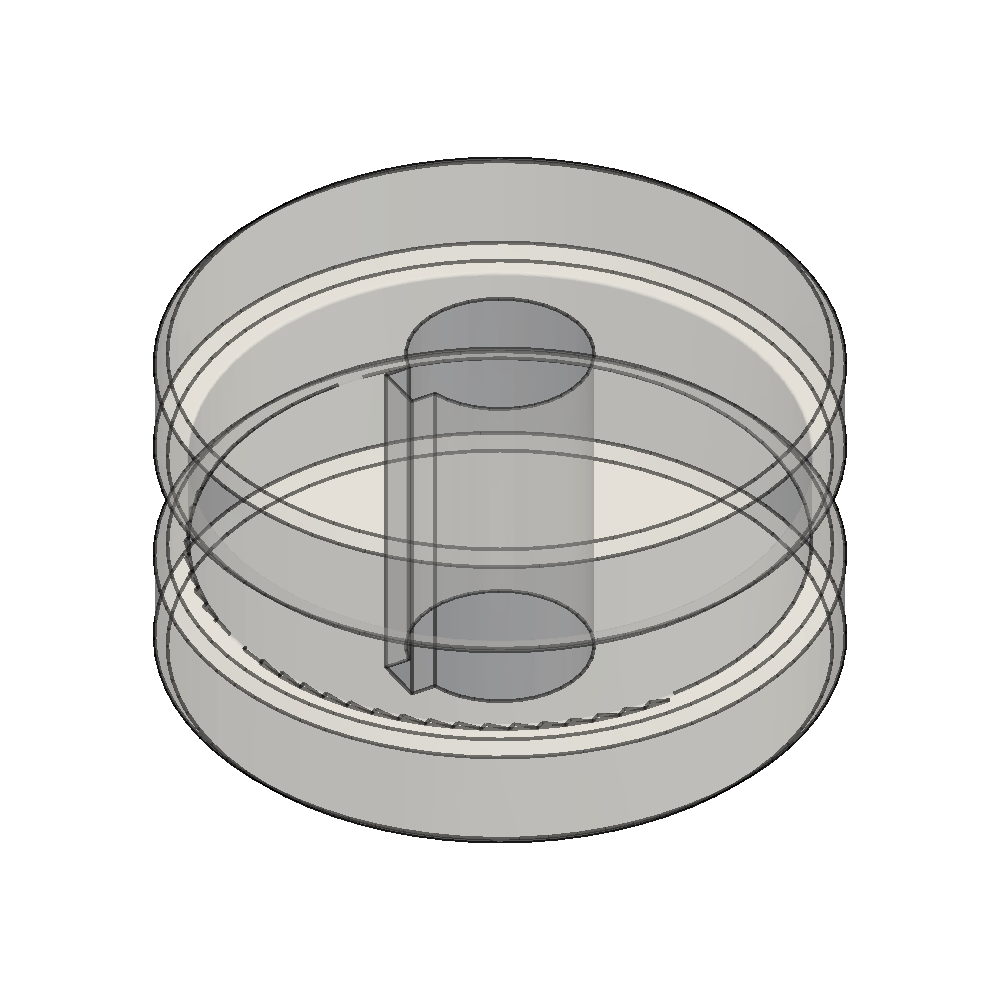}\hfill
    \includegraphics[width=0.117\linewidth]{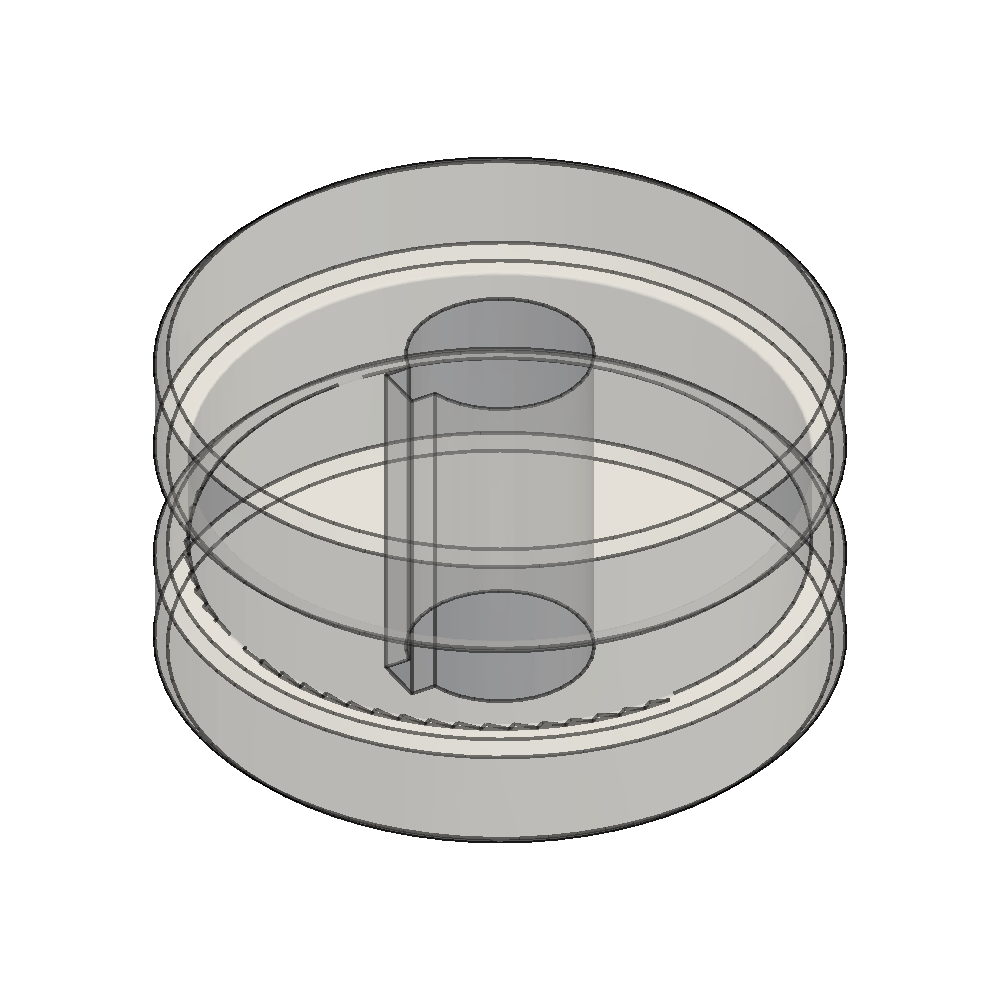}\hfill
    \includegraphics[width=0.117\linewidth]{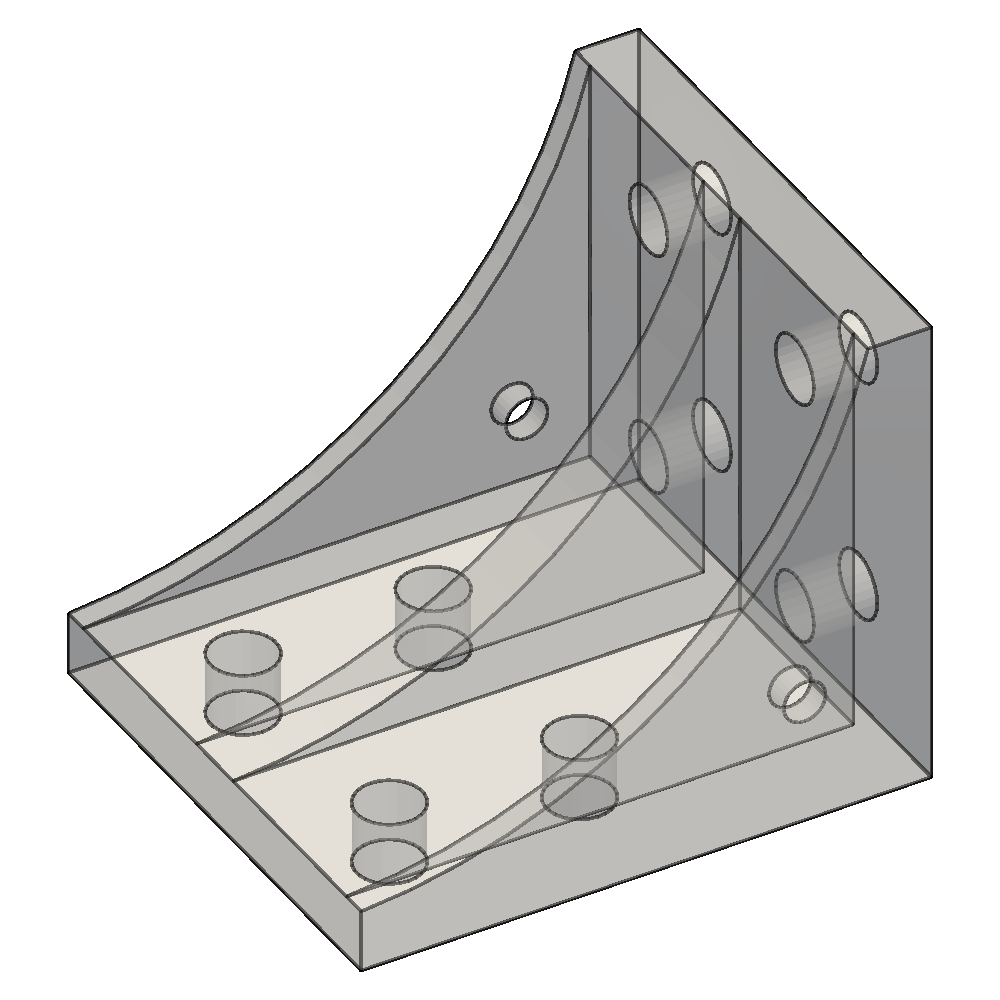}\hfill
    \includegraphics[width=0.117\linewidth]{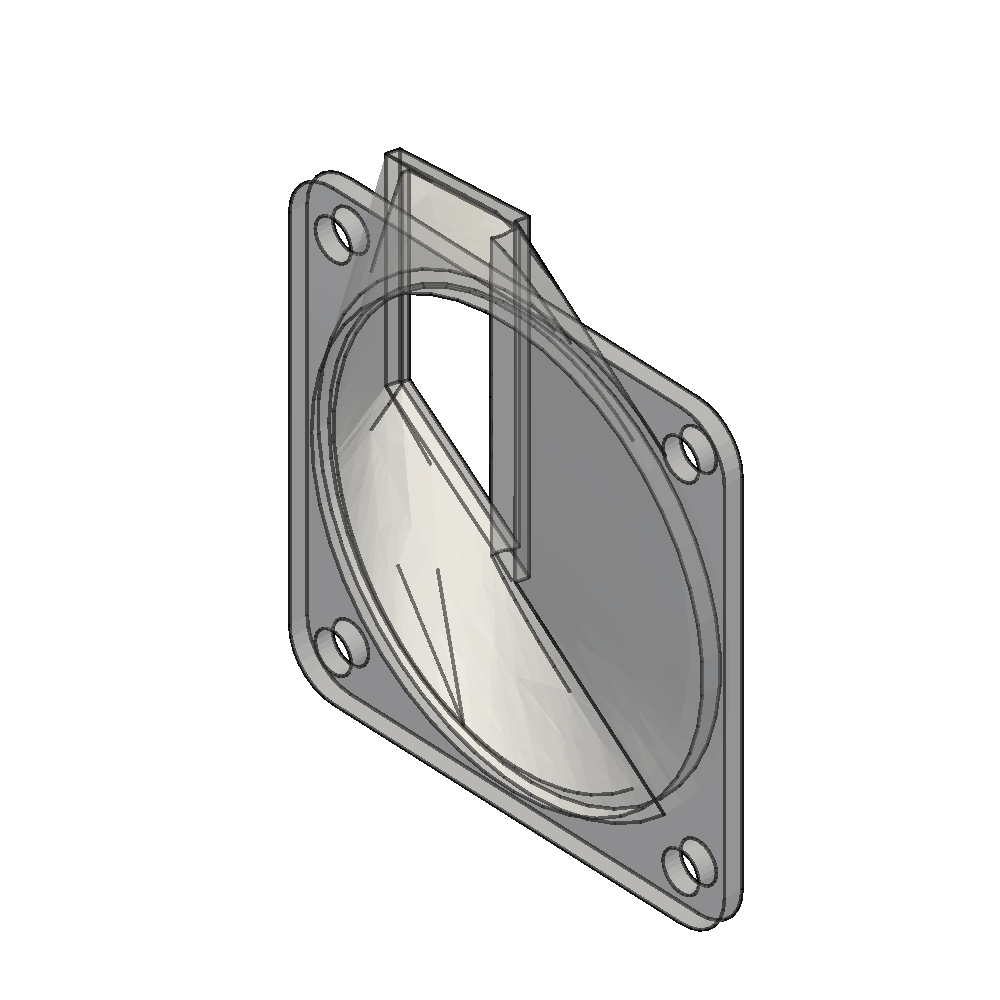}\hfill
    \includegraphics[width=0.117\linewidth]{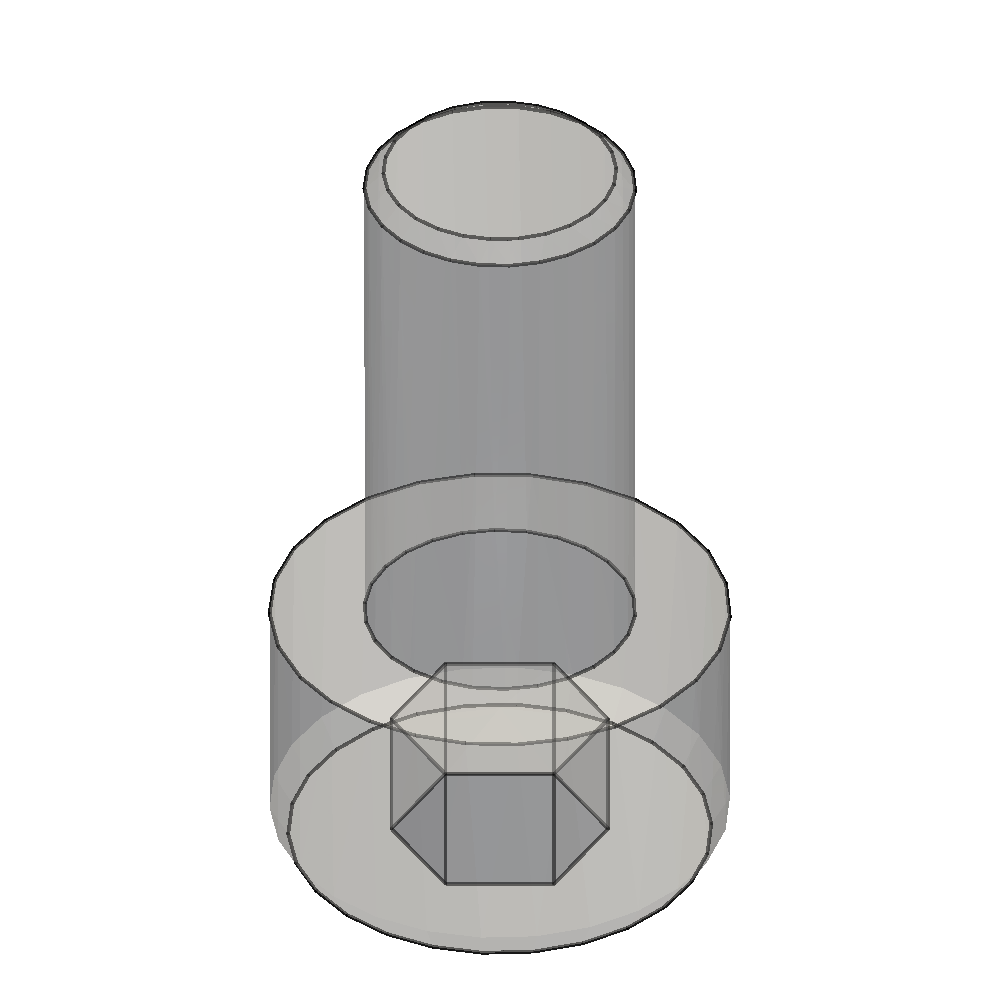}\hfill
    \includegraphics[width=0.117\linewidth]{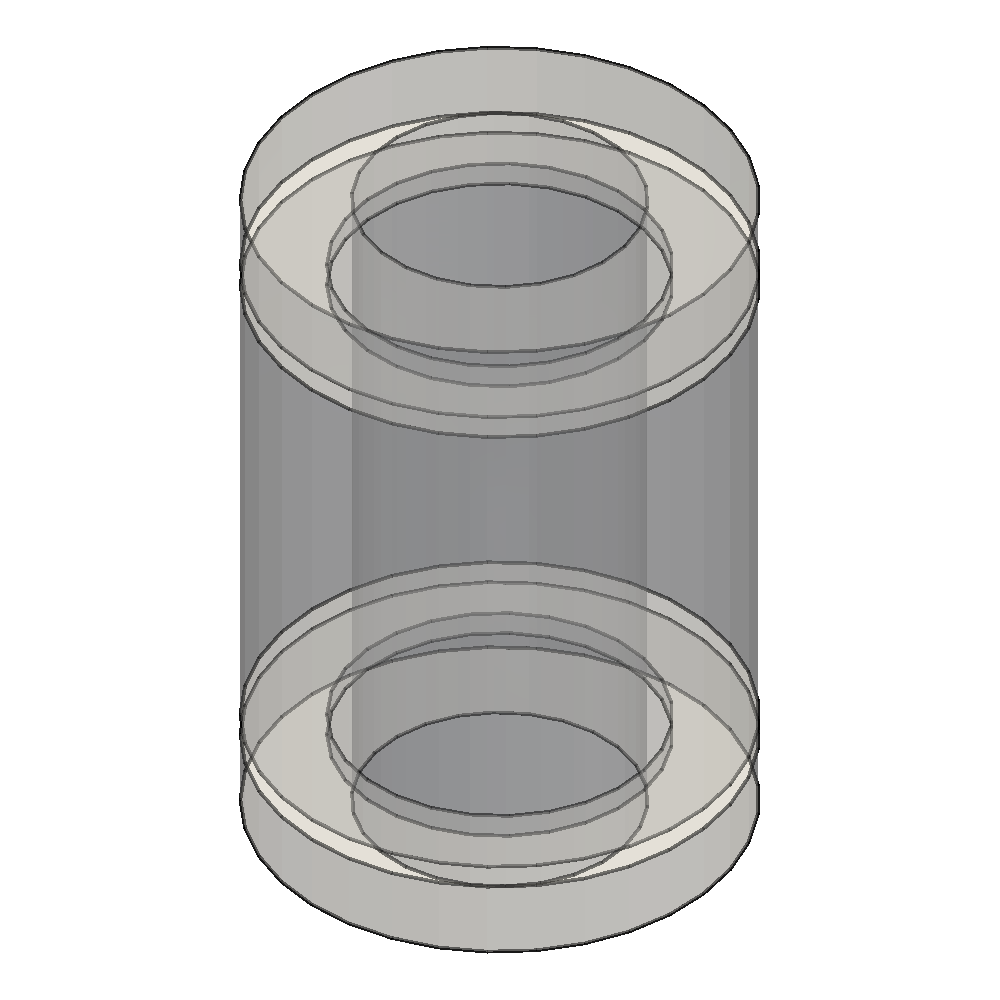}\hfill
    \includegraphics[width=0.117\linewidth]{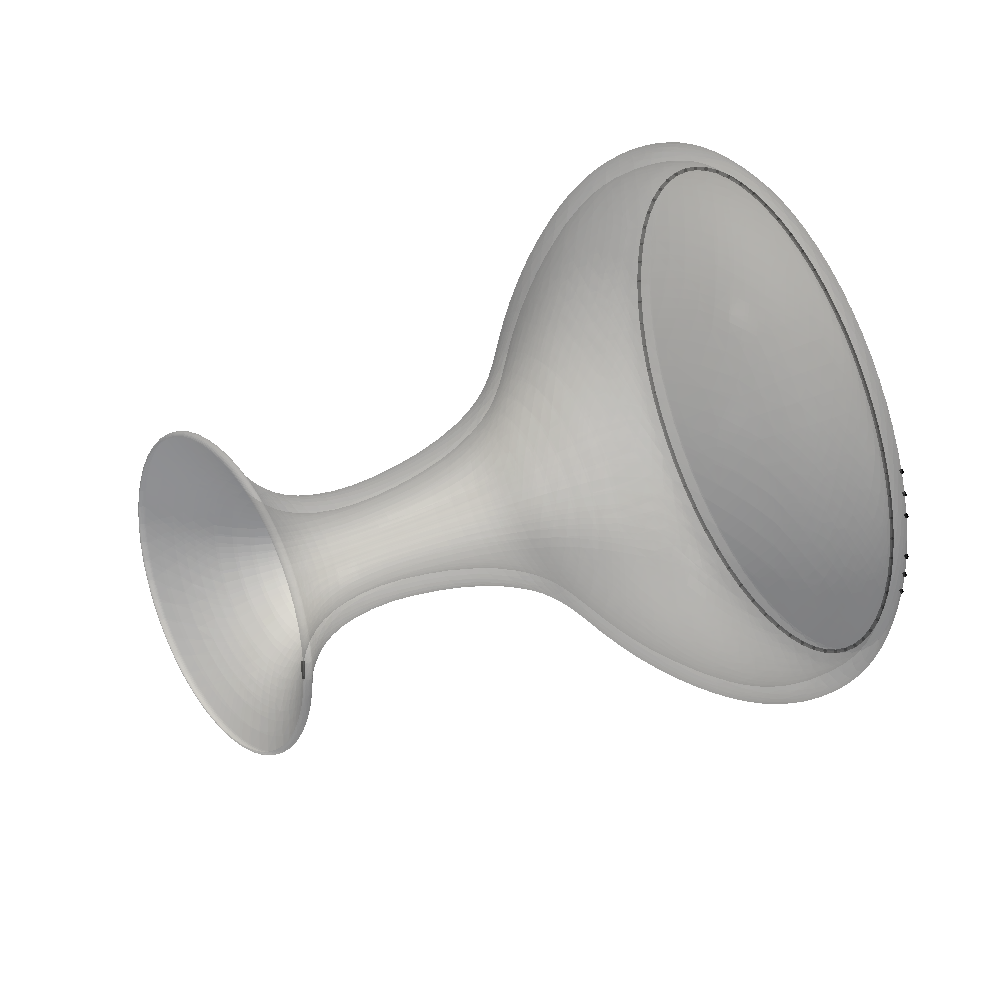}\\[\smallskipamount]
    \includegraphics[width=0.117\linewidth]{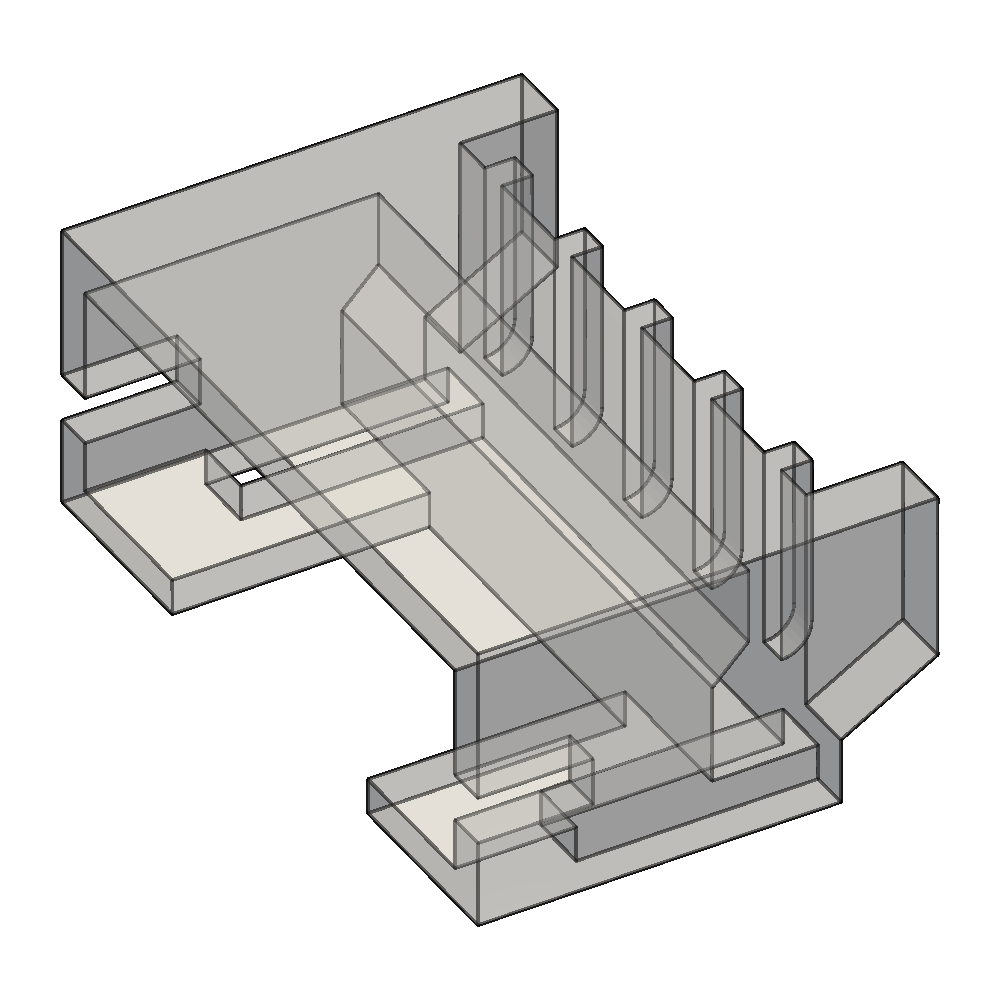}\hfill
    \includegraphics[width=0.117\linewidth]{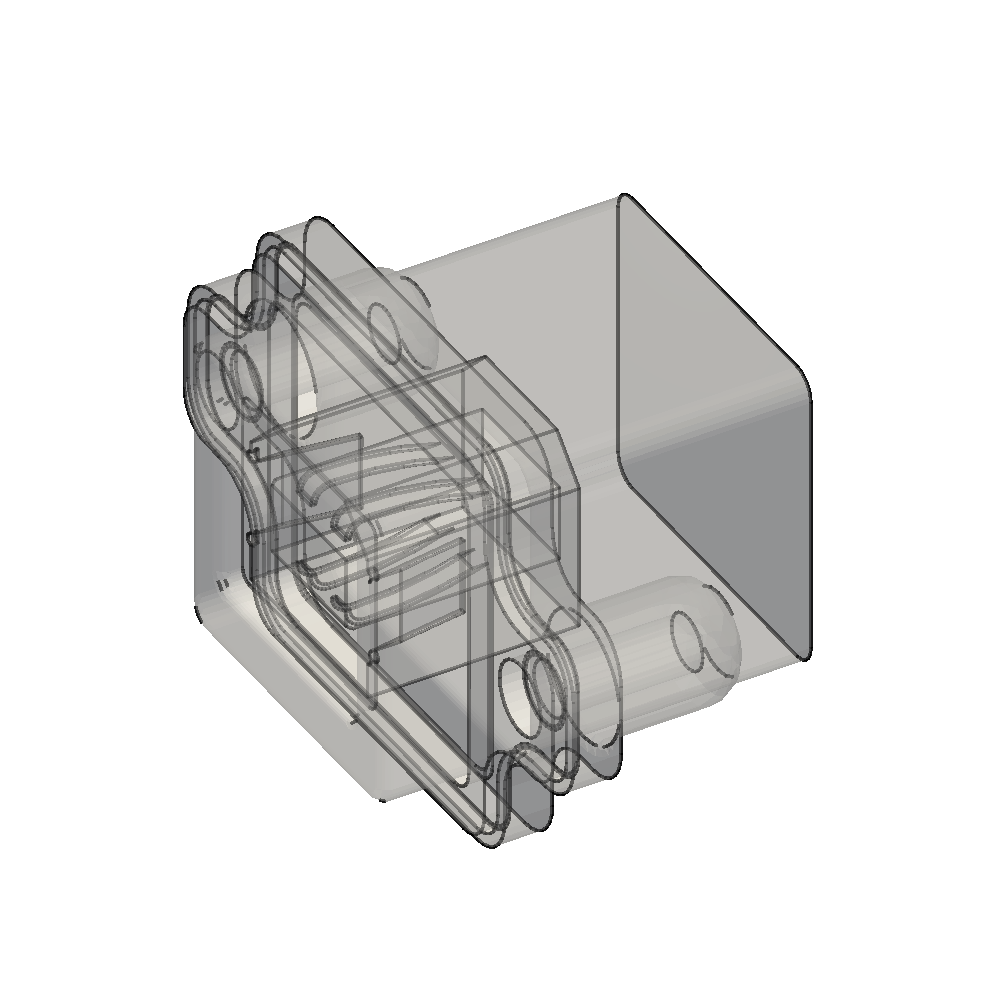}\hfill
    \includegraphics[width=0.117\linewidth]{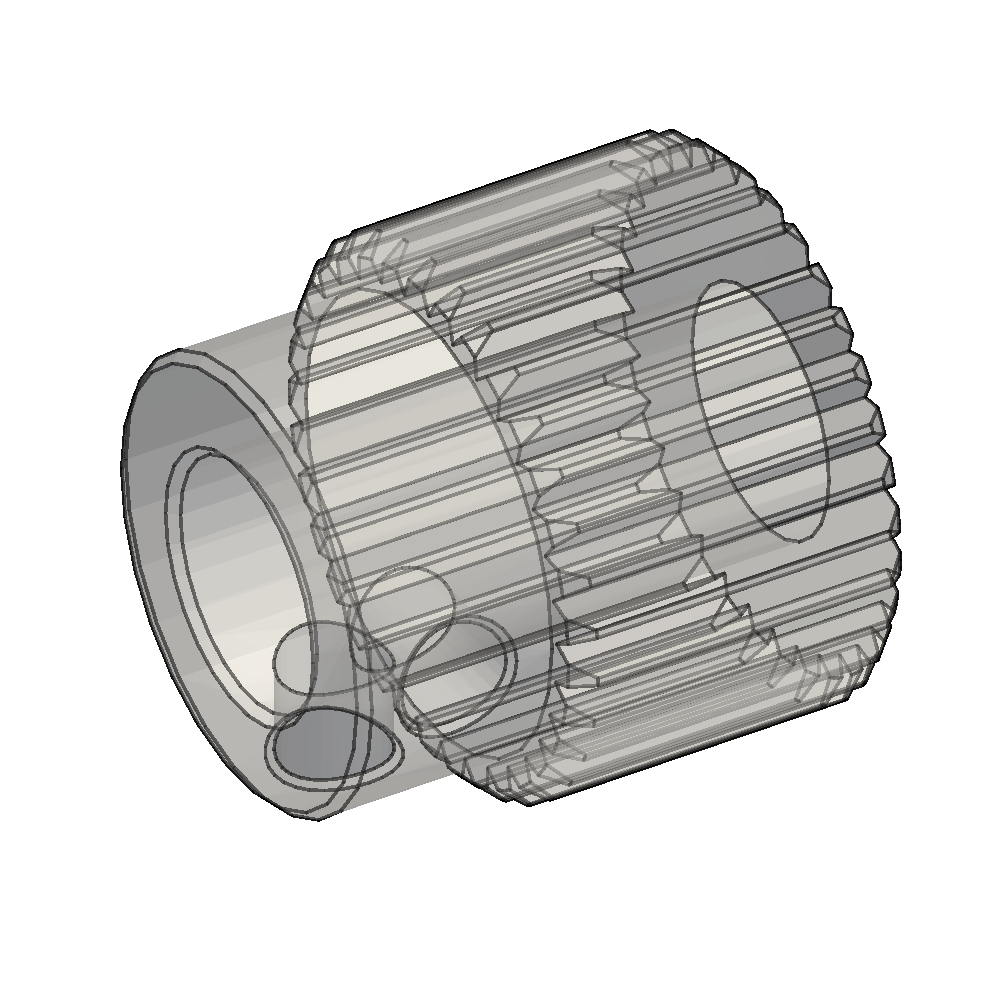}\hfill
    \includegraphics[width=0.117\linewidth]{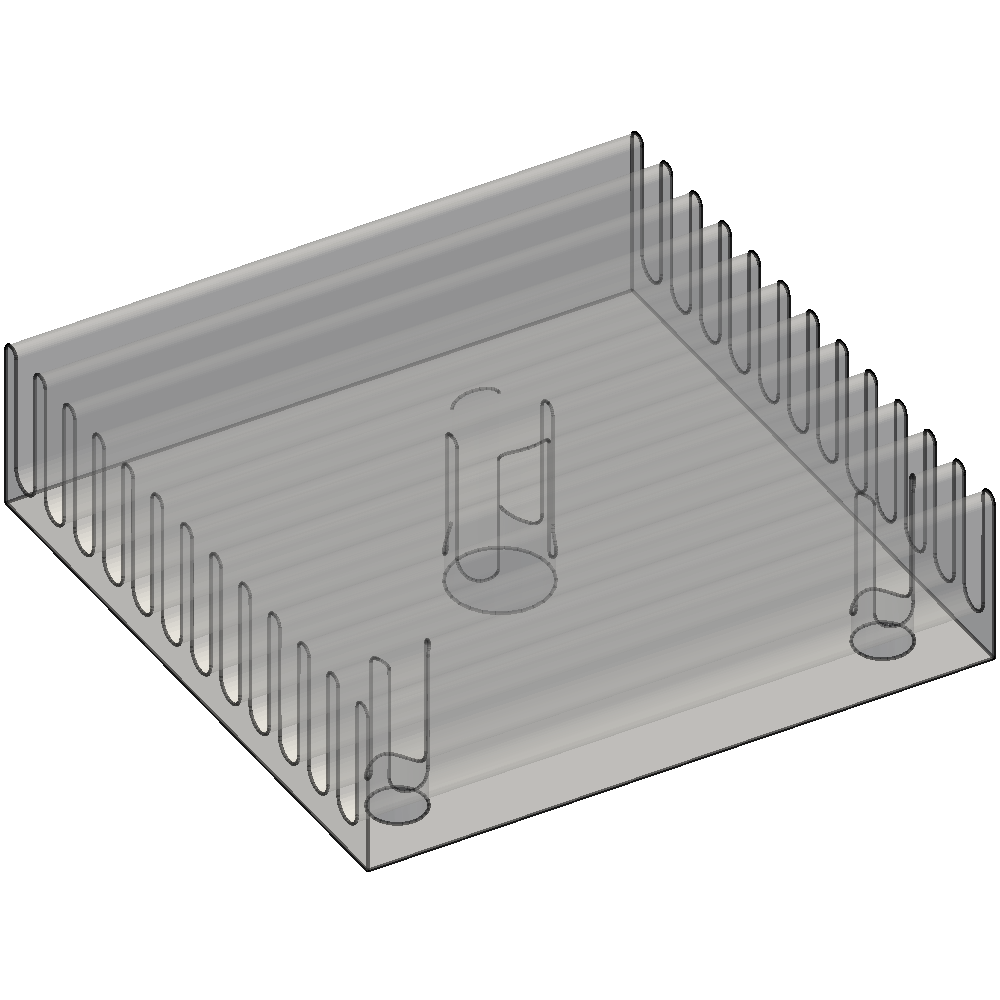}\hfill
    \includegraphics[width=0.117\linewidth]{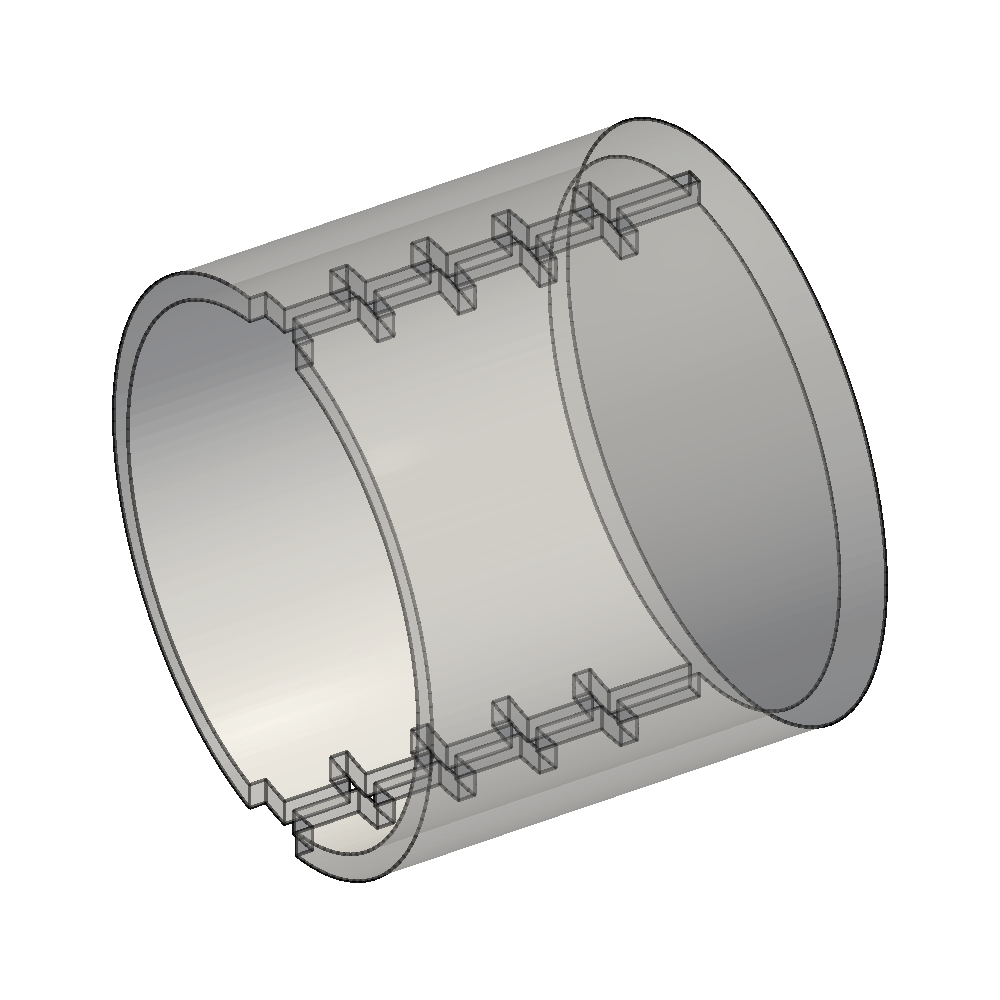}\hfill
    \includegraphics[width=0.117\linewidth]{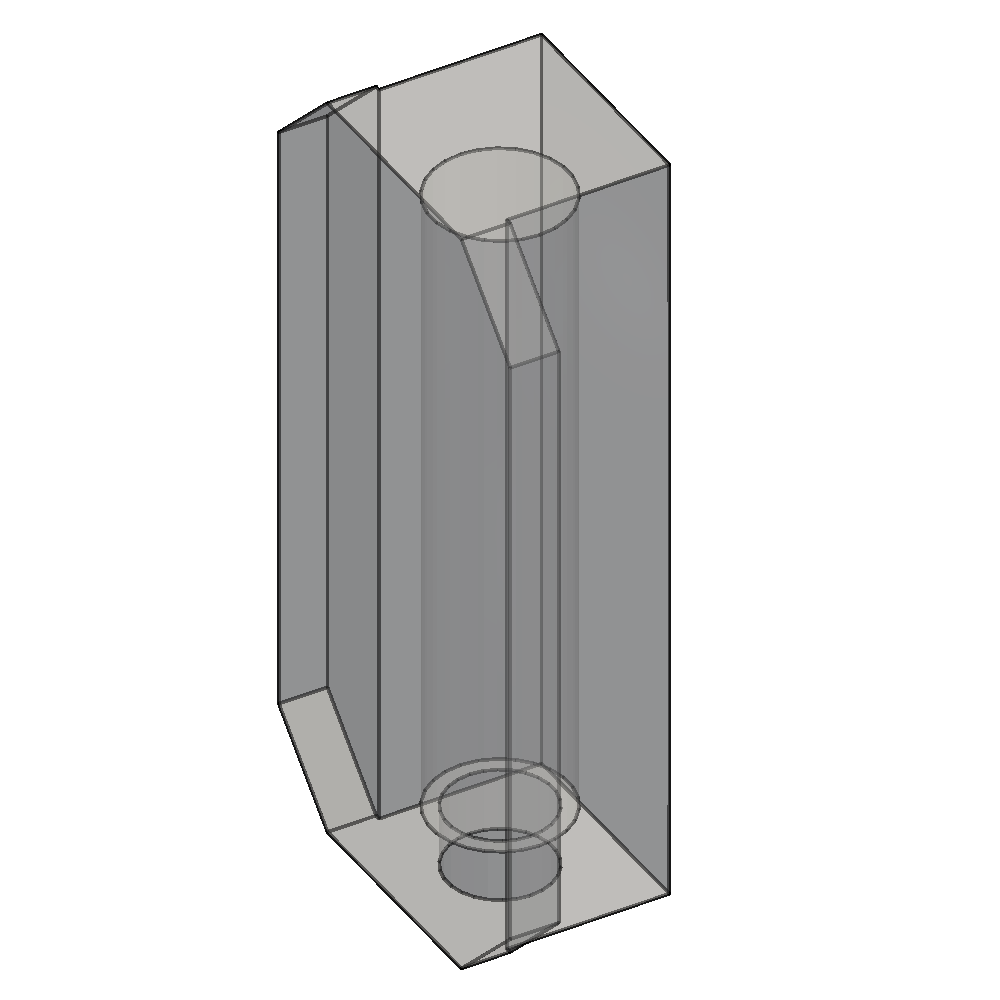}\hfill
    \includegraphics[width=0.117\linewidth]{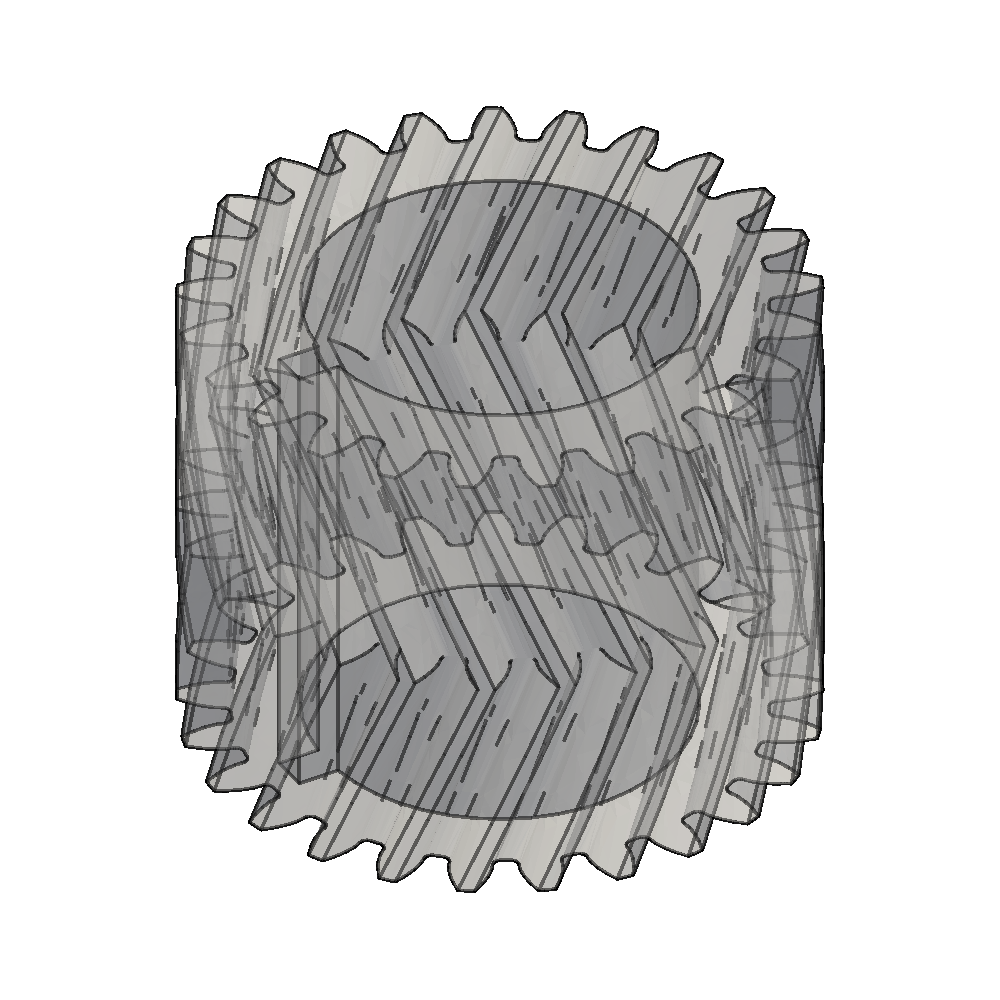}\hfill
    \includegraphics[width=0.117\linewidth]{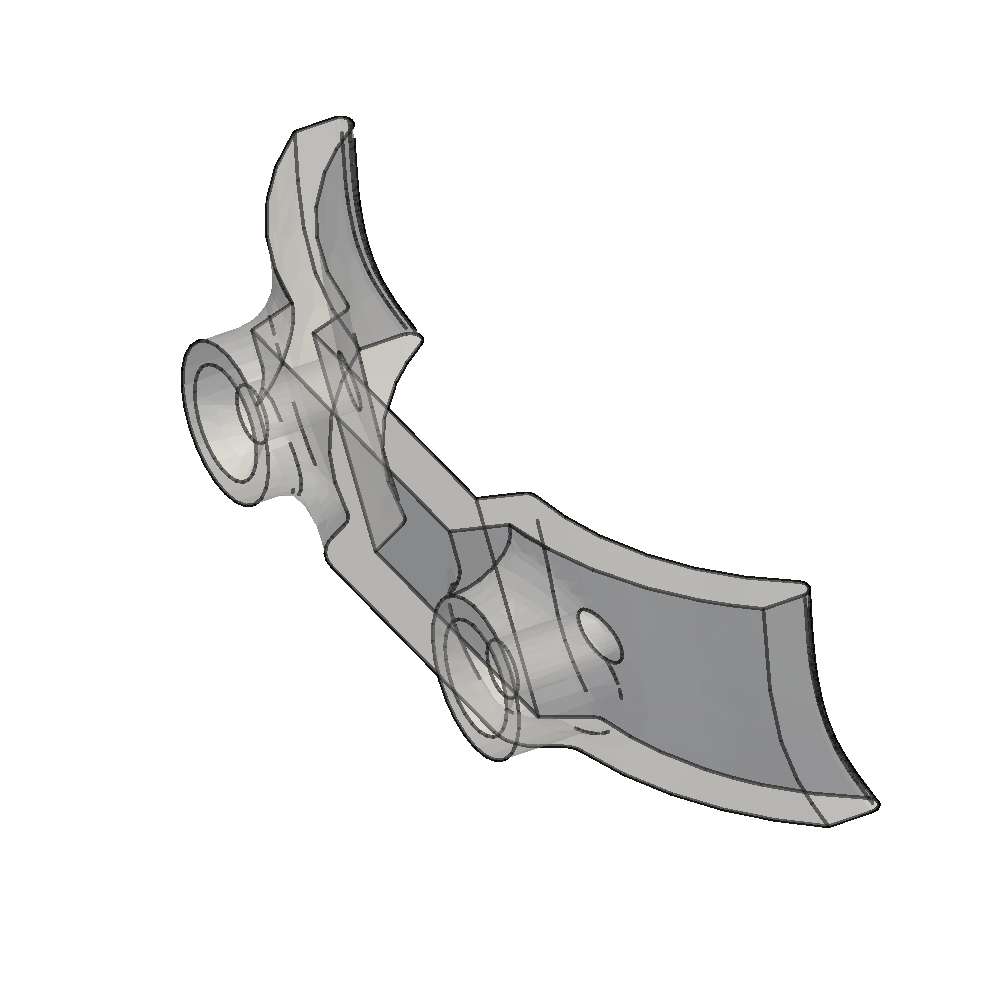}\\[\smallskipamount]
    \includegraphics[width=0.117\linewidth]{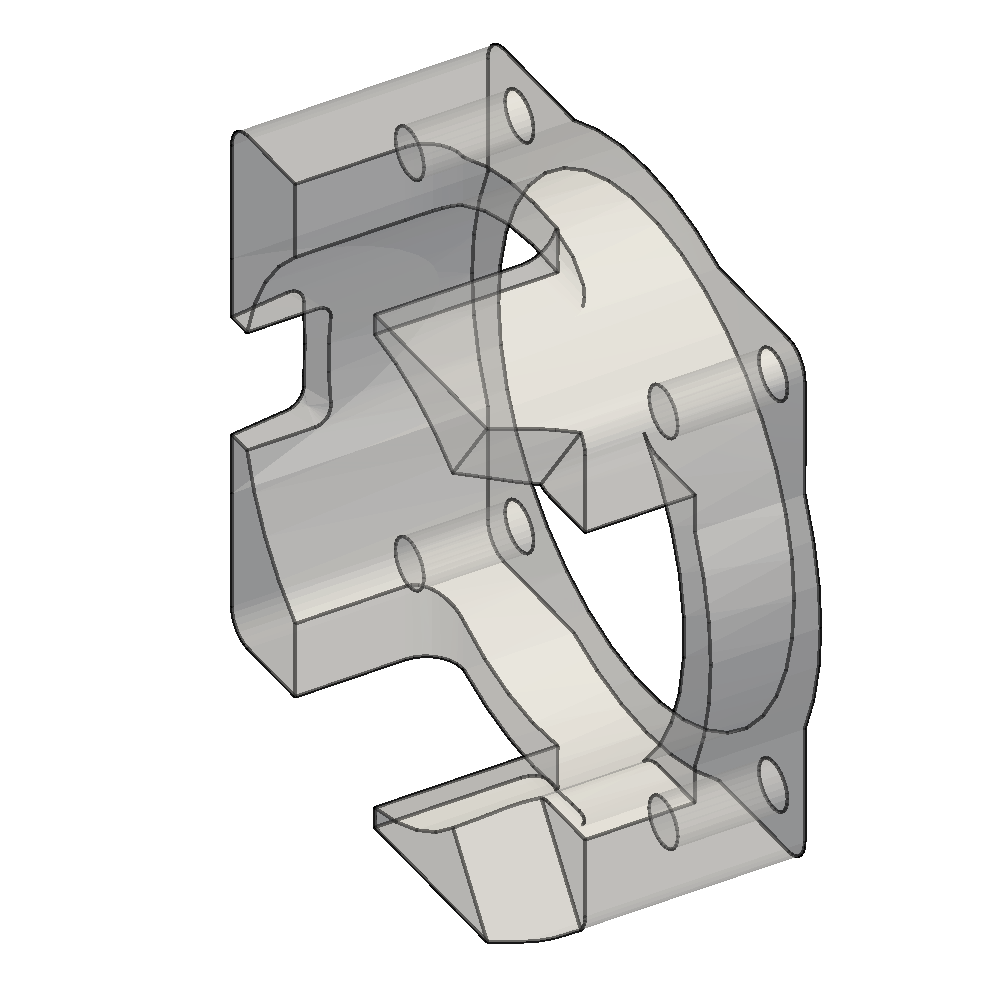}\hfill
    \includegraphics[width=0.117\linewidth]{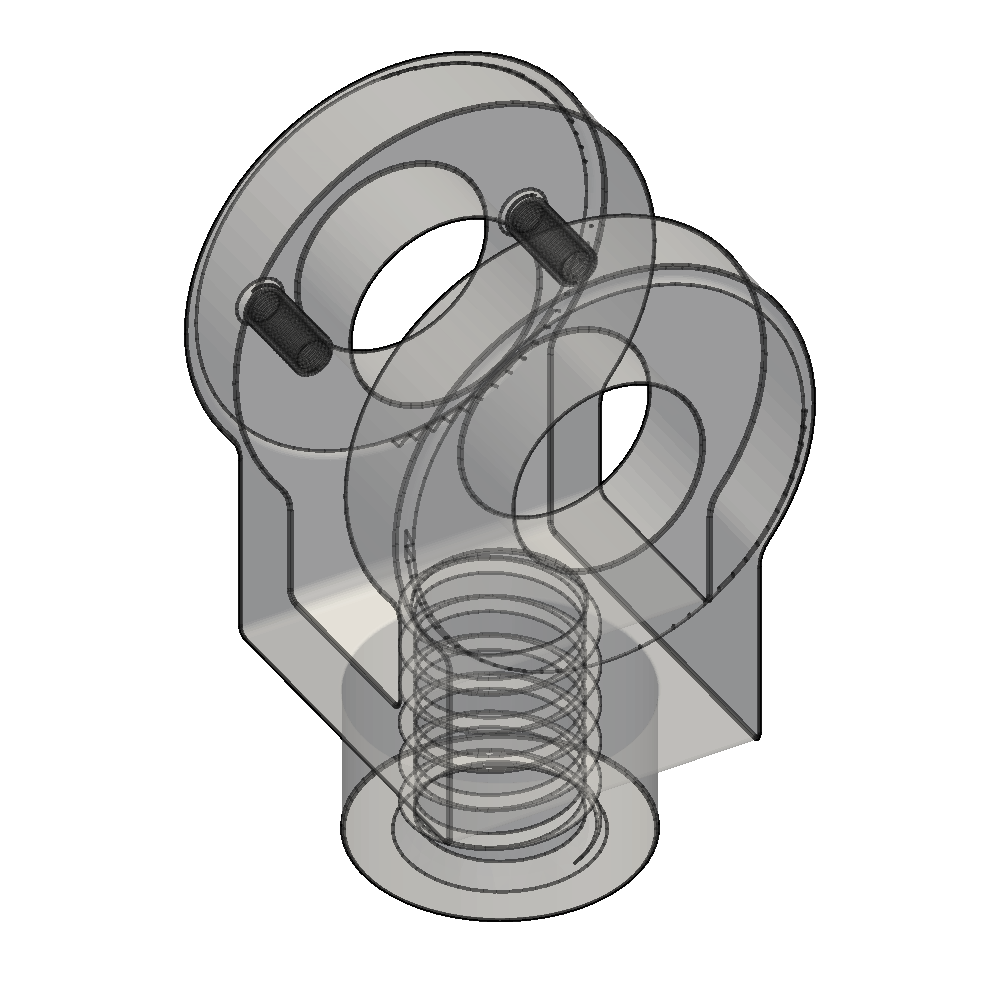}\hfill
    \includegraphics[width=0.117\linewidth]{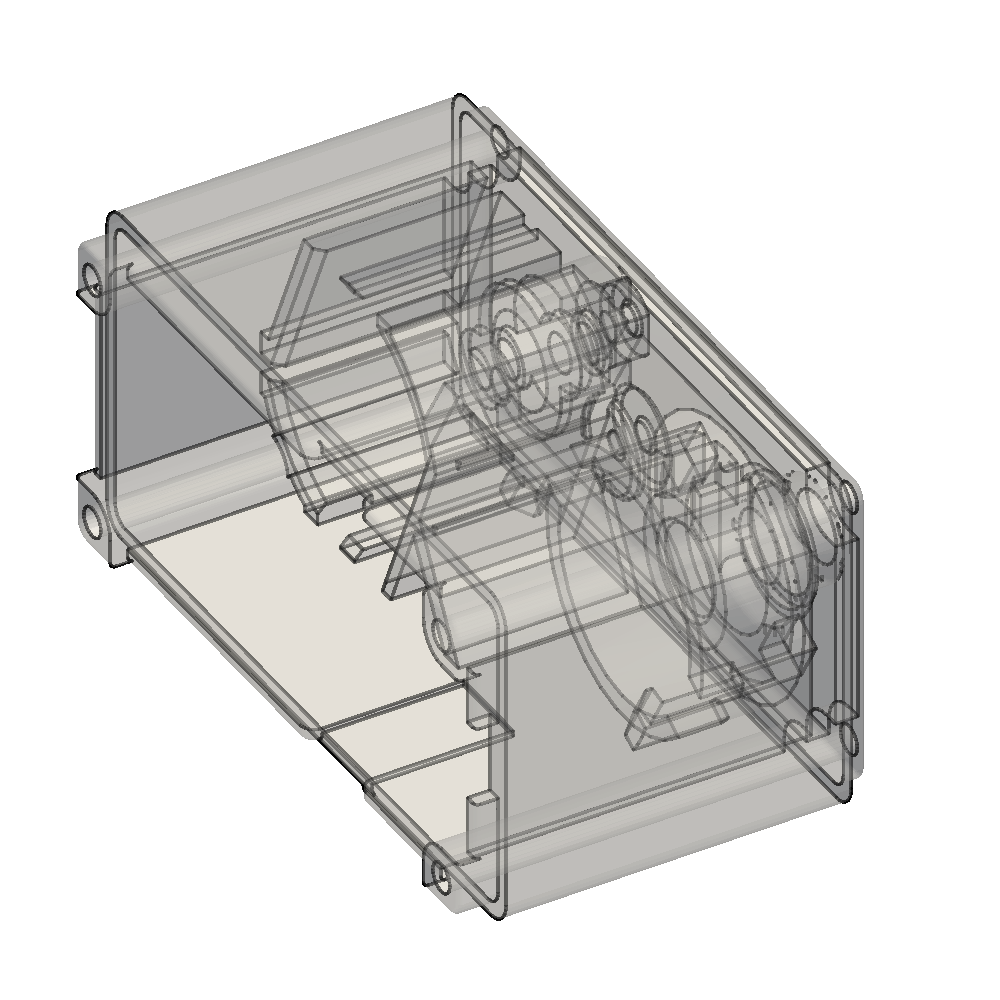}\hfill
    \includegraphics[width=0.117\linewidth]{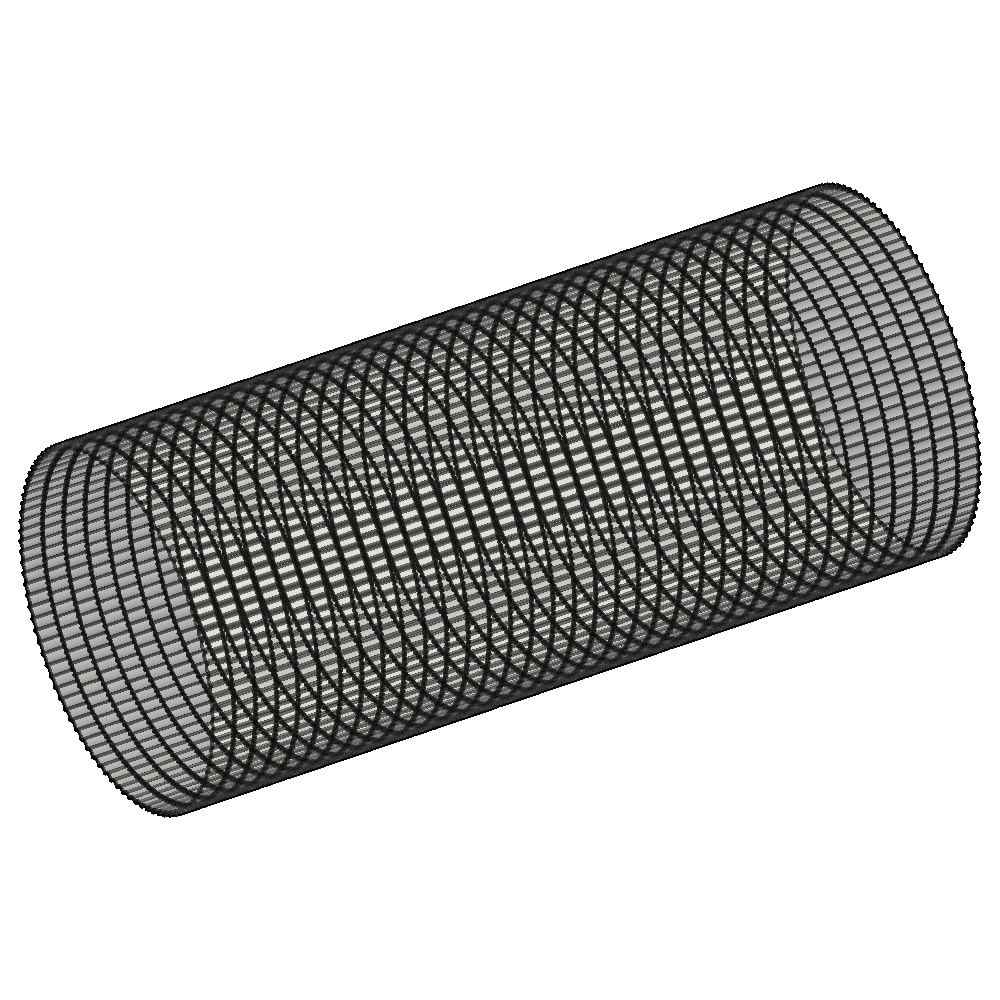}\hfill
    \includegraphics[width=0.117\linewidth]{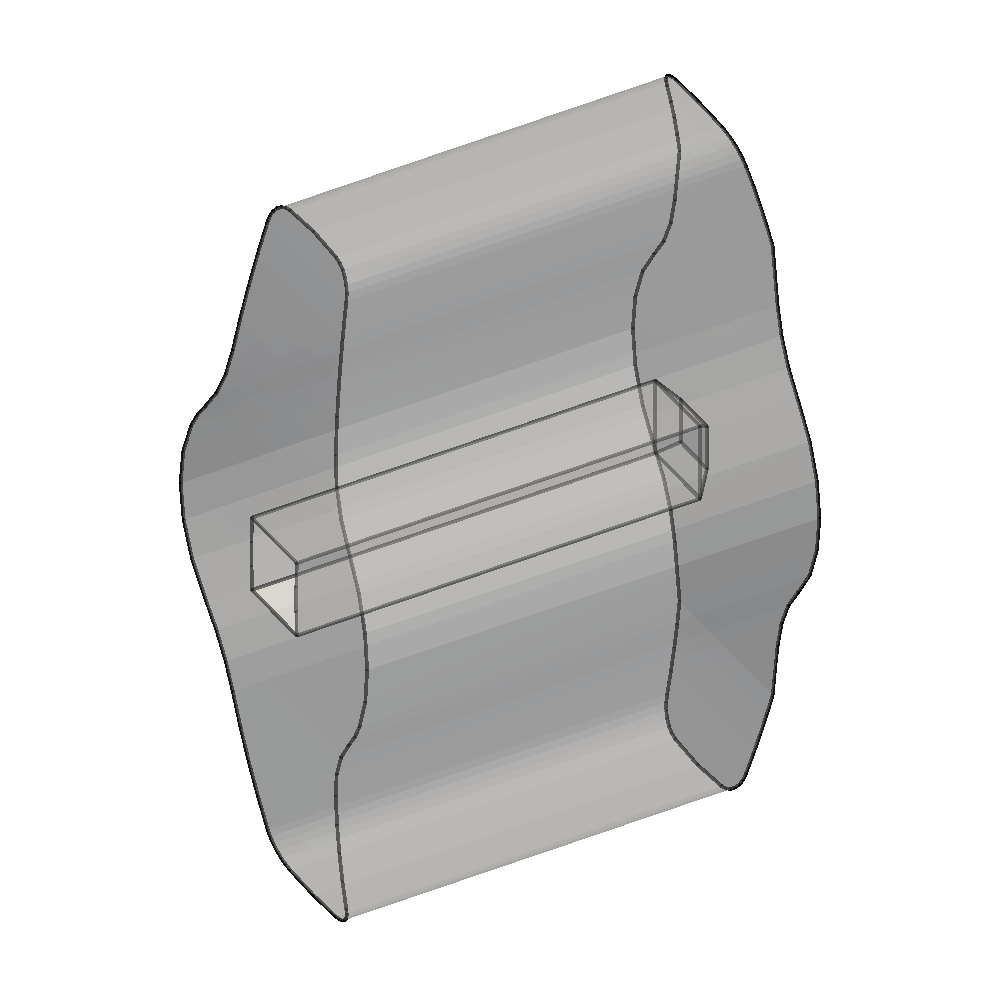}\hfill
    \includegraphics[width=0.117\linewidth]{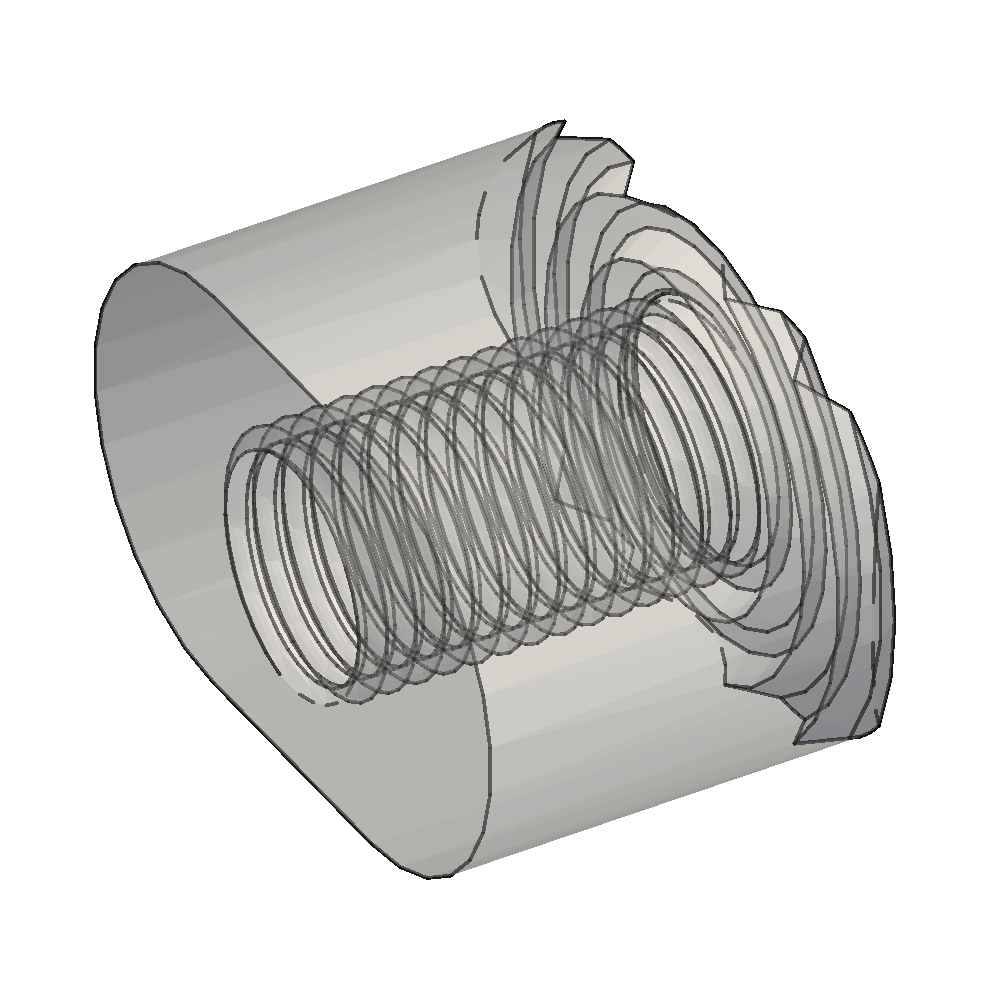}\hfill
    \includegraphics[width=0.117\linewidth]{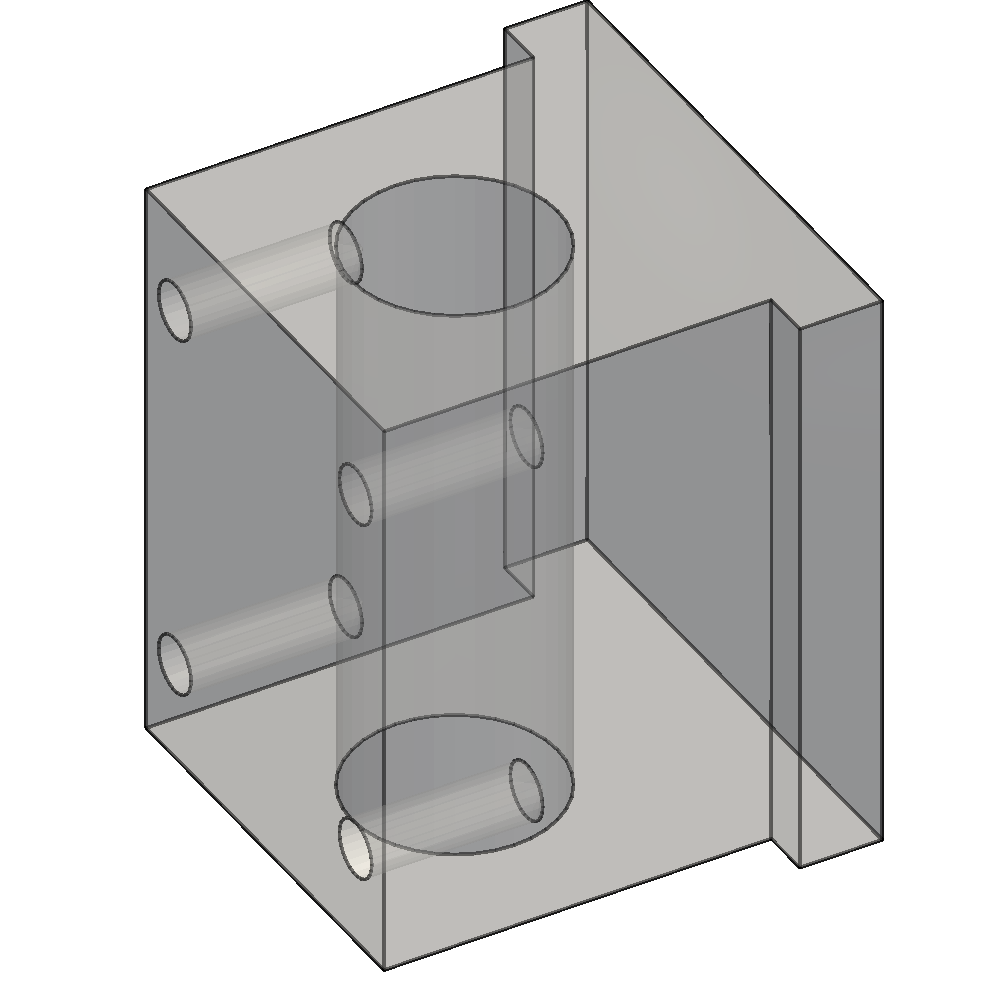}\hfill
    \includegraphics[width=0.117\linewidth]{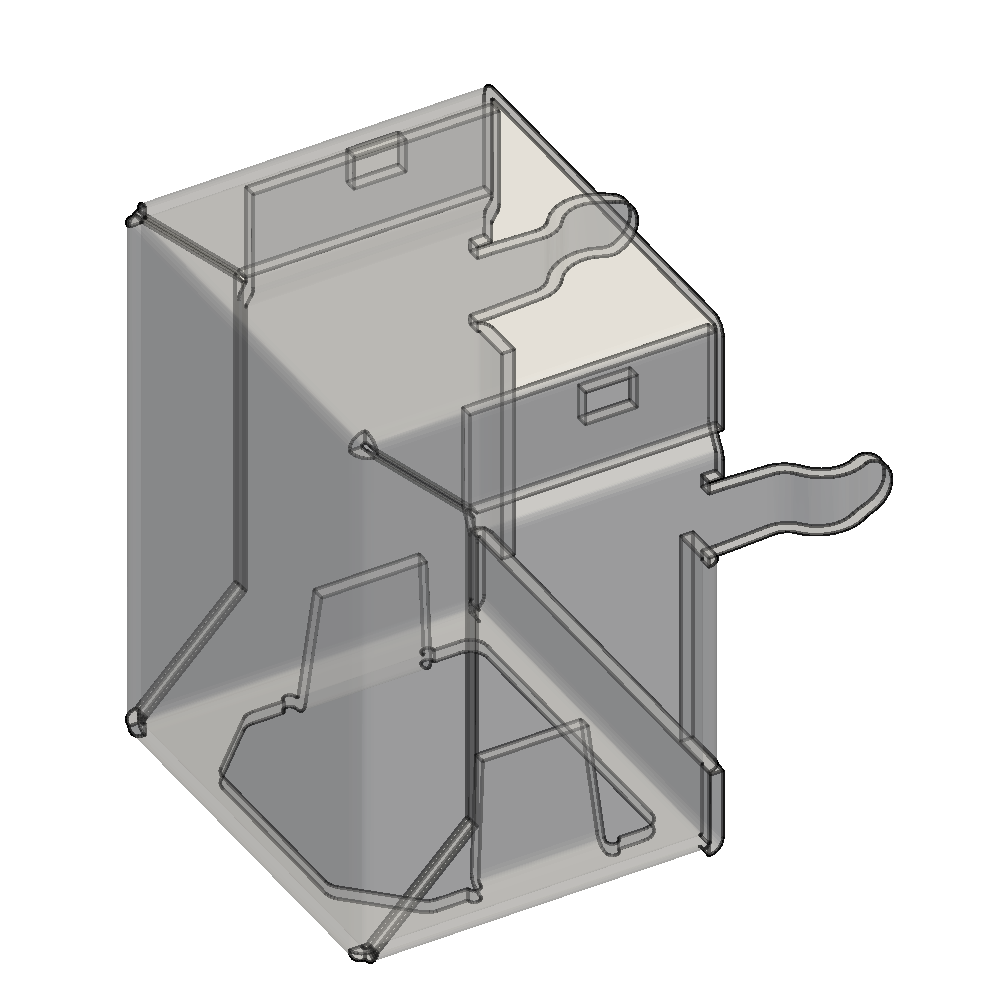}\\[\smallskipamount]
    \includegraphics[width=0.117\linewidth]{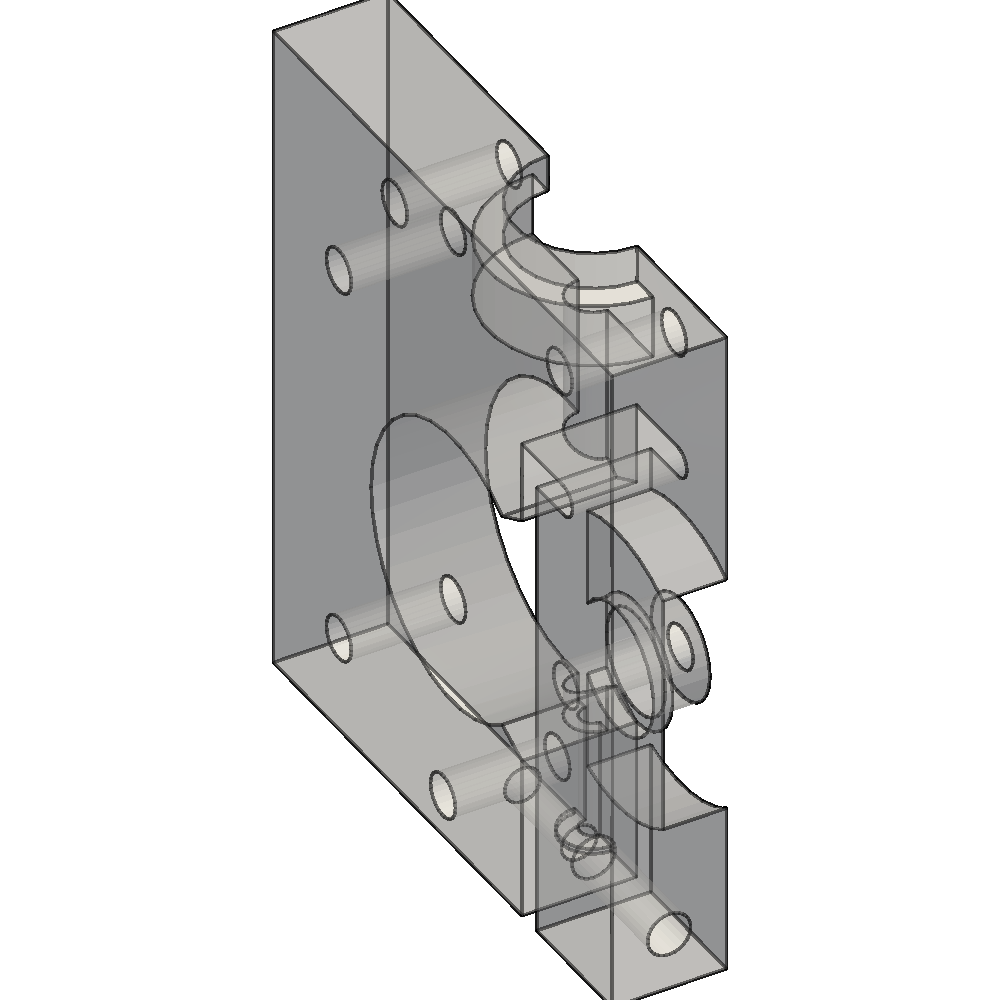}\hfill
    \includegraphics[width=0.117\linewidth]{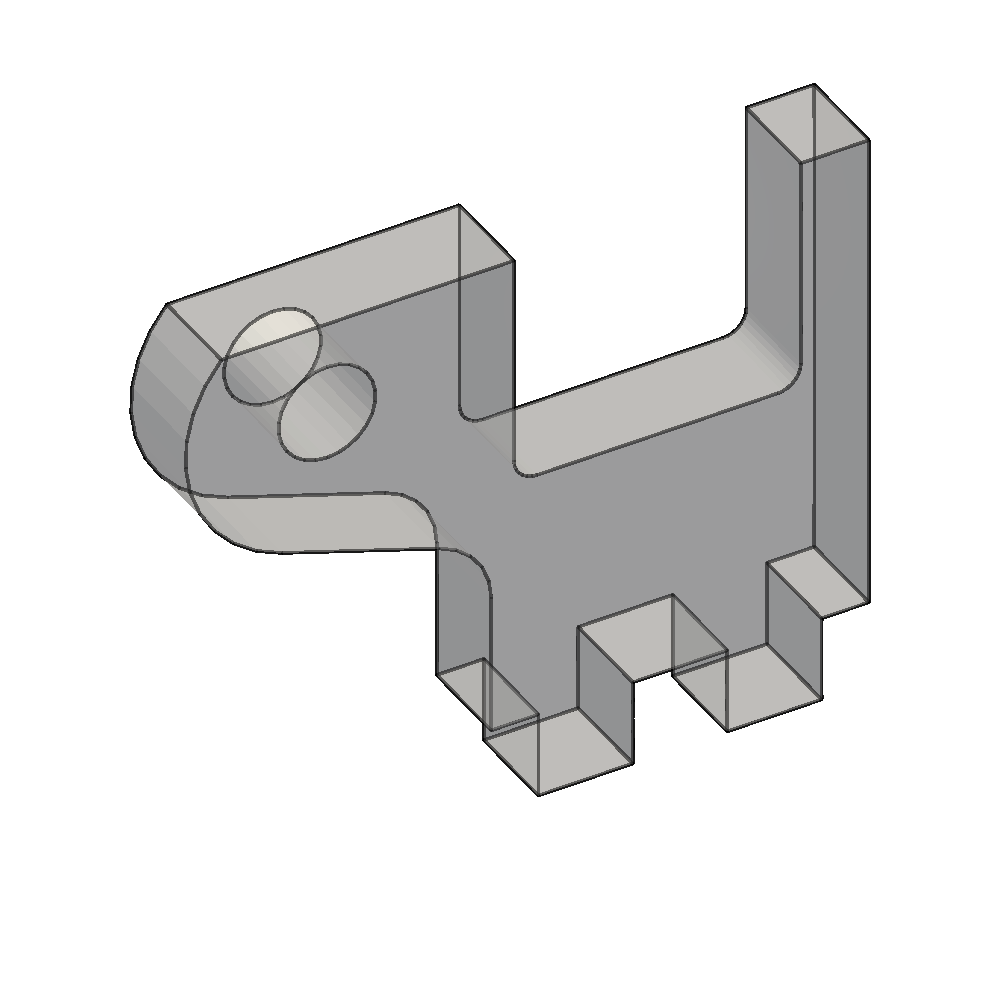}\hfill
    \includegraphics[width=0.117\linewidth]{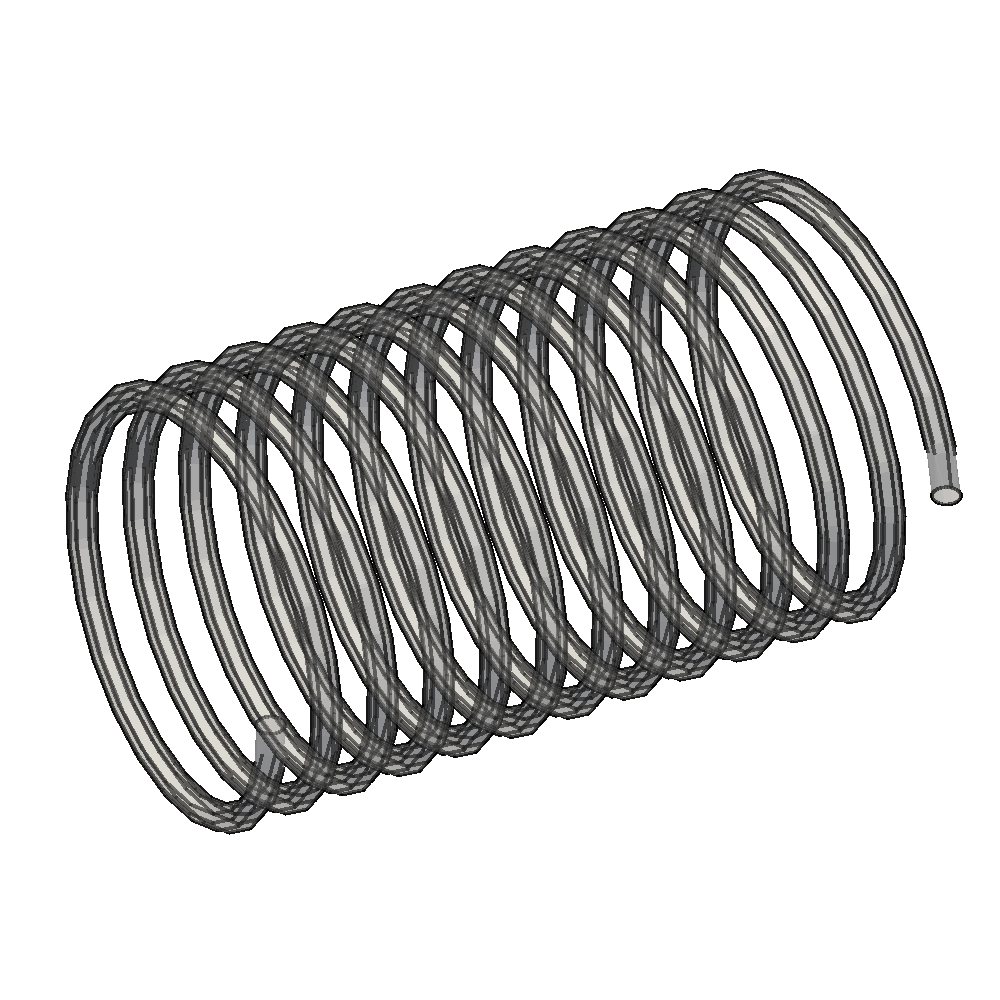}\hfill
    \includegraphics[width=0.117\linewidth]{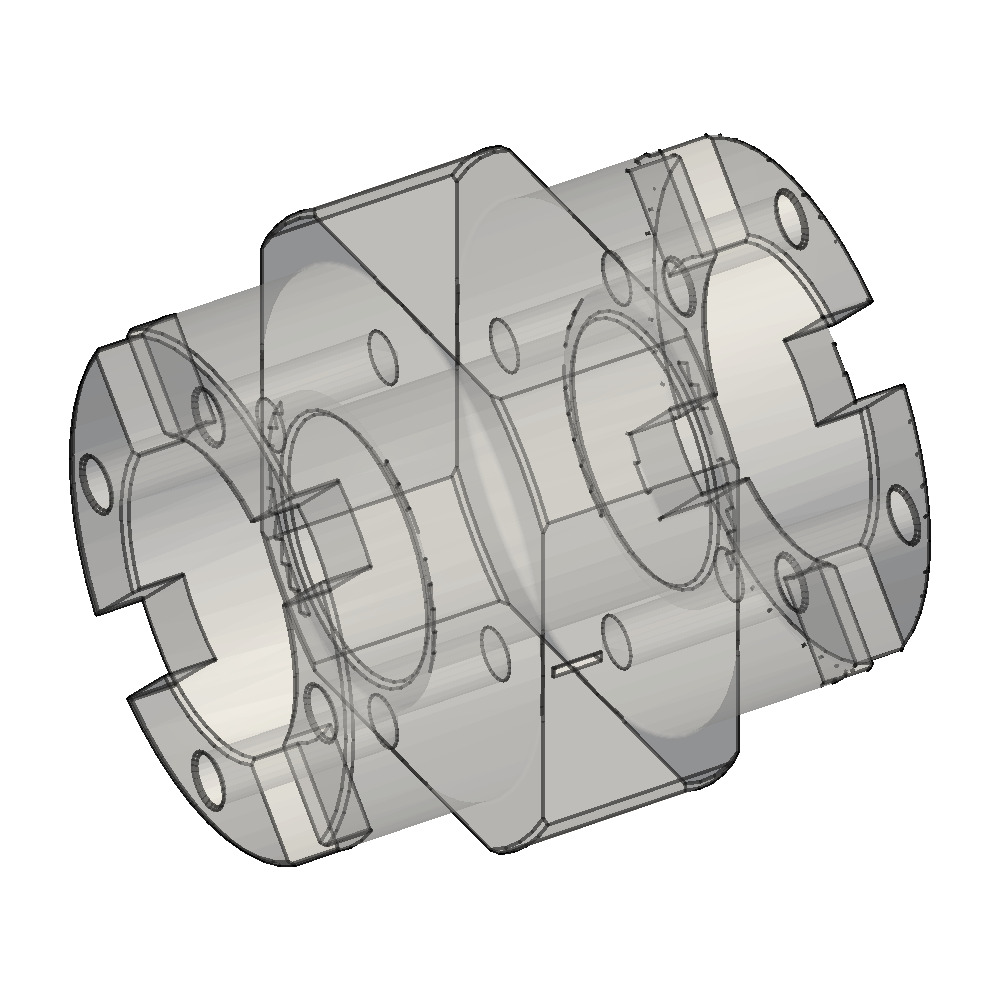}\hfill
    \includegraphics[width=0.117\linewidth]{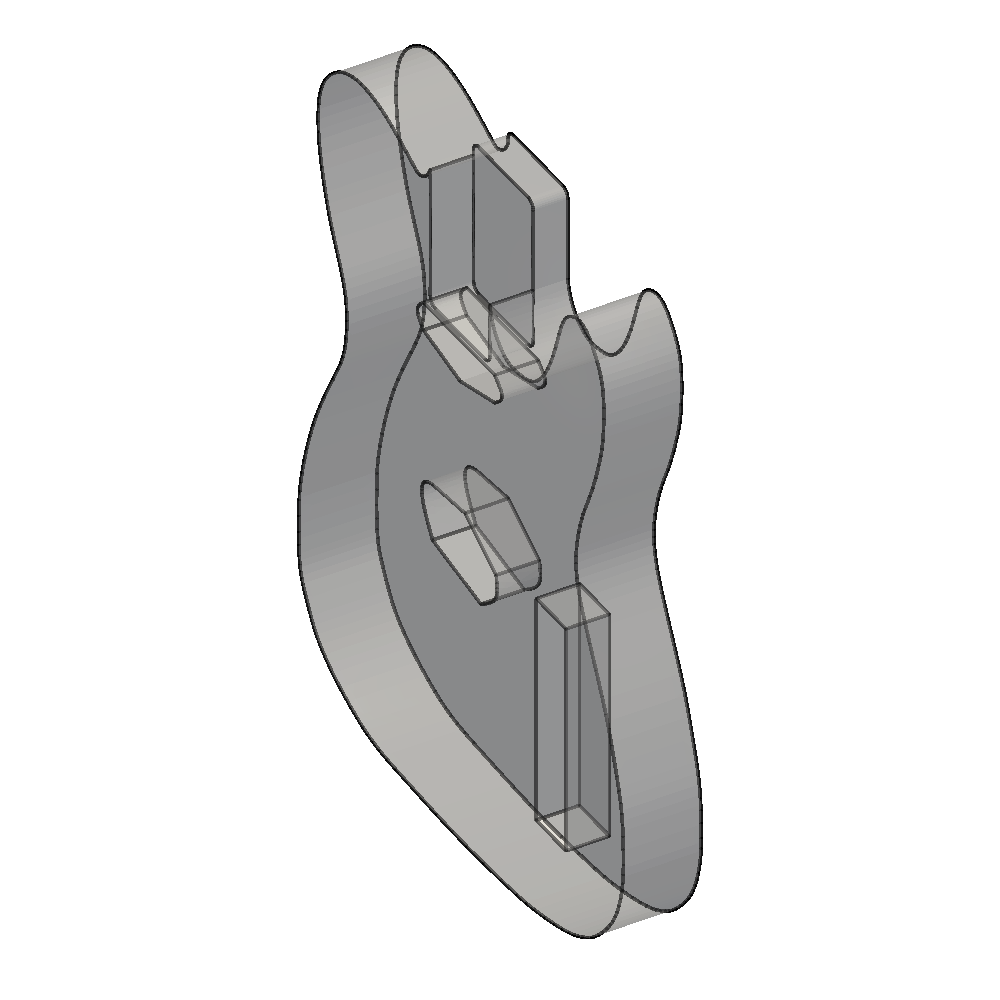}\hfill
    \includegraphics[width=0.117\linewidth]{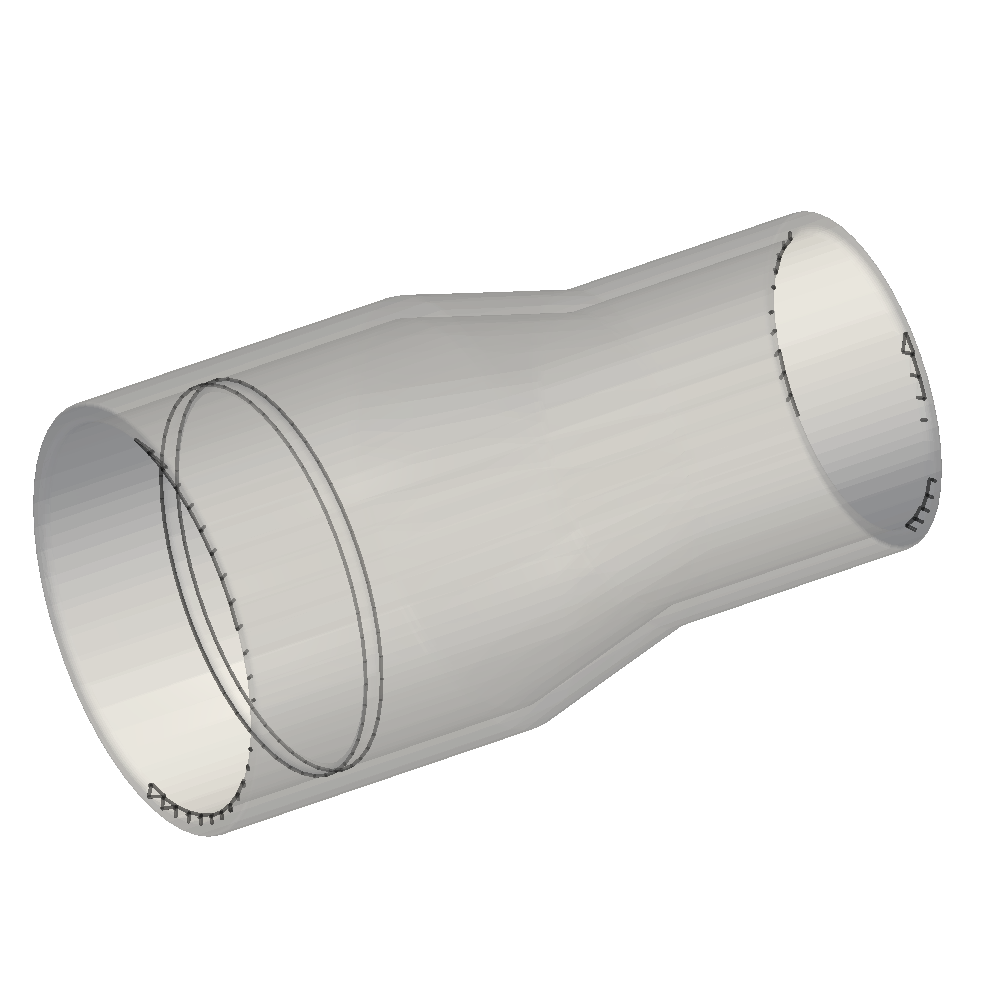}\hfill
    \includegraphics[width=0.117\linewidth]{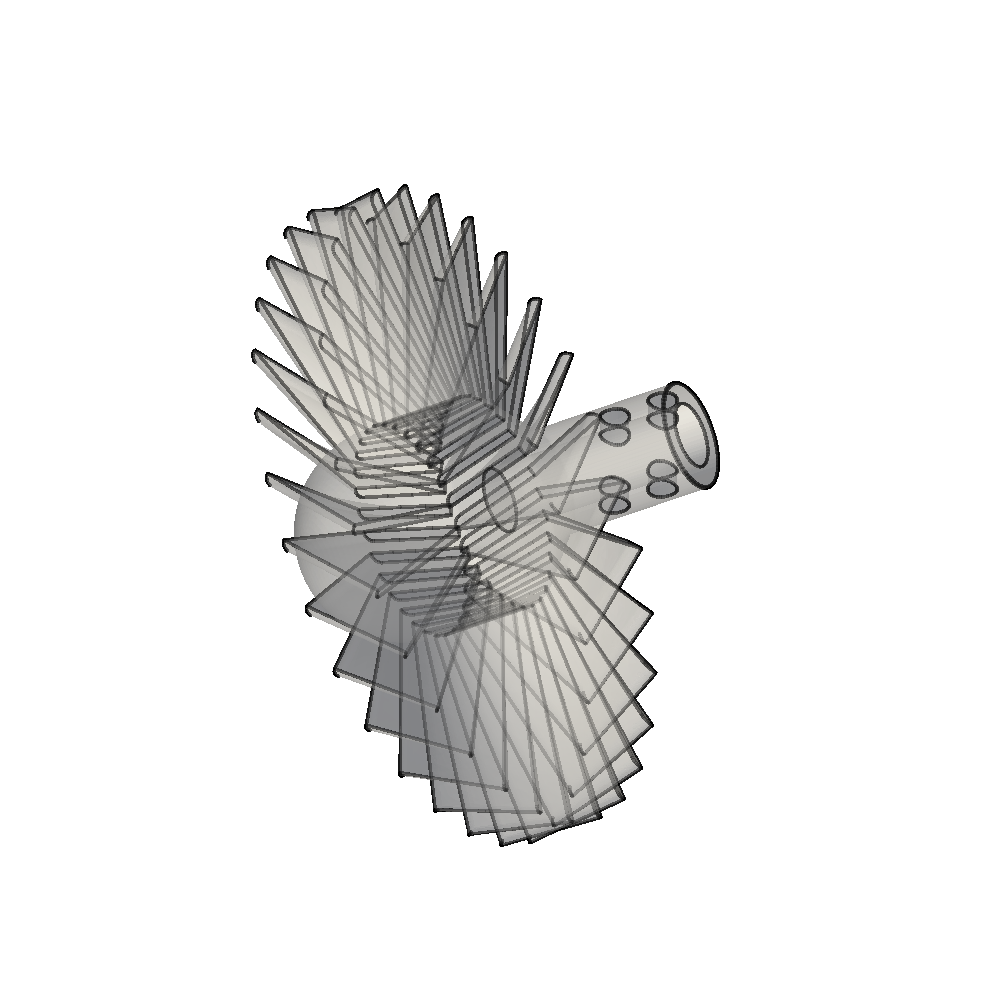}\hfill
    \includegraphics[width=0.117\linewidth]{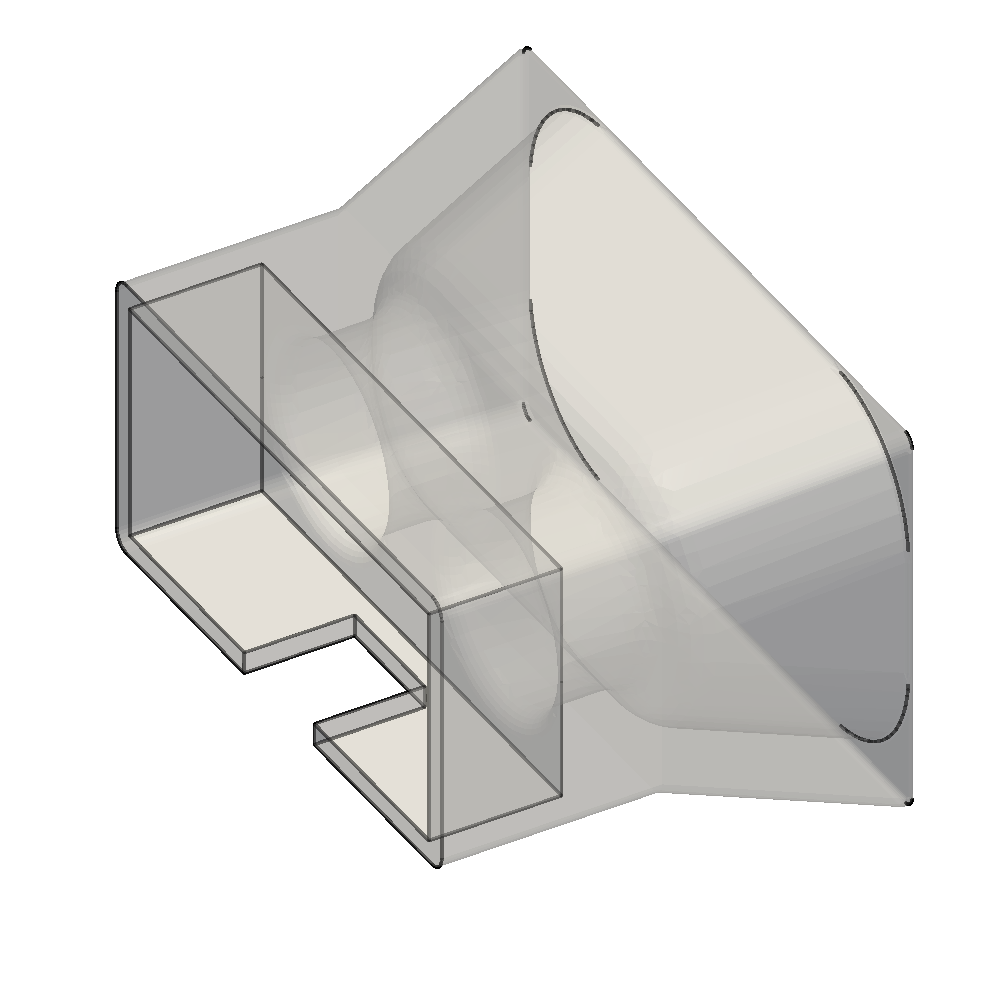}

    \caption{(continued, part 3 of 4) CAD parts of the 156-part evaluation benchmark.}
    \label{fig:cad_dataset_grid_3}
\end{figure}

\begin{figure}[h]
    \centering
    \includegraphics[width=0.117\linewidth]{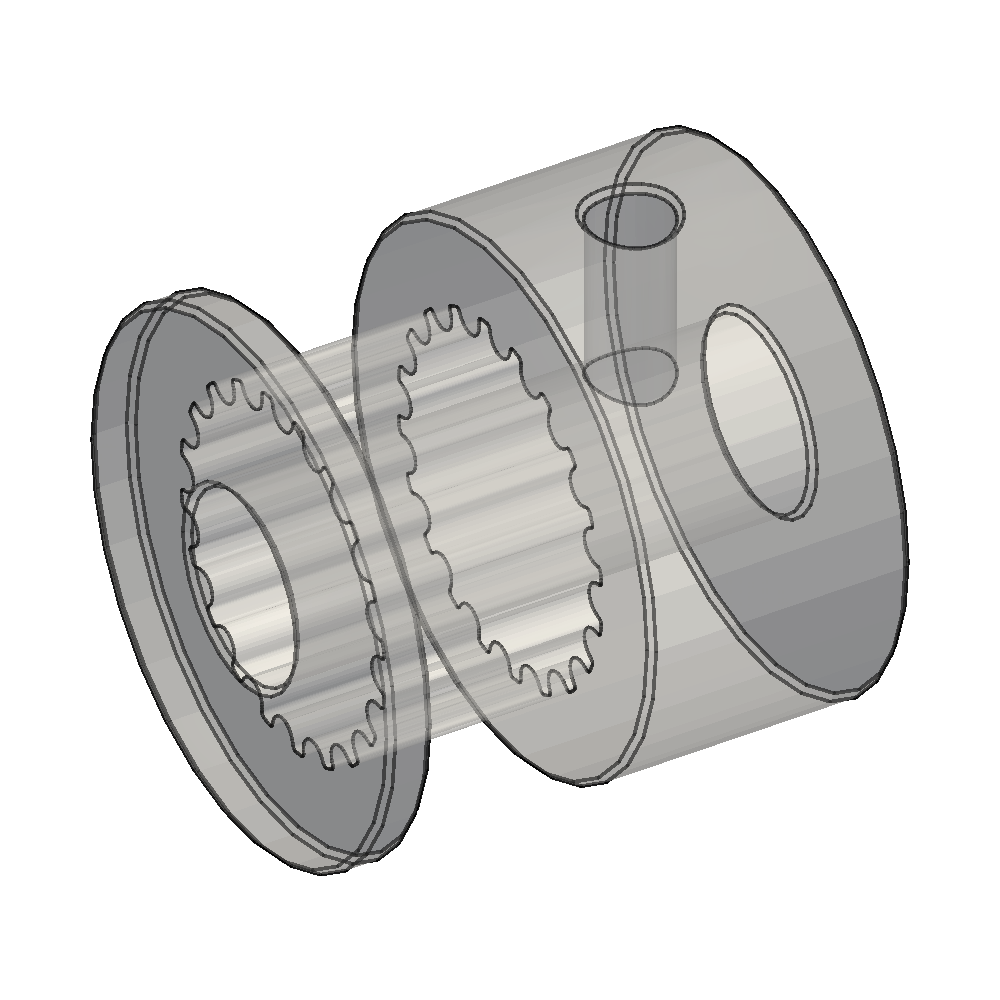}\hfill
    \includegraphics[width=0.117\linewidth]{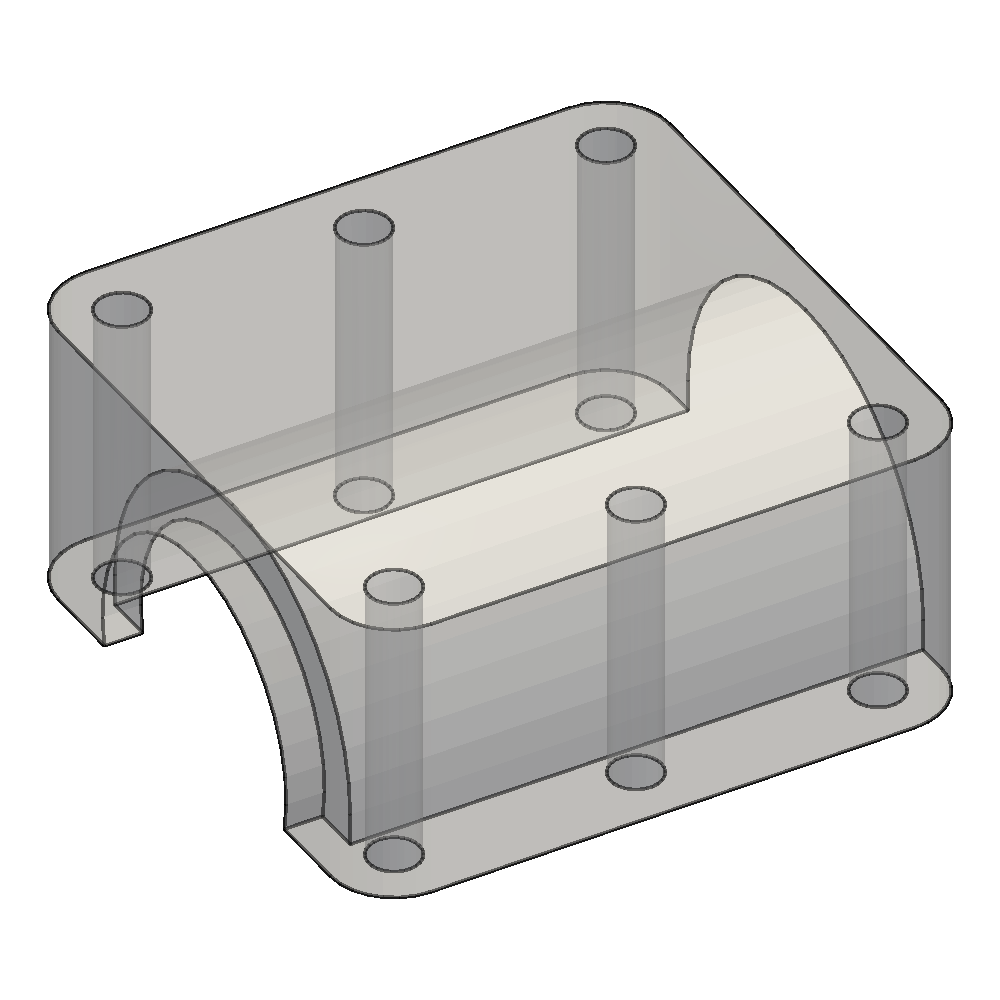}\hfill
    \includegraphics[width=0.117\linewidth]{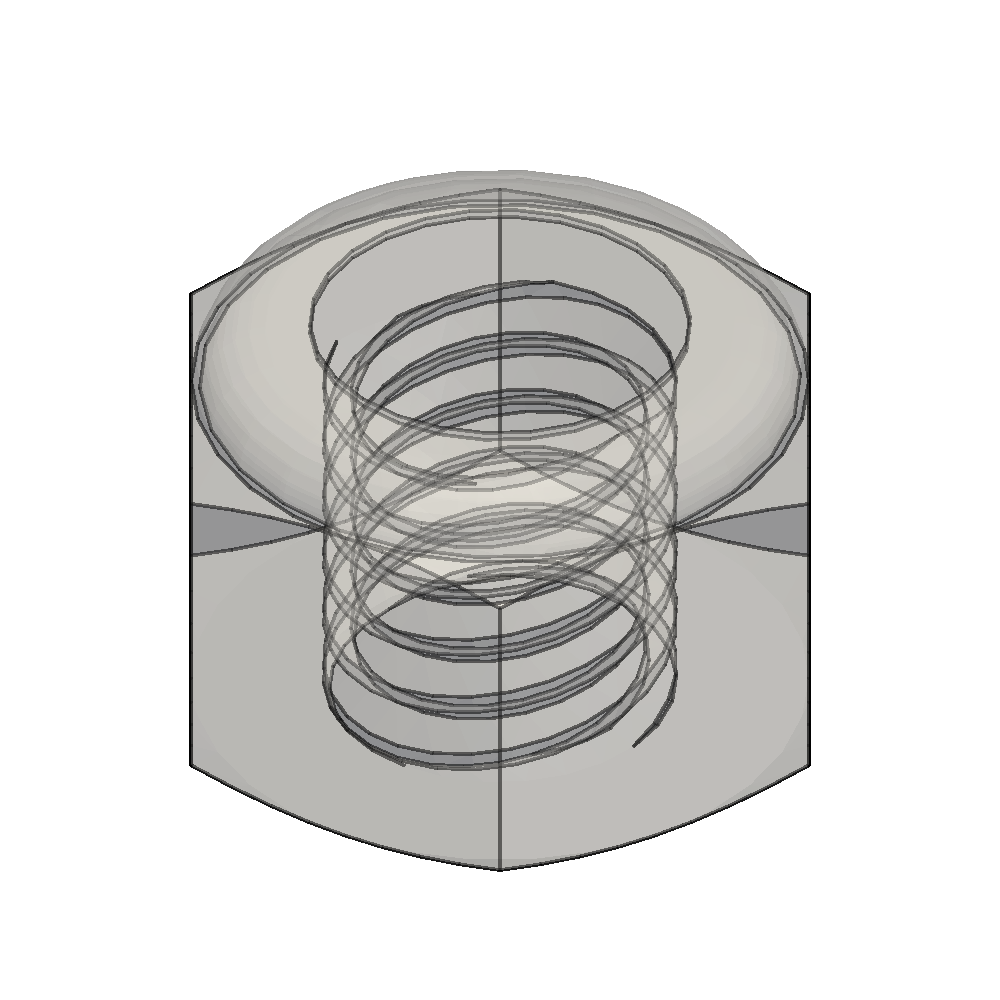}\hfill
    \includegraphics[width=0.117\linewidth]{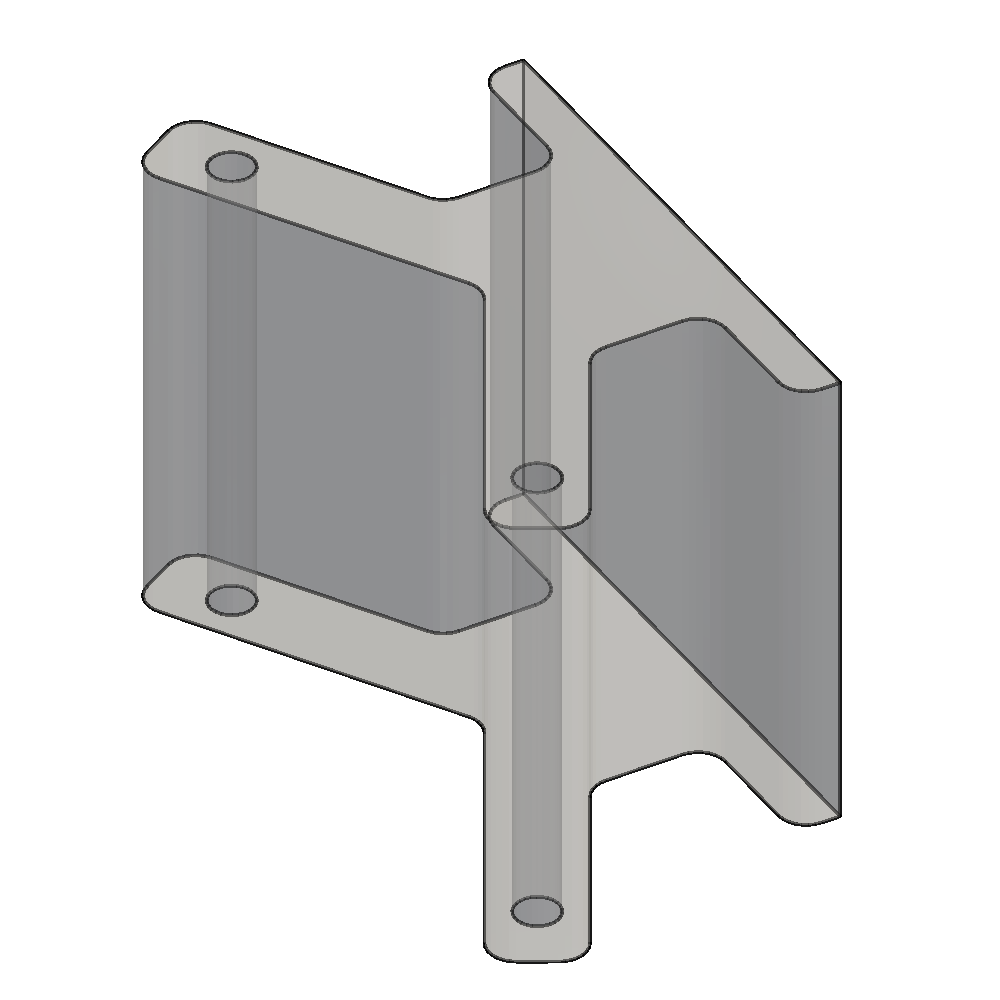}\hfill
    \includegraphics[width=0.117\linewidth]{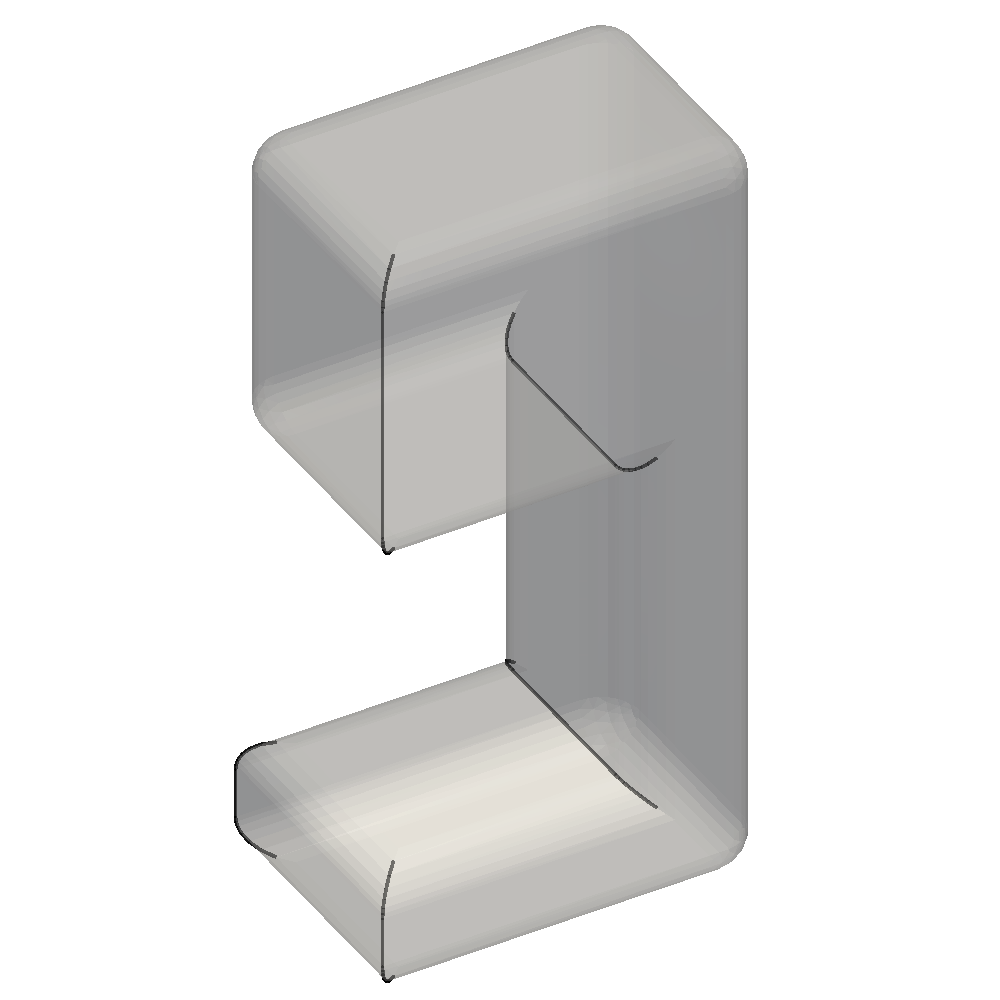}\hfill
    \includegraphics[width=0.117\linewidth]{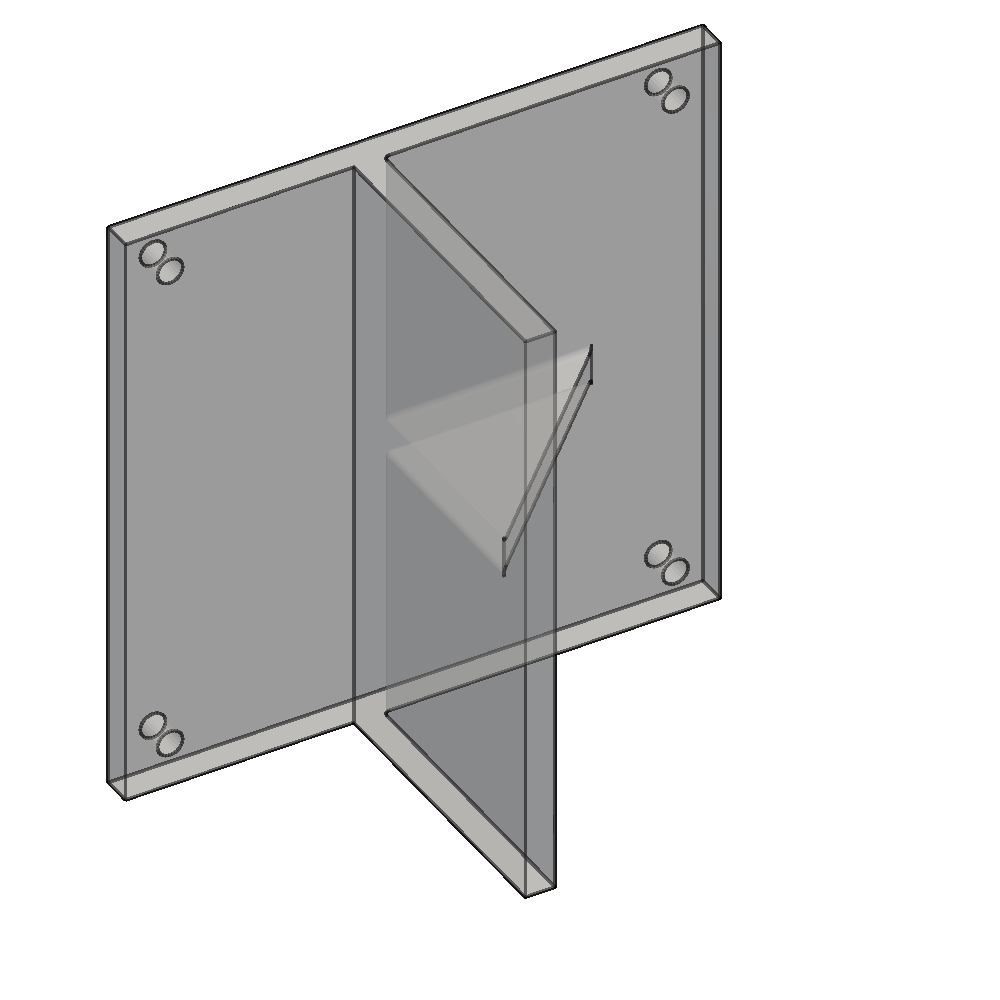}\hfill
    \includegraphics[width=0.117\linewidth]{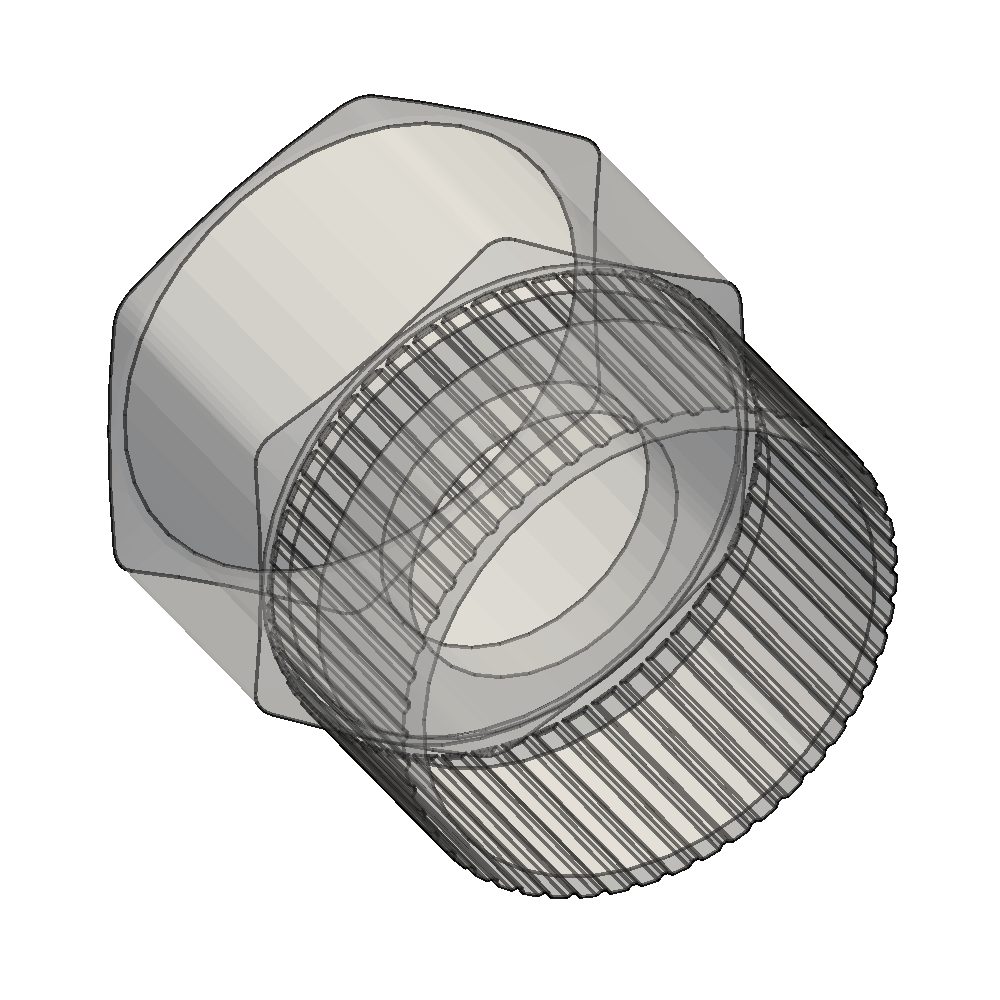}\hfill
    \includegraphics[width=0.117\linewidth]{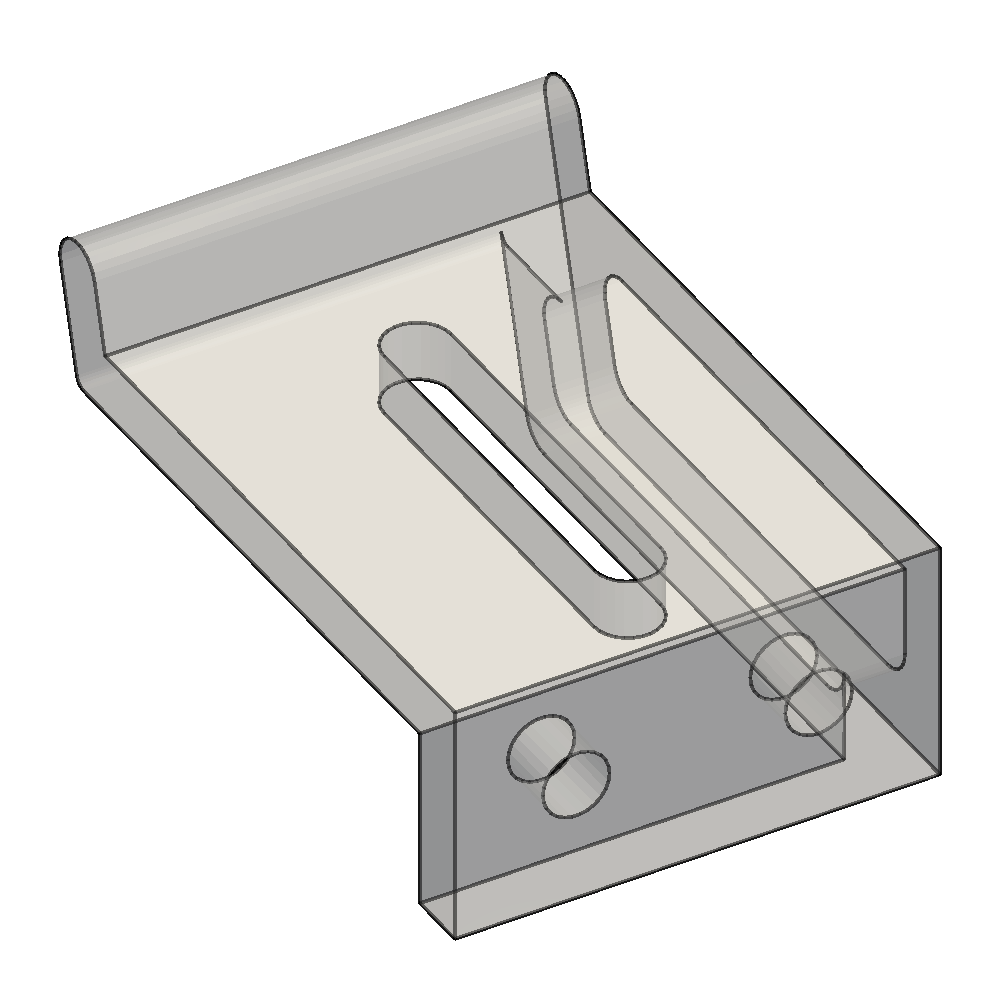}\\[\smallskipamount]
    \includegraphics[width=0.117\linewidth]{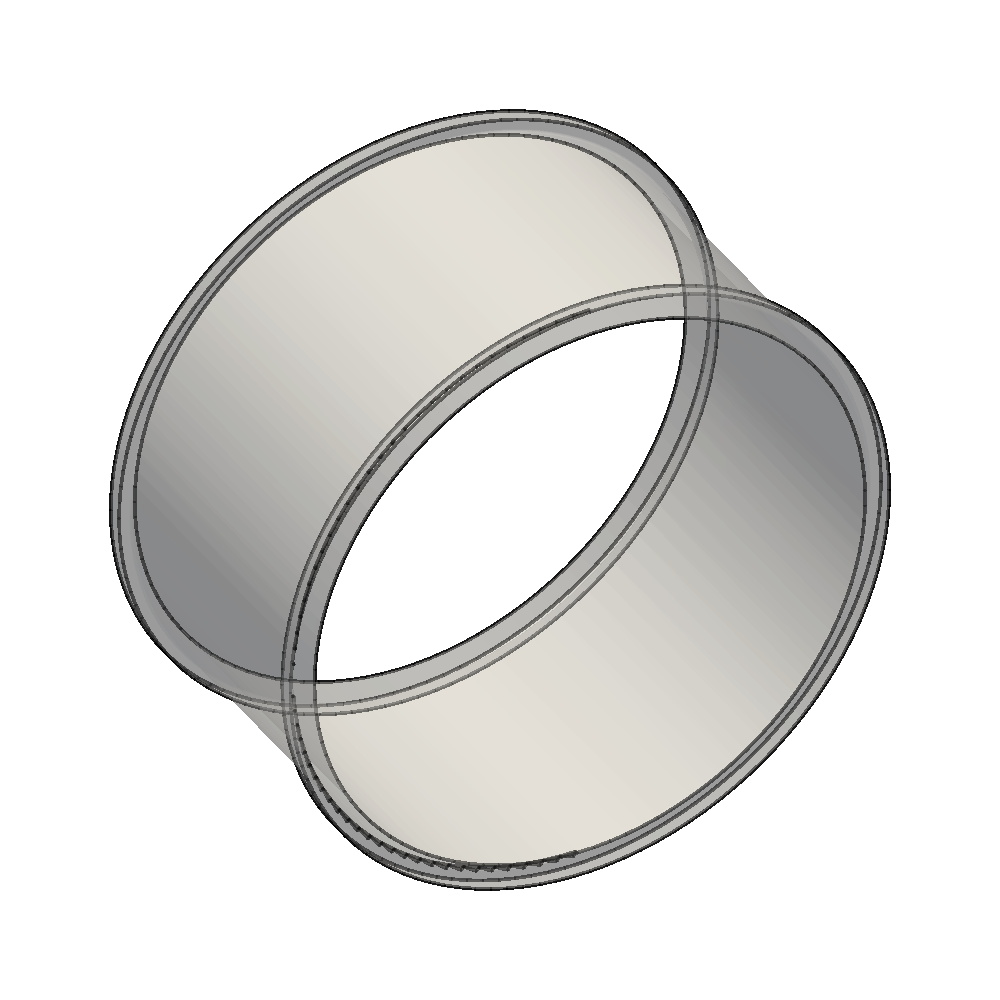}\hfill
    \includegraphics[width=0.117\linewidth]{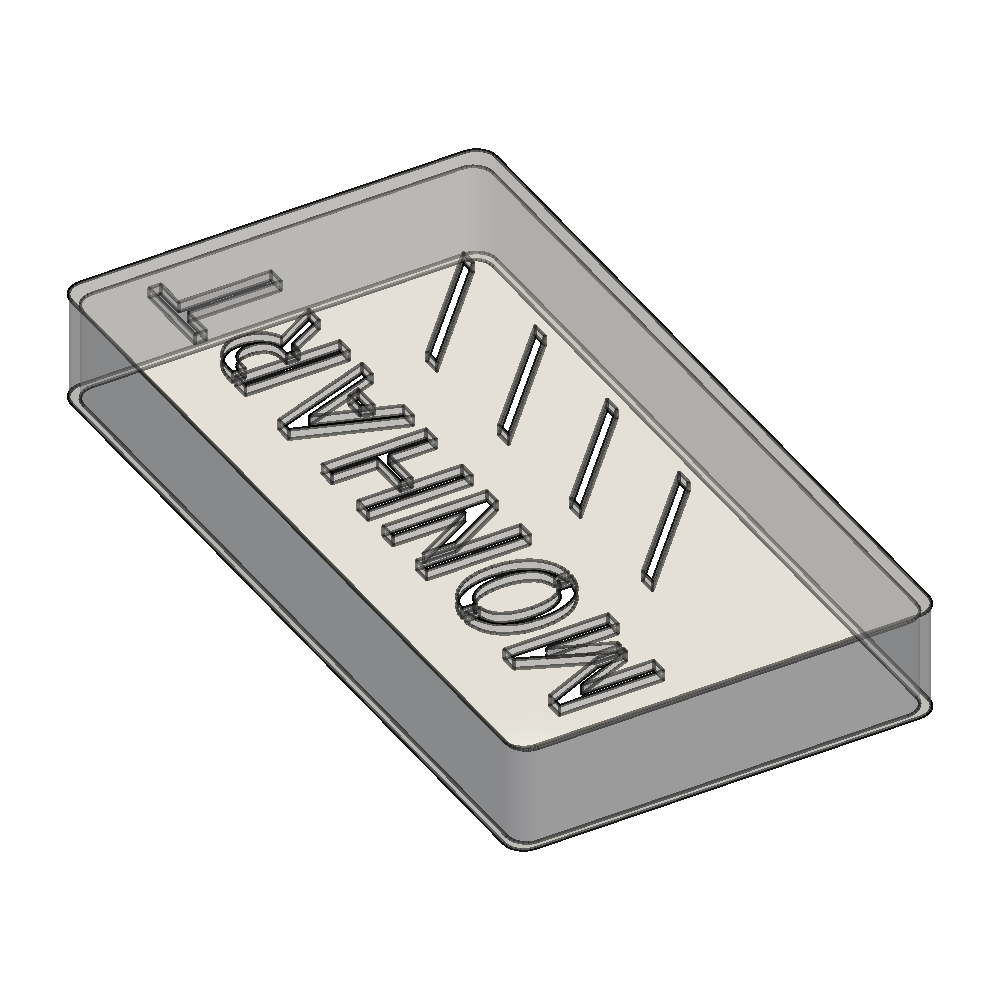}\hfill
    \includegraphics[width=0.117\linewidth]{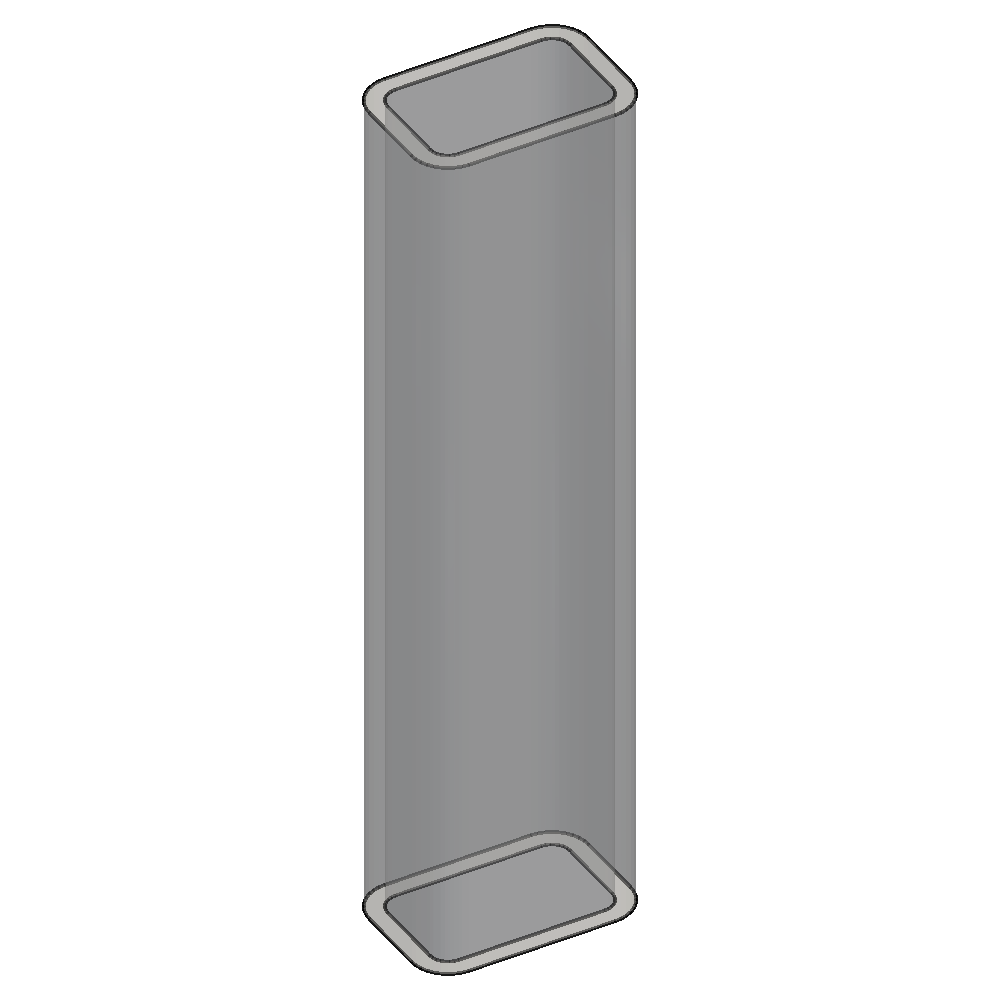}\hfill
    \includegraphics[width=0.117\linewidth]{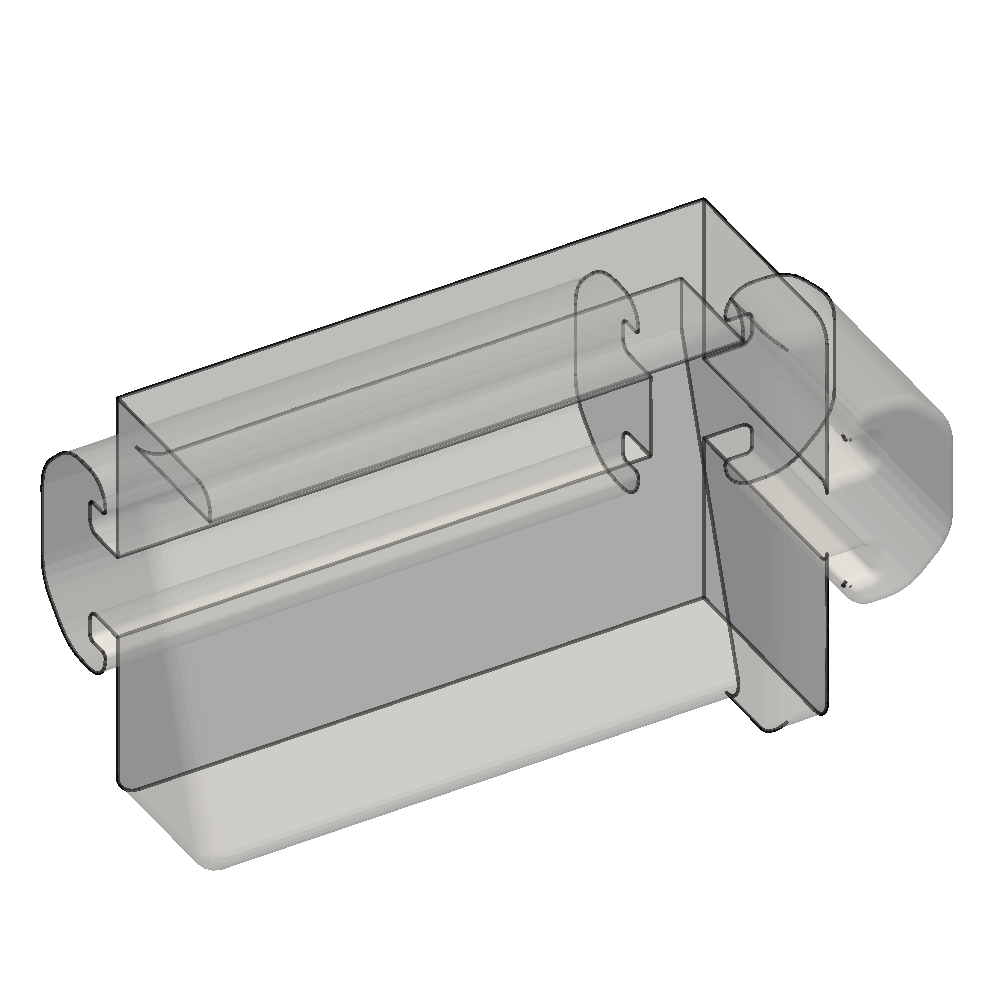}\hfill
    \includegraphics[width=0.117\linewidth]{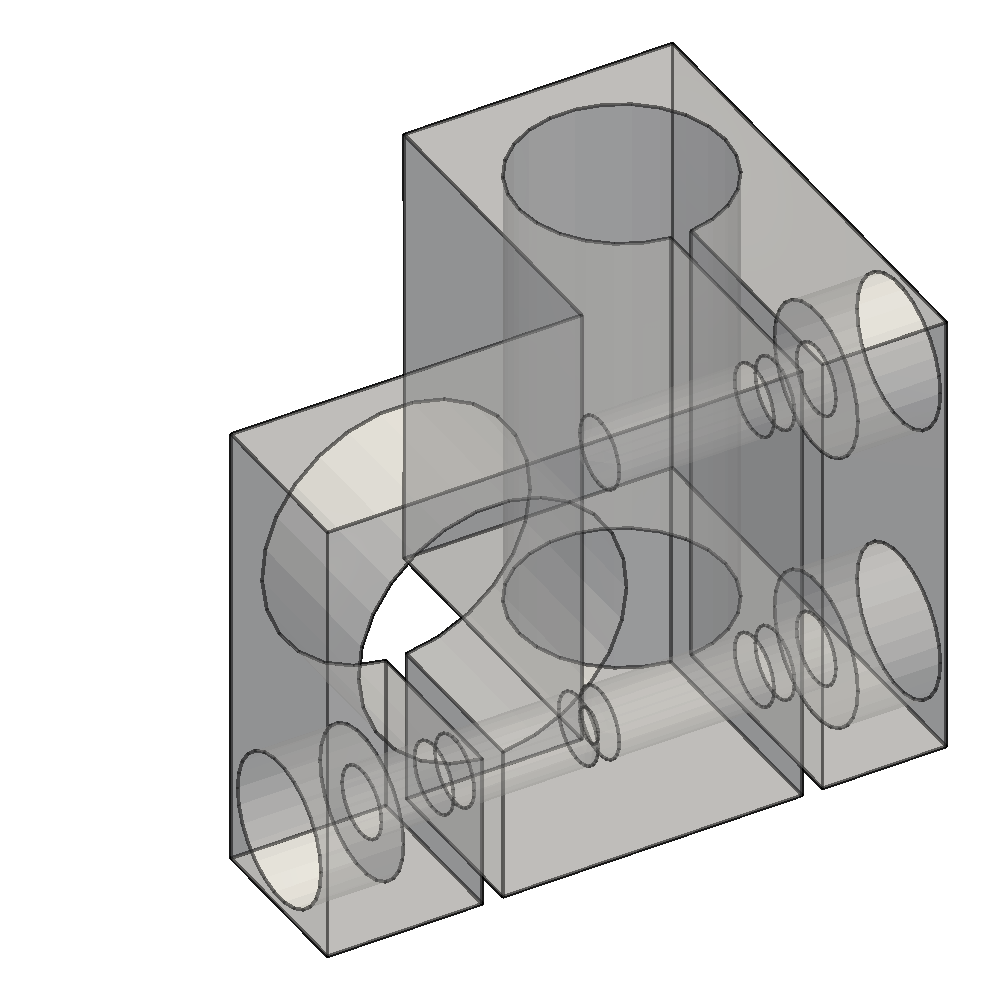}\hfill
    \includegraphics[width=0.117\linewidth]{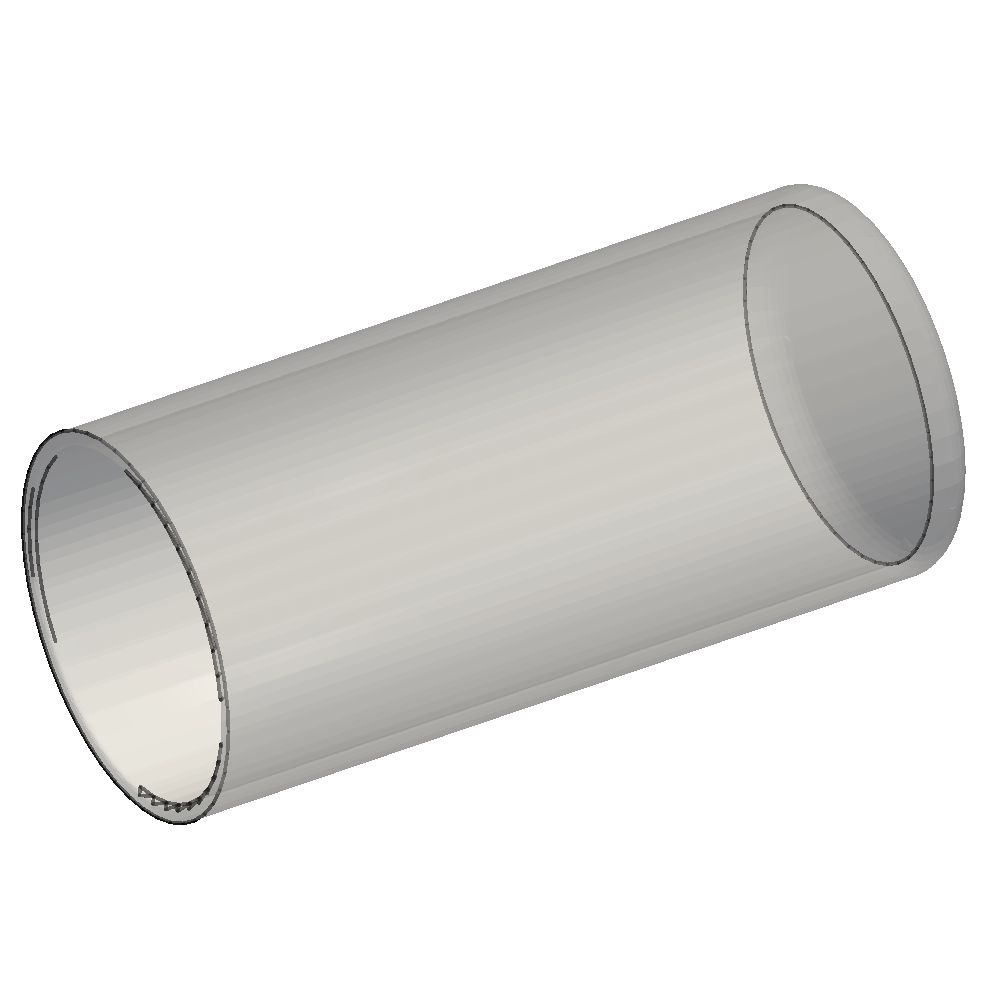}\hfill
    \includegraphics[width=0.117\linewidth]{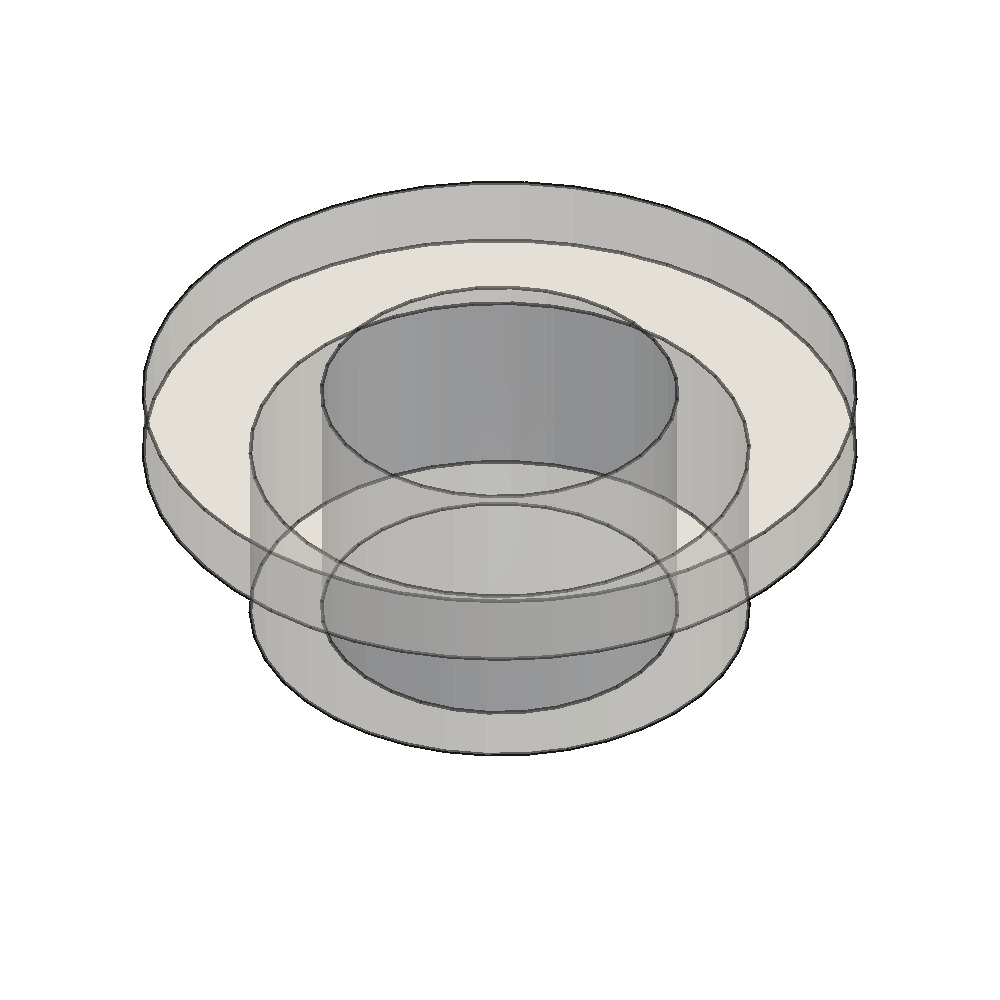}\hfill
    \includegraphics[width=0.117\linewidth]{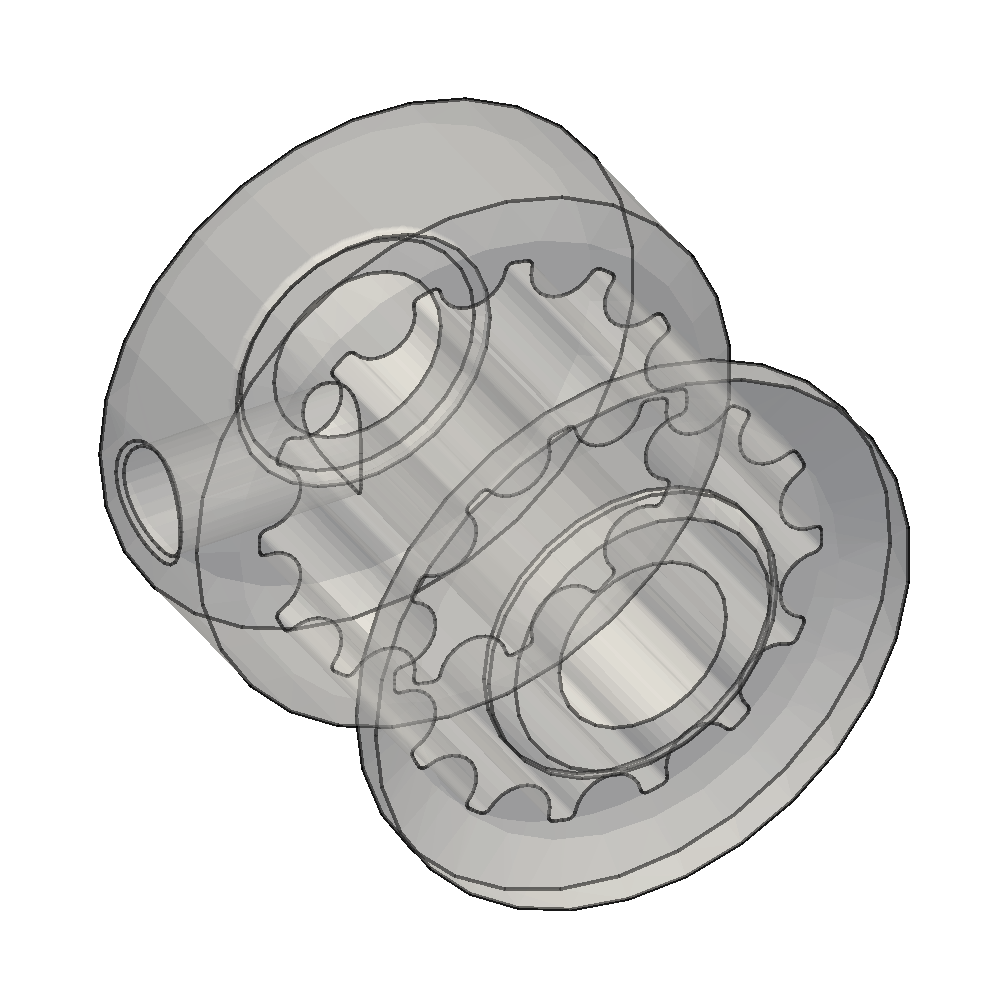}\\[\smallskipamount]
    \includegraphics[width=0.117\linewidth]{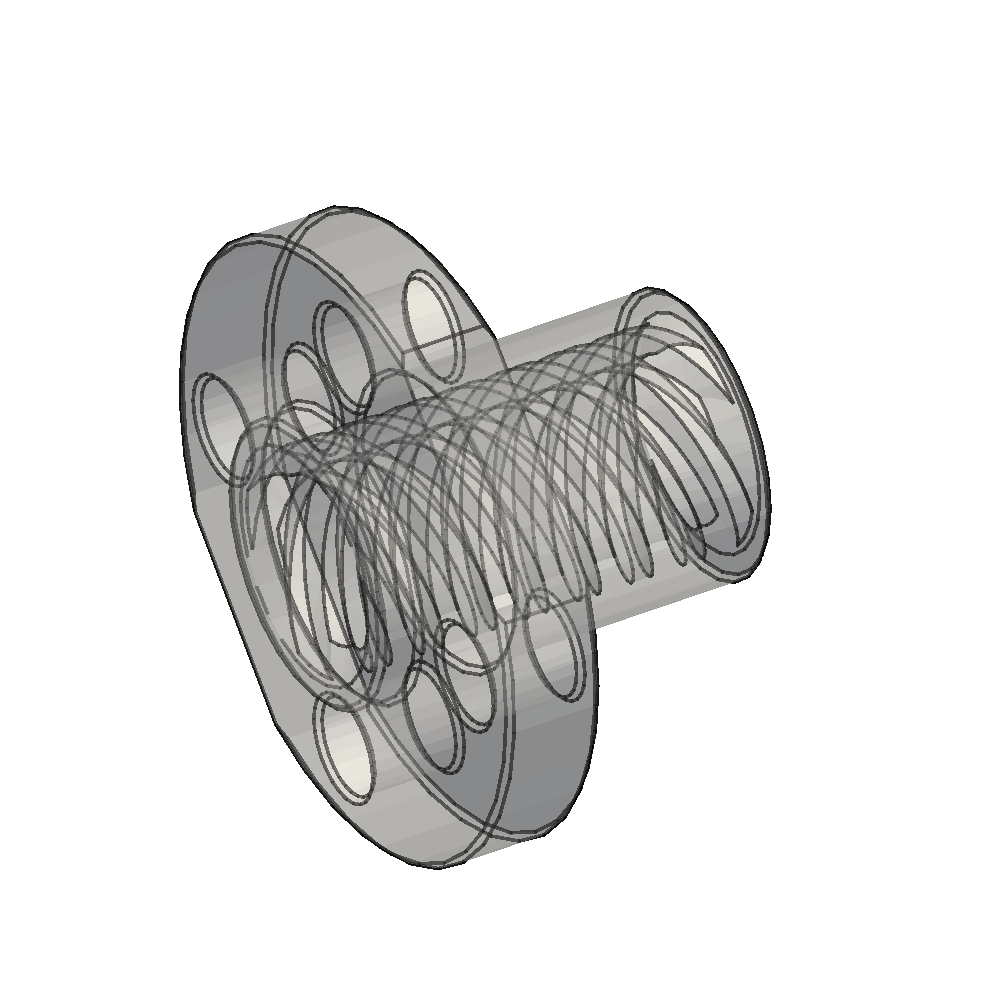}\hfill
    \includegraphics[width=0.117\linewidth]{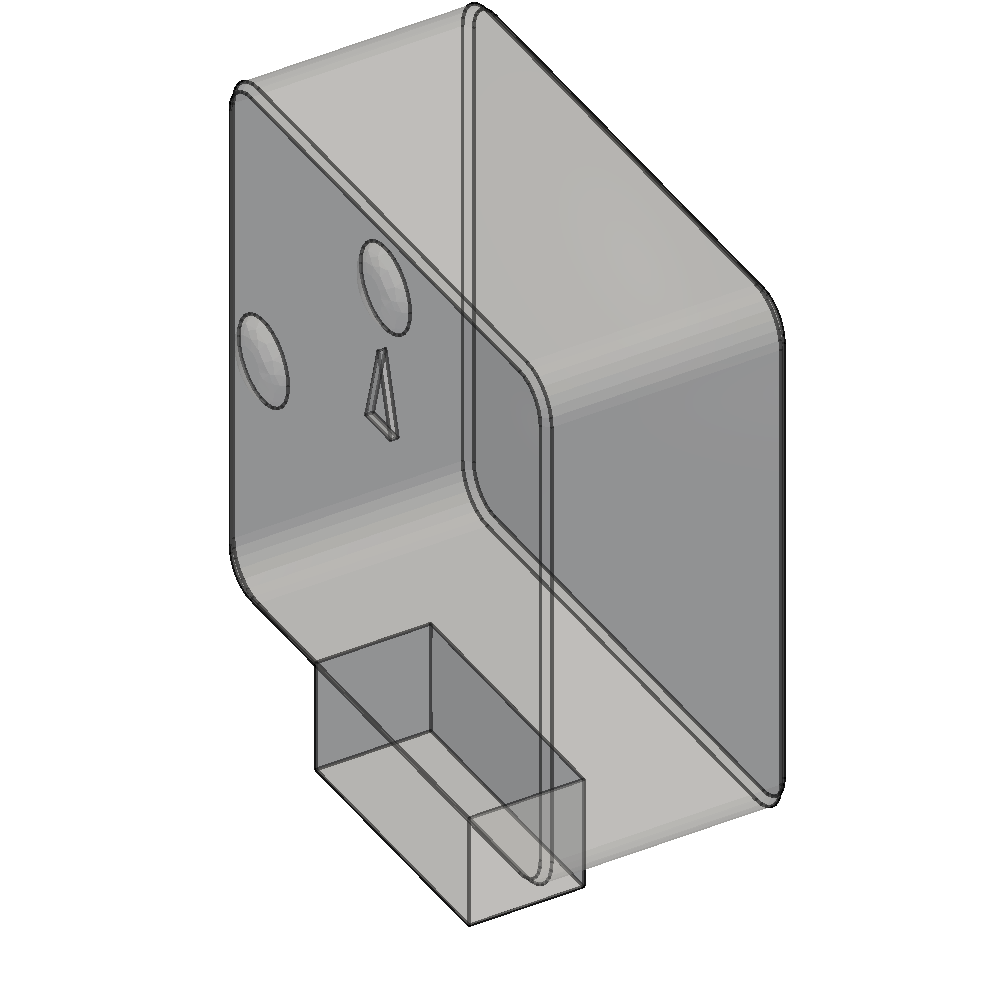}\hfill
    \includegraphics[width=0.117\linewidth]{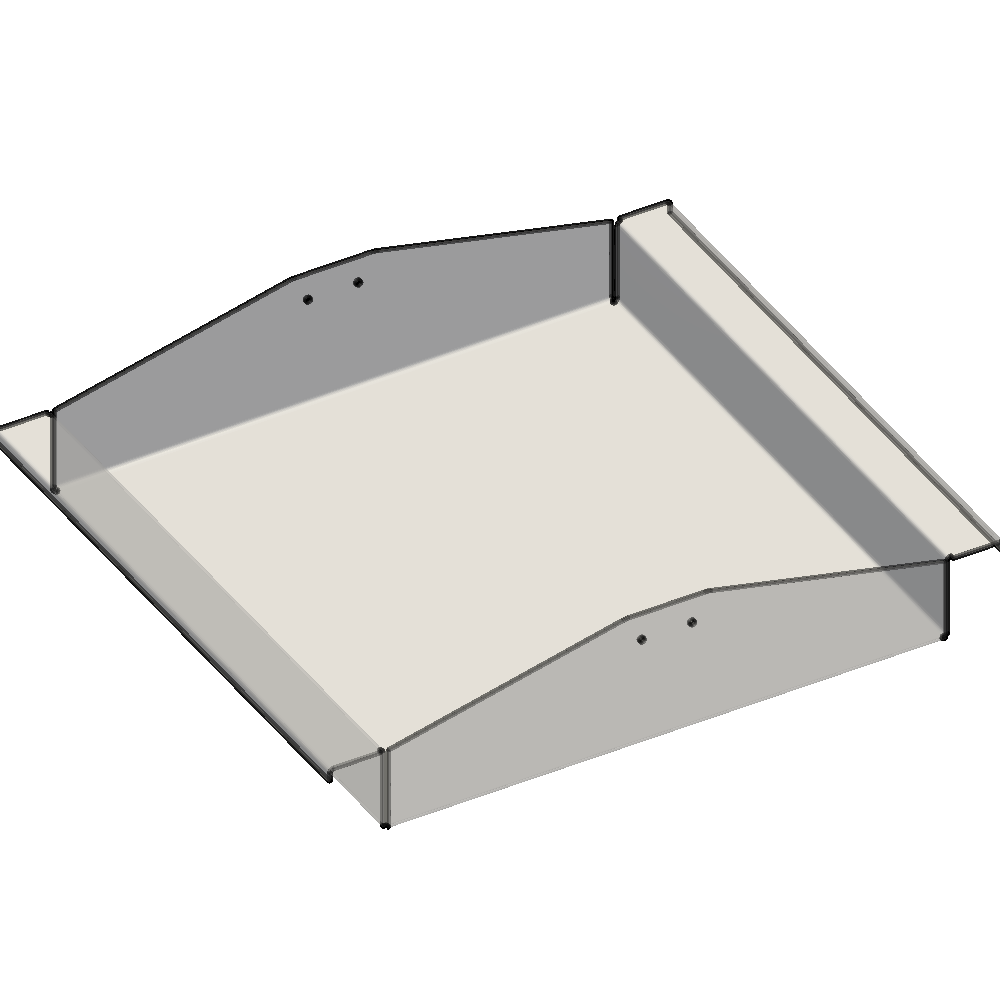}\hfill
    \includegraphics[width=0.117\linewidth]{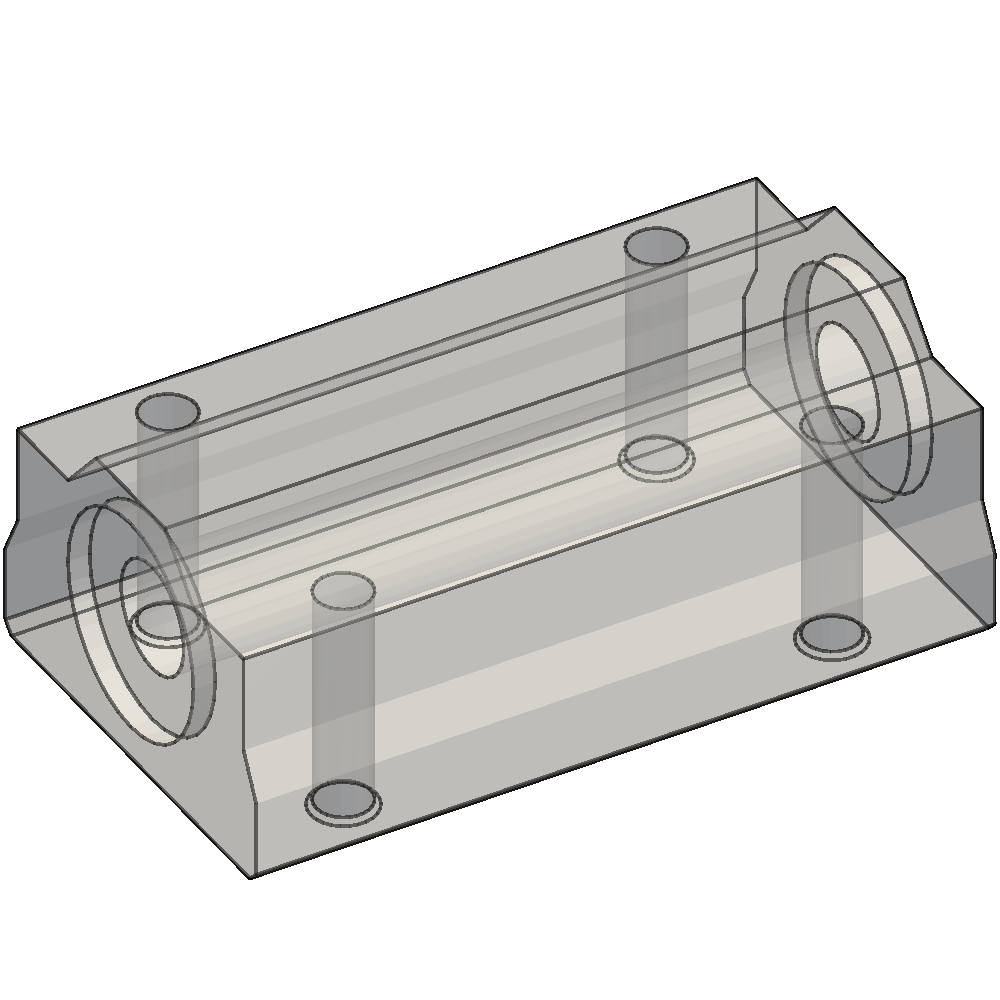}\hfill
    \includegraphics[width=0.117\linewidth]{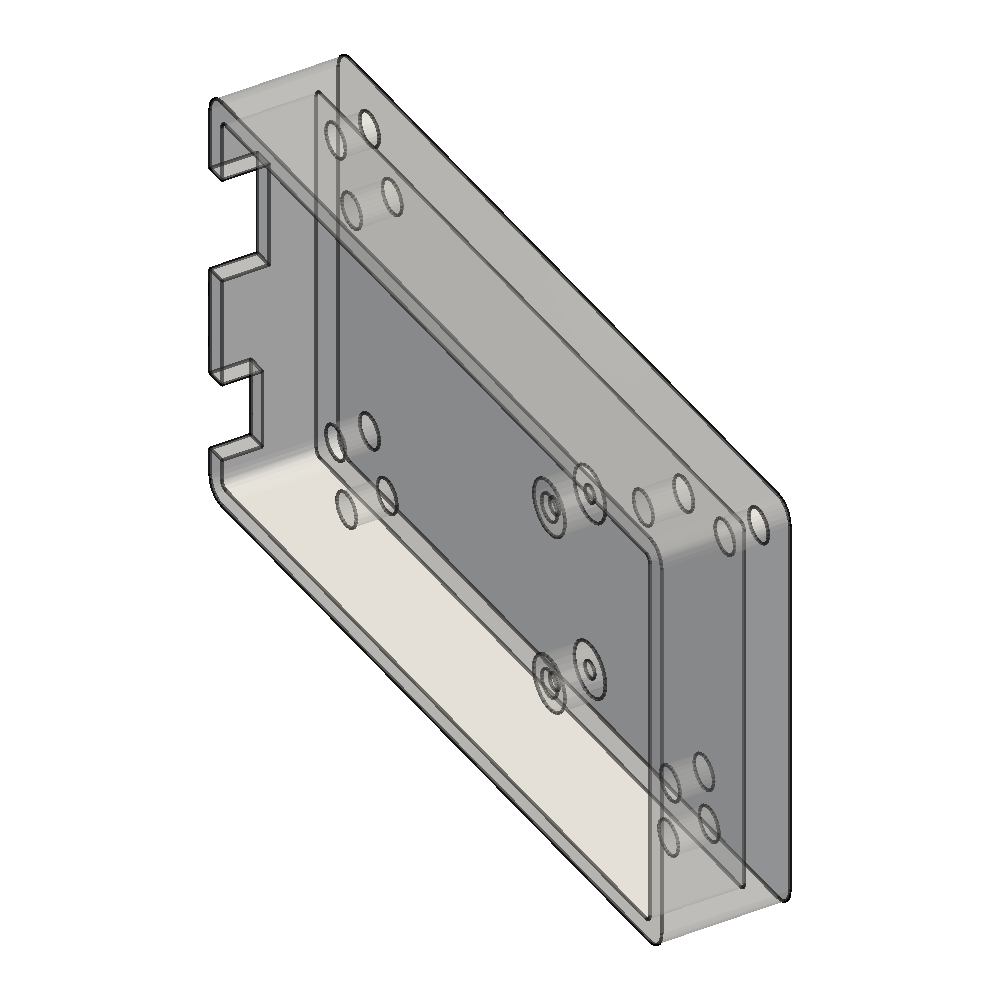}\hfill
    \includegraphics[width=0.117\linewidth]{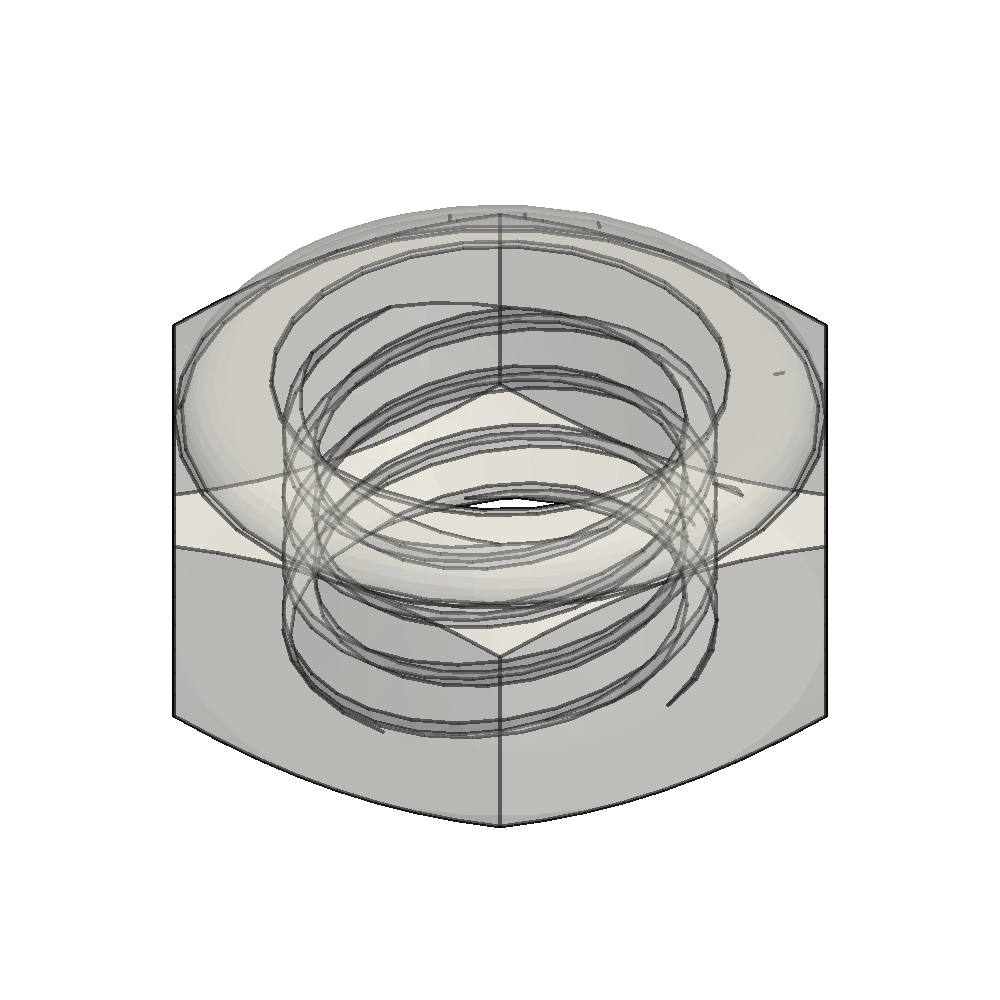}\hfill
    \includegraphics[width=0.117\linewidth]{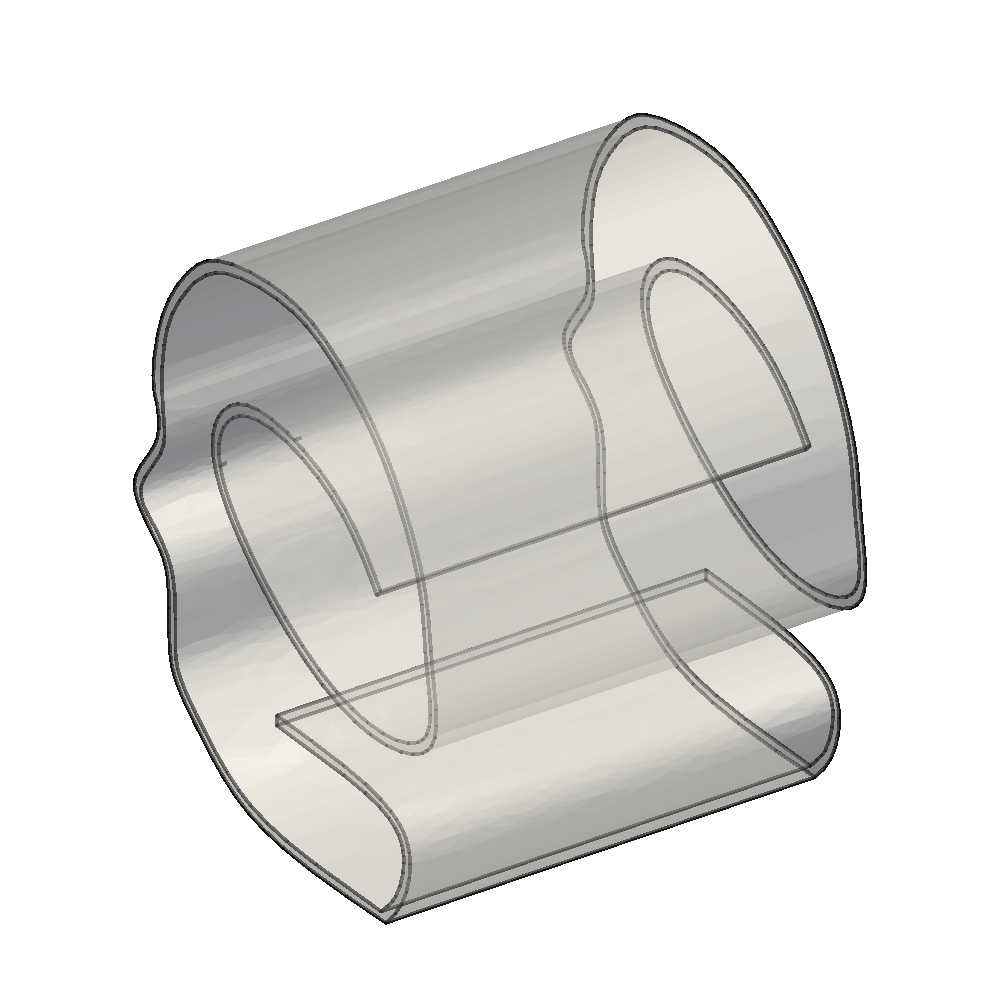}\hfill
    \includegraphics[width=0.117\linewidth]{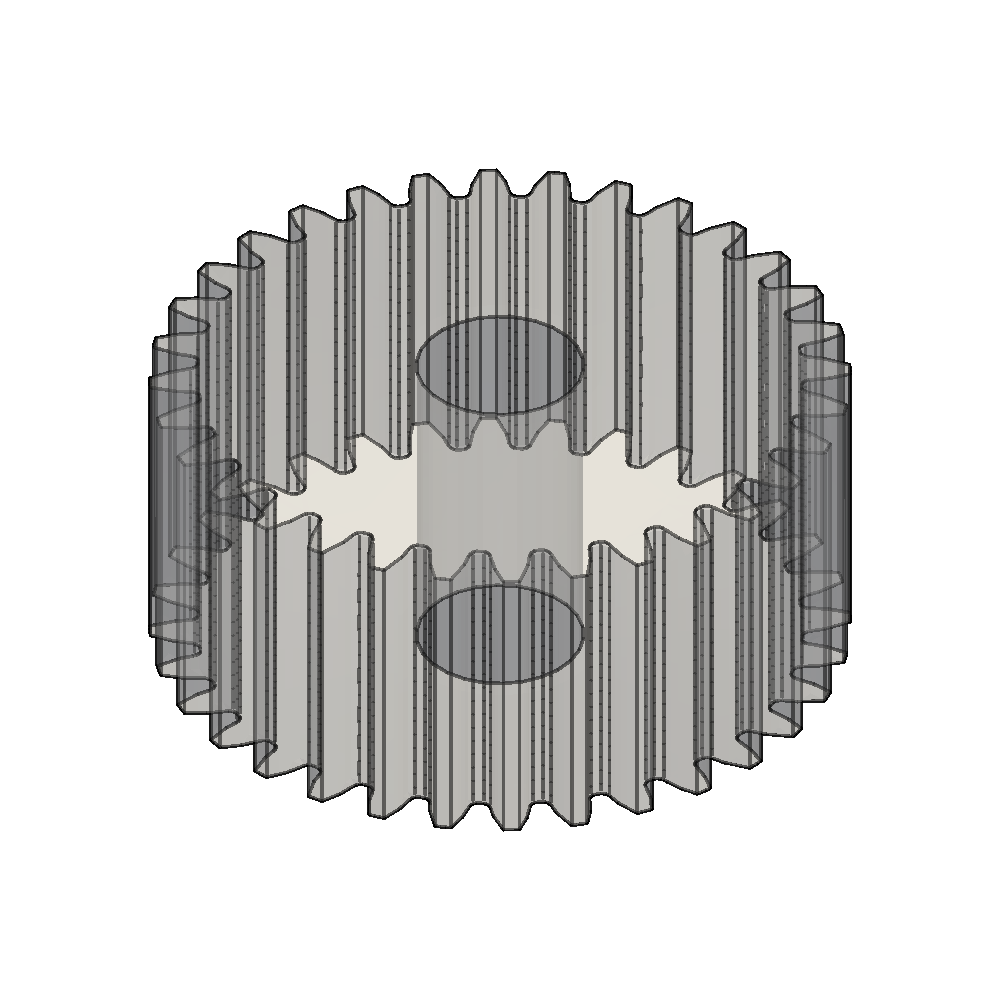}\\[\smallskipamount]
    \includegraphics[width=0.117\linewidth]{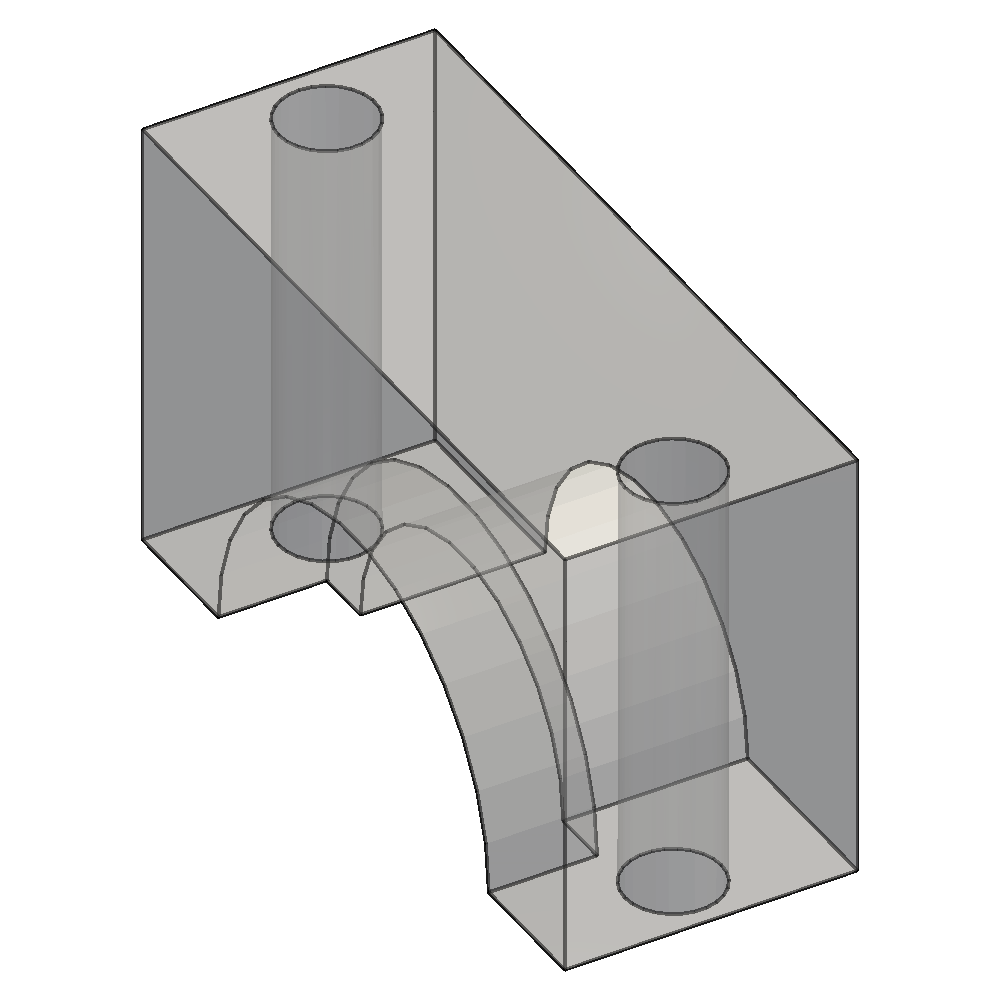}\hfill
    \includegraphics[width=0.117\linewidth]{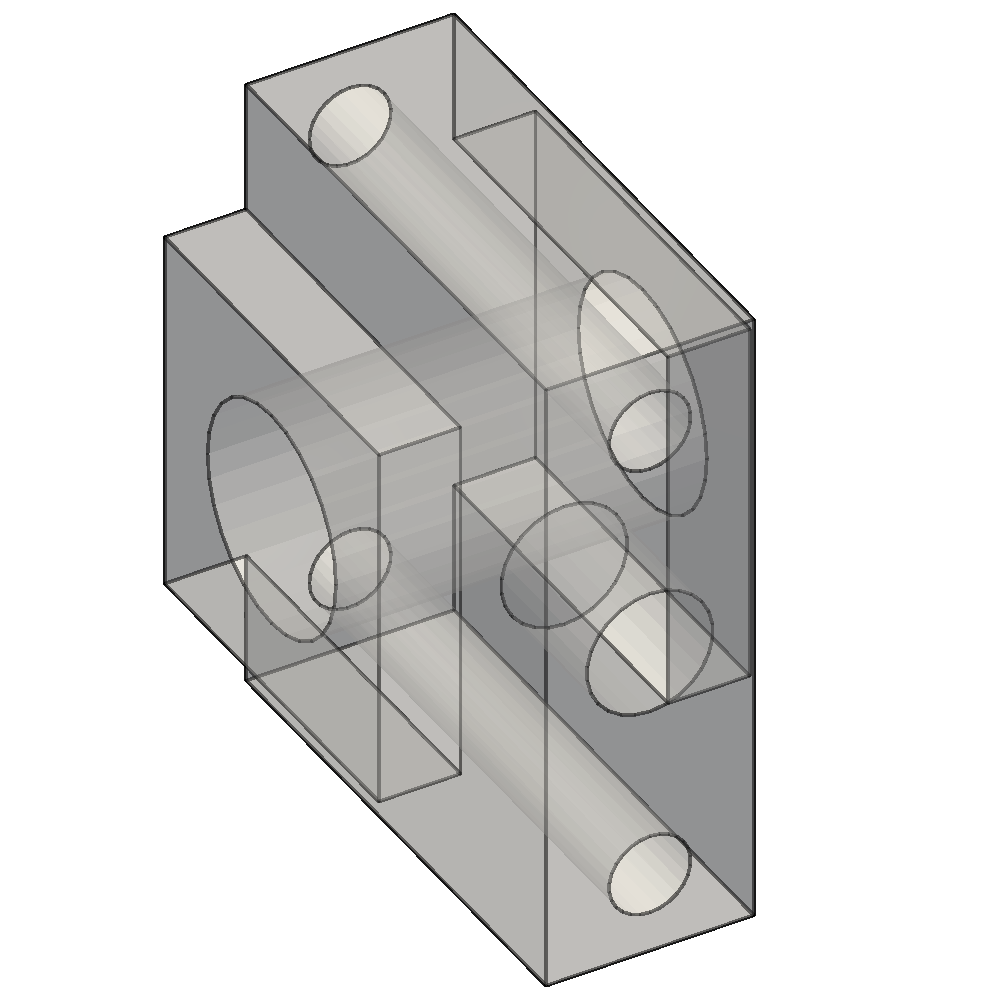}\hfill
    \includegraphics[width=0.117\linewidth]{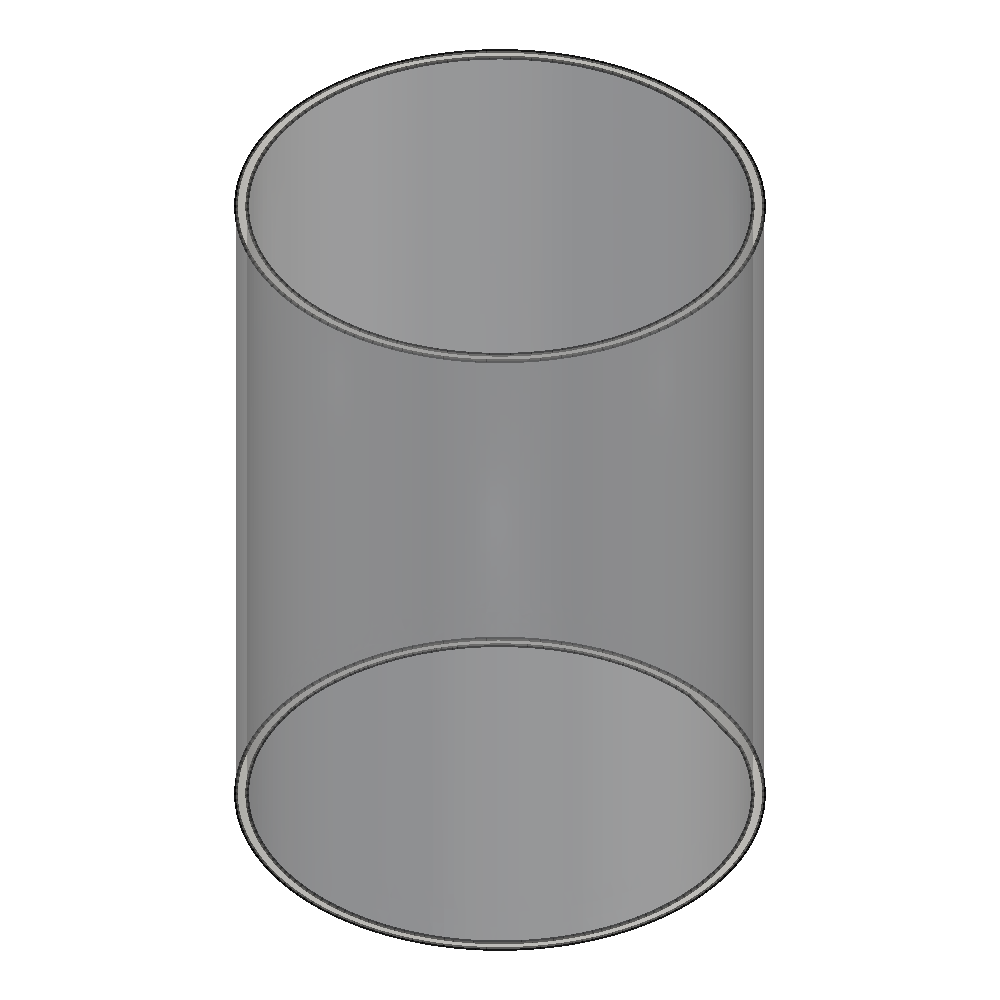}\hfill
    \includegraphics[width=0.117\linewidth]{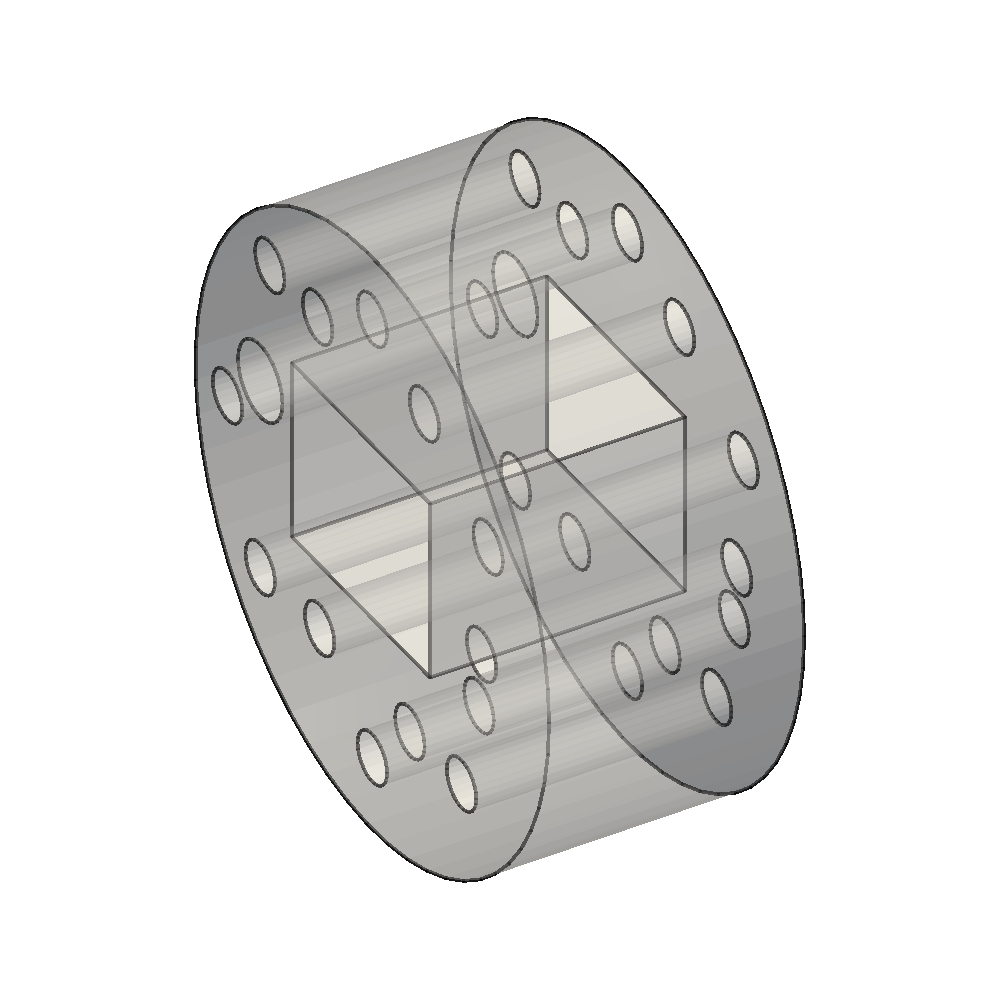}\hfill
    \includegraphics[width=0.117\linewidth]{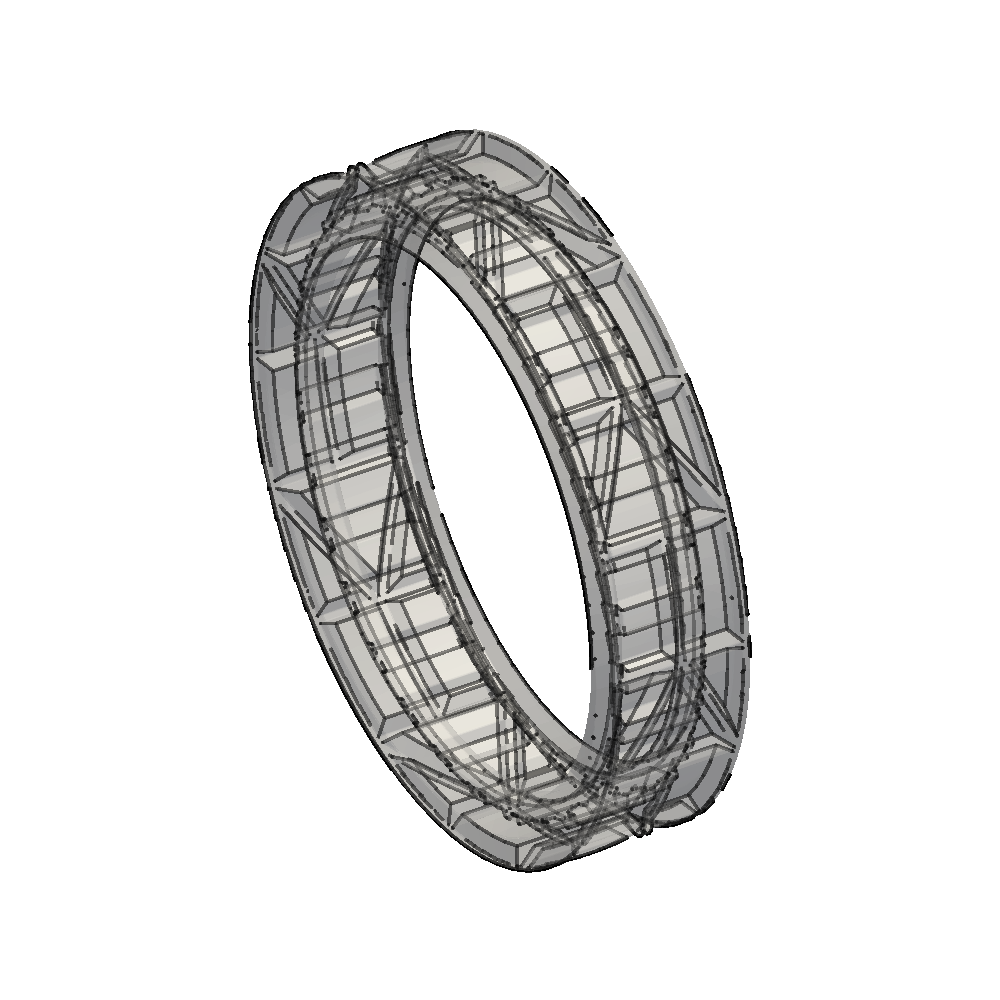}\hfill
    \includegraphics[width=0.117\linewidth]{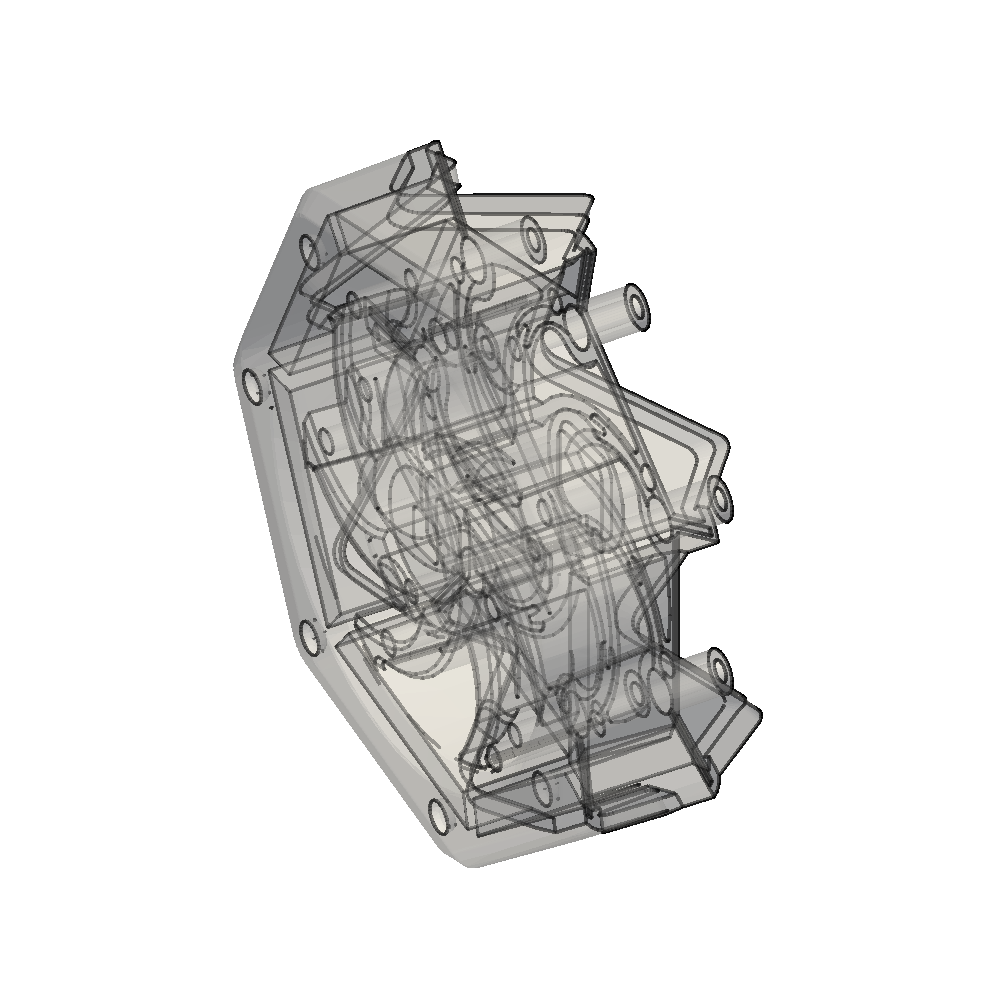}\hfill
    \includegraphics[width=0.117\linewidth]{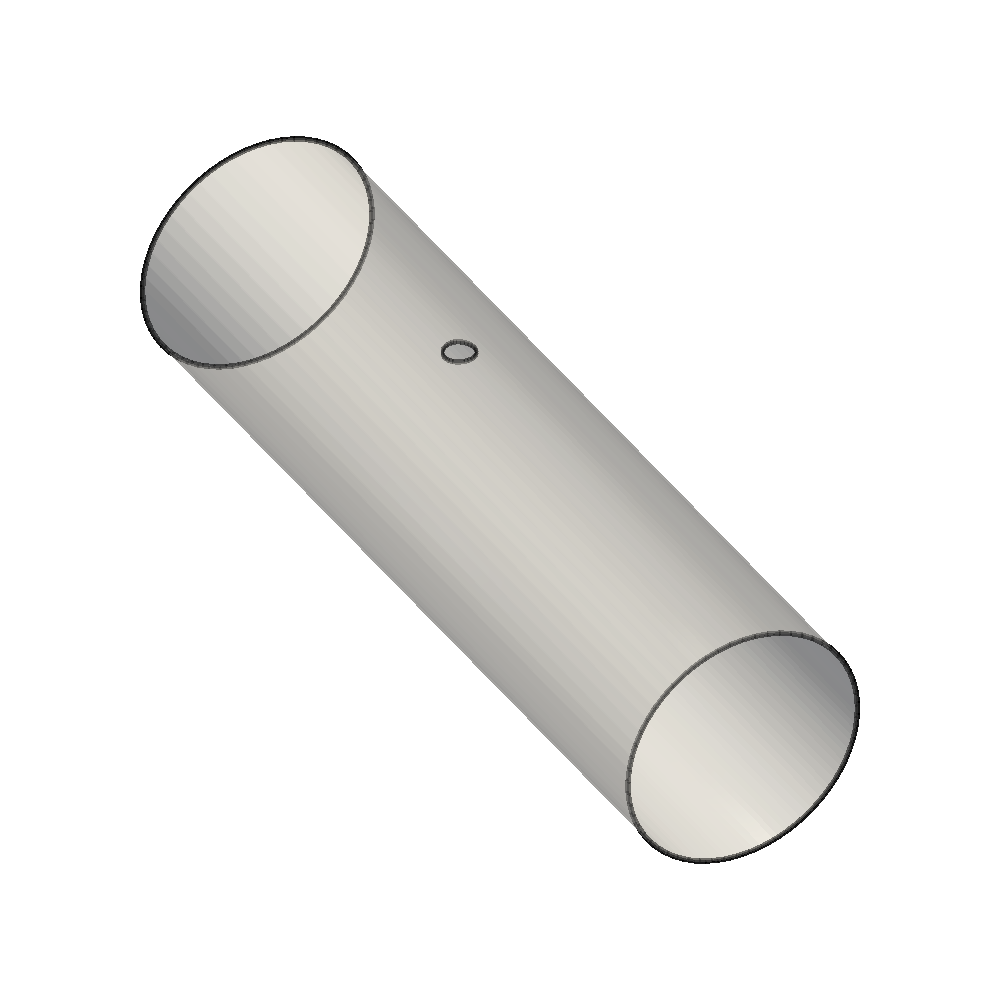}\hfill
    \includegraphics[width=0.117\linewidth]{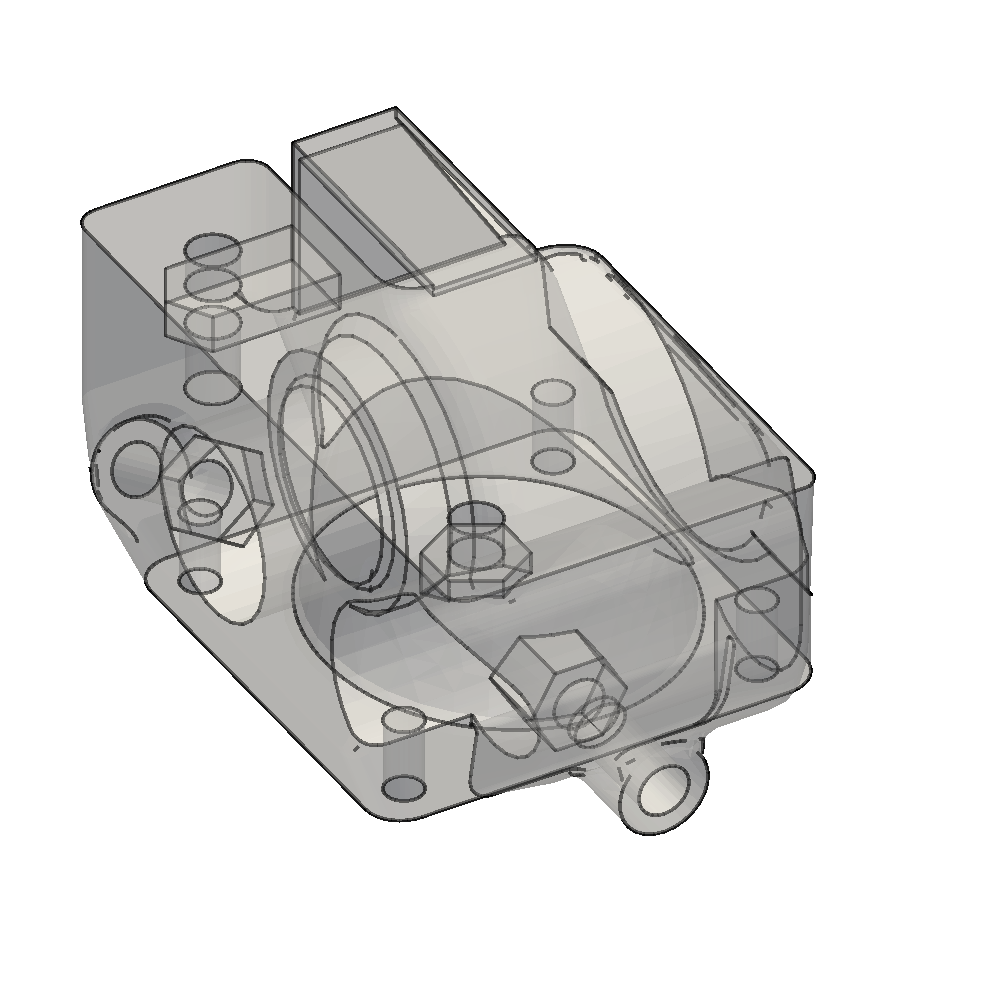}\\[\smallskipamount]
    \includegraphics[width=0.117\linewidth]{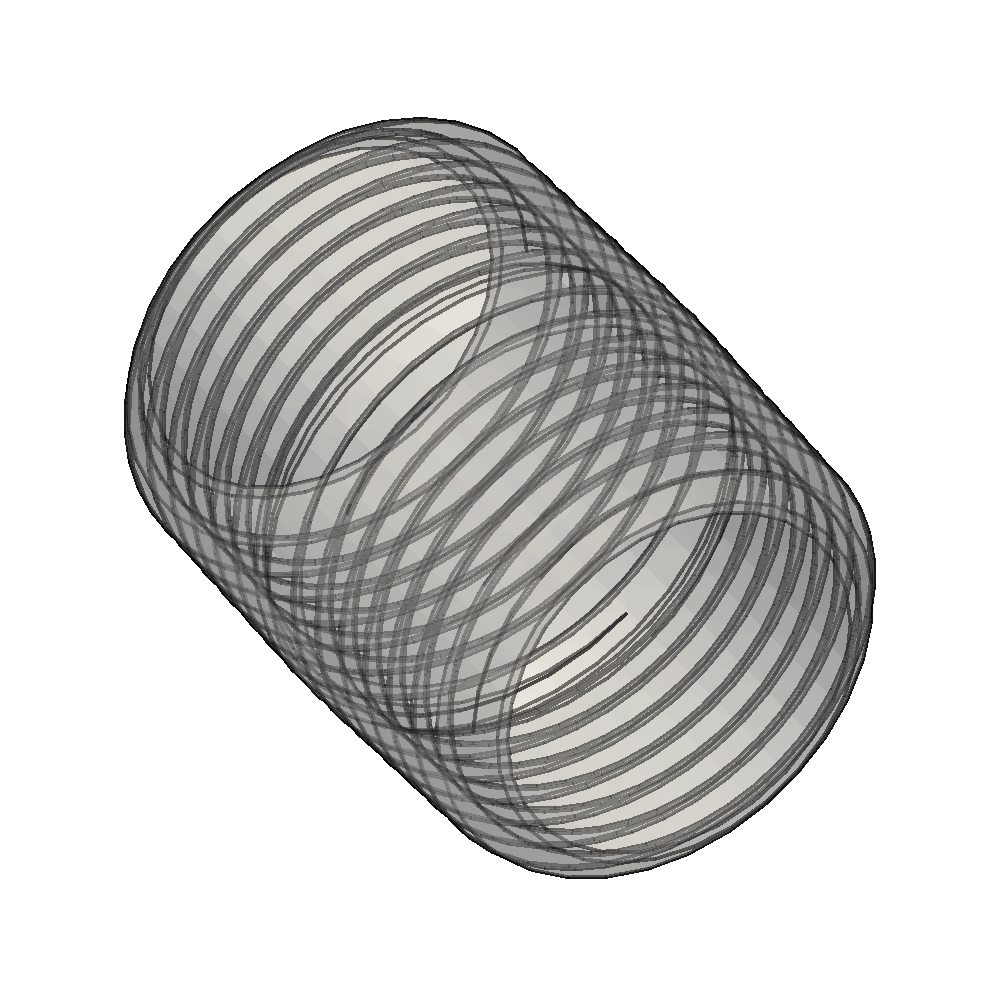}\hfill
    \includegraphics[width=0.117\linewidth]{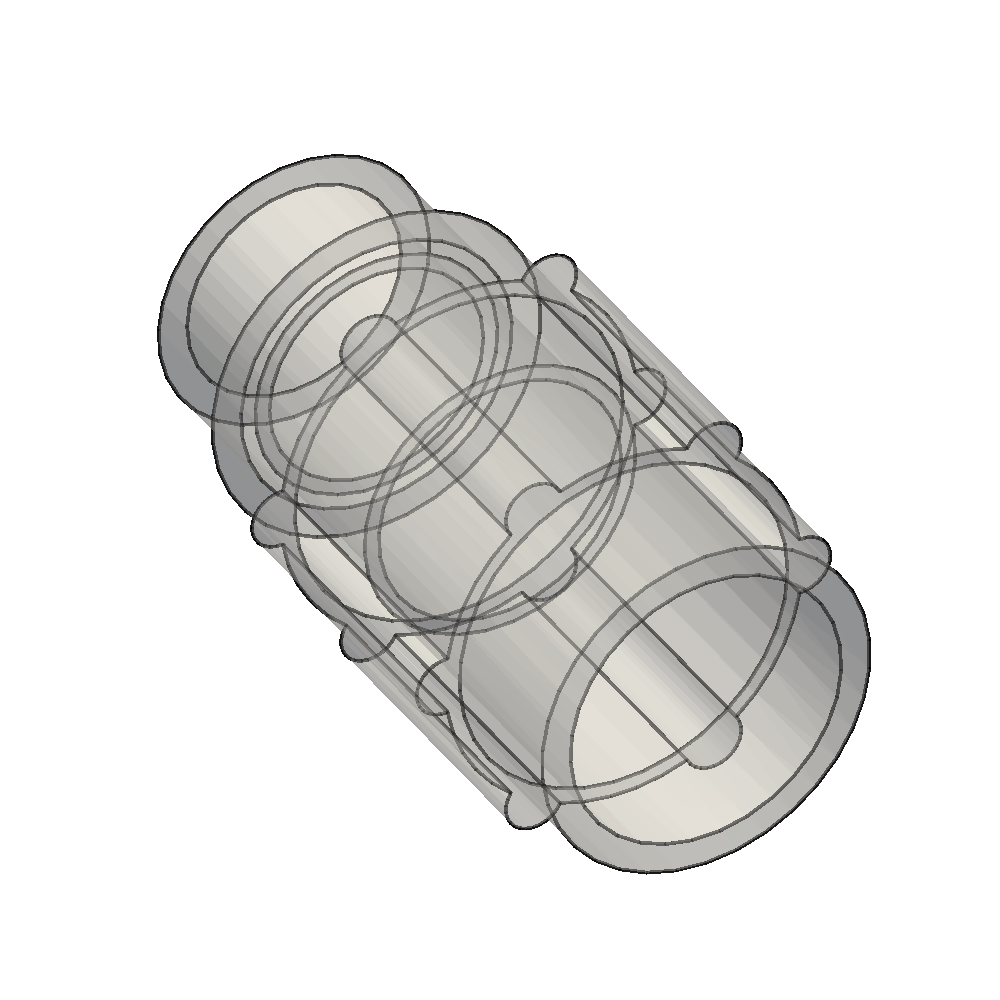}\hfill
    \includegraphics[width=0.117\linewidth]{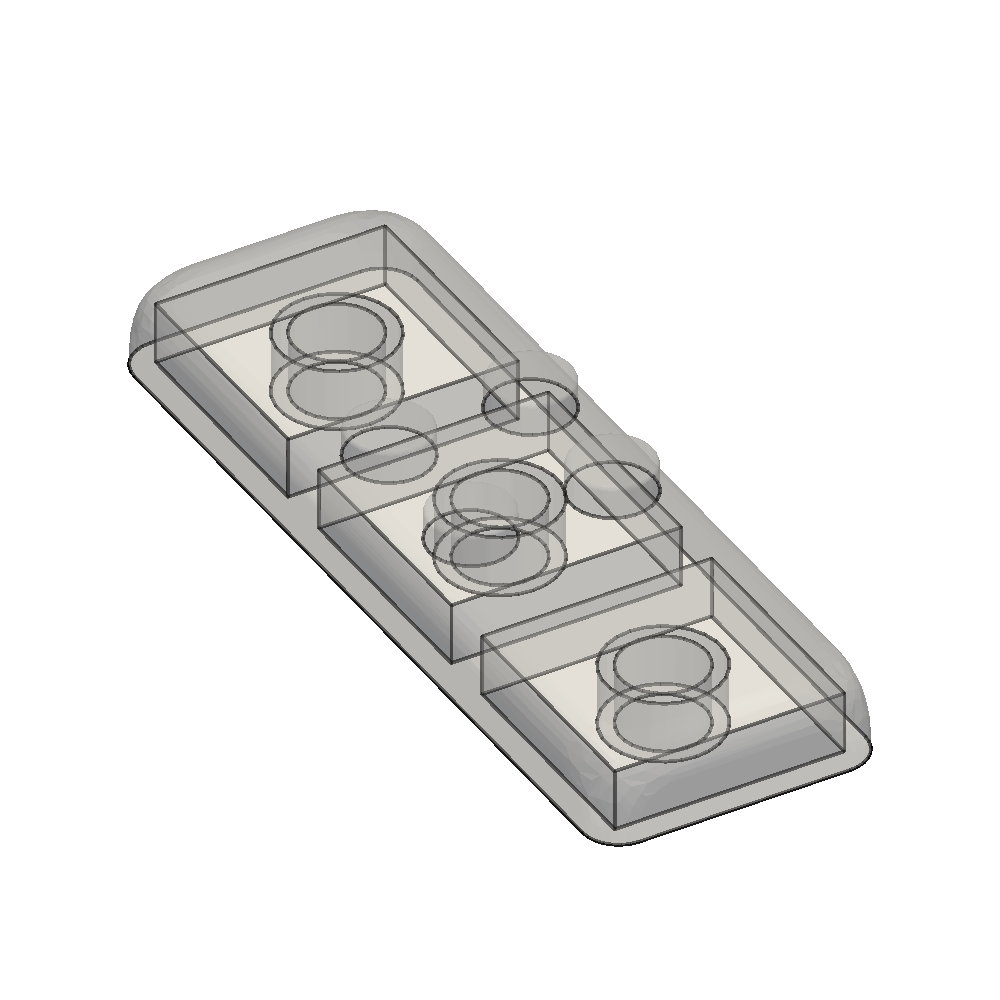}\hfill
    \includegraphics[width=0.117\linewidth]{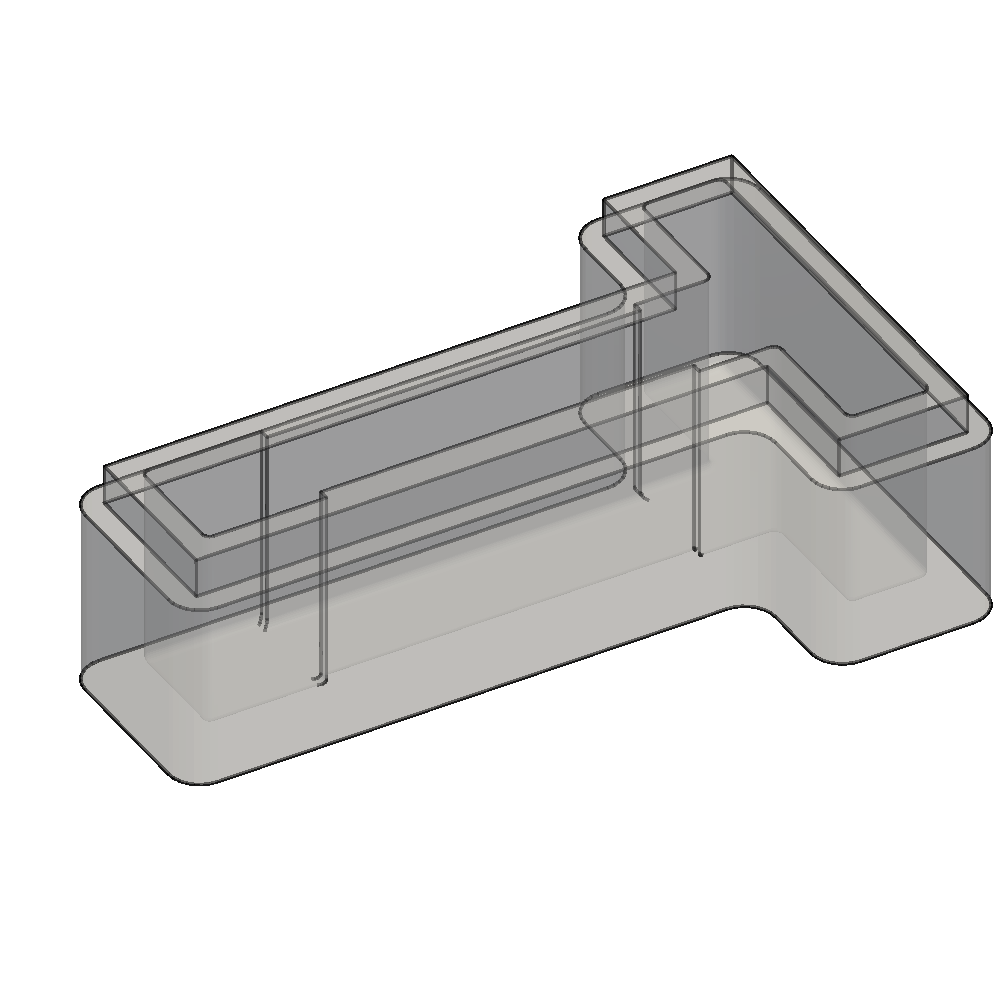}

    \caption{(continued, part 4 of 4) CAD parts of the 156-part evaluation benchmark.}
    \label{fig:cad_dataset_grid_4}
\end{figure}

\clearpage
\section{\textbf{GReFEM} Inference Pipeline}
\begin{algorithm}[ht]
  \caption{GReFEM Zero-Shot Inference Pipeline}
  \label{alg:grefem_pipeline}
  \begin{algorithmic}[1]
    \STATE {\bfseries Input:} 3D Surface Mesh $\mathcal{M}$, Load Case $L$, Point Density $\rho$, View count $K$
    \STATE {\bfseries Output:} 3D Refinement Anchor Points $\mathcal{P}_{3D}$

    \vspace{0.3em}
    \STATE \COMMENT{\textbf{Stage 1: View Selection (\textit{orthoViews})}}
    \STATE $\mathcal{V}_{\text{all}} \leftarrow \text{GenerateOrthographicViews}(\mathcal{M})$
    \STATE $\mathcal{V}_{\text{ranked}} \leftarrow \text{orthoViews}(\mathcal{V}_{\text{all}})$ \COMMENT{Rank via DINOv2 \& Self-Attention}
    \STATE $\mathcal{V}_{\text{top}} \leftarrow \text{SelectTopK}(\mathcal{V}_{\text{ranked}}, K)$
    \STATE $\mathcal{V}_{\text{grid}} \leftarrow \text{OverlayGrid}(\mathcal{V}_{\text{top}})$

    \vspace{0.3em}
     \STATE \COMMENT{\textbf{Stage 2: Physics-Guided Region Proposal (single MLLM call)}}
    \STATE $\text{Prompt} \leftarrow \text{ConstructPhysicsPrompt}(L, \text{PromptType})$
    \STATE $\mathcal{V}_{\text{persp}} \leftarrow \text{RenderPerspectiveViews}(\mathcal{M})$ \COMMENT{Non-gridded views for global 3D context}
    \STATE $\text{Response} \leftarrow \text{MLLM\_Inference}(\text{Prompt}, \mathcal{V}_{\text{persp}} \cup \mathcal{V}_{\text{grid}})$ \COMMENT{All views in one context}
    \FORALL{$V_k \in \mathcal{V}_{\text{grid}}$}
        \STATE $\mathcal{C}_{\text{pred}}[k] \leftarrow \text{ParseSemanticCells}(\text{Response}, k)$ \COMMENT{Feature type $\rightarrow$ cells for view $k$}
    \ENDFOR

    \vspace{0.3em}
    \STATE \COMMENT{\textbf{Stage 3: 2D Feature Extraction}}
    \STATE $\mathcal{Q}_{\text{2D}} \leftarrow \emptyset$ \COMMENT{Set of precise 2D pixel coordinates}
    \FORALL{$V_k \in \mathcal{V}_{\text{grid}}$}
        \STATE $\mathcal{E}_{\text{contours}} \leftarrow \text{ExtractGeometricContours}(V_k, \mathcal{M})$
        \FORALL{Feature Type $f \in \{\text{ICE, EC/PC, TH, F}\}$}
            \STATE $\text{Cells}_f \leftarrow \mathcal{C}_{\text{pred}}[k][f]$
            \STATE $\mathcal{Q}_f \leftarrow \text{IntersectContoursWithCells}(\mathcal{E}_{\text{contours}}, \text{Cells}_f)$
            \STATE $\mathcal{Q}_{\text{2D}} \leftarrow \mathcal{Q}_{\text{2D}} \cup \mathcal{Q}_f$
        \ENDFOR
    \ENDFOR

    \vspace{0.3em}
    \STATE \COMMENT{\textbf{Stage 4: 3D Spatial Grounding (MV-RaySeg)}}
    \STATE $\mathcal{P}_{\text{raw}} \leftarrow \emptyset$
    \FORALL{$V_k \in \mathcal{V}_{\text{grid}}$ and corresponding $\mathcal{Q}_{\text{2D}}$}
        \STATE $\mathbf{d}_v, \mathbf{c}_v \leftarrow \text{GetCameraParams}(V_k)$
        \FORALL{$\mathbf{q} \in \mathcal{Q}_{\text{2D}}$}
            \STATE $T_q \leftarrow \text{RayMeshIntersection}(\mathbf{q}, \mathbf{d}_v, \mathbf{c}_v, \mathcal{M})$
            \IF{Feature is Surface-Visible (e.g., EC/PC)}
                \STATE $\mathcal{P}_{\text{raw}} \leftarrow \mathcal{P}_{\text{raw}} \cup \{T_{q,0}\}$ \COMMENT{Keep first hit only}
            \ELSE
                \STATE \COMMENT{Deep volumetric feature (ICE, F, TH)}
                \FOR{$i=0$ to $|T_q|-1$ step $2$}
                    \STATE $\mathcal{P}_{\text{seg}} \leftarrow \text{SampleSegment}(T_{q,i}, T_{q,i+1}, \rho)$
                    \STATE $\mathcal{P}_{\text{raw}} \leftarrow \mathcal{P}_{\text{raw}} \cup \mathcal{P}_{\text{seg}}$
                \ENDFOR
            \ENDIF
        \ENDFOR
    \ENDFOR

    \vspace{0.3em}
    \STATE $\mathcal{P}_{3D} \leftarrow \text{DensityFilter}(\mathcal{P}_{\text{raw}}, \rho)$ \COMMENT{Remove redundant overlapping points}
    \RETURN $\mathcal{P}_{3D}$
  \end{algorithmic}
\end{algorithm}

\clearpage
\section{Mesh Geometric Heuristic Pipeline}
\begin{algorithm}[ht]
  \caption{Mesh Geometric Heuristic Baseline}
  \label{alg:geometric_baseline}
  \begin{algorithmic}[1]
    \STATE {\bfseries Input:} 3D Surface Mesh $\mathcal{M}=(\mathcal{V}, \mathcal{F})$, Angle Threshold $\theta_{\text{thresh}}$ (e.g., $30^\circ$), Spatial Density $\rho$
    \STATE {\bfseries Output:} Dense Geometric Refinement Points $\mathcal{P}_{\text{base}}$

    \vspace{0.3em}
    \STATE \COMMENT{\textbf{Stage 1: Process Mesh Connectivity}}
    \STATE Extract adjacent face pairs $(f_1, f_2)$ sharing an edge $e$
    \STATE Compute face normals $\mathbf{n}_1, \mathbf{n}_2$ and centroids $\mathbf{c}_1, \mathbf{c}_2$ for all faces in $\mathcal{F}$
    \STATE $\mathcal{P}_{\text{sharp}} \leftarrow \emptyset$, \quad $\mathcal{F}_{\text{smooth}} \leftarrow \emptyset$

    \vspace{0.3em}
    \STATE \COMMENT{\textbf{Stage 2: Extract and Upsample Sharp Concave Edges}}
    \FORALL{adjacent pairs $(f_1, f_2)$}
        \STATE $\theta \leftarrow \arccos(\mathbf{n}_1 \cdot \mathbf{n}_2)$
        \IF{$\theta > \theta_{\text{thresh}}$}
            \STATE $\mathbf{v}_{\text{dir}} \leftarrow \mathbf{c}_2 - \mathbf{c}_1$
            \IF{$\mathbf{v}_{\text{dir}} \cdot \mathbf{n}_1 > 0$}
                \STATE Retrieve vertices $\mathbf{v}_a, \mathbf{v}_b$ of edge $e$ \COMMENT{Edge is concave (inward bending)}
                \STATE $L_e \leftarrow \|\mathbf{v}_b - \mathbf{v}_a\|_2$
                \STATE $N_{\text{samples}} \leftarrow \max(2, \lceil L_e / \rho \rceil)$
                \STATE Sample $N_{\text{samples}}$ equidistant points along segment $[\mathbf{v}_a, \mathbf{v}_b]$
                \STATE $\mathcal{P}_{\text{sharp}} \leftarrow \mathcal{P}_{\text{sharp}} \cup \{ \text{sampled points} \}$
            \ENDIF
        \ENDIF
    \ENDFOR

    \vspace{0.3em}
    \STATE \COMMENT{\textbf{Stage 3: Extract and Sample Smooth Concave Surfaces (Fillets/Holes)}}
    \FORALL{adjacent pairs $(f_1, f_2)$}
        \STATE $\theta \leftarrow \arccos(\mathbf{n}_1 \cdot \mathbf{n}_2)$
        \IF{$1^\circ < \theta \leq \theta_{\text{thresh}}$}
            \STATE $\mathbf{v}_{\text{dir}} \leftarrow \mathbf{c}_2 - \mathbf{c}_1$
            \IF{$\mathbf{v}_{\text{dir}} \cdot \mathbf{n}_1 > 0$}
                \STATE $\mathcal{F}_{\text{smooth}} \leftarrow \mathcal{F}_{\text{smooth}} \cup \{f_1, f_2\}$ \COMMENT{Smooth concavity}
            \ENDIF
        \ENDIF
    \ENDFOR

    \vspace{0.3em}
    \STATE $A_{\text{total}} \leftarrow \sum_{f \in \mathcal{F}_{\text{smooth}}} \text{Area}(f)$
    \STATE $N_{\text{total\_samples}} \leftarrow \lceil A_{\text{total}} / \rho^2 \rceil$
    \STATE $\mathcal{P}_{\text{smooth}} \leftarrow \emptyset$
    \FOR{$i = 1$ to $N_{\text{total\_samples}}$}
        \STATE Select face $f \in \mathcal{F}_{\text{smooth}}$ with probability $\propto \text{Area}(f) / A_{\text{total}}$
        \STATE Generate random barycentric coordinates $(u, v, w)$ \COMMENT{Uniform triangle sampling}
        \STATE Compute 3D point $\mathbf{p} = u\mathbf{v}_0 + v\mathbf{v}_1 + w\mathbf{v}_2$ using vertices of $f$
        \STATE $\mathcal{P}_{\text{smooth}} \leftarrow \mathcal{P}_{\text{smooth}} \cup \{\mathbf{p}\}$
    \ENDFOR

    \vspace{0.3em}
    \STATE \COMMENT{\textbf{Stage 4: Combine and Filter}}
    \STATE $\mathcal{P}_{\text{base}} \leftarrow \mathcal{P}_{\text{sharp}} \cup \mathcal{P}_{\text{smooth}}$
    \STATE $\mathcal{P}_{\text{base}} \leftarrow \text{DensityFilter}(\mathcal{P}_{\text{base}}, \rho)$ \COMMENT{Remove redundant overlap}
    
    \RETURN $\mathcal{P}_{\text{base}}$
  \end{algorithmic}
\end{algorithm}

\clearpage

\end{document}